\documentclass[12pt,oneside]{mitthesis}
\usepackage{lgrind}
\usepackage{cmap}
\usepackage[T1]{fontenc}
\usepackage[utf8]{inputenc}
\usepackage[english]{babel}
\usepackage{lmodern}
\usepackage[babel=true]{microtype}
\usepackage{graphicx}
\usepackage[linesnumbered,ruled]{algorithm2e}
\usepackage{xspace} 
\usepackage{bm} 
\usepackage{array} 
\newcolumntype{L}[1]{>{\raggedright\let\newline\\\arraybackslash\hspace{0pt}}m{#1}}
\newcolumntype{C}[1]{>{\centering\let\newline\\\arraybackslash\hspace{0pt}}m{#1}}
\newcolumntype{R}[1]{>{\raggedleft\let\newline\\\arraybackslash\hspace{0pt}}m{#1}}
\usepackage{subcaption}
\usepackage{url} 
\usepackage{hhline}
\usepackage[normalem]{ulem}
\usepackage{booktabs}
\usepackage{makecell} 
\usepackage{titlesec} 
\usepackage{indentfirst} 
\usepackage{float} 
\usepackage{dblfloatfix} 
\usepackage{changepage} 

\newcommand{\incircle}[1]{\raisebox{0.5pt}{\protect \textcircled{\raisebox{-0.8pt}{#1}}}}

\newcommand{\cmdact}{\texttt{{ACTIVATE}}\xspace}
\newcommand{\cmdrd}{\texttt{{READ}}\xspace}

\usepackage{pifont}
\newcommand{\cmark}{\ding{51}}%
\newcommand{\xmark}{\ding{55}}%
\newcommand{\delete}[1]{{\color{red}\sout{}}}

\usepackage{multirow}
\usepackage[usenames,dvipsnames,svgnames,table]{xcolor}

\newcommand{\hcfirst}[0]{$HC_{\scalebox{0.7}{first}}$}  
\newcommand{\hcsecond}[0]{$HC_{\scalebox{0.7}{second}}$}  
\newcommand{\hcthird}[0]{$HC_{\scalebox{0.7}{third}}$}

\definecolor{amber}{rgb}{1.0, 0.49, 0.0}
\definecolor{darkgreen}{rgb}{0.0, 0.2, 0.13}
\definecolor{darkbyzantium}{rgb}{0.36, 0.22, 0.33}
\definecolor{darkseagreen}{rgb}{0.56, 0.74, 0.56}
\definecolor{darkspringgreen}{rgb}{0.09, 0.45, 0.27}
\definecolor{dollarbill}{rgb}{0.52, 0.73, 0.4}
\newcommand{\ap}[1]{{#1}}
\newcommand{\jk}[1]{{\color{black}#1}}

\newcommand{\jkz}[1]{{\color{black}#1}}
\newcommand{\jky}[1]{{\color{black}#1}}
\newcommand{\jkx}[1]{{\color{black}#1}}
\newcommand{\jkt}[1]{{\color{black}#1}}
\newcommand{\jkth}[1]{{\color{black}#1}}
\newcommand{\jkf}[1]{{\color{black}#1}}
\newcommand{\jks}[1]{{\color{black}#1}}
\newcommand{\jke}[1]{{\color{black}#1}}
\newcommand{\jkn}[1]{{\color{black}#1}}
\newcommand{\jktwo}[1]{{\color{black}#1}}
\newcommand{\jkthree}[1]{{\color{black}#1}}
\newcommand{\jkfour}[1]{{\color{black}#1}}
\newcommand{\jkfive}[1]{{\color{black}#1}}
\newcommand{\jksix}[1]{{\color{black}#1}}

\newcommand{\hasan}[1]{{\color{black}#1}}
\newcommand{\hh}[1]{{\color{black}#1}}
\newcommand{\hhtwo}[1]{{\color{black}#1}}
\newcommand{\hhthree}[1]{{\color{black}#1}}

\newcommand{\hhf}[1]{{\color{black}#1}}
\newcommand{\changes}[1]{{#1}}

\newcommand{\mpseven}[1]{{\color{black}#1}}

\newcommand{\mpsix}[1]{{\color{black}#1}}

\newcommand{\mpx}[1]{{\color{black}#1}}
\newcommand{\mpy}[1]{{\color{black}#1}}
\newcommand{\mpo}[1]{{\color{black}#1}}
\newcommand{\mpt}[1]{{\color{black}#1}}
\newcommand{\mpf}[1]{{\color{black}#1}}

\newcommand{\mechanism}[0]{D-RaNGe}

\newcommand{\squeezeme}{ \setlength{\itemsep}{0pt}
     \setlength{\parsep}{3pt}
     \setlength{\topsep}{3pt}
     \setlength{\partopsep}{0pt}
     \setlength{\leftmargin}{1.5em}
     \setlength{\labelwidth}{1em}
     \setlength{\labelsep}{0.5em} }

\usepackage{amsmath}
\usepackage[nolessnomore, italic]{mathastext}

\usepackage{fancyhdr}
\def\secondpage{\clearpage\null\vfill
\pagestyle{empty}
\begin{minipage}[b]{0.9\textwidth}
\footnotesize\raggedright
\setlength{\parskip}{0.5\baselineskip}
\center{Copyright \copyright\ \the\year\ Jeremie S. Kim\par} 
\center{All Rights Reserved} 
\end{minipage}
\vspace*{2\baselineskip}
\cleardoublepage
\rfoot{\thepage}}

\makeatletter
\g@addto@macro{\maketitle}{\secondpage}
\makeatother

\begin{document}

\title{Improving DRAM Performance, Security, \\and Reliability by Understanding and Exploiting \\DRAM Timing Parameter Margins} 

\author{Jeremie S. Kim} 
\department{Department of Electrical Engineering and Computer Science}

\degree{Master of Science in Electrical Engineering and Computer Science}

\degreemonth{June}
\degreeyear{1990}
\thesisdate{May 18, 1990}


\supervisor{William J. Dally}{Associate Professor}

\chairman{Arthur C. Smith}{Chairman, Department Committee on Graduate Theses}

\maketitle



\newpage
\newpage 
\setcounter{savepage}{\thepage}
\begin{abstractpage}

Characterization of real DRAM devices has enabled findings in DRAM device properties, which has led to proposals that significantly improve overall system performance by reducing DRAM access latency and power consumption. In addition to improving system performance, a deeper understanding of DRAM technology via characterization can also improve device reliability and security. These can be seen with the recent discoveries of 1) DRAM-based true random number generators (TRNGs), a method for generating true random numbers using DRAM devices which can be used in many applications, 2) DRAM-based physical unclonable functions (PUFs), a method for generating unique device-dependent keys for identification and authentication, and 3) the RowHammer vulnerability, a phenomenon where repeatedly accessing a DRAM row can cause failures in unaccessed neighboring DRAM rows. 
 
To advance DRAM-based discoveries and mechanisms, this dissertation rigorously characterizes many modern commodity DRAM devices and shows that by exploiting DRAM access timing margins within manufacturer-recommended DRAM timing specifications, we can significantly improve system performance, reduce power consumption, and improve device reliability and security. First, we characterize DRAM timing parameter margins and find that certain regions of DRAM can be accessed faster than other regions due to DRAM cell process manufacturing variation. We exploit this by enabling variable access times depending on the DRAM cells being accessed, which not only improves overall system performance, but also decreases power consumption. Second, we find that we can uniquely identify DRAM devices by the locations of failures that result when we access DRAM with timing parameters reduced below specification values. Because we induce these failures with DRAM accesses, we can generate these unique identifiers significantly more quickly than prior work. Third, we propose a random number generator that is based on our observation that timing failures in certain DRAM cells are randomly induced and can thus be repeatedly polled to very quickly generate true random values. Finally, we characterize the RowHammer security vulnerability on a wide range of modern DRAM chips while violating the DRAM refresh requirement in order to directly characterize the underlying DRAM technology without the interference of refresh commands. We demonstrate with our characterization of real chips, that existing RowHammer mitigation mechanisms \jky{either} are not scalable or suffer from prohibitively large performance overheads in projected future devices and it is critical to research more effective solutions to RowHammer. Overall, our studies build a new understanding of modern DRAM devices to improve computing system performance, reliability and security all at the same time.

\end{abstractpage}


\cleardoublepage

\section*{Acknowledgments}

I have many people to thank for their support during my PhD journey. First and
foremost, I am extremely grateful to my advisor, Onur Mutlu, who has generously
mentored and guided me since my sophomore year of college. His passion for
Computer Architecture and his research initially caught my interest during a
seminar course lecture, and his exemplary feedback, encouragement, and ongoing
support helped me to adopt his passion into my own work as well. I am grateful
to have experienced, first-hand, the thought and care Onur puts into both his
teaching and his research, and I have learned greatly through many iterations
of paper submissions, conference talks, course exams, and lectures with him.
Onur has also provided countless opportunities for collaboration within the
SAFARI research group and industrial collaborators, which provided me with many
new experiences during my exciting and unique PhD experience, as well as made
this thesis possible. 

I am grateful to the members of my PhD committee, James Hoe, Derek Chiou, and
Saugata Ghose, for their valuable feedback and stimulating discussions. 

I am immensely grateful to have found many great friends in the SAFARI research
group. I could not have done it without Minesh Patel, a great friend that truly
made my PhD experience enjoyable. He brought an immense wealth of knowledge,
enforced high research standards, and fostered my enjoyment of hard liquors.  I
am very thankful for Can Firtina, an endless source of entertainment,
unforgettable experiences, and shelter. I am grateful for Hasan Hassan who on
many occasions kept me company in the office late into the night. I am thankful
for Damla Senol Cali, my first friend in SAFARI with whom I had to navigate the
new field of bioinformatics and learn Onur's research process. I also want to
thank Giray Yaglikci for his friendship and company on hikes around the world. 

I would like to acknowledge all past and current members of our research group
for being both great friends and colleagues. I want to especially thank Yoongu
Kim, Hongyi Xin, Donghyuk Lee, Rachata Ausavarungnirun, Yixin Luo, Saugata
Ghose, Vivek Seshadri, and Kevin Chang, for their mentorshop during my
formative years in SAFARI. I thank Can Firtina, Giray Yaglikci, Nastaran
Hajinazar, Geraldo De Oliveira, and Ivan Fernandez-Vega for gracefully
welcoming my invasion of their office space and providing a fun working
environment. I also thank all others for their discussions, feedback,
collaboration, and support: Arash Tavakkol, Jawad Haj-Yahya, Mohammed Alser,
Roknoddin Azizibarzoki, Nika Mansourighiasi, Lois Orosa, Juan Gomez Luna,
Amirali Boroumand, Jisung Park, Nandita Vijaykumar, Ivan Puddu, Ataberk Olgun,
Nisa Bostanci, Rahul Bera, Konstantinos Kanellopoulos, and Taha Shahroodi. 

I am especially grateful to have met and worked with Tyler Huberty, Stephan
Meier, Jared Zerbe, Jung-Sik Kim, Heonjae Ha, Seung Lee, Gihong Kim, Taehyun
Kim, Augustin Hong, Can Alkan, and countless others during my PhD journey. They
have all provided great insight, mentorship, and support.  

I would also like to thank my internship mentors, Stefan Saroiu and Alec
Wolman, during my time at Microsoft Research, who provided me with a
stimulating environment and an industrial perspective on system security. I
sincerely thank Microsoft for this opportunity.

I would like to thank the National Science Foundation (grants 1212962 and
1320531), the National Institutes of Health (grant HG006004) and SAFARI
Research Group's industrial partners for respectively the financial support and
the gift funding they have provided that have contributed to works during my
PhD.

To my many friends and family, your support and encouragement throughout my
journey were worth more than I can express on paper. I want to particularly
thank Jimmy Lee, Ho-Gyun Choi, Noelle Jung, Stephanie Chen, Justine Kim, and
Matt Yin for always being there. 

Finally, I want to thank my parents Hyong and Anita for their unwavering
support, encouragement, and love.


\tableofcontents
\newpage
\listoffigures
\newpage
\listoftables

\chapter{Introduction}

\section{Problem and Thesis Statement} 
\label{sec:introduction} 

Characterization of real DRAM devices has enabled findings in DRAM device
properties, which has lead to proposals that significantly improve overall
system performance by reducing DRAM access latency and power consumption. In
addition to improving system performance, a deeper understanding of DRAM
technology via characterization can also improve device reliability and
security. These can be seen with the recent discoveries of 1) DRAM-based true
random number generators (TRNGs)~\jky{\cite{keller2014dynamic, sutar2018d,
hashemian2015robust, tehranipoor2016robust, eckert2017drng, pyo2009dram,
talukder2019exploiting, orosa2019dataplant}}, a method for generating true
random numbers using DRAM devices which can be used in many applications, 2)
DRAM-based physical unclonable functions (PUFs)~\jky{\cite{sutar2016d,
keller2014dynamic, hashemian2015robust, tehranipoor2015dram,
tehranipoor2017investigation, rahmati2015probable, tang2017dram,
sutar2017memory, karimian2019generate, muelich2019channel,
talukder2019prelatpuf, orosa2019dataplant}}, a method for generating unique
device-dependent keys for identification and authentication, and 3) the
RowHammer vulnerability~\cite{kim2014flipping, cojocar2019exploiting,
gruss2018another, gruss2016rowhammer, lipp2018nethammer, qiao2016new,
razavi2016flip, seaborn2015exploiting, tatar2018defeating, van2016drammer,
xiao2016one, frigo2020trrespass, ji2019pinpoint, kwong2020rambleed,
mutlu2017rowhammer, van2018guardion}, a phenomenon where repeatedly accessing a
DRAM row can cause failures in unaccessed neighboring DRAM rows. To advance
this collection of discoveries and mechanisms, we rigorously characterize many
modern commodity DRAM devices and show that by exploiting DRAM access timing
margins and specifications, we can significantly improve system performance,
reduce power consumption, and improve device reliability and security. First,
we characterize DRAM timing parameter margins and find that certain regions of
DRAM can be accessed faster than other regions due to DRAM cell process
manufacturing variation. We exploit this by enabling variable access times
depending on the DRAM cells \jky{that are} being accessed, which not only
improves overall system performance, but also decreases power consumption.
Second, with further characterization, we find that we can uniquely identify
DRAM devices by the locations of failures that result when we access DRAM with
timing parameters reduced below specification values. Because we induce these
failures with DRAM accesses, we can generate these unique identifiers
significantly quicker than prior work. Third, we propose a random number
generator that is based on our observation that timing failures in certain DRAM
cells are randomly induced and can thus be repeatedly polled to very quickly
generate true random values.  Finally, we characterize the RowHammer security
vulnerability on a wide range of modern DRAM devices while violating the DRAM
refresh requirement in order to directly characterize the underlying DRAM
technology without the interference of refresh commands. 

\medskip
\jky{Our thesis statement is as follows:} 
\textbf{\emph{By rigorously understanding and exploiting DRAM device
characteristics, we can significantly improve system performance and enhance
system security and reliability.}}

\section{Overview of Our Approach} 
\label{sec:overview} 

In line with our thesis statement, we use rigorous characterization of real
DRAM chips to make novel observations on chip properties and use these
observations to improve system performance and enhance system security and
reliability. \jkz{We demonstrate across four works, that by understanding
per-chip error characteristics using a profiling mechanism, we can develop
mechanisms that exploit chip-dependent error profiles to improve system
performance or enhance system security and reliability.} 

The first mechanism that we develop based on our observations,
\emph{\underline{S}ubarray-\underline{o}ptimized Access \underline{La}tency
\underline{R}eduction DRAM (Solar-DRAM)}, builds on our detailed experimental
characterization and exploits each of our novel observations to significantly
and robustly reduce DRAM access latency.  The key ideas of Solar-DRAM are to
issue 1) DRAM reads with reduced $t_{RCD}$ (i.e., by 39\%) unless the requested
DRAM cache line contains weak DRAM cells that are likely to fail under reduced
$t_{RCD}$, and 2) all DRAM writes with reduced $t_{RCD}$ (i.e., by 77\%).
Solar-DRAM determines whether a DRAM cell is weak using a \jkz{per-chip}
\emph{static profile} of local bitlines, which we experimentally find to be
\emph{reliable across time}.  Compared to state-of-the-art LPDDR4 DRAM,
Solar-DRAM provides significant system performance improvement while
maintaining data correctness. 

The second mechanism, \emph{the DRAM latency PUF}, \jkz{exploits \jky{our} novel
observation that reducing DRAM read access latency below the reliable datasheet
specifications using software-only system calls results in error patterns that
can be used as unique identifiers.  We demonstrate that users can further
enhance the reliability of the unique identifiers using an error profile of the
DRAM chip, which enables users to select regions of DRAM that are better suited
for evaluating PUFs.} We experimentally demonstrate, using 223 modern LPDDR4
DRAM chips, that the DRAM latency PUF satisfies all of the requirements of an
effective runtime-accessible PUF.  In particular, a DRAM latency PUF can be
evaluated in 88.2ms on average across all devices at all operating
temperatures. We show that, for a constant DRAM capacity overhead of 64KiB, the
DRAM latency PUF's average (minimum, maximum) evaluation time speedup over the
DRAM retention PUF~\cite{sutar2016d, keller2014dynamic, liu2014trustworthy,
xiong2016run} is 152x (109x, 181x) at 70$^{\circ}$C and 1426x (868x, 1783x) at
55$^{\circ}$C, with exponentially increasing speedups at even lower
temperatures.

The third mechanism, D-RaNGe, \jkz{exploits the novel observation from our
characterization results that true random numbers can be extracted from access
latency failures with high throughput.} D-RaNGe~consists of two steps: 1)
identifying specific DRAM cells that are vulnerable to activation failures
using a \emph{low-latency} profiling step and 2) generating a continuous stream
(i.e., constant rate) of random numbers by repeatedly inducing activation
failures in the previously-identified vulnerable cells.  D-RaNGe~runs entirely
in software and is capable of immediately running on any commodity system that
provides the ability to manipulate DRAM timing parameters within the memory
controller~\cite{bkdg_amd2013, opteron_amd}. For most other devices, a simple
software API must be exposed without any hardware changes to the commodity DRAM
device (e.g., similarly to SoftMC~\cite{hassan2017softmc,
softmc-safarigithub}), which makes D-RaNGe suitable for implementation on most
existing systems today. 

Finally, we demonstrate via characterization of many DRAM chips and technology
node generations, that the DRAM-based security vulnerability, RowHammer, is
getting worse as device feature size reduces. This means that the number of
activations needed to induce a RowHammer bit flip also reduces, to as few as
9.6k in the most vulnerable chip we tested. We then use our characterization
results to demonstrate how five state-of-the-art RowHammer mitigation
mechanisms do not scale to support the degrees of RowHammer vulnerability that
we expect to see in future devices. We conclude by discussing \jky{various methods for
improving RowHammer mitigation for future DRAM devices.} 


\section{Contributions} 

This dissertation makes the following \textbf{key contributions}: 
\begin{enumerate}
\squeezeme
\item Using 282 LPDDR4 DRAM modules from three major DRAM manufacturers,
we extensively characterize the effects of multiple testing conditions (e.g.,
DRAM temperature, DRAM access latency parameters, data patterns written
in DRAM) on activation failures. We demonstrate the viability of mechanisms that exploit variation in
access latency of DRAM cells by showing that cells that operate correctly at
reduced latency continue to operate correctly at the same latency over time.
That is, a DRAM cell's activation failure probability is \emph{not} vulnerable
to significant variation over short time intervals. 
    \begin{enumerate}
    \squeezeme
    \item We present data across our DRAM \jks{modules}, that activation failures
    exhibit high spatial locality and are tightly constrained to a small number of
    \emph{columns} (i.e., on average 3.7\%/2.5\%/2.2\% per bank for DRAM chips of
    manufacturers A/B/C) at the granularity of a DRAM subarray. 
    \item We demonstrate that $t_{RCD}$ can be greatly reduced (i.e., by 77\%) for
    DRAM \emph{write} requests while \emph{still} maintaining data integrity. This
    is because $t_{RCD}$ defines the amount of time required for data within DRAM
    cells to be amplified to a \emph{readable} voltage level, which does \emph{not}
    govern DRAM \emph{write} operations. 
    \item We find that across SPEC CPU2006 benchmarks, DRAM accesses to closed rows
    typically request the \emph{$0^{th}$ cache line} in the row, with a maximum
    (average) probability of 22.2\% (6.6\%). \jkt{This is much greater than the
    expected probability (i.e., 3.1\%) assuming that DRAM accesses to closed rows
    access each cache line with an equal probability.} Since $t_{RCD}$ affects only
    DRAM accesses to closed DRAM rows, we find that simply reducing $t_{RCD}$ for
    all accesses to the $0^{th}$ cache lines of all DRAM rows improves overall
    system performance by up to 6.54\%. 

    \item We propose Solar-DRAM, a mechanism that exploits our three key
    observations on reliably reducing the $t_{RCD}$ timing parameter. Solar-DRAM
    selectively reduces $t_{RCD}$ for 1) reads to DRAM cache lines containing
    ``weak'' or ``strong'' cells, and 2) writes to all of DRAM. We evaluate
    Solar-DRAM on a variety of multi-core workloads and show that compared to
    \emph{state-of-the-art} LPDDR4 DRAM, Solar-DRAM \jks{improves performance by
    4.97\% (8.79\%) on} heterogeneous and \jke{by} 4.31\% (10.87\%) \jks{on}
    homogeneous workloads.
    \end{enumerate} 

\item We introduce the DRAM latency PUF, a new class of DRAM PUFs, that
is based on the deliberate violation of manufacturer-specified DRAM latency
parameters. DRAM latency PUFs can be implemented \emph{with no additional
hardware overhead} on any commodity off-the-shelf (COTS) system that
\mpsix{permits} software-controlled manipulation of DRAM access latencies at
the memory controller (e.g., \cite{lee2015adaptive, AMD_opteron,
bkdg_amd2013}).

    \begin{enumerate}
    \squeezeme 
    \item Using experimental data from 223 real LPDDR4 DRAM chips, we extensively
    analyze both DRAM latency PUFs and DRAM retention PUFs.  We show that DRAM
    latency PUFs 1) satisfy all characteristics of an effective PUF, and 2) are
    suitable for use as runtime-accessible PUFs across a \emph{wide range} of
    temperatures. We also present an extensive characterization of DRAM retention
    PUFs under a wide range of temperatures. We show that while DRAM retention PUFs
    can be evaluated faster at higher temperatures, their evaluation time at
    temperatures \mpsix{even as high as 70$^\circ$C} is \emph{prohibitively} slow.

    \item We experimentally show that the DRAM latency PUF significantly
    outperforms DRAM retention PUFs, achieving \jkfour{an average speedup of
    152x/1426x at 70$^{\circ}$C/55$^{\circ}$C} when evaluating PUFs with a constant
    DRAM capacity overhead of 64KiB. We also find that while DRAM retention PUFs
    suffer from \emph{temperature-dependent evaluation times}, the DRAM latency PUF
    provides a \mpsix{consistently low average evaluation time of 88.2ms at} all
    operating temperatures.
    \end{enumerate} 

\item We introduce \mechanism, a new methodology for extracting true random
numbers from a commodity DRAM device at \hhtwo{high} throughput \hhtwo{and low
latency}. The key idea of \mechanism~is to use DRAM cells as entropy sources to
generate true random numbers by accessing them with a latency
that is \hhtwo{lower than}
manufacturer-recommended specifications.

    \begin{enumerate} 
    \squeezeme 
    \item Using experimental data from 282 state-of-the-art LPDDR4 DRAM devices
    from three major DRAM manufacturers, we present a rigorous characterization of 
    \hhtwo{randomness in errors induced by accessing DRAM with low latency}.
    Our analysis demonstrates that \mechanism~is able
    to maintain high-quality random number generation both over 15 days of testing
    and across the entire reliable testing temperature range of our
    infrastructure (55$^{\circ}$C-70$^{\circ}$C). We verify our observations from
    this study with prior works' observations on DDR3 DRAM
    devices~\cite{chang2016understanding, lee-sigmetrics2017, lee2015adaptive,
    kim2018solar}. Furthermore, we experimentally demonstrate on four DDR3 DRAM
    devices, from a single manufacturer, that \mechanism~is suitable for implementation
    in a \hhtwo{wide} range of commodity DRAM devices.

    \item We evaluate the quality of \mechanism's output bitstream using the
    \mpo{standard} NIST statistical test suite for
    randomness~\cite{rukhin2001statistical} and find that it successfully passes
    every test.  We also compare \mechanism's performance to four previously
    proposed DRAM-based TRNG designs (Section~\ref{comparison}) and show that
    \mechanism\ outperforms the best prior DRAM-based TRNG design by over two orders
    of magnitude in terms of maximum and average throughput.
    \end{enumerate} 


\item We provide the first rigorous RowHammer failure characterization study of a broad range of real modern DRAM chips across different DRAM types, technology node generations, and manufacturers. We experimentally study 1580 DRAM chips (408$\times$ DDR3, 652$\times$ DDR4, and 520$\times$ LPDDR4) from 300 DRAM modules (60$\times$ DDR3, 110$\times$ DDR4, and 130$\times$ LPDDR4) and present our RowHammer characterization results for both aggregate RowHammer failure rates and the behavior of individual cells while sweeping the hammer count ($HC$) and stored data pattern. 

    \begin{enumerate} 
    \squeezeme 
    \item Via our rigorous characterization studies, we definitively demonstrate that the RowHammer vulnerability significantly worsens (i.e., the number of hammers required to induce a RowHammer bit flip, {\hcfirst}, greatly reduces) in newer DRAM chips (e.g., {\hcfirst} reduces from $69.2k$ to $22.4k$ in DDR3, $17.5k$ to $10k$ in DDR4, and $16.8k$ to $4.8k$ in LPDDR4 chips across multiple technology node generations).
    \item We demonstrate, based on our rigorous evaluation of five state-of-the-art RowHammer mitigation mechanisms, that even though existing RowHammer mitigation mechanisms are reasonably effective at mitigating RowHammer in today's DRAM chips (e.g., 8\% average performance loss on our workloads when {\hcfirst} is 4.8k), they will cause significant overhead in future DRAM chips with even lower {\hcfirst} values (e.g., 80\% average performance loss with the most scalable mechanism when {\hcfirst} is 128).  
    \item We evaluate an ideal refresh-based mitigation mechanism that selectively refreshes a row only just before it is about to experience a RowHammer bit flip, and find that in chips with high vulnerability to RowHammer, there is still significant opportunity for developing a refresh-based RowHammer mitigation mechanism with low performance overhead that scales to low {\hcfirst} values. We conclude that it is critical to research more effective solutions to RowHammer, and we provide promising directions for future research. 
    \end{enumerate} 
\end{enumerate} 
 
\section{Dissertation Outline}

This thesis is organized into 7 chapters. Chapter 2 describes necessary
background on DRAM organization, operations, and failure mechanisms. Chapter 3
presents solar-DRAM, a mechanism for reducing DRAM access latency. Chapter 4
presents the DRAM Latency PUF, a fast and efficient method for generating
unique identifiers in commodity DRAM. Chapter 5 presents D-RaNGe, a method for
quickly and efficiently generating true random numbers in DRAM. Chapter 6
presents our experimental study of RowHammer on several technology node
generations of DRAM chips. Finally, Chapter 7 presents conclusions and future
research directions that are enabled by this dissertation.

\chapter{Background} 

We describe the DRAM organization and operation necessary for
understanding our observations and mechanism for reducing DRAM access
latencies. We refer the reader to prior
works~\cite{lee2015adaptive, khan2016parbor,
seshadri2016simple, hassan2016chargecache, hassan2017softmc, zhang2014half,
lee2013tiered, kim2012case, seshadri2013rowclone, chang2016understanding,
lee2016reducing, chang2017thesis, chang2016low, lee-sigmetrics2017,
seshadri2017ambit, lee2015decoupled, kim2016ramulator, kim2014flipping,
bhati2015flexible, chang_understanding2017, liu2013experimental,
khan2014efficacy, qureshi2015avatar, zhang2014half, khan2017detecting,
patel2017reaper, mukundan2013understanding, chang2014improving, raidr,
kim2018dram, kim2018solar, kim2019d, vampire2018ghose, ghose2019demystifying} for more
detail.

\section{DRAM System Organization}
\label{subsubsec:dram_org}

\begin{figure}[h] \centering 
    \includegraphics[width=0.65\linewidth]{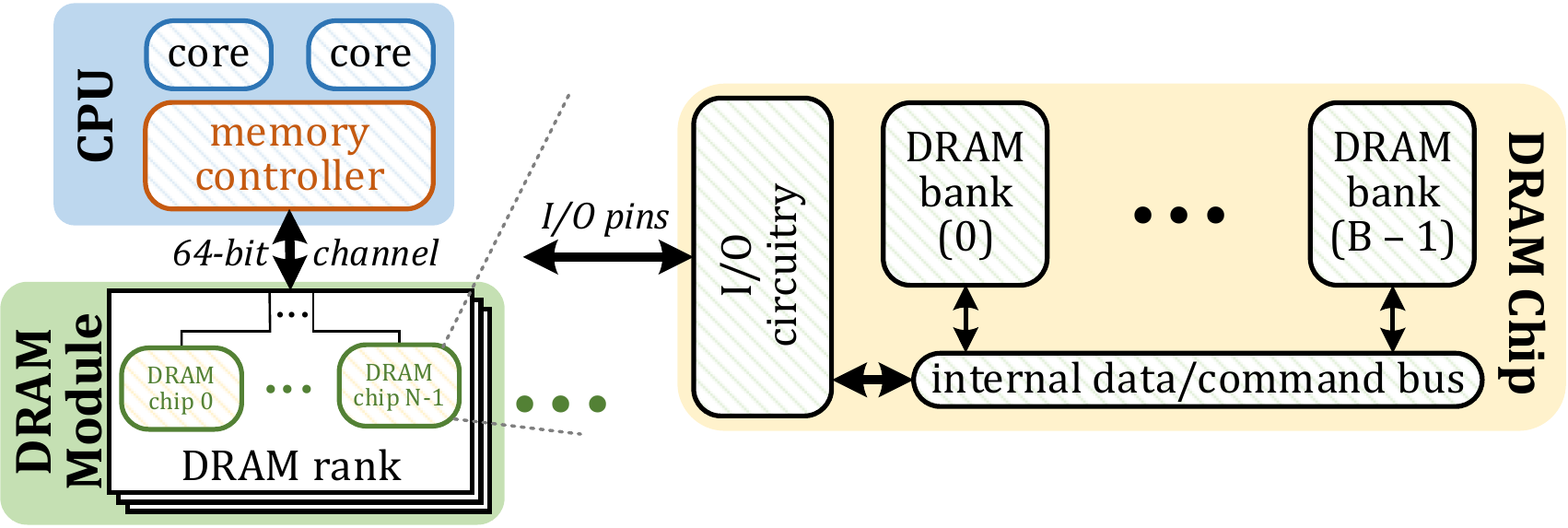} 
    \caption{A typical DRAM-based system~\cite{kim2018solar}.}
    \label{fig:dram_org}
\end{figure}

~In a typical system configuration, a CPU chip includes a set of memory
controllers, where each memory controller interfaces \hhtwo{with} a DRAM channel to
perform read and write operations. As we show in Figure~\ref{fig:dram_org}
\hh{(left)}, a DRAM channel has its own I/O bus and operates independently
of other channels in the system. \hh{To achieve high memory capacity, a
channel can host multiple DRAM modules by sharing the I/O bus between the
modules. A DRAM module implements a single or multiple DRAM ranks.}
Command and data transfers are serialized between ranks in the same channel due
to the shared I/O bus. A DRAM rank consists of multiple DRAM chips that operate
in lock-step, i.e., all chips simultaneously perform the same operation\hhtwo{,
but they do so on different bits}. The number of DRAM chips per rank depends on
the data bus width of the DRAM chips and the channel width. For example, a
typical system has a 64-bit wide DRAM channel. Thus, four 16-bit or eight 8-bit
DRAM chips are needed to build a DRAM rank.

\section{DRAM Chip Organization}
\label{subsubsec:dram_bank}

~At a high-level, a DRAM chip consists of \hh{billions} of DRAM cells that are
hierarchically organized to maximize storage density and performance. We
describe each level of the hierarchy of a modern DRAM chip.


A modern DRAM chip is composed of multiple DRAM banks (shown in
Figure~\ref{fig:dram_org}\hh{, right}). The chip communicates with the memory controller
through the \emph{I/O circuitry}. The I/O circuitry is connected to the
\emph{internal command and data bus} that is shared among all banks in the
chip.

\begin{figure}[h]
    \centering
    \begin{subfigure}[b]{.380\linewidth}
        \centering
        \includegraphics[width=\linewidth]{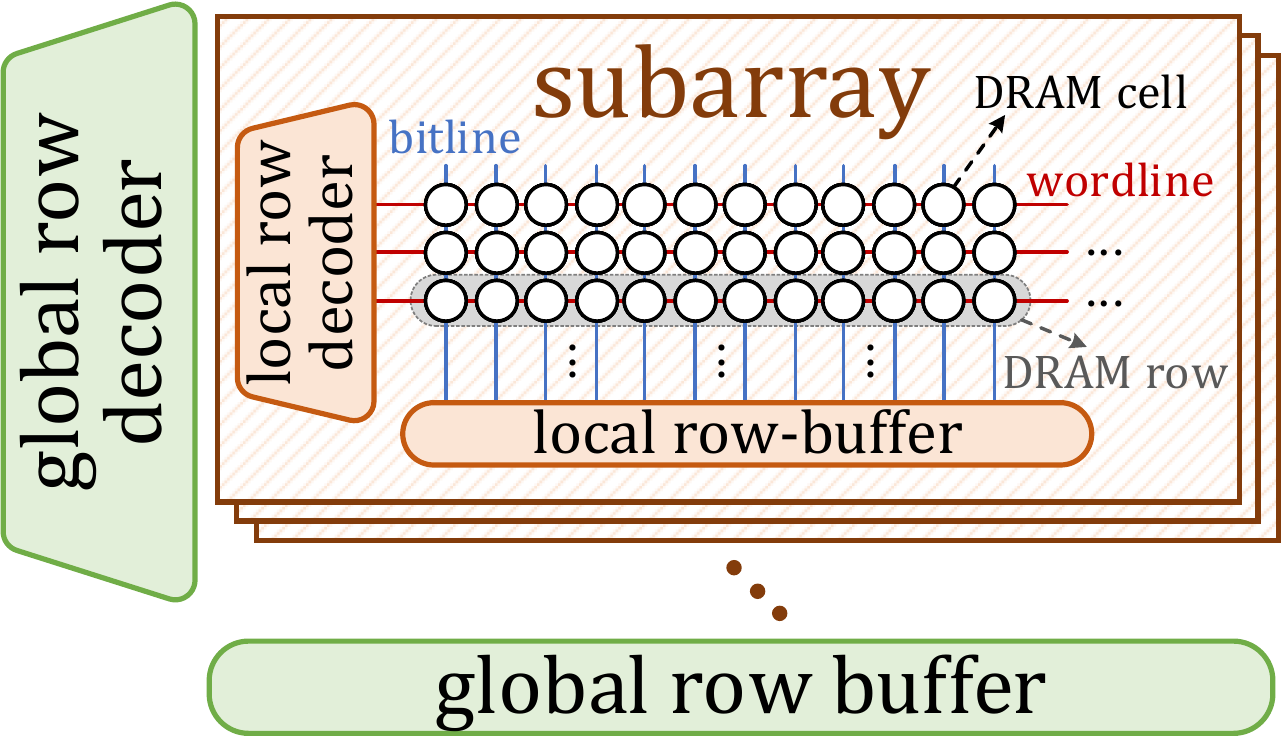}
        \caption{DRAM bank.}
        \label{subfig:dram_bank}
    \end{subfigure}
    \quad
    \begin{subfigure}[b]{.24\linewidth}
        \centering
        \includegraphics[width=\linewidth]{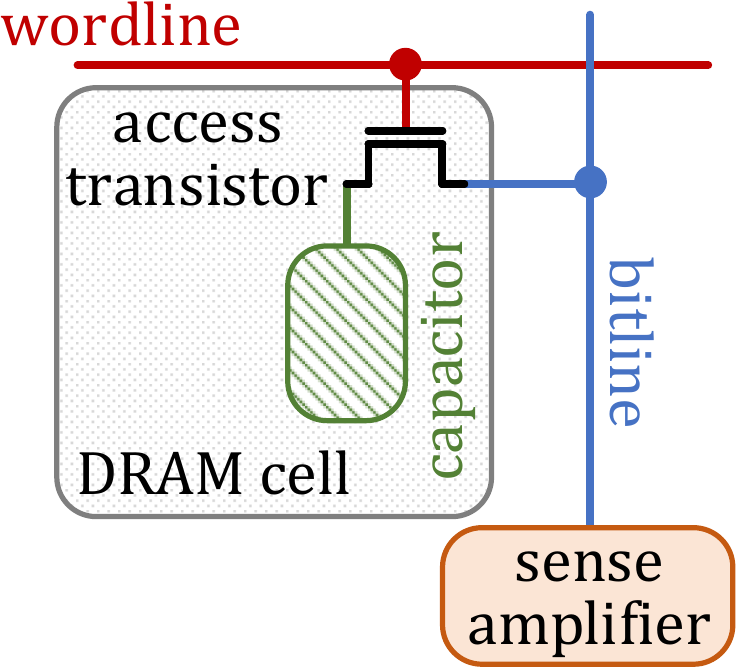}
        \caption{DRAM cell.}
        \label{subfig:dram_cell}
    \end{subfigure}

        \caption{DRAM bank and cell architecture~\cite{kim2018solar}.}
        \label{figure:dram_bank_and_cell}
\end{figure}

Figure~\ref{subfig:dram_bank} illustrates the \hasan{organization} of a DRAM bank. In a
bank, the \emph{global row decoder} partially decodes the address of the
accessed \emph{DRAM row} to select the corresponding \emph{DRAM subarray}. A
DRAM subarray is a 2D array of DRAM cells, where cells are horizontally
organized into multiple DRAM rows. A DRAM row is a set of DRAM
cells that share a wire called the \emph{wordline}, which the \emph{local row
decoder} of the subarray drives after fully decoding the row address. In a
subarray, a column of cells shares a wire, referred to as the \emph{bitline},
that connects the column of cells to a \emph{sense amplifier}.  The sense
amplifier is the circuitry used to read and modify the data of a DRAM cell.
\hh{The row} of sense amplifiers in the subarray is referred to as the \emph{local
row-buffer}.  To access a DRAM cell, the corresponding DRAM row first needs to
be copied into the local row-buffer, which connects to the internal I/O bus via
the \emph{global row-buffer}.

Figure~\ref{subfig:dram_cell} illustrates a DRAM cell, which is composed of a
\emph{storage capacitor} and \emph{access transistor}. A DRAM cell stores a
single bit of information based on the charge level of the capacitor. The data
stored in the cell is interpreted as a ``1'' or ``0'' depending on whether the
charge stored in the cell is above or below a certain threshold. Unfortunately,
the capacitor and the access transistor are not ideal circuit components and
have \emph{charge leakage paths}. Thus, to ensure that the cell does not leak
charge to the point where the bit stored in the cell flips, the cell needs to
be periodically \emph{refreshed} to fully restore its original charge.

\section{DRAM Commands}
\label{subsubsec:dram_cmds}

~The memory controller issues a set of DRAM commands to access data in the DRAM
chip. To perform a read or write operation, the memory controller first needs
to \emph{open} a row, i.e., copy the data of the cells in the row to the
row-buffer. To open a row, the memory controller issues an \emph{activate
(ACT)} command to a bank by specifying the address of the row to open. The
memory controller can issue \emph{ACT} commands to different banks in
consecutive DRAM bus cycles to operate on \emph{multiple banks in parallel}.
After opening a row \hh{in a bank}, the memory controller issues either a \emph{READ} or a
\emph{WRITE} command to read or write a DRAM word (which is typically equal to
64 bytes) \hh{within the open row}. 
\hh{An open row can serve multiple \emph{READ} and \emph{WRITE} requests
without incurring precharge and activation delays.}
A DRAM row typically contains 4-8 KiBs of data. To access data from
another DRAM row \hh{in} the same bank, the memory controller must first close the
currently open row by issuing a \emph{precharge (PRE)} command. The memory
controller also periodically issues \emph{refresh (REF)} commands to prevent
data loss due to charge leakage.

\section{DRAM Cell Operation}
\label{subsubsec:dram_operation}


~We describe DRAM operation by explaining the steps involved in reading data
from \hh{a DRAM cell}.\footnote{Although we focus only on reading data, steps
involved in a write operation are similar.} The memory controller initiates
each step by issuing a DRAM command. Each step takes a certain amount of time
to complete, and thus, a DRAM command is typically associated with one or more
timing constraints known as \emph{timing parameters}. It is the responsibility
of the memory controller to satisfy these timing parameters in order to ensure
\emph{correct} DRAM operation.

\begin{figure}[h] \centering 
\includegraphics[width=0.8\linewidth]{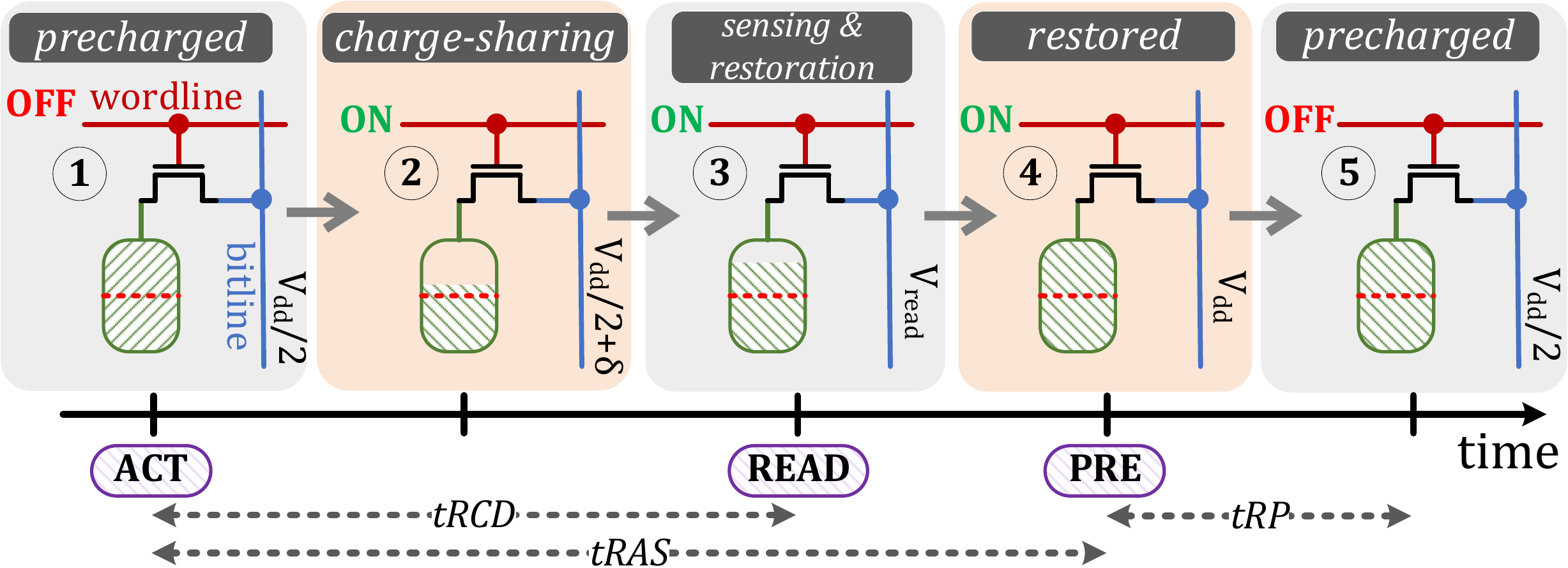} 
\caption{Command sequence for reading data from DRAM and the state of a DRAM cell during each \hhtwo{related} step.}  
\label{fig:dram_op}
\end{figure}

In Figure~\ref{fig:dram_op}, we show how the state of a DRAM cell changes
during the steps involved in a read operation. Each DRAM cell diagram
corresponds to the state of the cell at exactly the tick \hh{mark} on the time
axis. Each command (\jkfour{shown} in purple boxes below the \hhtwo{time} axis)
is issued by the memory controller at the \hh{corresponding} tick \hh{mark}.
Initially, the cell is in a \emph{precharged} state~\incircle{1}. When
precharged, the capacitor of the cell is disconnected from the bitline since
the wordline is not asserted and thus the access transistor is off. The bitline
voltage is stable at $\frac{V_{dd}}{2}$ and is ready to \hh{be perturbed}
towards the \hh{voltage level} of the cell \hh{capacitor} upon enabling the
access transistor. 

To read data from a cell, the memory controller first needs to perform
\emph{row activation} by issuing an \emph{ACT} command. During row activation
\hh{(\incircle{2}),} the row decoder asserts the wordline that connects the
storage capacitor of the cell to the bitline by enabling the access transistor.
At this point, the capacitor charge perturbs the bitline \hh{via the}
\emph{charge sharing} \hh{process}. Charge sharing continues until the
capacitor and bitline voltages reach an equal value of $\frac{V_{dd}}{2} +
\delta$. After charge sharing \hh{(\incircle{3}),} the sense amplifier begins
driving the bitline towards \hh{either} $V_{dd}$ or $0V$ depending on the
direction of the perturbation in the charge sharing step.  This step\hh{, which
amplifies} the voltage level \hh{on the bitline as well as the cell} is called
\emph{charge restoration}. Although charge restoration continues until the
original capacitor charge is fully replenished (\incircle{4}), the memory
controller can issue a \emph{READ} command to safely read data from the
activated row \hh{before the capacitor charge is fully replenished}. A
\emph{READ} command can \hasan{reliably} be issued when the bitline voltage
reaches \hh{the} voltage level $V_{read}$.  To ensure that the read occurs
after the bitline reaches $V_{read}$, the memory controller inserts a time
interval $t_{RCD}$ between the \emph{ACT} and \emph{READ} commands. It is the
responsibility of the DRAM \hh{manufacturer} to ensure that their DRAM chip
operates safely as long as the \hh{memory controller obeys the} $t_{RCD}$
timing parameter, which is defined in the DRAM standard~\cite{2014lpddr4}. If
the memory controller issues a \emph{READ} command before $t_{RCD}$ elapses,
the bitline voltage may be below $V_{read}$, which can lead to \hh{the} reading
\hh{of} a wrong value. 

To return a cell to its precharged state, the voltage in the cell must first be
fully restored. A cell is expected to be fully \hh{restored when the memory
controller satisfies} \hh{a} time \hh{interval dictated by} $t_{RAS}$
\hh{after} issuing the \emph{ACT} command. Failing to satisfy $t_{RAS}$ may
lead to \hh{insufficient amount of charge \hh{to be} restored} in the cells of
the accessed row. A subsequent activation of the row can then result in the
reading \hh{of} incorrect data from the cells.

Once the cell is successfully \emph{restored} \hh{(\incircle{4}),} the
memory controller can issue a \emph{PRE} command to close the
\hh{currently-open} row to prepare the bank for an access to another row.
The cell returns to the precharged state (\incircle{5}) after waiting for
\hh{the} timing parameter $t_{RP}$ following the \emph{PRE} command.
Violating $t_{RP}$ \hh{may prevent} the sense amplifiers from fully driving
the bitline back to $\frac{V_{dd}}{2}$, \hh{which may later result in the row
to be activated with too small amount of charge in its cells,} \hhtwo{potentially
preventing the sense amplifiers to read the data correctly}.

For correct DRAM operation, it is critical for the memory controller to ensure
that the DRAM timing parameters defined in the DRAM specification are
\emph{not} violated. Violation of the timing parameters may lead to
\hh{incorrect data to be read from the DRAM,} and \hh{thus cause} unexpected
program behavior~\cite{lee2015adaptive, khan2016parbor, hassan2017softmc,
chang2016understanding, chang_understanding2017, chang2017thesis,
lee-sigmetrics2017}. In this work, we study \hhtwo{the} failure modes \hhtwo{due to
violating DRAM timing parameters} and explore their application \hh{to}
reliably generating true random numbers.

\section{DRAM Failure Modes} 
\label{background:failure_modes} 

As we describe in Section~\ref{subsubsec:dram_operation}, the memory controller
must satisfy timing parameters associated with DRAM commands for correct
operation. We define \emph{access latency failures} as failures that occur due
to accessing a DRAM module with \emph{any} reduced timing parameter. In this
dissertation, we focus on \emph{activation failures}, which is a special case of
access latency failures, caused by reducing the $t_{RCD}$ timing parameter. 

An \emph{activation failure} occurs due to insufficient time for the sense
amplifier to drive the bitline to $V_{access}$. Depending on the reduction in
the $t_{RCD}$ parameter, there are two modes of \emph{activation failure}.
The first mode of activation failure results in transient
failures in the returned data, but no failures in the data stored in the DRAM
cells. In this case, the \emph{next access} to the same row that satisfies the
timing parameters would return correct data. Such a failure may happen when the
bitline does \emph{not} reach $V_{access}$ prior to the read operation but the
sense amplifier continues to drive the bitline towards the same direction
(i.e., full 0 or 1) as the charge-sharing phase has already started.  

The second mode of activation failure \emph{destroys} the data stored in a DRAM
cell such that future accesses (with default timing parameters) return
failures. Such a failure may happen when, at the time the \cmdrd is issued, the
bitline voltage level is sufficiently low.  In this case, the read operation
could significantly disturb the bitline such that the sense amplifier starts
driving the bitline towards the opposite of the original direction. We observe
both of the \emph{activation failure} modes in our experiments with real LPDDR4
DRAM modules.

\section{Violating Manufacturer-Specified Timing Parameters} \hspace{2pt}
\label{section:out_of_spec} 
Different cells in the same DRAM chip have different reliable operation
latencies (for each timing parameter) due to two major reasons: 1) design
(architectural) differences~\cite{lee-sigmetrics2017}, and 2) process
variation~\cite{lee2015adaptive}. For example, a cell located closer to
the sense amplifiers than an otherwise-equivalent cell can operate correctly
with a lower $t_{RCD}$ constraint~\cite{lee-sigmetrics2017} because the
inherent latency to access a cell close to the sense amplifiers is lower.
Similarly, a cell that happens to have a larger capacitor (due to
manufacturing process variation) can operate reliably with tighter timing
constraints than a smaller cell elsewhere in the same
chip~\cite{lee2015adaptive}.

Because manufacturing process variation occurs in random and unpredictable
locations within and across chips~\cite{desai2012process, zhang2015exploiting,
li2011dram, kim2009new, lee2015adaptive,lee-sigmetrics2017,
chang_understanding2017, chang2016understanding, chang2017thesis,
lee2016reducing, kim2014flipping}, the manufacturer-published timing parameters
are chosen to ensure reliable operation of the \emph{worst-case} cell in any
acceptable device at the \emph{worst-case} operating conditions (e.g., highest
supported temperature, lowest supported voltage). This results in a large
\emph{safety margin} (or, \emph{guardband}) for each timing parameter, which
prior work shows can often be reliably reduced at \emph{typical} operating
conditions~\cite{chang2016understanding, lee2015adaptive,
chandrasekar2014exploiting}.

Prior work also shows that decreasing the timing parameters \emph{too
aggressively} results in failures, with increasing error rates observed for
larger reductions in timing parameter values~\cite{liu2013experimental,
qureshi2015avatar, khan2014efficacy, lee2015adaptive, patel2017reaper,
chang2016understanding, hassan2016chargecache, chang_understanding2017,
hassan2017softmc, khan2016parbor, khan2017detecting, khan2016case}. Errors
occur because, with reduced timing parameters, the internal DRAM circuitry is
\emph{not} allowed \emph{enough time} to properly perform its 
functions and stabilize outputs
before the memory controller issues the next command
(Section~\ref{subsubsec:dram_operation}).

\chapter{Solar-DRAM: Reducing DRAM Access Latency by Exploiting the Variation in Local Bitlines}
\label{ch3-solar}

DRAM latency is a major bottleneck for many applications in modern computing
systems. In this chapter, we rigorously characterize the effects of reducing DRAM
access \jkf{latency} on 282 state-of-the-art LPDDR4 DRAM \jks{modules}.  As found in
prior work on older DRAM generations (DDR3), we show that regions of
LPDDR4 DRAM \jks{modules} can be accessed with latencies that are significantly lower
than manufacturer-specified values \emph{without} causing failures.  We present
novel data that 1) further supports the viability of such latency reduction
mechanisms and 2) exposes a variety of new cases in which access latencies can
be effectively reduced. Using our observations, we propose a new low-cost
mechanism, Solar-DRAM, that 1) identifies failure-prone regions of DRAM at
reduced latency and 2) robustly reduces average DRAM access \jkf{latency} while
maintaining data correctness, by issuing DRAM requests with reduced access
latencies to non-failure-prone DRAM regions.  We evaluate Solar-DRAM on a wide
variety of multi-core workloads and show that for 4-core homogeneous workloads,
Solar-DRAM provides an average (maximum) system performance improvement of
4.31\% (10.87\%) compared to using \jkf{the default fixed DRAM access latency.} 

\section{Motivation and Goal} 
\label{solar:sec:motivation} 

Many prior works~\cite{hassan2016chargecache, kim2010thread, lee2009improving,
muralidhara2011reducing, mutlu2007stall, mutlu2008parallelism,
zhang2000permutation, kim2012case, lee2013tiered, lee-sigmetrics2017} show that various important
workloads exhibit \emph{low} access locality and thus are unable to effectively
exploit \emph{row-buffer locality}. In other words, these workloads issue a
significant number of DRAM accesses that result in bank (i.e., row buffer)
conflicts, which \emph{negatively} impact overall system performance. Each
access that causes a bank conflict requires activating a closed row, \jkth{a
process whose latency is dictated by the $t_{RCD}$ timing parameter.  The
memory controller must wait for $t_{RCD}$ before issuing any other command} to
that bank. To reduce the overhead of bank conflicts, we aim to reduce the
$t_{RCD}$ timing parameter while maintaining data correctness. 

\textbf{Prior Observations.} In a recent publication, Chang et
al.~\cite{chang2016understanding} observe that activation failures 1) are
\emph{highly} constrained to specific columns of DRAM cells across an entire
DRAM bank, i.e., global bitlines, and regions of memory that are closer to the
row decoders, 2) can \emph{only} affect cells within the cache line granularity
of bits that is first requested in a closed row, and 3) propagate back into
DRAM cells and become \emph{permanent} failures in the stored data. 

Based on these observations, Chang et al. propose FLY-DRAM, which
\emph{statically} profiles DRAM \emph{global bitlines} as \emph{weak} or
\emph{strong} using a one-time profiling step.  During execution,
FLY-DRAM relies on this \emph{static} profile to access \emph{weak} or
\emph{strong} global bitlines with \emph{default} or \emph{reduced} $t_{RCD}$,
respectively. 

Unfortunately, \cite{chang2016understanding} falls short in three aspects.
First, the paper lacks \jkf{analysis of} whether a \emph{strong} bitline will
ever become a \emph{weak} bitline or vice versa. This analysis is necessary to
demonstrate the viability of relying on a static profile of global bitlines
to guarantee data integrity. Second, the authors present a
characterization of activation failures on an older generation of DRAM (DDR3).
Third, the proposed mechanism, FLY-DRAM, does not \emph{fully} take advantage
\jkf{of all opportunities} to reduce $t_{RCD}$ in modern DRAM \jks{modules} (as we
show in Section~\ref{sec:characterization}). 

Given the shortcomings of prior work~\cite{chang2016understanding}, \textbf{our
goal} is to 1) present a more rigorous characterization of activation failures
on \emph{state-of-the-art LPDDR4} DRAM \jks{modules}, 2) demonstrate the
viability of mechanisms that rely on a static profile of weak cells to reduce
DRAM access latency, and \jkf{3)} devise new mechanisms that exploit \emph{more
activation failure characteristics} on \emph{state-of-the-art LPDDR4} DRAM
\jks{modules} to further reduce DRAM latency.

\section{Testing Methodology} 

To analyze DRAM behavior under reduced $t_{RCD}$ values, we developed an
infrastructure to characterize state-of-the-art LPDDR4 DRAM
chips~\cite{2014lpddr4} in a thermally-controlled chamber.  Our testing
environment gives us precise control over DRAM commands and $t_{RCD}$, as
verified via a logic analyzer probing the command bus. In addition, we
determined the address mapping for internal DRAM row scrambling so that we
could study the spatial locality of activation failures in the physical DRAM
chip. We test for activation failures across a DRAM \jkf{module} using
Algorithm~\ref{alg:testing_lat}. The key idea is to access every cache line
across DRAM, and open a closed row on each access.  This \emph{guarantees} that
\jkf{we test} every DRAM cell's propensity for activation failure. 
\begin{algorithm}[h]\footnotesize
    \SetAlgoNlRelativeSize{0.7}
    \SetAlgoNoLine
    \DontPrintSemicolon
    \SetAlCapHSkip{0pt}
    \caption{DRAM Activation Failure Testing}
    \label{alg:testing_lat}


    \textbf{DRAM\_ACT\_fail\_testing($data\_pattern$, $reduced\_t_{RCD}$):} \par
    ~~~~write $data\_pattern$ (e.g., solid 1s) into all DRAM cells \par
    ~~~~\textbf{foreach} $col$ in DRAM \jkf{module}: \par
    ~~~~~~~~\textbf{foreach} $row$ in DRAM \jkf{module}: \par
    ~~~~~~~~~~~~$refresh(row)$ \textcolor{gray}{~~~~~~~~~~~~~~~~~~~~~~~~~~~// replenish cell voltage } \par
    ~~~~~~~~~~~~$precharge(row)$ \textcolor{gray}{~~~~~~~~~~~~~~~~~~~~~~// ensure next access activates row} \par
    ~~~~~~~~~~~~$read(col)$ with $reduced\_t_{RCD}$ \textcolor{gray}{~~~// induce activation failures on col} \par
    ~~~~~~~~~~~~find and record activation failures \par
\end{algorithm}

We first write a known data pattern to DRAM (Line~2) for consistent testing
conditions.  The \emph{for loops} (Lines~3-4) ensure that we test all DRAM
cache lines. For each cache line, we 1) refresh the row containing it (Line 5)
to induce activation failures in cells with similar levels of charge, 2)
precharge the row (Line 6), and 3) activate the row again with a \emph{reduced
$t_{RCD}$} (Line 7) to induce activation failures. We then find and record the
activation failures in the row (Line 8), by comparing the read data to the data
pattern the row was initialized with. \jkf{We experimentally determine that
Algorithm~\ref{alg:testing_lat} takes approximately 200ms to test a single
bank.} 

Unless otherwise specified, we perform all tests using 2y-nm LPDDR4 DRAM
chips from three major manufacturers in a thermally-controlled chamber held at
55$^{\circ}$C. We control the ambient temperature precisely using heaters
and fans. A microcontroller-based PID loop controls the heaters and fans
to within an accuracy of 0.25$^{\circ}$C and a reliable range of 40$^{\circ}$C
to 55$^{\circ}$C. We keep the DRAM temperature at 15$^{\circ}$C above
ambient temperature using a separate local heating source. This local heating
source probes local on-chip temperature sensors to smooth out temperature
variations due to self-induced heating.

%
 
\section{Activation Failure Characterization} 
\label{sec:characterization} 

We present our extensive characterization of activation failures in modern
LPDDR4 DRAM \jks{modules} from three major DRAM manufacturers. We make a number of
key observations that 1) support the viability of a mechanism that uses a
\emph{static profile} of weak cells to exploit variation in access latencies of
DRAM cells, and 2) enable us to devise new mechanisms that exploit \emph{more
activation failure characteristics} to \jkf{further} reduce DRAM latency. 


\subsection{Spatial Distribution of Activation Failures}
\label{subsec:spatial_obs} 

\begin{figure}[h]
    \centering
    \includegraphics[width=0.55\linewidth]{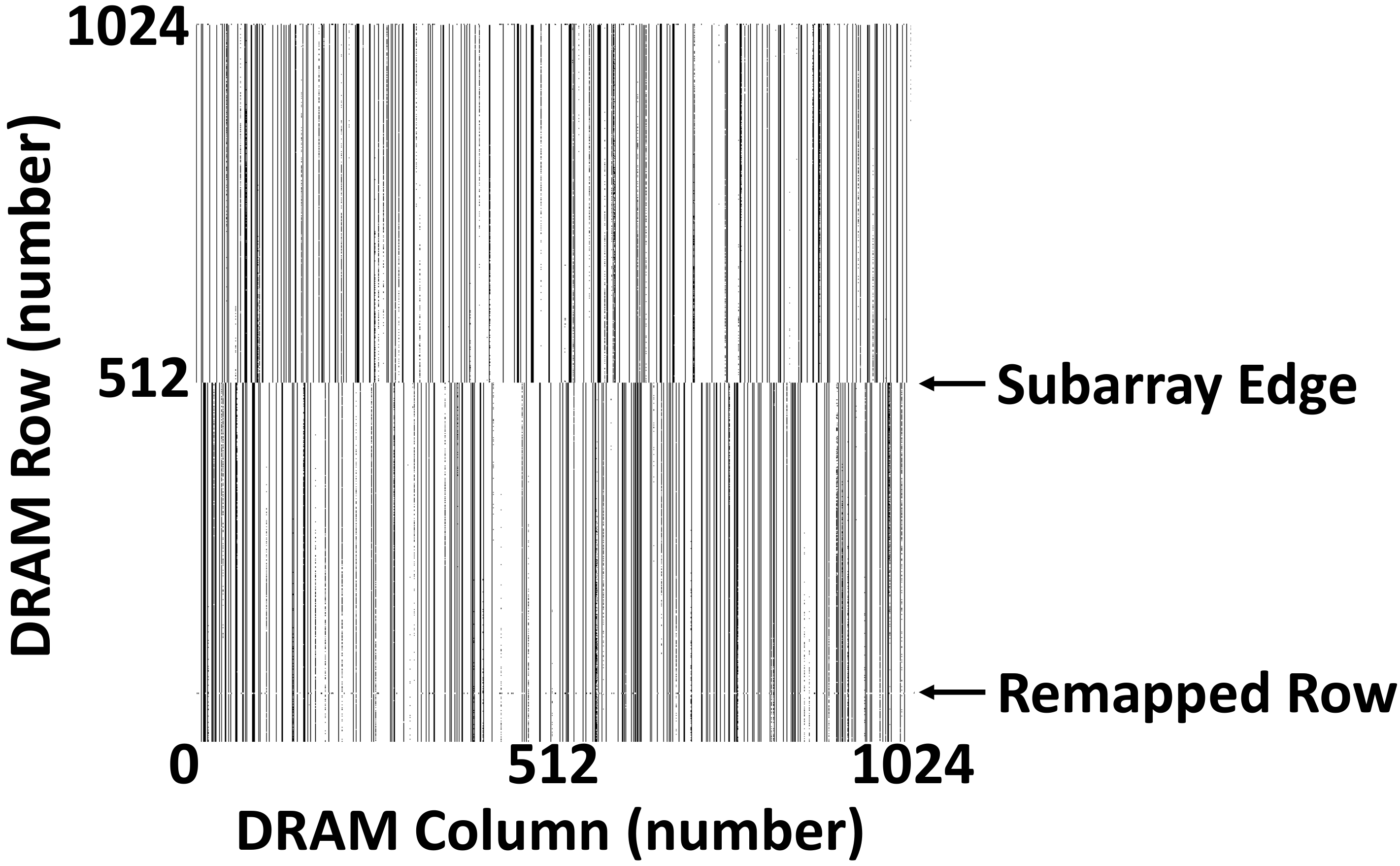}
    \caption{Activation failure bitmap in 1024x1024 cell array.} 
    \label{fig:act_fail_bitmap}
\end{figure}

We first analyze the spatial distribution of activation failures across DRAM
\jks{modules} by visually inspecting bitmaps of activation failures across many DRAM
banks. A \emph{representative} 1024x1024 array of DRAM cells with a significant number
of activation failures is shown in Figure~\ref{fig:act_fail_bitmap}. Using
these bitmaps, we make three key observations. \textbf{Observation 1:}
Activation failures are highly
constrained to \emph{local bitlines}. We infer that the granularity at which
we see bitline-wide \jkf{activation} failures is a subarray. This is because
the number of consecutive rows with \jkf{activation failures on} the same
bitline falls within the range of expected modern subarray sizes of 512 to
1024~\cite{kim2012case, lee-sigmetrics2017}.  We hypothesize that this occurs
as a result of process manufacturing variation at the level of the local sense
amplifiers. Some sense amplifiers are manufactured ``weaker'' and \emph{cannot}
amplify data on the local bitline as quickly.  This results in a higher
probability of activation failures in DRAM cells attached to the same
``weak'' local bitline.  While manufacturing process variation dictates
the local bitlines that contain errors, the manufacturer design decisions for
subarray size dictates the number of cells attached to the same local bitline,
and thus, the number of consecutive rows that contain activation failures in
the same local bitline.  \textbf{Observation 2:} Subarrays from Vendor B and
C's DRAM \jks{modules} consist of 512 DRAM rows, while subarrays from Vendor A's DRAM
\jks{modules} consist of 1024 DRAM rows.  \textbf{Observation 3:} We find that within
a set of subarray rows, very few rows \jkf{(<0.001\%)} exhibit a significantly
different set of cells that experience activation failures compared to the
expected set of cells. We hypothesize that the rows with significantly
different failures are rows that are \emph{remapped} to redundant rows
\jkf{(see~\cite{khan2016parbor, liu2013experimental})} after the \jks{DRAM
module} was manufactured (indicated in Figure~\ref{fig:act_fail_bitmap}).

We next study the granularity at which activation failures can be induced when
accessing a row. We make two observations (also seen in prior
work~\cite{chang2016understanding}).  \textbf{Observation 4:} When accessing a
row with low $t_{RCD}$, the errors in the row are constrained to the DRAM cache
line granularity (typically 32 or 64 bytes), and only occur in the
aligned 32 bytes that is first accessed in a closed row (i.e., up to 32 bytes
are affected by a single low $t_{RCD}$ access). Prior
work~\cite{chang2016understanding} also observes that failures are
constrained to cache lines on a system \jkf{with 64 byte cache lines}.
\textbf{Observation 5:} The first cache line accessed in a closed DRAM row is
the \emph{only} cache line in the row that we observe to exhibit
activation failures. We hypothesize that DRAM cells that are subsequently
accessed in the same row have enough time to have their charge amplified
and completely restored for correct \jkf{sensing}.

\jkth{We next study the proportion of weak subarray columns per bank across
many DRAM banks from all 282 of our \jks{DRAM modules}. We collect the
proportion of weak subarray columns per bank across two banks from each of our
DRAM \jks{modules} across all three manufacturers. For a given bank, we aggregate the
subarray columns that contain activation failures when accessed with reduced
$t_{RCD}$ across our full range of temperatures. \textbf{Observation 6:} We
observe that banks from manufacturers A, B, and C have an \jkf{average/maximum
(standard deviation) proportion of weak subarray columns of 3.7\%/96\% (12\%),
2.5\%/100\% (6.5\%), and 2.2\%/37\% (4.3\%)}, respectively. We find that on
average, banks have a \emph{very low proportion of weak subarray columns},
which means \jkf{that} the memory controller can issue DRAM accesses to
\emph{most} subarray columns with reduced $t_{RCD}$.} 

\begin{figure}[ht]
    \centering
    \includegraphics[width=0.8\linewidth]{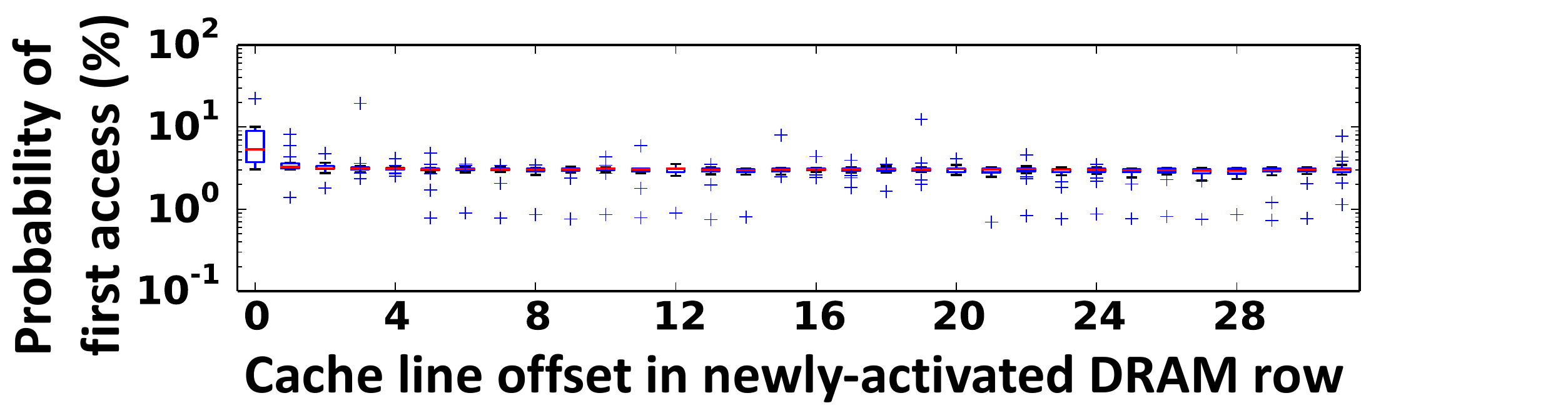}
	\caption{Probability of \jkf{the first} access to a \jkf{newly-activated} row going to a particular cache line \jkth{offset} within the row.} 
    \label{fig:cacheline_motivation} 
\end{figure}

We next study how a real workload might be affected by reducing $t_{RCD}$.
We use Ramulator~\cite{kim2016ramulator, ramulatorgithub} to analyze the
spatial distribution of accesses immediately following an \cmdact (i.e., accesses
that can induce activation failures) across 20 workloads from \jkf{the} 
SPEC CPU2006 benchmark suite~\cite{spec2006}.
Figure~\ref{fig:cacheline_motivation} shows the probability that the first
access to a newly-activated row \jke{is} to a particular cache line offset within
the row. For a given cache line offset (x-axis value), the probability is
presented as a distribution of probabilities, found across the SPEC CPU2006
workloads. Each distribution of probabilities is shown as a box-and-whisker
plot\footnote{A box-and-whisker plot emphasizes the important metrics of a
dataset's distribution. The box is lower-bounded by the first quartile (i.e.,
the median of the first half of the ordered set of data points) and
upper-bounded by the third quartile (i.e., the median of the second half of the
ordered set of data points). The median falls within the box. The \emph
{inter-quartile range} (IQR) is defined as the distance between the first and
third quartiles, or the size of the box.  Whiskers extend an additional $1.5
\times IQR$ on either side of the box. We indicate outliers, or data points
outside of the whiskers, with pluses.} where the probability (y-axis) is
logarithmically scaled.  \textbf{Observation 7:} A significant proportion of
\jkf{first accesses to a newly-activated} DRAM row requests the $0^{th}$ cache
line in the row, with a maximum (average) proportion of 22.2\% (6.6\%). This
indicates that simply reducing $t_{RCD}$ for all accesses to \emph{only} the
$0^{th}$ cache line of each DRAM row can significantly improve overall system
performance. We hypothesize that the $0^{th}$ cache line is accessed with a
significantly higher probability due to a significant number of streaming
accesses to DRAM rows in our evaluated workloads. Streaming accesses would
result in accesses \jke{first} to the $0^{th}$ cache line of a newly-activated
row followed by accesses to the remaining cache lines in the row \jke{in a
consecutive manner}. 

\subsection{Data Pattern Dependence} 
\label{solar:subsec:dpd}

To understand the effects of DRAM data patterns on activation failures in local
bitlines, we analyze the number of local bitlines containing activation
failures \jkf{with} different data patterns written to the DRAM array.  Similar
to prior works~\cite{patel2017reaper, liu2013experimental} that extensively
describe \jkf{DRAM} data patterns, we study a total of 40 unique data patterns:
solid 1s, checkered, row stripe, column stripe, 16 walking 1s, \emph{and} the
inverses of all 20 aforementioned data patterns.

\begin{figure}[ht]
    \centering
    \includegraphics[width=\linewidth]{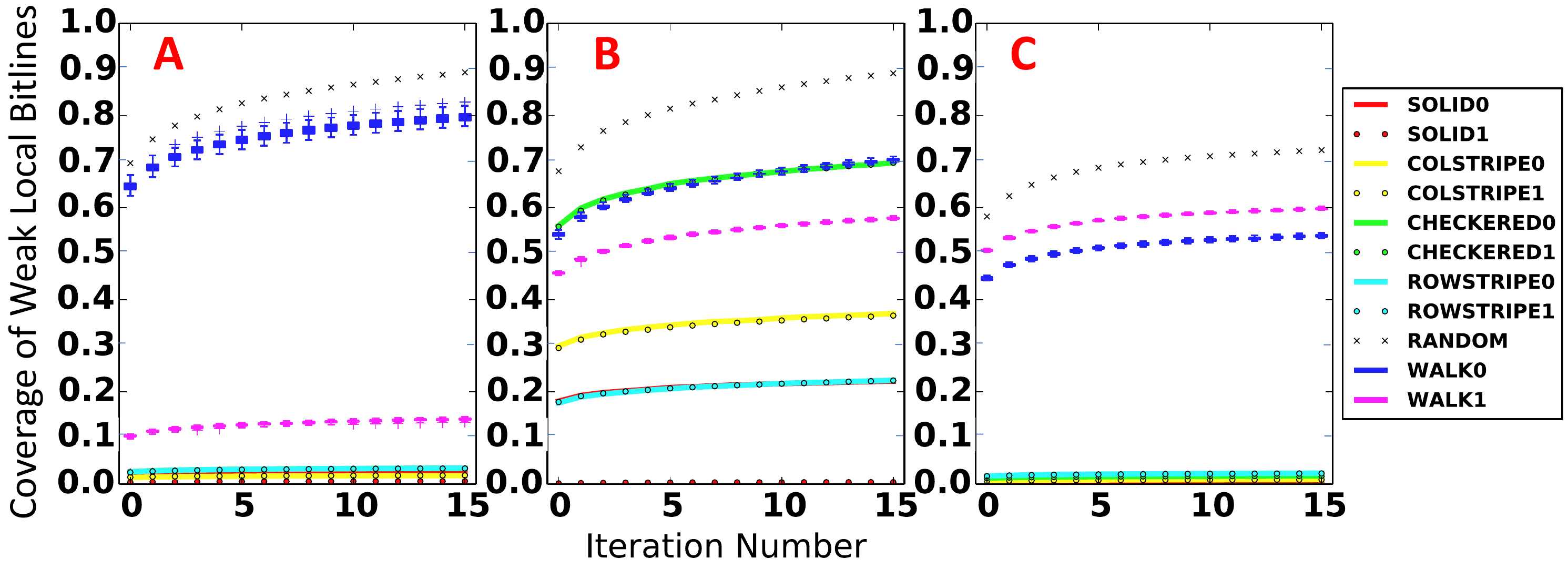}
    \caption{Data pattern dependence of the \jkf{proportion} of local bitlines \jkf{with activation failures found} over 16 iterations.} 
    \label{fig:data_pattern_dependence} 
\end{figure}

Figure~\ref{fig:data_pattern_dependence} shows the \emph{cumulative} number of
unique local bitlines containing activation failures over 16 iterations with
different data patterns across representative DRAM \jks{modules} from three DRAM
manufacturers.  This data was gathered with 100 iterations of
Algorithm~\ref{alg:testing_lat} per data pattern, but we \jkf{present only} the
first 16 iterations to highlight the accumulation rate of local bitlines with
failures in earlier iterations. For a given iteration, we calculate the
\jkf{\emph{coverage of each data pattern}} as: 
\begin{equation} \label{eq:dp_coverage}
\frac{\sum_{n=1}^{x} unique\_local\_bitlines(data\_pattern, iteration_n)}{total\_local\_bitlines\_with\_failures}
\end{equation}
where $unique\_local\_bitlines()$ is the number of local bitlines observed to
contain failures \jkf{in} a given iteration but \jkf{\emph{not}} observed to
contain failures in any prior iteration when using a specific data pattern, and
$total\_local\_bitlines\_with\_failures$ is the total number of unique local
bitlines observed to contain failures at \jkf{\emph{any}} iteration, with
\jkf{\emph{any}} data pattern.  The \emph{\jkf{coverage}} of a single data
pattern indicates the effectiveness of that data pattern to identify the full
set of local bitlines containing activation-failure-prone DRAM cells.
\textbf{Observation 8:} Each walking pattern in a set of WALK1s or WALK0s
(i.e., 16 walking 1 patterns and \jkf{their} inverses) finds a similar
coverage of local bitlines over many iterations.  Given Observation~8, we
\jks{have already simplified} Figure~\ref{fig:data_pattern_dependence} by
grouping the set of 16 walking 1 patterns and plotting the distribution of
coverages of the patterns as a box-and-whisker-plot \jkf{(WALK1)}. We \jks{have
done} the same for the set of 16 walking 0 patterns \jkf{(WALK0)}.
\textbf{Observation 9:} The random data pattern exhibits the highest coverage
of \jkf{activation-failure-prone} local bitlines across all three DRAM
manufacturers.  We hypothesize that the random data results in, on average
across DRAM cells, the worst-case coupling noise of a DRAM cell and its
neighbors. This is consistent with prior works' \jks{experimental} observations
that the random data pattern causes the highest rate of charge leakage in
cells~\cite{patel2017reaper, liu2013experimental, khan2014efficacy}.  


\subsection{Temperature Effects} 
\label{subsec:temperature_characterization}

We next study the effect of \jks{DRAM temperature (at the granularity of
$5^{\circ}C$)} on the number of activation failures across a DRAM \jks{module}
(at reduced $t_{RCD}$). We make similar observations as prior
work~\cite{chang2016understanding} and see no clear correlation between the
\jke{\emph{total number}} of activation failures across a DRAM device and DRAM
temperature. However, when we analyze the activation failure rates at the
granularity of a \emph{local bitline}, we \jke{observe} correlations between
DRAM temperature and the number of activation failures in a \jke{\emph{local bitline}}. 

To determine the effect of temperature on a local bitline's probability to
contain cells with activation failures, we study activation failures on a local
bitline granularity with a range of temperatures. For a set of $5^{\circ}C$
intervals of DRAM temperature between $55^{\circ}C$ and $70^{\circ}C$, we run
100 iterations of Algorithm~\ref{alg:testing_lat}, recording each cell's
probability of failure across all our DRAM \jks{modules}. We indicate a \emph{local
bitline's probability \jkf{of} failure ($F_{prob}$)} as: 
\begin{equation} \label{eq:bitline_fail_prob} 
F_{prob} = 
\sum_{n=1}^{cells\_in\_SA\_bitline} \frac{num\_iters\_failed_{cell_n}}{num\_iters\times cells\_in\_SA\_bitline}
\end{equation}
where $cells\_in\_SA\_bitline$ indicates the number of cells in a local
bitline, $num\_iters\_failed_{cell_n}$ indicates the number of iterations out
of the 100 tested iterations \jkf{in which} $cell_n$ fails, and $num\_iters$ is
the total number of iterations that the DRAM \jks{module} is tested for. 

\begin{figure}[ht]
    \centering
    \includegraphics[width=0.8\linewidth]{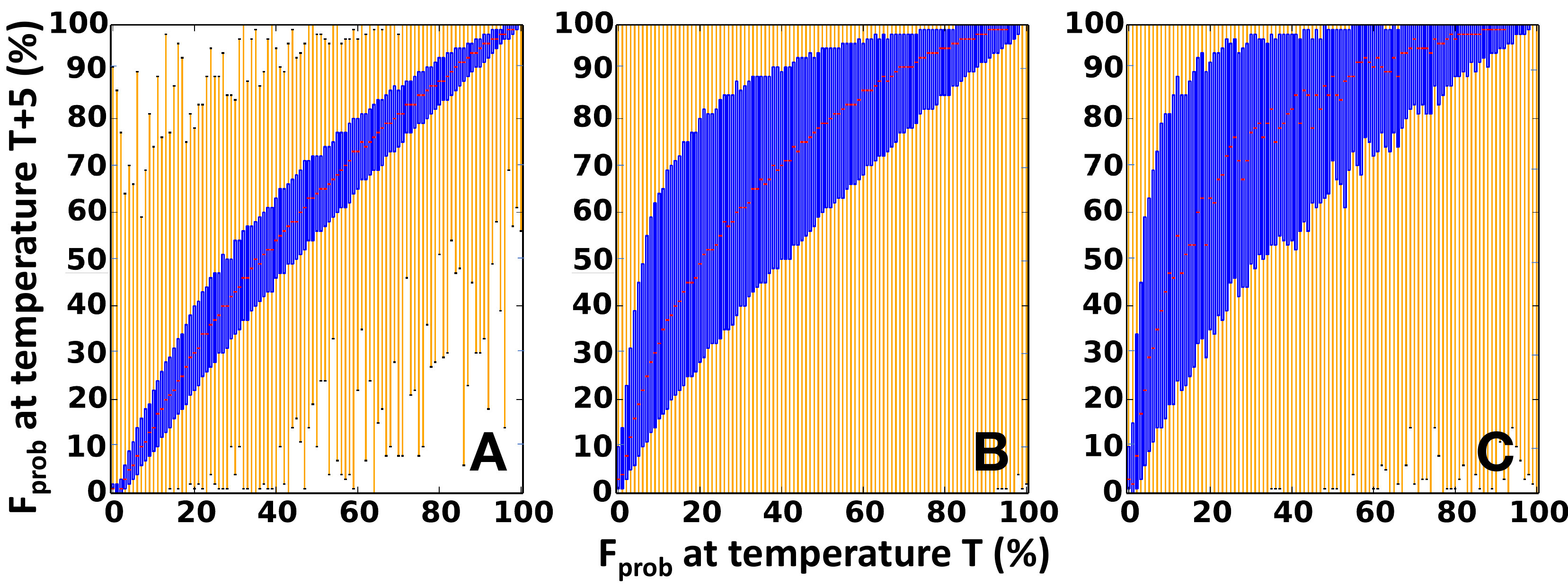}
    \caption{Temperature effects on a local bitline's $F_{prob}$.}
    \label{fig:temperature_correlation} 
\end{figure}

Figure~\ref{fig:temperature_correlation} aggregates our data across 30 DRAM
\jks{modules} from each DRAM manufacturer. Each point in the figure represents the
$F_{prob}$ of a local bitline at temperature $T$ on the x-axis (i.e., the
baseline temperature) and the $F_{prob}$ of the same local bitline at
temperature $T+5$ on the y-axis (i.e., $5^{\circ}C$ above the baseline
temperature).
The $F_{prob}$ values at the baseline temperature are binned at the granularity
of 1\% and represent the range of $F_{prob}\pm0.5\%$. We aggregate the
$F_{prob}$ values at temperature $T+5$ for every local bitline whose $F_{prob}$
at temperature $T$ falls within the same bin on the x-axis. We aggregate each
set of $F_{prob}$ values with box-and-whisker plots \jkf{to show how the
$F_{prob}$ is generally affected by increasing the temperature.}  We draw each
box-and-whisker plot with a blue box, orange whiskers, black whisker ends, and
red medians. \textbf{Observation 10:} \jkf{We observe that $F_{prob}$ at
temperature $T+5$ tends to be higher than $F_{prob}$ at temperature $T$
\jke{(i.e., the blue region of the figure is above the $x=y$ line)}.
\jks{Thus,} $F_{prob}$ tends to increase with increased temperature.} However,
there are cases \jke{(i.e., <25\% of all data points)} where the $F_{prob}$
decreases with an increased temperature.  We conclude that in order to find a
comprehensive set of weak subarray columns, we must profile for activation
failures with a range \jkf{(e.g., 40$^{\circ}$C to 55$^{\circ}$C)} of DRAM
temperatures.

\subsection{Latency Effects}


We next study the effects of changing the value of $t_{RCD}$ on
activation failures. We sweep $t_{RCD}$ between 2ns and 18ns (default) at the
coarse granularity of 2ns, and we study the correlation of $t_{RCD}$ with the
total number of activation failures. We make \emph{two} observations analogous
to those made by Chang et al.~\cite{chang2016understanding}.
\textbf{Observation 11:} We observe \emph{no} activation failures when using
$t_{RCD}$ values above 14ns regardless of the temperature. The first $t_{RCD}$
at which activation failures occur is 4ns below manufacturer-recommended
values. This demonstrates the additional \emph{guardband} that manufacturers
place to account for process variation. \textbf{Observation 12:}
We observe that a small reduction (i.e., by 2ns) in $t_{RCD}$ results in a
significant increase (>10x) in the number of activation failures.

In addition to repeating analyses on older generation
\jks{modules}~\cite{chang2016understanding}, we \jkf{are the first to} study the
effects of changing the $t_{RCD}$ value on the failure probability of an
individual cell.  \textbf{Observation 13:} \jkf{We observe that, if a DRAM cell
fails 100\% of the time when accessed with a reduced $t_{RCD}$ of $n$, the same
cell will likely fail between 0\% and 100\% when $t_{RCD}$ is set to $n+2$, and
0\% of the time when $t_{RCD}$ is set to $n+4$. We hypothesize that the large
changes in activation failure probability is due to the coarse granularity with
which we can change $t_{RCD}$ (i.e., 2ns\jks{; due to experimental
infrastructure limitations}). For this very reason, we cannot observe gradual
changes in the activation failure probability \jks{that} we expect would occur
at smaller intervals of $t_{RCD}$. We leave the exploration of correlating
finer granularity changes of $t_{RCD}$ with the probability of activation
failure of a DRAM cell to future work.} 

\subsection{Short-term Variation} 
\label{subsec:time_obs}

\begin{figure}[h] \centering
\includegraphics[width=0.4\linewidth]{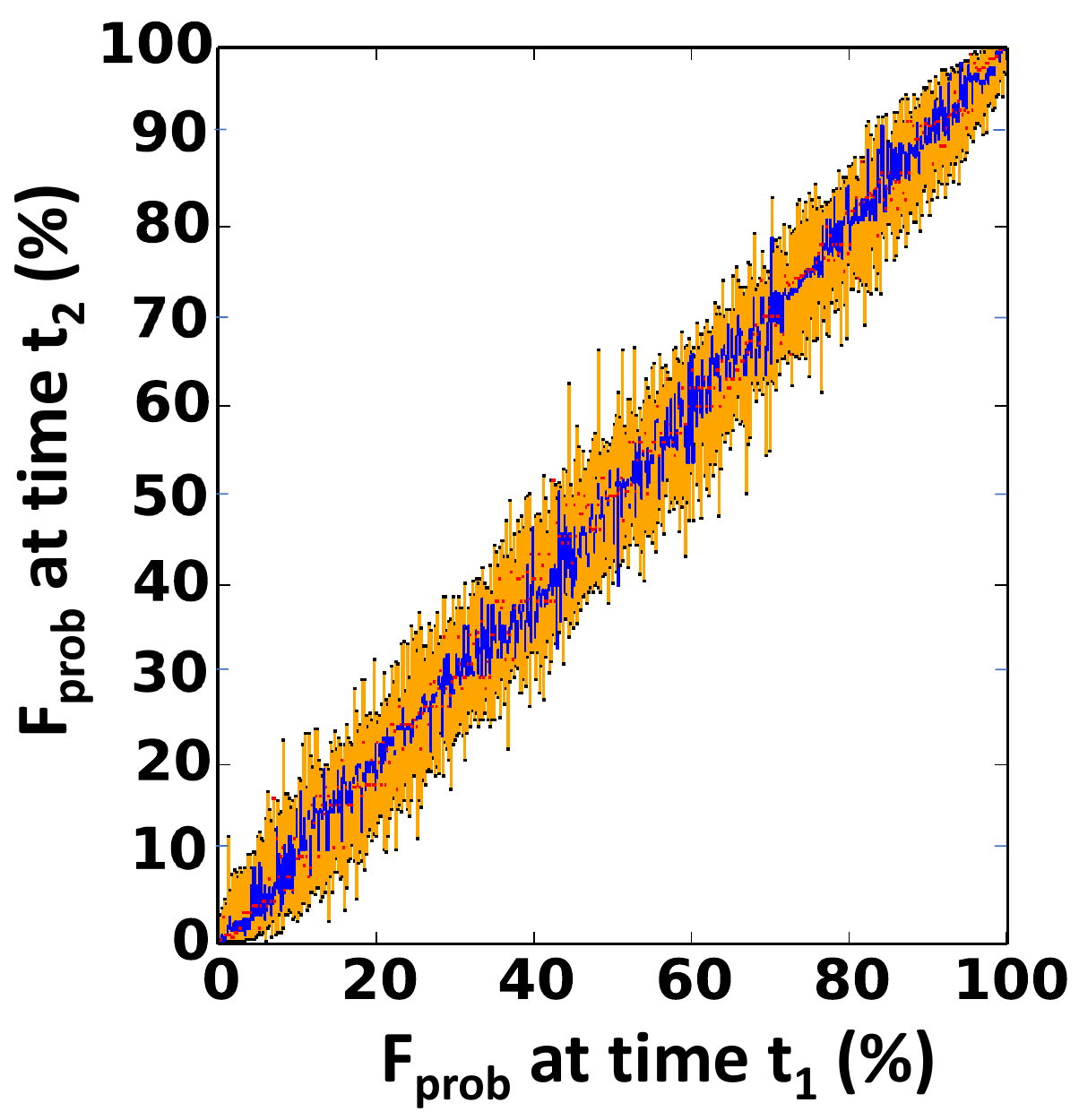}
\caption{$F_{prob}$ of local bitlines across time.}
\label{fig:constant_failure_probabilities} 
\end{figure}

Many previous DRAM retention characterization works~\cite{baek2014refresh,
khan2014efficacy, liu2013experimental, qureshi2015avatar,
venkatesan2006retention, hassan2017softmc, chang_understanding2017,
lee-sigmetrics2017, schroeder2009dram, sridharan2012study, kim2014flipping,
jung2016reverse, khan2016case, khan2016parbor, yaney1987meta, restle1992dram,
kang2014co, mori2005origin, patel2017reaper, khan2017detecting} \jkf{have
shown} that there is a well-known phenomenon called variable retention time
(VRT), where variation occurs \emph{over time} in DRAM circuit elements that
\jkf{results} in significant and sudden changes in the leakage rates of charge from a
DRAM cell.  This affects the retention time of \jkf{a DRAM cell} over short-term
intervals, resulting in varying retention failure probabilities for a given
DRAM cell over the span of minutes or hours.  To see if a similar
time-based variation phenomenon affects the probability of an \emph{activation
failure}, we sample the $F_{prob}$ of many local bitlines every six hours over
14 days and study how $F_{prob}$ changes across the samples for a given local
bitline.  Figure~\ref{fig:constant_failure_probabilities} plots the change in
$F_{prob}$ of a given local bitline from one \jkf{time} sample to another. For a given
local bitline, every pair of sample $F_{prob}$ values (across the 14 day study)
are plotted as (x,y) pairs. We collect these data points across all local
bitlines in 30 DRAM \jks{modules} (10 of each DRAM manufacturer) and plot the points.
All points sharing the same $F_{prob}$ on the x-axis, are aggregated into
box-and-whisker plots.
\textbf{Observation 14:} We find that the box-and-whisker plots show a tight
distribution around the diagonal axis (where x equals y). This indicates that
the $F_{prob}$ of a given local bitline remains highly similar
($correlation~r~=~0.94$) across time.  \emph{This means that a weak local bitline
is very likely to remain weak and a strong local bitline is very likely to
remain strong across time.} Thus, we can identify the set of weak local
bitlines \emph{once} and that set would remain \emph{constant} across time.  To
determine the number of iterations we expect to profile for to find a
comprehensive set of weak local bitlines, we run iterations of
Algorithm~\ref{alg:testing_lat} \jkf{for each bank} until we \jkf{only} observe
either zero or one failing bit in a local bitline that has never been observed
to fail before \jkf{in the tested bank}. At this point, we say that we have
found \jkf{the \emph{entire}} set of local bitlines containing activation
failures.  \textbf{Observation 15:} We find that the required number of
iterations to find \jkf{the entire} set of local bitlines containing activation
failures differs significantly across chips and manufacturers.  The
\jkf{average/maximum (standard deviation) number of iterations required to find
the entire set of local bitlines for manufacturers A, B, and C is 843/1411
(284.28), 162/441 (174.86), and 1914/1944 (26.28), respectively.} 

\subsection{DRAM Write Operations} 
\label{subsec:reduced_writes}

We next study the effects of reduced $t_{RCD}$ on \emph{\jkf{write}}
operations.  We hypothesize that $t_{RCD}$ is \emph{\jke{mostly unnecessary}}
for DRAM write operations, because $t_{RCD}$ dictates the time required for the
sense amplifiers to amplify the data in DRAM cells to an \emph{I/O readable
value} ($V_{access}$) such that reads can be correctly serviced. To determine
the effects of reducing $t_{RCD}$ on DRAM write operations, we run two
experiments with our DRAM \jks{modules}.  First, we sweep the value of
$t_{RCD}$ between 2ns and 18ns, and write a known data pattern across DRAM. We
then read every value in the DRAM array with the default $t_{RCD}$ and compare
each read value with the expected value. We repeat this process 100 times using
the \emph{random} data pattern for each of our \jks{DRAM modules}. We observe
activation failures only when $t_{RCD}$ is set below 4ns. We conclude that we
can reliably issue DRAM write operations to our LPDDR4 DRAM \jks{modules} with
a \emph{significantly} reduced $t_{RCD}$ \jkf{(i.e., 4ns; a reduction of 77\%)}
without loss of data integrity.

\section{Exploiting Activation Latency Variation} 
\label{sec:mechanism}

Based on our key observations from our extensive characterization of
activation latency failures in DRAM (Section~\ref{sec:characterization}), we
propose \emph{\underline{S}ubarray-\underline{o}ptimized Access
\underline{La}tency \underline{R}eduction DRAM (Solar-DRAM)}, a mechanism that
robustly reduces $t_{RCD}$ for both DRAM read and write requests. 

\subsection{Solar-DRAM} 
\label{subsec:soar}

Solar-DRAM consists of three components that exploit various observations on
activation failures and memory access patterns. These three components are pure
hardware approaches implemented within the memory controller without any DRAM
changes and are invisible to applications. 

\textbf{Component I: Variable-latency cache lines (VLC).} The first key
observation that we exploit is that activation failures are highly constrained
to \emph{some} (or few) \emph{local bitlines} (i.e., \jkf{only}
3.7\%/2.5\%/2.2\% of subarray columns per bank are weak on average for DRAM
manufacturers A/B/C respectively\jkf{. See Section~\ref{subsec:spatial_obs})}, and
the local bitlines with activation-failure-prone cells are
\emph{randomly} distributed across the chip \jkf{(not shown)}.  Given the known
spatial distribution of activation failures, the memory controller can issue
memory requests with varying activation latency depending on whether or not the
access is to data contained in a ``weak'' local bitline.  To enable such a
mechanism, Solar-DRAM requires the use of a \emph{weak subarray column profile}
that identifies local bitlines as either \emph{weak} or \emph{strong}.
However, since activation failures affect DRAM \jkf{only} at the granularity of a
\jkf{cache line} (Section~\ref{subsec:spatial_obs}), Solar-DRAM needs to \jkf{only}
store whether or not a column of \jkf{cache-line-aligned} DRAM cells within a
subarray, i.e., a \emph{subarray column}, contains a weak local bitline.

The second key observation that we exploit is that the failure probability of a
cell, when accessed with a reduced $t_{RCD}$, is \emph{not} vulnerable to
short-term time variation, i.e., we did not observe significant changes within
14 days of testing (Section~\ref{subsec:time_obs}). This novel observation is
necessary to ensure that a profile of weak local bitlines will \emph{not change
over time} and thus allows Solar-DRAM to rely on a static profile.\footnote{We
acknowledge that we do \emph{not} consider long-term variation that may arise
from aging or wearout of circuit components. We leave this exploration to
future work. Such long-term effects can have implications for a static profile
(as discussed in DIVA-DRAM~\cite{lee-sigmetrics2017}), but one can devise a
mechanism that updates the profile at regular long time intervals with low
overhead, e.g., as in prior work~\cite{patel2017reaper,
qureshi2015avatar}\jke{.}}  

Given a static profile of weak subarray columns, we can safely access the weak
subarray columns with the default $t_{RCD}$, and \emph{all other} subarray
columns with a reduced $t_{RCD}$. We observe that after finding the initial set
of failing columns there is still a very low probability \jks{(i.e.,
$<5\times10^{-7}$)} that a strong column will result in a single error.
Fortunately, we find this probability to be low enough such that employing
error correction codes (ECC)~\cite{kang2014co, nair2016xed, hamming1950error,
cha2017defect}, which are already present in modern DRAM chips, would
transparently mitigate low-probability activation failures in strong columns. 

\textbf{Component II: Reordered subarray columns (RSC).} We observe in
Section~\ref{subsec:spatial_obs}, that the memory controller accesses the
$0^{th}$ cache line of a newly-activated DRAM row with the highest probability
compared to the rest of the cache lines. Thus, we would like to devise a
mechanism that reduces access latency (i.e., $t_{RCD}$) \jke{\emph{specifically to the
$0^{th}$ cache line in each row}} because the first accessed cache line in a
newly-activated row is most affected by $t_{RCD}$.  To this end, we propose a
mechanism that scrambles column addresses such that the $0^{th}$ cache line in
a row is \emph{unlikely} to get mapped to weak subarray columns.  Given a weak
subarray column profile, we identify the \emph{global column} (i.e., the
column of cache-line-aligned DRAM cells across a full DRAM bank) containing the
fewest weak subarray columns\jke{, called the \emph{strongest global column}.} 
We then scramble the column address bits such that the $0^{th}$ cache line for
each bank maps to the \emph{strongest global column} in the bank. We
perform this scrambling \jke{by changing the DRAM address mapping} at the
granularity of the global column, \jks{in order to} to reduce the overhead in
address scrambling.

\textbf{Component III: Reduced latency for writes (RLW).} The final observation
that we exploit in Solar-DRAM is that write operations do \emph{not} require
the default $t_{RCD}$ value (Section~\ref{subsec:reduced_writes}).  To exploit
this observation, we use a reliable, reduced $t_{RCD}$ (i.e., 4ns\jkf{,} as
measured with our experimental infrastructure) for \emph{all} write operations
to DRAM. 

\subsection{Static Profile of Weak Subarray Columns} 

To obtain the static profile of weak \jkf{subarray} columns, we run multiple
iterations of Algorithm~\ref{alg:testing_lat}, recording all subarray columns
containing observed activation failures. As we observe in
Section~\ref{sec:characterization}, there are various factors that affect a
local bitline's probability of failure ($F_{prob}$). We use these factors to
determine a method for identifying a comprehensive profile of weak subarray
columns for a given DRAM \jks{module}. First, we use our observation on the
accumulation rate of finding weak local bitlines
(Section~\ref{subsec:time_obs}) to determine the number of iterations we expect
to test each \jks{DRAM module}. However, since there is such high variation
across each \jks{DRAM module} (as seen in the standard deviations of the
distributions in Observation 11), we can only provide \jks{the expected number
of iterations needed to find a comprehensive profile} for \jks{DRAM modules} of a
manufacturer, and the time to profile depends on the \jks{module}. We show in
Section~\ref{solar:subsec:dpd} that no \emph{single} data pattern alone finds a high
coverage of weak local bitlines.  This indicates that we must test each data
pattern (40 data patterns) for the \jks{expected number of iterations needed to
find a comprehensive profile} of a \jks{DRAM module} for a range of temperatures
(Section~\ref{subsec:temperature_characterization}). While this could result in
many iterations of testing (on the order of a few thousands\jkf{; see
Section~\ref{subsec:time_obs})}, this is a one-time process \jkf{on the order
of half a day per bank} that results in a reliable profile of weak subarray
columns. \jke{The required one-time profiling can be performed} in two ways: 1)
\jke{t}he system running Solar-DRAM can profile a DRAM \jks{module} when the
memory controller \jke{detects} a new DRAM \jks{module} at bootup, or 2) the DRAM
manufacturer can profile each DRAM \jks{module} and provide the profile within
the \emph{Serial Presence Detect} (SPD) circuitry (a Read-Only Memory present
in each DIMM)~\cite{SPD_DDR4}. 

To minimize the storage overhead of the weak subarray column profile in the
memory controller, we encode each subarray column with a bit indicating whether
or not to issue accesses to it with a reduced $t_{RCD}$. \jks{After profiling
DRAM, the memory controller loads the weak subarray column profile once into a}
small lookup table in the DRAM channel's memory controller.\footnote{To
store the lookup table for a DRAM channel\jks{,} we require $\jkf{num\_banks}\times
num\_subarrays\_per\_bank\times\frac{row\_size}{cacheline\_size}$ \jkf{bits,
where $num\_subarrays\_per\_bank$ is the number of subarrays in a bank,
$row\_size$ is the size of a DRAM row in bits, and $cacheline\_size$ is the
size of a cache line in bits. For a 4GB DRAM \jks{module} with 8 banks, 64 subarrays
per bank, 32-byte cache lines, and 2KB per row, the lookup table requires 4KB
of storage.}} For any DRAM request, the memory controller references the lookup
table with the subarray column that is being accessed.  The memory controller
\jks{determines} the $t_{RCD}$ timing parameter according to the value of the
bit found in the lookup table.

\section{Solar-DRAM Evaluation} 
\label{sec:evaluation} 

We first discuss our evaluation methodology and evaluated system
configurations. We then present our multi-core simulation results for our
chosen system configurations. 

\subsection{Evaluation Methodology} 

\noindent
\textbf{System Configurations.} We evaluate the performance of Solar-DRAM on a
4-core system using Ramulator~\cite{kim2016ramulator, ramulatorgithub}, an
open-source cycle-accurate DRAM simulator, in CPU-trace-driven mode. We analyze
various real workloads with traces from the SPEC CPU2006
benchmark~\cite{spec2006} that we collect using Pintool~\cite{luk2005pin}.
Table~\ref{table:system_config} shows the configuration of our evaluated
system. We use the standard LPDDR4-3200~\cite{2014lpddr4} timing parameters
as our baseline. To give a conservative estimate of Solar-DRAM's performance
improvement, we simulate with a 64B cache line and a subarray size of 1024
rows.\footnote{Using the typical upper-limit values for these
configuration variables reduces the total number of subarray columns that
comprise DRAM (\jks{to} 8,192 subarray columns per bank). A smaller number of
subarray columns reduces the granularity at which we can issue DRAM accesses
with reduced $t_{RCD}$, which reduces Solar-DRAM's potential for performance
benefit\jks{. This is because a single activation failure requires the memory
controller to access \emph{\jke{larger}} regions of DRAM with default $t_{RCD}$.} } 

\begin{table}[h!] 
    \centering \footnotesize \renewcommand{\arraystretch}{1.4}
    \resizebox{0.7\linewidth}{!}{
    \begin{tabular}{|m{2.7cm}
        | m{7.5cm}|}
        \hline \textbf{Processor} & 4 cores, 4 GHz, 4-wide issue, 8
        MSHRs/core, OoO 128-entry window\\
        \hline
        \textbf{LLC} & $8~MiB$ shared, 64B cache line, 8-way associative\\
        \hline
        \textbf{Memory \hspace{2mm} \newline Controller} & 64-entry R/W queue,
        {FR-FCFS}~\cite{frfcfs, zuravleff1997controller} \\
        \hline
        \textbf{DRAM} & LPDDR4-3200~\cite{2014lpddr4}, 2 channels, 1 rank/channel,
        8 banks/rank, 64K rows/bank, 1024 rows/subarray, $8~KiB$ row-buffer, 
        Baseline: $t_{RCD}$/$t_{RAS}$/$t_{WR}$~=~29/67/29 cycles (18.125/41.875/18.125 ns)\\
        \hline
		\textbf{Solar-DRAM} & \jks{reduced} $t_{RCD}$ for \jke{requests to} \jkf{strong} cache lines: 18 cycles (11.25ns) \newline\jks{reduced} $t_{RCD}$ for write requests: 7 cycles (4.375ns)\\ 
		\hline
    \end{tabular}
    }
\caption{Evaluated system configuration.}
\label{table:system_config}
\end{table}

\noindent
\textbf{Solar-DRAM Configuration.} 
To evaluate Solar-DRAM and FLY-DRAM~\cite{chang2016understanding} on a variety
of different \jks{DRAM modules} with unique properties, we simulate varying 1)
the number of weak subarray columns \jkf{per bank between $n = 1~to~512$}, and
2) the chosen weak subarray columns in each bank. For a given \jkf{$n$, i.e.,}
weak subarray column count, we generate 10 \emph{unique} profiles with $n$
randomly chosen weak subarray columns per bank. \jkf{The profile indicates
whether a subarray column} should be accessed with \jkf{the} default $t_{RCD}$
(29 cycles; 18\jke{.13} ns) or the reduced $t_{RCD}$ (18 cycles; 11\jke{.25}
ns).  We use these profiles to evaluate 1) Solar-DRAM's three components
(described in Section~\ref{subsec:soar}) independently, 2) Solar-DRAM with all
its three components, 3) FLY-DRAM~\cite{chang2016understanding}, and 4) our
baseline LPDDR4 DRAM. 

\emph{Variable latency cache lines} (VLC), directly uses a weak subarray column
profile to determine whether an access should be issued with a reduced or
default $t_{RCD}$ value. \emph{Reordered subarray columns} (RSC) takes a
profile and maps the $0^{th}$ cache line to the \jke{\emph{strongest global
column}} in each bank. For a given profile, this maximizes the probability that
any access to the $0^{th}$ cache line of a row will be issued with a reduced
$t_{RCD}$. \emph{Reduced latency for writes} (RLW) reduces $t_{RCD}$ to 7
cycles (4\jke{.38} ns) (Section~\ref{subsec:reduced_writes}) for \emph{all}
write operations to DRAM.  \emph{Solar-DRAM} (Section~\ref{subsec:soar})
combines all three components (\emph{VLC}, \emph{RSC}, and \emph{RLW}).  Since
\emph{FLY-DRAM}~\cite{chang2016understanding} issues read requests at the
granularity of the \emph{\jkf{global} column} depending on whether a
\jkf{global column} contains weak bits, we evaluate FLY-DRAM by taking a weak
subarray column profile and extending each weak subarray column to \jkf{the
global column containing it.} Baseline LPDDR4 uses a fixed $t_{RCD}$ of 29
cycles (18\jke{.13} ns) for all accesses. We present performance improvement of
the different mechanisms over this \jkf{LPDDR4} baseline.

\begin{figure*}[!b]
    \centering
    \includegraphics[width=\linewidth]{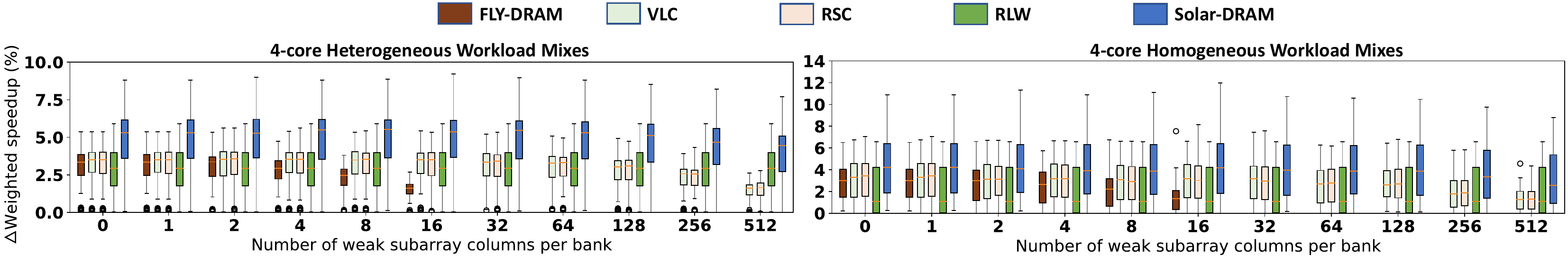}
    \caption{Weighted speedup improvements of Solar-DRAM, its three individual components, and FLY-DRAM over baseline LPDDR4 DRAM\jkf{,} evaluated over various 4-core workload mixes from the SPEC CPU2006 benchmark suite.} 
    \label{fig:evaluation} 
\end{figure*}

\subsection{Multi-core Evaluation Results}

Figure~\ref{fig:evaluation} plots the improvement \jkf{in} weighted
speedup\jks{\cite{snavely2000symbiotic}}, which corresponds to system
throughput~\cite{eyerman2008system}, over the baseline on 20 homogeneous mixes
of 4-core workloads and 20 heterogeneous mixes of 4-core workloads randomly
combined from the set of workloads in the SPEC CPU2006 benchmark
suite~\cite{spec2006}. For each configuration of \jke{\emph{<weak subarray
column count, weak subarray column profile, mechanism, workload mix>}}, we
aggregate all weighted speedup improvement results into a box-and-whisker
plot. 

We make four key observations. First, Solar-DRAM provides significant weighted
speedup improvement. Even when half of the subarray columns are classified as
weak \jkf{(which is very unrealistic and conservative, as our experiments on real
\jks{DRAM modules} show)}, Solar-DRAM improves performance by 4.03\% (7.71\%) for
heterogeneous and 3.36\% (8.80\%) for homogeneous workloads. In the ideal case,
where there are 0 weak subarray columns per bank and thus, the memory
controller issues \jkf{\emph{all}} memory accesses with a reduced $t_{RCD}$,
Solar-DRAM \jkf{improves performance by} 4.97\% (8.79\%) for heterogeneous and
4.31\% (10.87\%) for homogeneous workloads.  Second, \jkf{e}ach individual
component of Solar-DRAM improves system performance\jkf{. \emph{RLW} is the
best alone: it improves performance by} 2.92\% (5.90\%) for heterogeneous and
2.25\% (6.59\%) for homogeneous workloads.  Because \emph{RLW} is independent
of the number of weak subarray columns in \jkf{a bank, its} weighted speedup
improvement is constant regardless of the number of weak subarray columns per
bank. \jkf{Third,} Solar-DRAM provides higher performance improvement than each
of its components, demonstrating that the combination of \emph{VLC},
\emph{RSC}, and \emph{RLW} \jkf{is} \emph{synergistic}. \jkf{Fourth,}
Solar-DRAM provides much higher performance improvement \jks{than} FLY-DRAM.  This is
because \jkf{Solar-DRAM 1)} exploits the observation that \emph{all write
requests} can be issued with a greatly reduced $t_{RCD}$ (i.e., by 77\%), and
2) issues read requests with reduced $t_{RCD}$ at the granularity of the
\emph{local} bitline rather than the global bitline. This means that for a
single weak cache line in a subarray, Solar-DRAM issues \jkf{read requests with
default $t_{RCD}$ \emph{only}} to cache lines in the \emph{subarray column}
containing the weak cache line, while FLY-DRAM would issue \jkf{read requests
with default $t_{RCD}$} to \emph{all} cache lines in the column across the
\emph{full bank}. For this \jkf{very} same reason, we also observe that
\emph{VLC} alone \jkf{outperforms} FLY-DRAM. Fourth, Solar-DRAM enables
significantly higher performance improvement on DRAM \jks{modules} with a high
rate of activation failures, \jkf{where} FLY-DRAM provides no benefit.  Because
FLY-DRAM categorizes columns across the \emph{\jks{entire} bank} as strong or
weak, even a low activation failure rate across the DRAM chip \jkf{results} in
a high number of cache lines requiring the default $t_{RCD}$ timing parameter
\jkf{in FLY-DRAM}. We experimentally observe the average proportion of weak
subarray columns per bank to be 3.7\%/2.5\%/2.2\% for DRAM manufacturers A/B/C
(Section~\ref{subsec:spatial_obs}).  Even at such a low proportion of weak
subarray columns (i.e., 38/26/23 subarray columns out of 1024 subarray columns
in our evaluated DRAM configuration), we \jks{expect the performance benefit of
FLY-DRAM to be well below 1.6\% (i.e., the median performance benefit when we
evaluate FLY-DRAM with 16 weak subarray columns in Figure~\ref{fig:evaluation}
across all workload mixes) for DRAM manufacturers B and C, and 0\% for DRAM
manufacturer A.} We conclude that Solar-DRAM's three components provide
significant performance improvement on modern LPDDR4 DRAM \jks{modules} over
LPDDR4 DRAM and FLY-DRAM.

\section{Related Work} 

Many works seek to improve DRAM access latency. They can be classified
according to the mechanisms they take advantage of, as follows\jkf{.} 

\noindent
\textbf{Static Variation.} We have already described these
works~\cite{chang2016understanding} in detail in
Section~\ref{solar:sec:motivation} and compared to
FLY-DRAM~\cite{chang2016understanding} in Section~\ref{sec:evaluation}.
\jkf{Solar-DRAM outperforms FLY-DRAM.} \jkn{Das et al.~\cite{das2018vrl}
propose a method to reduce \emph{refresh latency}, which is orthogonal to
Solar-DRAM.}

\noindent
\textbf{Operational Factors.} Prior works improve DRAM latency by controlling
or taking advantage of changes in operational factors such as
temperature~\cite{lee2015adaptive} and voltage~\cite{chang_understanding2017}.
These works are orthogonal to Solar-DRAM since they reduce latency in response
to changes in factors \jkf{that are} independent of latency variations inherent
to the \jks{DRAM module}. 

\noindent
\textbf{Access Locality.} Some work exploits locality in DRAM access
patterns~\jkf{\cite{hassan2016chargecache, wang2018reducing, shin2014nuat}} and
reorganizes DRAM accesses to allow for higher
locality~\cite{mutlu2008parallelism, lee2010dram, seshadri2014dirty,
seshadri2015gather} in order to reduce average DRAM access latency. These can
be combined with Solar-DRAM for further latency reduction.

\noindent
\textbf{Modifications to DRAM Architecture.} Various
works~\jky{\cite{chang2014improving, chang2016low, hidaka1990cache, kim2012case,
lee2013tiered, lu2015improving, seongil2014row, seshadri2013rowclone,
seshadri2016buddy, seshadri2017ambit, seshadri2015fast, son2013reducing,
zhang2014half, luo2020clr, wang2018reducing}} propose mechanisms that change
the structure of DRAM to reduce latency. Solar-DRAM requires \emph{no} changes
to the DRAM chip. 

\noindent
\textbf{\jkf{Software} Support.} Several works~\cite{ding2014compiler,
jeong2012balancing, lee2009improving, pai1999code} propose using compile-time
optimizations to improve DRAM access locality and thus, decrease overall DRAM
access latency. Solar-DRAM reduces the latency of the average memory access and
would provide added benefits to \jkf{software} optimizations.  If the profile
of weak subarray columns \jkf{is exposed to the compiler or the system
software, the software} could potentially use this device-level information to
allocate latency-critical data at \emph{stronger} locations in DRAM, while
decreasing the hardware overhead of storing weak subarray column profiles in
the memory controller.

\section{\jky{Limitations}} 

\jkz{Solar-DRAM has a few limitations that must be considered when analyzing the
viability of such a mechanism on a real system. } 

\jkz{First, our characterization in Solar-DRAM demonstrates that failures due
to \jky{low-latency} accesses are generally localized to specific local bitlines.
However, there are low probability bit flips that may occur outside of the
"weak" local bitlines less predictably. Solar-DRAM relies on Error Correcting
Code (ECC) hardware to handle such low probability bit flips that are not
encapsulated by the Solar-DRAM profile. In DRAM devices without ECC hardware or
margin in their ECC capability for additional bit flips, applications using
Solar-DRAM on such devices will likely see failures in their data. We expect
future work to enhance Solar-DRAM by enabling the ability to either 1)
comprehensively predict all weak subarray cache lines with a better profiling
methodology, 2) predict low probability bit flips and mitigate them before they
occur, or 3) understand why these low probability bit flips occur and eliminate
them completely.} 

\jkz{Second, a number of unknowns prevents us from providing sampling statistics
for our data. These unknowns include 1) the percentage of DRAM modules that we
characterize compared to the total number of available modules (billions), 2)
the total number of DRAM chips available for a given DRAM type, technology, and
process size, and 3) the inability to randomly sample DRAM modules. Without
this knowledge, we are unable to provide confidence intervals, or margins of
sampling error. Therefore, the profile storage sizes and performance benefit
distributions may not be representative of the average DRAM device. However,
given that we were able to see the overarching trend of weak subarray bitlines
in each DRAM chip that we tested, we are confident that this effect is
prevalent across chips of the DRAM types and process nodes that we tested and
all chips should be able to benefit from Solar-DRAM with low storage overhead.} 

\jkz{Third, we did not account for long-term DRAM aging effects during
characterization. Thus, we are not able to confidently determine how long a
profile will remain viable, and how often re-profiling will need to occur in
order to reliably issue low latency accesses according to the Solar-DRAM
profile. However, we do show that a profile will likely remain viable after 14
days of usage. In the worst case, the system may require profiling the DRAM
chip every 14 days in order to create a reliable profile. However, we believe
this is a conservative estimate, and there may be more intelligent ways for
identifying gradual changes to the profile during normal execution. }

\section{Summary} 

We introduced 1) a rigorous characterization of activation failures across 282
\emph{real state-of-the-art LPDDR4} DRAM \jks{modules}, 2) Solar-DRAM, whose
key idea is to exploit our observations and issue DRAM accesses with variable
latency depending on the target DRAM location's propensity to fail with reduced
access latency, and 3) an evaluation of Solar-DRAM and its three individual
components, with comparisons to the
state-of-the-art~\cite{chang2016understanding}.  We find that Solar-DRAM
provides significant performance improvement over the state-of-the-art DRAM
latency reduction mechanism across a wide variety of workloads,
\jkf{\emph{without} requiring any changes to DRAM chips or software.}



\chapter{The DRAM Latency PUF: Quickly Evaluating Physical Unclonable Functions by Exploiting the Latency-Reliability Tradeoff in Modern Commodity DRAM Devices} 
\label{ch4-dlpuf} 

Physically Unclonable Functions (PUFs) are commonly used in cryptography to
identify devices based on the uniqueness of their \emph{physical
microstructures}. DRAM-based PUFs have numerous advantages over PUF designs
that exploit alternative substrates: DRAM is a major component of many modern
systems, and a DRAM-based PUF can generate many unique identifiers. However,
none of the prior DRAM PUF proposals provide implementations suitable for
runtime-accessible PUF evaluation on commodity DRAM \jkfive{devices}. Prior
DRAM PUFs exhibit unacceptably high latencies, especially at low temperatures
(e.g., >125.8s on average for a 64KiB memory segment below 55$^{\circ}$C), and
they cause high system interference by keeping part of DRAM unavailable during
PUF evaluation.

In this chapter, we introduce the DRAM latency PUF, a new class of fast,
reliable DRAM PUFs. The key idea is to reduce DRAM read access latency below
the reliable datasheet specifications using software-only system calls.  Doing
so results in error patterns that reflect the compound effects of manufacturing
variations in various DRAM structures (e.g., capacitors, wires, sense
amplifiers). Based on a rigorous experimental characterization of 223 modern
LPDDR4 DRAM chips, we demonstrate that these error patterns 1) satisfy
runtime-accessible PUF requirements, and 2) are quickly generated
\jkfour{(i.e., at 88.2ms) irrespective of operating temperature using a real
system with no additional hardware modifications.} We show that, for a constant
DRAM capacity overhead of 64KiB, our implementation of the DRAM latency PUF
enables \jkfive{an average (minimum, maximum) PUF} evaluation time speedup of
152x (109x, 181x) at 70$^{\circ}$C and 1426x (868x, 1783x) at
55$^{\circ}$C when compared to a DRAM retention PUF and achieves
greater speedups at even lower temperatures.

\section{Physical Unclonable Functions} 

A Physically Unclonable Function (PUF) maps a set of \emph{input parameters} to
unique, device-specific signatures that can be generated \emph{repeatably} and
\emph{reliably}. We refer to the process of generating a signature using a
given set of input parameters as the \emph{evaluation} of a PUF. The resulting
signature reflects a device's inherent, random physical variations introduced
during manufacturing. This property guarantees that the signature is
practically impossible to predict or replicate without access to the device
itself~\cite{yan2015novel, gassend2002silicon}. These characteristics enable
PUFs to be frequently used in security applications such as low-cost
authentication mechanisms against system security attacks and prevention of
integrated circuit (IC) counterfeiting~\cite{sutar2016d, xiong2016run}.

PUFs are generally used in a challenge-response (CR)
protocol~\cite{sutar2016d}, in which a trusted server gives a device a
\emph{challenge} (i.e., a set of input parameters and conditions with which to
evaluate a PUF), and verifies the device's \emph{PUF response} (i.e., the
signature generated by the PUF). A CR protocol generally consists of two
phases: \emph{enrollment} and \emph{authentication}.  Enrollment is a one-time
setup phase in which a given device is analyzed, and all possible PUF responses
are stored in the trusted server. Authentication occurs when an application
running on the enrolled device requests escalated permissions from the trusted
server to perform a secure action. The server provides a challenge to the
application, which then evaluates the PUF with the requested parameters and
returns the PUF response. If the response matches with the previously-enrolled
response for the challenge, i.e., the \emph{golden key}, authentication is
successful. The CR can be done \emph{statically}, where the PUF is evaluated
only once before runtime (e.g., at bootup) or at \emph{runtime}, where an
application running on the enrolled device can evaluate a PUF
on-demand~\cite{xiong2016run}.

PUFs for silicon devices were first introduced as a method for integrated
circuit (IC) identification, exploiting manufacturing process variation among
devices for \emph{disambiguating} different devices~\cite{lofstrom2000ic}.
Since then, many prior works have proposed PUF evaluation techniques for
different substrates (e.g., ASICs, FPGAs, memories), exploiting manufacturing
variation in different components such as emerging memory
technologies~\cite{rose2013foundations,koeberl2013memristor,iyengar2014dwm,vatajelu2015stt},
flash memory~\cite{wang2012flash}, Application Specific Integrated Circuit
(ASIC) logic~\cite{helinski2009physical, tuyls2006read, guajardo2007physical,
lee2004technique, lim2005extracting, majzoobi2008testing,
ruhrmair2009foundations, ozturk2008physical, ozturk2008towards,
hammouri2008unclonable, gassend2003physical, gassend2002silicon,
suh2007physical, kumar2008butterfly, su20071, maes2008intrinsic,
van2010hardware, yin2013design}, Static Random Access Memory
(SRAM)~\cite{guajardo2007fpga, holcomb2007initial, holcomb2009power,
cortez2013adapting, bhargava2012reliability, zheng2013resp, xiao2014bit,
bacha2015authenticache}, and Dynamic Random Access Memory
(DRAM)~\cite{sutar2016d, keller2014dynamic, hashemian2015robust,
tehranipoor2015dram, tehranipoor2017investigation, rahmati2015probable,
tang2017dram, sutar2017memory}.

PUFs must satisfy \emph{five} key characteristics to be effective in security
applications~\cite{tehranipoor2015dram, xiong2016run, hashemian2015robust,
sutar2016d, maes2010physically}. We describe these characteristics in detail in
Section~\ref{subsection:ideal_puf}. PUFs satisfying these characteristics
\jkfive{1)} guarantee a level of robustness for \emph{disambiguating} many
devices and \jkfive{2)} are practically impossible for an attacker to duplicate
\emph{without} access to the physical device itself. In addition to these
properties, a \emph{runtime-accessible} PUF, \jkfive{i.e., a PUF that is
accessible online} to an application running on an enrolled device, must 1) be easily
evaluated with \emph{low latency} to prevent unnecessary slowdown of the
application requesting authentication, and 2) provide \emph{low system
interference}, i.e., minimize the disturbance PUF evaluation causes to other
applications running on the same system.
Section~\ref{subsection:ideal_runtime_puf} describes the characteristics of
\emph{ideal runtime-accessible PUFs}.

\section{Motivation and Goal} 

DRAM-based PUFs, henceforth called \emph{DRAM PUFs}, have recently
attracted significant interest for two key reasons: 1) DRAM is already widely
used in a wide variety of modern systems~\cite{mutlu2013memory,
mutlu2014research}, ranging from embedded to server, and 2) DRAM's large
address space, which is on the order of Giga- or Tera-bytes, makes it
naturally suitable for CR applications by providing a greater CR space
relative to smaller components (e.g., SRAMs)~\cite{guajardo2007fpga,
holcomb2007initial, guajardo2007physical, holcomb2009power,
cortez2013adapting, bhargava2012reliability, zheng2013resp, xiao2014bit,
bacha2015authenticache}.  Prior DRAM PUF proposals exploit variations in DRAM
start-up values~\cite{tehranipoor2015dram}, DRAM write access
latencies~\cite{hashemian2015robust}, and DRAM cell retention
failures~\cite{sutar2016d, keller2014dynamic, liu2014trustworthy,
xiong2016run} to generate reliable PUF responses.

Unfortunately, these prior DRAM PUF proposals have significant drawbacks that
make them unsuitable as \emph{runtime-accessible} PUFs. PUFs that use DRAM
start-up values~\cite{tehranipoor2015dram} preclude runtime-accessible PUF
evaluation by requiring a DRAM power cycle for \emph{every} authentication.
This requires either interrupting other applications using DRAM or restarting
the entire system, which is likely infeasible at runtime. On the other hand,
PUFs that \mpsix{exploit} variation in write access
latencies~\cite{hashemian2015robust} \emph{can be} evaluated at runtime.
However, \cite{hashemian2015robust}'s proposal requires additional circuitry in
a DRAM chip to allow fine-grained manipulation of write
latency~\cite{hashemian2015robust}. This requires changes to DRAM chips,
\jksix{rendering} such proposals inapplicable to devices used in the field
today.  In this chapter, we would like to design a new runtime-accessible PUF
\emph{without} modifying \mpsix{commodity DRAM chips}.

Using cell charge retention \emph{failures} and their resulting \emph{error
patterns}~\cite{raidr, hamamoto1998retention, liu2013experimental,
khan2014efficacy, patel2017reaper, qureshi2015avatar} is the best candidate for
runtime-accessible DRAM PUF evaluation in commodity devices today, since it
does \emph{not} require a power cycle or any modifications to DRAM chips.
Unfortunately, such \emph{DRAM retention PUFs} impose two major overheads.
\jkfive{First, due to the 1) wide distribution of charge retention times across
DRAM cells~\cite{khan2014efficacy, liu2013experimental, patel2017reaper,
qureshi2015avatar, hamamoto1998retention} and 2) roughly-uniform spatial
distribution of retention failures across a chip~\cite{shirley2014copula,
baek2014refresh}, we find that the evaluation time of a DRAM retention PUF
\jksix{takes} on the order of \emph{minutes} at $55^\circ$C to identify enough
retention failures. The evaluation time increases exponentially as temperature
decreases.}  Second, this means that DRAM refresh \emph{must} be disabled for
long periods of time. Because DRAM refresh can \emph{only} be disabled for
large regions of DRAM~\cite{chang2014improving}, evaluating a DRAM retention
PUF on a small region of memory, i.e., a \emph{PUF memory segment}, requires
disabling refresh on the \emph{entire} large memory region containing the PUF
memory segment. However, to maintain the integrity of data inside the large
region but outside
\jksix{of} the PUF memory segment, \emph{all} such data must be
\emph{continuously} refreshed with additional DRAM commands, which results in
significant system interference~\cite{xiong2016run}. Based on extensive
experimental analysis using 223 state-of-the-art LPDDR4 DRAM devices, we find
that DRAM retention PUFs are too slow for reasonable runtime operation, e.g.,
\jksix{they have} average evaluation times of 125.8s at 55$^{\circ}$C and 13.4s
at 70$^{\circ}$C using a 64KiB memory segment
\jksix{(Section~\ref{ret_puf_analysis}).}

\textbf{Our goal} in this work is to develop a new runtime-accessible PUF that
1) uses \emph{existing} commodity DRAM devices, 2) satisfies all
characteristics of an effective \emph{runtime-accessible} PUF, and 3) provides
\emph{low-latency} evaluation with \emph{low system interference} across
\emph{all operating conditions}.

\section{Properties of a Runtime-Accessible PUF}
\label{section:PUF_metrics}

In this section, we examine the desired properties of an \emph{effective 
runtime-accessible PUF}. Prior works present various different metrics
for defining the effectiveness of a PUF~\cite{tehranipoor2015dram,
xiong2016run, hashemian2015robust, sutar2016d, maes2010physically}. We
consolidate these metrics into five key properties below. We then discuss
two properties that we consider necessary for an effective
\emph{runtime-accessible} PUF. We refer to these seven properties when
analyzing \jkfour{DRAM PUFs} (Section~\ref{ret_puf_analysis}
and~\ref{section:key_idea}). In Section~\ref{section:key_idea}, we show
how DRAM latency PUFs overcome the weaknesses of DRAM retention PUFs
based on a comparison of these properties between the two types of
PUFs.

\subsection{Characteristics of a Desirable PUF}
\label{subsection:ideal_puf} 
The following five key properties must be provided by any effective \emph{PUF}
\jkfour{that can be evaluated across a set of devices:}

\begin{enumerate}
    \itemsep0em
    \item \emph{Diffuseness}: a single device is able to generate many unique
    and independent responses to different input
        parameters~\cite{bacha2015authenticache,gao2015memristive,danger2016pufs,
        hori2010quantitative}. 
    \item \emph{Uniqueness}: a single device can be uniquely identified among the set of 
        devices~\cite{tehranipoor2015dram, xiong2016run, hashemian2015robust,
        wang2017current, liu2017acro, hori2010quantitative}. 
    \item \emph{Uniform Randomness}: all possible PUF responses must
        be equally different from each other~\cite{tehranipoor2015dram,
        xiong2016run, liu2017acro, hori2010quantitative}.
    \item \emph{Unclonability}: it should be practically impossible for
    an adversary to construct a device that exhibits the same properties as
        another~\cite{maes2010physically, katzenbeisserpufs,
        wang2017current}.
	\item \emph{Repeatability}: given a set of input parameters, PUF evaluation 
	results in the same PUF regardless of internal and external conditions (e.g., temperature,
        aging)~\cite{tehranipoor2015dram, xiong2016run,
        hashemian2015robust, wang2017current, liu2017acro,
        hori2010quantitative}. 
\end{enumerate}

These five properties ensure that a PUF can be used effectively for
challenge-response authentication.

\subsection{Characteristics of a \emph{Runtime-Accessible} PUF}
\label{subsection:ideal_runtime_puf} 

There are many important use cases for runtime-accessible PUFs.  Examples
include \jkfour{1) systems that employ remote communication protocols to access
devices via remote direct memory access (RDMA~\cite{rdmaconsortium}) or to
perform functions on remote devices (e.g., remote servers), 2) systems that
have interchangeable/broken system components (e.g., SSD drives, external
sensors, peripheral devices). In each of these systems, a connection/component
can be maliciously swapped out \emph{during runtime} so that a \emph{malicious}
device can be swapped in.  One way to avoid such an attack is to utilize a
low-overhead runtime-accessible PUF-based challenge-response mechanism that
frequently authenticates the communicating devices.  This enables
re-authentication of the system components during each step of communication
rather than just once at bootup time. More generally, a} fast
\emph{runtime-accessible} PUF enables the protection of a system from attacks
that exploit the fact that the time of check is different from the time of
use~\cite{xiong2016run}. 

In order to be useful for \emph{runtime} authentication, a PUF must be
effectively \emph{usable} while the system is running \emph{without}
significantly interfering with application execution and system operation.
\jkfour{Thus, a runtime-accessible PUF must possess the following two key
properties:} 

\begin{enumerate}
    \itemsep0em
    \item \emph{Low Latency}: PUF evaluation must be fast so that the
    application requesting authentication stalls for the \emph{smallest
    possible amount of time.}
    \item \emph{Low System Interference}: PUF evaluation must \emph{not}
    significantly slow down concurrently-running applications.
\end{enumerate}

\section{Testing Environment}
\label{methodology} 

To analyze DRAM behavior with \jkfour{both reduced refresh rates and} reduced
timing parameters, we developed an infrastructure to characterize modern
LPDDR4~\cite{2014lpddr4} DRAM chips.  Our testing environment gives us precise
control over the DRAM commands and DRAM timing parameters as verified
with a logic analyzer probing the command bus. 

We perfom all tests, unless otherwise specified, using a total of 223 2y-nm
LPDDR4 DRAM chips from three major manufacturers in a thermally-controlled
chamber held at 45$^{\circ}$C. For consistency across results, we stabilize the
ambient temperature precisely using heaters and fans controlled via a
microcontroller-based proportional-integral-derivative (PID) loop to within an
accuracy of 0.25$^{\circ}$C and a reliable range of 40$^{\circ}$C to
55$^{\circ}$C.  We maintain DRAM temperature at 15$^{\circ}$C above ambient
temperature using a separate local heating source. We utilize temperature
sensors to smooth out temperature variations caused by self-induced heating.

\section{DRAM Retention PUFs: Analysis}
\label{ret_puf_analysis} 

\sloppypar 
Recent works~\cite{sutar2016d, xiong2016run, rahmati2015probable,
sutar2017memory, tehranipoor2017investigation} propose DRAM retention
PUFs\jkfour{, which} require no modifications to commodity DRAM chips. These
works evaluate their proposals using DDR3 DRAM modules and find that while the
use of charge retention times in DRAM cells can result in repeatable PUFs,
delays on the order of \emph{minutes} are required to produce enough failures
for uniquely identifying many devices. 

In this section, we evaluate prior proposals using our own infrastructure with
223 modern LPDDR4 DRAM modules. Our experimental results
(Section~\ref{subsection:rpufgt}) confirm that DRAM retention PUFs can be
effectively implemented with commodity LPDDR4 DRAM devices. However, similarly
to prior work~\cite{sutar2016d, xiong2016run}, we find that the time required
to evaluate retention PUFs is prohibitively long (e.g., on the order of
minutes) \emph{at temperatures} that are likely encountered under common-case
operating conditions \jkfive{(e.g.,
35$^\circ$C-55$^\circ$C)}~\cite{choi2007modeling, lee2015adaptive,
el2012temperature}.

\subsection{Evaluating Retention PUFs}
\label{subsection:genretpuf}

We evaluate DRAM retention PUFs on our \jkfour{modern LPDDR4} devices, as shown
in Algorithm~\ref{alg:ret_puf}. The DRAM retention PUF disables refresh for a
period indicated by \jkfour{the $wait\_time$ input} parameter on a memory segment
indicated by the segment ID ($seg\_id$) input parameter (line~3). In order to
constrain retention failures to the PUF memory segment, the user must refresh
the rows contained in the DRAM rank, but \emph{not} in the PUF memory segment
during the $wait\_time$ interval (line~5-8). The resulting data in the memory
segment after the $wait\_time$ interval is the PUF response that is returned
for authentication (line~10).  \jkfour{This PUF response is uniquely
represented by the pattern of DRAM cells that fail in the memory segment
after \emph{not} being refreshed during the $wait\_time$ interval.} 
    
The memory controller can disable refresh only at the granularity of
\jkfour{DRAM ranks or banks}~\cite{2014lpddr4}. Therefore, in order to prevent
potential data loss, evaluation of a \emph {runtime-accessible} DRAM retention
PUF using a given DRAM memory segment \emph{requires} continuous refreshing of
all rows that are within the same \jkfour{rank or bank} but outside of the
\jkfive{PUF} memory segment. Doing so results in high system interference (see,
e.g.,~\cite{chang2014improving, raidr}) that is greatly exacerbated by the long
refresh intervals (e.g., \emph{60s} vs. the standard \emph{64ms}) required for
repeatable retention PUF evaluation at \jkfive{common-case temperatures (e.g.
35$^{\circ}$C-55$^{\circ}$C).}
\begin{algorithm}\footnotesize 
    \SetAlgoNlRelativeSize{0.7}
    \SetAlgoCaptionSeparator{\kern-0.3em{ }:\kern-0.1em{ }}
    \newcommand{\bfnarrow}[1]{\scalebox{.985}[1.0]{\textbf{#1}}}
    \newcommand{\normnarrow}[1]{\scalebox{.985}[1.0]{#1}}
    \SetAlCapSty{bfnarrow}
    \SetAlCapNameSty{normnarrow}
    \SetAlCapHSkip{0pt}
    \SetAlgoNoLine
    \DontPrintSemicolon
    \captionsetup[algorithm2e]{singlelinecheck=on}
    \caption{Evaluate\kern-0.1em{ }Retention\kern-0.1em{ }PUF\kern-0.1em{~}\cite{sutar2016d, xiong2016run, rahmati2015probable, sutar2017memory, tehranipoor2017investigation}}
    \label{alg:ret_puf}

    \textbf{evaluate\_DRAM\_retention\_PUF($seg\_id$, $wait\_time$):} \par
    ~~~~$rank\_id$ $\gets$ DRAM rank containing $seg\_id$ \par	
	~~~~disable refresh for Rank[$rank\_id$] \par
	~~~~$start\_time$ $\gets$ $current\_time()$ \par
	~~~~\textbf{while} $current\_time()$ - $start\_time$ < $wait\_time$: \par
    ~~~~~~~~\textbf{foreach} $row$ \textbf{in} Rank[$rank\_id$]: \par
	~~~~~~~~~~~~\textbf{if} $row$ \textbf{not in} Segment[$seg\_id$]: \par
	~~~~~~~~~~~~~~~~~issue refresh to $row$ \textcolor{gray}{~~~~~~~~~~~~~~~~~~~~~~~~~// refresh all other rows} \par
	~~~~enable refresh for Rank[$rank\_id$] \par
	~~~~\textbf{return} data at Segment[$seg\_id$]
\end{algorithm} 

\subsection{Evaluation Times of Retention PUFs} 
\label{subsection:rpufgt}

In this section, we explore the effects of DRAM temperature during DRAM
retention PUF evaluation on the DRAM retention PUF evaluation time. Based on
extensive experimental data from 223 LPDDR4 DRAM chips, we find that the
evaluation time of a DRAM retention PUF exhibits a strong dependence on DRAM
temperature during evaluation. With even just a 10$^{\circ}$C decrease in DRAM
temperature, the evaluation time for the \emph{same} PUF memory segment
increases by 10x~\cite{sutar2016d, xiong2016run}. This is due to the
\jkfour{direct correlation between} retention failure rate and temperature. We
reproduce the \emph{bit error rate} (BER) vs.  temperature relationship studied
for DDR3~\cite{liu2013experimental} and LPDDR4~\cite{patel2017reaper} chips
using our own LPDDR4 chips. We find that below refresh intervals of 30s,
there is an exponential dependence of BER on temperature with an average
exponential growth factor of 0.23 per 10$^{\circ}$C. This results in
approximately a 10x decrease in the retention failure rate with every
10$^{\circ}$C decrease in temperature and is consistent with prior work's
findings with older DRAM chips~\cite{liu2013experimental, sutar2016d,
patel2017reaper}. Due to the sensitivity of DRAM retention PUFs to temperature,
a \emph{stable temperature} is required to generate a \emph{repeatable} PUF
response. 

\begin{figure}[h]
    \centering
    \includegraphics[width=0.7\linewidth]{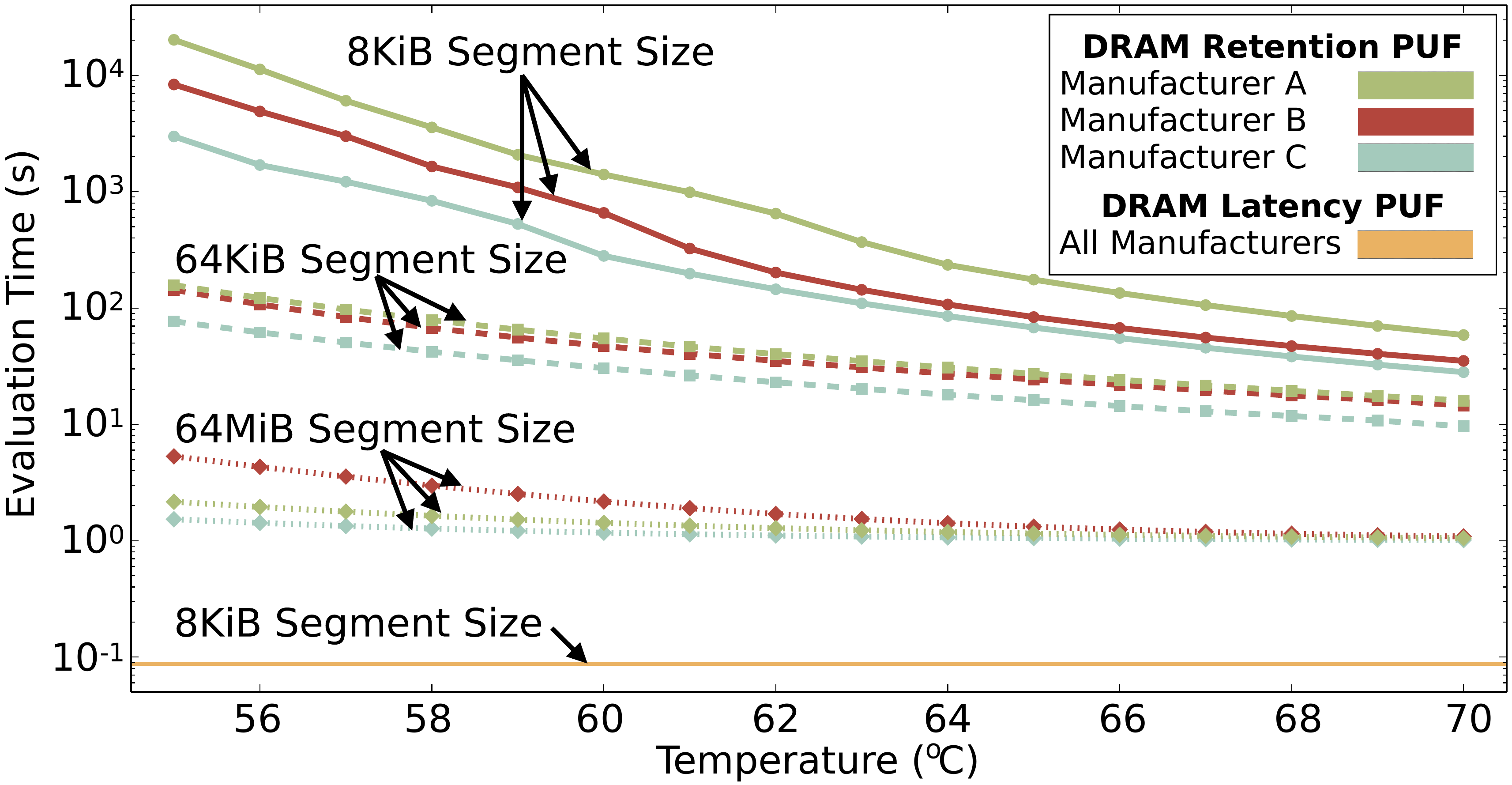} 
    \caption{Average DRAM retention PUF evaluation time vs. temperature
    shown for three selected memory segment sizes for each manufacturer.
    Average DRAM latency PUF evaluation time
    (Section~\ref{subsubsec:low_latency}) is shown as a comparison point.\vspace{5pt}}
    \label{fig:raw_runtime}
\end{figure}

To find the evaluation time of DRAM retention PUFs, we use a similar
methodology to prior works on DRAM retention PUFs, which disable DRAM refresh
and wait for at least 512 retention failures to accumulate across a memory
segment~\cite{sutar2016d, keller2014dynamic}. Figure~\ref{fig:raw_runtime}
shows the results of DRAM retention PUF evaluation times for three different
memory segment sizes (8KiB, 64KiB, 64MiB) across our testable DRAM
temperature range (i.e., 55$^{\circ}$C-70$^{\circ}$C). Results are shown for
the average across all tested chips from each manufacturer in order to isolate
manufacturer-specific variation~\cite{raidr, liu2013experimental,
khan2014efficacy, patel2017reaper}. Figure~\ref{fig:raw_runtime} also shows,
for comparison, the DRAM latency PUF evaluation time, which is experimentally
determined to be 88.2ms on average for \emph{any} DRAM device \jkfour{at}
\emph{all} operating temperatures (see
Section~\ref{subsubsec:low_latency}).

We find that at our maximum testing temperature of 70$^{\circ}$C, the
average DRAM retention PUF across all manufacturers can be evaluated on
\jkfour{average (minimum, maximum) in 40.6s (28.1s, 58.6s)} using an 8KiB
segment size. By increasing the memory segment size from 8KiB to 64KiB, we can
evaluate a DRAM retention PUF in \jkfour{13.4s (9.6s, 16.0s)}, and at 64MiB, in
\jkfour{1.05s (1.01s, 1.09s)}. However, at our lowest testable temperature
(i.e., 55$^{\circ}$C), DRAM retention PUF evaluation time increases to
\jkfour{2.9 \emph{hours} (49.7 minutes, 5.6 \emph{hours})} using an 8KiB
segment, \jkfour{125.8s (76.6s, 157.3s)} using a 64KiB segment, and
\jkfour{3.0s (1.5s, 5.3s)} using a 64MiB segment.\footnote{These evaluation
times are consistent with prior work on DRAM retention PUFs~\cite{sutar2016d,
keller2014dynamic, xiong2016run}, which find that evaluation times on the order
of minutes or longer are required to induce enough retention failures in a
128KiB memory segment to generate a PUF response at 20$^{\circ}$C.} 

A DRAM retention PUF evaluation time on the order \jkfour{of even seconds or
minutes} is \emph{prohibitively high} for \jkfour{at least three reasons: 1)
such high latency leads to very long application stall times and very high
system interference, 2) since DRAM refresh intervals can be modified only at a
rank/bank granularity, the memory controller must continuously issue
\emph{extra accesses}, during PUF evaluation, to each row inside the rank/bank
but \jkfive{outside of the PUF memory segment, which causes significant
bandwidth performance and energy overhead}, and 3) such a long evaluation time
allows ample opportunity for temperature to fluctuate, which would result in a
PUF response with low similarity to the golden key, and thus, an unreliable
PUF.} 

In general, DRAM retention PUF evaluation time increases with \emph{decreasing}
temperature. This is due to the temperature dependence of charge leakage in
DRAM cell capacitors, and is a \emph{fundamental limitation} of \jkfour{using
DRAM retention failures as a PUF mechanism.} Therefore, any devices operating
at common-case operating temperatures
(35$^{\circ}$C-55$^{\circ}$C)~\cite{lee2015adaptive, el2012temperature,
liu2011hardware} or below will have great difficulty adopting DRAM retention
PUFs for runtime accessibility.  In Sections~\ref{section:metrics_evaluation}
and~\ref{section:generating_reliable_lpuf}, we describe the DRAM latency PUF in
detail and show how it 1) provides a much lower evaluation time than the DRAM
retention PUF, and 2) enables a reliably short evaluation time across
\emph{all} operating temperatures.

\subsection{Optimizing Retention PUFs} 
\label{subsection:retention_time_optimization}

\jkfour{We explore if it is possible to make DRAM retention PUFs
runtime-accessible (i.e., significantly faster) at common-case operating
temperatures by increasing \jkfive{the rate at which retention failures are
induced}. Given that ambient (i.e., environmental) temperature is fixed, we can
increase the rate of induced retention failures in two ways: 1) using a larger
PUF memory segment in DRAM, or 2) accelerating the rate of charge leakage using
means other than increasing ambient temperature.} 

\textbf{Larger PUF memory segments.} Using a larger PUF memory segment results
in additional DRAM capacity overhead that does \emph{not} \jkfive{scale
favorably with decreasing} temperatures.  As shown in
Section~\ref{subsection:rpufgt}, the number of retention failures drops
exponentially with temperature, so the PUF memory segment size required to
compensate for the decreasing retention failure rate \emph{increases
exponentially}.  Our experimental analysis in Figure~\ref{fig:raw_runtime}
shows that at 55$^\circ$C, even using a PUF memory segment size on the order of
tens of megabytes, a DRAM retention PUF \emph{cannot} be evaluated in under 1
second.  Assuming the exponential growth factor of 0.23 for DRAM BER as a
function of temperature (found in Section~\ref{subsection:rpufgt}), a
corresponding PUF evaluation time of \textasciitilde1s at 20$^\circ$C would
require a PUF memory segment over a thousand times larger (i.e., hundreds of
gigabytes)\jkfour{.  Thus, it is not cost-effective (i.e., scalable) to
na{\"i}vely increase the PUF memory segment size.} 

\textbf{Accelerating charge leakage.} Accelerating charge leakage given
a fixed temperature can be done \jkfour{by either 1)} making hardware
modifications or 2) exploiting factors other than temperature that affect
charge leakage.  Unfortunately, as we discuss in this section, there is no easy
way to achieve these using commodity off-the-shelf (COTS) systems.

In-DRAM hardware modifications proposed in prior work can be leveraged to
increase the number of retention failures observed at a fixed ambient
temperature. For example, partial restoration of DRAM
cells~\cite{lee2015adaptive, zhang2016restore} can be used to prepare the PUF
memory segment with reduced charge levels in order to exacerbate the number of
retention failures observed with a given refresh interval. Similarly, other
mechanisms in prior work (e.g.,~\cite{shin2014nuat, hassan2017softmc}) can be
used to decrease DRAM retention PUF evaluation time at \jkfour{common-case}
temperatures where DRAM retention PUFs are otherwise infeasible. However, these
approaches require modifications in DRAM or the memory controller, and thus,
cannot be used in COTS DRAM.

System-level hardware modifications, such as adding a heating source local to
the DRAM chip~\cite{govindavajhala2003using}, could be used to exacerbate the
occurrence of retention failures at low ambient temperatures.  However, these
approaches require custom system architectures, which contradicts our goal of
designing a PUF for COTS systems. \jkfive{They may also open up system security and
reliability concerns.} 

Experimental studies on DRAM have shown that charge leakage rates are dependent
on factors such as supply voltage~\cite{chang_understanding2017}, data pattern
effects~\cite{liu2013experimental, lee2010mechanism, khan2016parbor,
khan2016case, khan2017detecting, patel2017reaper, khan2014efficacy}, and random
charge fluctuations known as \emph{variable retention time}
(\emph{VRT})~\cite{yaney1987meta, restle1992dram, liu2013experimental,
qureshi2015avatar, khan2014efficacy, patel2017reaper}. Analogously to
temperature control, any of these quantities could be intelligently manipulated
to exacerbate the number of retention failures observed.  Unfortunately, these
effects are either \emph{relatively weak} to significantly increase the number
of observed retention failures (e.g., data pattern dependence), require
\emph{system modifications} to implement (e.g., voltage
control~\cite{david2011memory, chang_understanding2017}), or are inherently
\emph{difficult to control} (e.g., VRT effects).

In order to reduce the number of extra row refresh operations necessary
to prevent data loss throughout retention PUF evaluation
(Section~\ref{subsection:genretpuf}), DRAM refresh optimizations proposed in
prior work~\cite{radar, raidr, liu2013experimental, lin2012secret,
nair2013archshield, cui2014dtail, baek2014refresh, wang2014proactivedram,
qureshi2015avatar, bhati2016dram, patel2017reaper} can be used to increase the
granularity of the refresh operation. While this approach could potentially
eliminate the extra refresh operations altogether, these mechanisms
come with their own hardware and runtime overheads that may diminish the benefits
of \emph{not} having to issue the extra refresh commands during PUF
evaluation. Many such mechanisms also require hardware modifications to either
DRAM chips or memory controllers or both. 

We conclude that there is no good known way to optimize DRAM retention PUF
evaluation time for COTS DRAM \jkfive{devices} today. While many approaches to
improve evaluation time exist, they are all impractical in COTS systems due
\jkfour{to 1) lack of applicability and scalability to common-case
temperatures, 2) need for DRAM modification, or 3) inherent difficulties in
control. This motivates the need for a runtime-accessible PUF that is suitable
across all temperature conditions and can be implemented on COTS DRAM devices
today.}

\section{DRAM Latency PUFs} 
\label{section:key_idea} 

Our goal is to develop a DRAM PUF that can be evaluated 1) with low latency and
low system interference across all operating temperatures, and 2) without any
modification to DRAM chips. To this end, we present the DRAM latency PUF, a new
class of DRAM PUFs \jkfive{with these} characteristics. In particular, a DRAM
latency PUF provides low evaluation time \jkfive{at a wide range of operating
temperatures (0$^{\circ}$C-70$^{\circ}$C), which includes common-case
temperatures (35$^{\circ}$C-55$^{\circ}$C)~\cite{lee2015adaptive,
el2012temperature, liu2011hardware}.} 

\textbf{Key Idea.} The key idea of the DRAM latency PUF is to 
provide unique device signatures using the error pattern resulting from
accessing DRAM with \emph{reduced} timing parameters. These \emph{latency
failures} are inherently related to chip-specific random process variation
introduced during manufacturing (Section~\ref{section:out_of_spec}), which allows
us to use the failures as unique identifiers for each DRAM chip. To evaluate a
DRAM latency PUF, we write known data into a fixed-size \emph{memory segment}
(e.g., 4 DRAM rows $\approx$ 8KiB in our LPDDR4 DRAM chips) and read
it back with reduced timing parameters. The resulting failures \jkfour{form a}
pattern of bits unique to the tested device. 

\textbf{Probabilistic Nature.} Inducing latency failures is a
stochastic process in which the probability of cell failure
is based on random variations in both the cell itself and any
peripheral circuitry used to access \jkfour{the cell}. This is due to the
probabilistic behavior of circuit elements when timing requirements are
violated. To find a repeatable set of latency-failure-prone DRAM
cells, each cell should be accessed \emph{multiple} times with reduced timing
parameters. In the case of reduced $t_{RCD}$, we require multiple
\emph{iterations} of reading each cell to accumulate a reliable set of latency
failures. Fortunately, as we show in Section~\ref{subsubsec:low_latency},
finding a reliable set of latency failures is a relatively fast process (i.e.,
it takes 88.2ms on average).

\textbf{Key Variables.} We identify \jkfour{three} key variables to optimize for when
designing the DRAM latency PUF. These variables define the tradeoffs between
the DRAM latency PUF's evaluation time and its effectiveness.

1) \emph{Memory segment ID}. \jkfive{DRAM} PUFs can be evaluated using memory
segments from different parts of DRAM. Each segment results in unique error
patterns and can therefore be used for \emph{different} challenge-response
pairs. In Section~\ref{subsection:among_memseg_variation}, we discuss how
variation in process manufacturing causes some chips to have fewer memory
segments (where fewer is worse) that are viable for \jkfive{DRAM latency PUF}
evaluation than others.

2) \emph{Memory segment size}. Larger memory segments allow more devices to be
uniquely identified at the cost of higher PUF evaluation time because more
memory accesses are required to induce latency failures across the memory
segment. \jksix{With an experimental analysis of memory segment size based on}
data from 223 real DRAM chips (Section~\ref{section:metrics_evaluation}), we
find that a memory segment size of 8KiB is sufficient to find enough latency
failures for an effective \jkfour{DRAM latency PUF}.

3) \emph{DRAM timing parameters}. Both using different timing parameters and
changing the amount of reduction in the chosen timing parameter result in
\emph{different} error patterns (Section~\ref{section:out_of_spec}).
This is because \jkfour{1) different timing parameters guard against different
underlying error mechanisms~\cite{chang_understanding2017, lee-sigmetrics2017,
lee2015adaptive}, and 2) different amounts of latency reduction exercise
different failure-prone bits~\cite{lee2015adaptive}. These two dimensions of
control add more degrees of freedom to the DRAM latency PUF}, further
increasing its diffuseness (Section~\ref{subsubsec:diffuseness}).

Throughout the rest of this section, we first demonstrate that the
DRAM latency PUF satisfies all requirements for 1) a reliable PUF
(Section~\ref{subsection:ideal_puf}) and 2) runtime-accessible PUF evaluation
(Section~\ref{subsection:ideal_runtime_puf}) across all temperatures. We focus
on $t_{RCD}$-induced DRAM read errors in this work, but DRAM latency PUFs also
work with any other timing parameter whose timing violation results in failures 
(e.g., $t_{RP}$, $t_{RAS}$, $t_{WR}$), thereby enabling a potentially larger
challenge-response space than obtained by using a single timing
parameter alone.


\subsection{PUF Characteristics: Experimental Analysis} 
\label{section:metrics_evaluation}

This section shows, with experimental results from 223 state-of-the-art LPDDR4
DRAM chips, that the DRAM latency PUF satisfies each of the five
characteristics of a desirable PUF discussed in
Section~\ref{subsection:ideal_puf}. 

\subsubsection{Diffuseness}\hspace{2pt}
\label{subsubsec:diffuseness}
Different memory segments within the same device result in different error
patterns~\cite{lee2015adaptive, lee-sigmetrics2017, chang_understanding2017,
chang2016understanding}. Given the large address space provided by modern DRAM,
\jkfive{different memory segments provide different challenge-response pairs.}
For example, our selected segment size of 8KiB
(Section~\ref{subsection:among_memseg_variation}) in a 2GiB DRAM, offers up to
256K ($\frac{2GiB}{8KiB}$) different \jkfive{challenge-response pairs}, which
is on the same order of magnitude as prior DRAM PUFs~\cite{xiong2016run,
sutar2016d}. 

\subsubsection{Uniqueness and Uniform Randomness}\hspace{2pt}
\label{subsubsec:uniqueness} 
To show the uniqueness and uniform randomness of DRAM latency PUFs evaluated
across different memory segments, we study a large number of different memory
segments from each of \jkfive{our} 223 LPDDR4 DRAM chips (as specified in
Table~\ref{Tab:mem_segs}). 
\begin{table}[h!]
\footnotesize
\begin{center}
\begin{tabular}{ c|c|c| } 
\cline{2-3}
 & \#Chips & \#Tested Memory Segments \\ 
\cline{2-3}
A & 91 & 17,408 \\
B & 65 & 12,544 \\
C & 67 & 10,580 \\ 
\cline{2-3}
\end{tabular} 
\caption{The number of tested PUF memory segments across the tested chips from each of the three manufacturers.} 
\label{Tab:mem_segs} 
\end{center} 
\end{table} 
\vspace{-8pt}

For each memory segment, we evaluate the PUF 50 times at 70$^{\circ}$C. To
measure the uniqueness of a PUF, we use the notion of a \emph{Jaccard
index}~\cite{jaccard1901etude}, as suggested by prior work~\cite{xiong2016run,
schaller2017intrinsic, aysu2017new}. \jkfour{We use the Jaccard index to
measure the similarity of two PUF responses. The Jaccard index is} determined
by taking the two sets of latency failures ($s_1$, $s_2$) from two \jkfour{PUF
responses} and computing the ratio \jkfive{of the size of} \jkfive{the shared
set of failures over the total number of unique errors} in the two sets
$\frac{|s_1 \cap s_2|}{|s_1 \cup s_2|}$. \jkfour{A Jaccard index value closer
to 1 indicates a high similarity between the two PUF responses, and a value
closer to 0 indicates uniqueness \jkfive{of the two}. Thus, a unique PUF should
have Jaccard index values close to 0 across all pairs of \emph{distinct} memory
segments.} 

We choose to employ the Jaccard index instead of the \emph{Hamming
distance}~\cite{hamming1950error} as our metric for evaluating the
\jkfour{similarity between PUF responses because the Jaccard index places} a
heavier emphasis on the differences between two large bitfields. This is
especially true in the case of devices that exhibit inherently lower failure 
rates. In the case of Hamming distance, calculating similarity between two PUF
responses depends heavily on the number of failures found, and we find this to be
an unfair comparison due to the large variance in the number of failures across
distinct memory segments.  For example, consider the case where two memory
segments each generate PUF responses consisting of a single failure in
different locations of a bitfield comprised of 100 cells. The Hamming distance
between these PUF responses would be 1, which could be mistaken for a match,
but the Jaccard index would be calculated as a 0, which would guarantee a
mismatch.  Because we are more interested in the locations \emph{with} failures
than without, we use the Jaccard index, which discounts locations without
failures.  Throughout the rest of this chapter, \jkfour{we use the terms 1)
\emph{Intra-Jaccard}~\cite{xiong2016run, schaller2017intrinsic} to refer to the
Jaccard index of two PUF responses from the \emph{same} memory segment and
2) \emph{Inter-Jaccard}~\cite{xiong2016run, schaller2017intrinsic} to refer to
the Jaccard index of two PUF responses from \emph{different} memory segments.}

A PUF must exhibit uniqueness and uniform randomness across any memory
segment from any device from any manufacturer. To show that these
characteristics hold for the DRAM latency PUF, we ensure that the
distribution of Inter-Jaccard indices are distributed near 0. This demonstrates
that 1) the error patterns are unique such that no two distinct memory
segments would generate PUF responses with high similarity, and 2) the error
patterns are distributed uniform randomly across the DRAM \jkfive{chip(s) such
that the likelihood of two chips (or two memory segments)} generating the same
error pattern is exceedingly low.

\begin{figure}[h]
    \centering
    \includegraphics[width=0.7\linewidth]{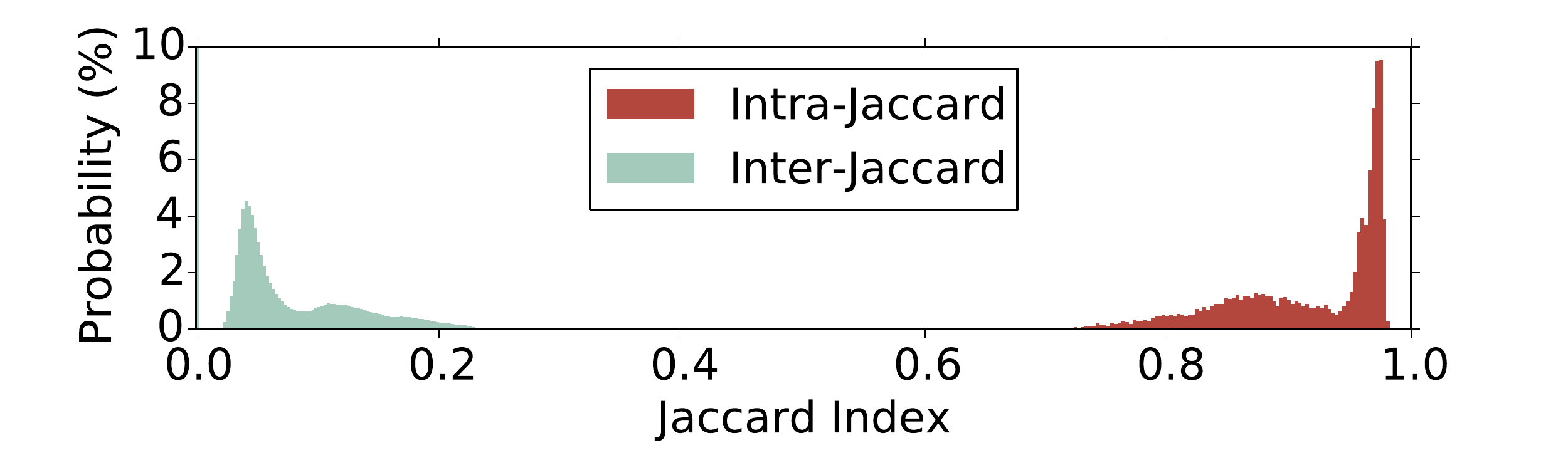} 
    \caption{Distributions of Jaccard indices calculated across every possible pair of PUF responses across all tested PUF memory segments from each of 223 LPDDR4 DRAM chips.\vspace{5pt}}
    \label{fig:key_idea_uniqueness}
\end{figure}

Figure~\ref{fig:key_idea_uniqueness} plots, in blue, the distribution of
Inter-Jaccard indices calculated between \emph{all possible pairs} of PUF
responses generated at the same operating temperature (70$^{\circ}$C) from
all tested memory segments across \jkfour{all} chips from three manufacturers.
The distribution of the Intra-Jaccard indices are also shown in red (discussed
later in this section).  The x-axis shows the Jaccard indices and the y-axis
marks the probability of any pair of memory segments (either within the same
device or across two different devices) resulting in the Jaccard index
indicated by the x-axis. We observe that the distribution of the Inter-Jaccard
indices is multimodal, \jkfour{but the Inter-Jaccard index \emph{always} remains below
0.25 for \emph{any pair} of distinct memory segments. This means that PUFs from
different memory segments have low similarity.  Thus,} we conclude that
latency-related error patterns approximate the behavior of a desirable PUF with
regard to both uniqueness and uniform randomness.

\begin{figure}[h]
    \centering
    \includegraphics[width=0.7\linewidth]{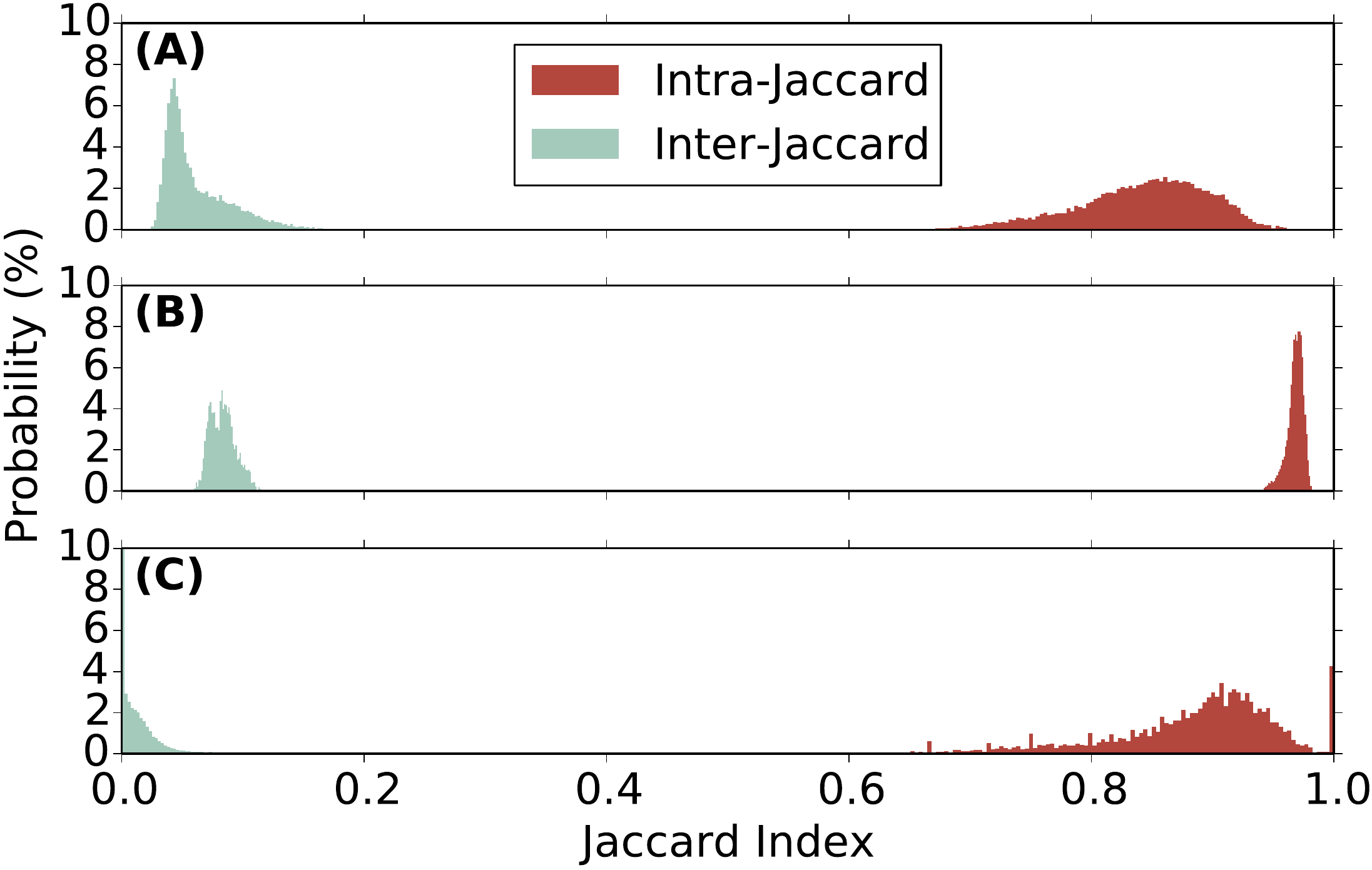} 
    \caption{\jkfive{Distributions of Jaccard indices calculated between PUF responses of DRAM chips from a single manufacturer.\vspace{5pt}}}
    \label{fig:uniqueness_jaccards_vendor_split} 
\end{figure}

To understand manufacturer-related effects,
Figure~\ref{fig:uniqueness_jaccards_vendor_split} separately plots the Intra-
and Inter-Jaccard distributions of PUF responses from chips of a
\emph{single} manufacturer in subplots.  Each subplot indicates the manufacturer
encoding in the top left corner (A, B, C). From these per-manufacturer
distributions, we make three major observations: 1) Inter-Jaccard values are
quite low, per-manufacturer, which shows uniqueness and uniform randomness, 2)
there is variation across manufacturers, as expected, and 3)
Figure~\ref{fig:key_idea_uniqueness}'s multimodal behavior for Inter- and
Intra-Jaccard index distributions can be explained by the mixture of
per-manufacturer distributions. We also find that the distribution of
Inter-Jaccard indices calculated between two PUF responses from chips of
distinct manufacturers are tightly distributed close to 0 (not shown).

\subsubsection{Unclonability} \hspace{2pt}
\label{subsubsec:unclonability} 
We attribute the probabilistic behavior of latency failures to physical
variation inherent to the chip (discussed in
Section~\ref{section:out_of_spec}). Chips of the same design contain physical
differences due to manufacturing process variation which \jkfive{occurs as a
result of} imperfections in manufacturing~\cite{lee2015adaptive,
lee-sigmetrics2017, chang_understanding2017, chang2016understanding,
chang2017thesis, lee2016reducing, kim2014flipping}. The exact physical
variations are inherent to each individual chip, as shown by previous
work~\cite{lee2015adaptive, lee-sigmetrics2017, chang_understanding2017,
chang2016understanding, chang2017thesis, lee2016reducing, kim2014flipping} and
confirmed by our experiments (not shown), and the pattern of variations is very
difficult to replicate as it is created entirely unintentionally.

\subsubsection{Repeatability}\hspace{2pt}
\label{subsubsec:repeatability} 
To demonstrate that the DRAM latency PUF exhibits repeatability, we show how
well a PUF memory segment can result in the \emph{same} PUF response 1) at
different times or 2) under different operating temperatures. For each of many
different memory segments, we evaluate a PUF multiple times and calculate all
possible \emph{Intra-Jaccard} indices (i.e., Jaccard indices between two PUF
responses generated from the \emph{same} exact memory segment). Because a
highly-repeatable PUF generates very similar PUF responses during each
evaluation, we expect the Intra-Jaccard indices between PUF responses of a
highly-repeatable PUF to be tightly distributed near a value of 1.
Figure~\ref{fig:key_idea_uniqueness} plots the distribution of Intra-Jaccard
indices across every PUF memory segment we tested in red. We observe that while
the distribution is multimodal, the Intra-Jaccard \jkfour{indices are clustered
very close to 1.0 and} \emph{never} drop below 0.65.

Similarly to the Inter-Jaccard index distributions (discussed in
Section~\ref{subsubsec:uniqueness}), we find that the different modes of the
Intra-Jaccard index distribution shown in Figure~\ref{fig:key_idea_uniqueness}
arise from combining the Intra-Jaccard index distributions from all three
manufacturers. We plot the Intra-Jaccard index distributions for each
manufacturer alone in Figure~\ref{fig:uniqueness_jaccards_vendor_split} as
indicated by (A),(B), and (C).  We observe from the higher distribution mean of
Intra-Jaccard indices in Figure~\ref{fig:uniqueness_jaccards_vendor_split}
for manufacturer B that DRAM latency PUFs evaluated on chips from
manufacturer B exhibit higher repeatability than those from manufacturers A or
C. We conclude from the high Intra-Jaccard indices in
Figures~\ref{fig:key_idea_uniqueness}
and~\ref{fig:uniqueness_jaccards_vendor_split}, that DRAM latency PUFs exhibit
high repeatability.

\textbf{Long-term Repeatability.} We next study the repeatability of DRAM
latency PUFs on a subset of chips over a 30-day period to show that the
repeatability property holds for longer periods of time (i.e., a memory segment
generates a PUF response similar to its previously-enrolled golden key
irrespective of the time since its enrollment). We examine a total of more than
a million 8KiB memory segments \emph{across} many chips from each of the three
manufacturers as shown in Table~\ref{Tab:mem_segs_long_term}. \jkfive{The right
column} indicates the number of memory segments across $n$ devices, where $n$
is indicated \jkfive{in the left column}, and the rows indicate the different
manufacturers of the chips containing the memory segments. 
\begin{table}[h!]
\footnotesize
\begin{center}
\begin{tabular}{ c|c|c|}
\cline{2-3}
 & \#Chips & \#Total Memory Segments \\
\cline{2-3}
A & 19 & 589,824\\
B & 12 & 442,879\\
C & 14 & 437,990\\
\cline{2-3}
\end{tabular} 
\caption{\jkfour{Number} of PUF memory segments tested for 30 days.} 
\label{Tab:mem_segs_long_term} 
\end{center} 
\end{table} 
\vspace{-8pt}

In order to demonstrate the repeatability of evaluating a DRAM latency PUF over
long periods of time, we continuously evaluate our DRAM latency PUF across a
30-day period using each of our chosen memory segments.  For each memory
segment, we calculate the Intra-Jaccard index between the first PUF response
and each subsequent PUF response.  We find the \emph{Intra-Jaccard index range}, or
the range of values ($max\_value~-~min\_value$) found across the Jaccard
indices calculated for every pair of PUF responses from a memory segment. If a
memory segment exhibits a low \jkfour{Intra-Jaccard index range}, the memory
segment generates highly-similar PUF responses during each evaluation over our
testing period.  Thus, memory segments that exhibit low Intra-Jaccard index ranges
demonstrate high repeatability. 
    
\begin{figure}[h]
    \centering
    \includegraphics[width=0.7\linewidth]{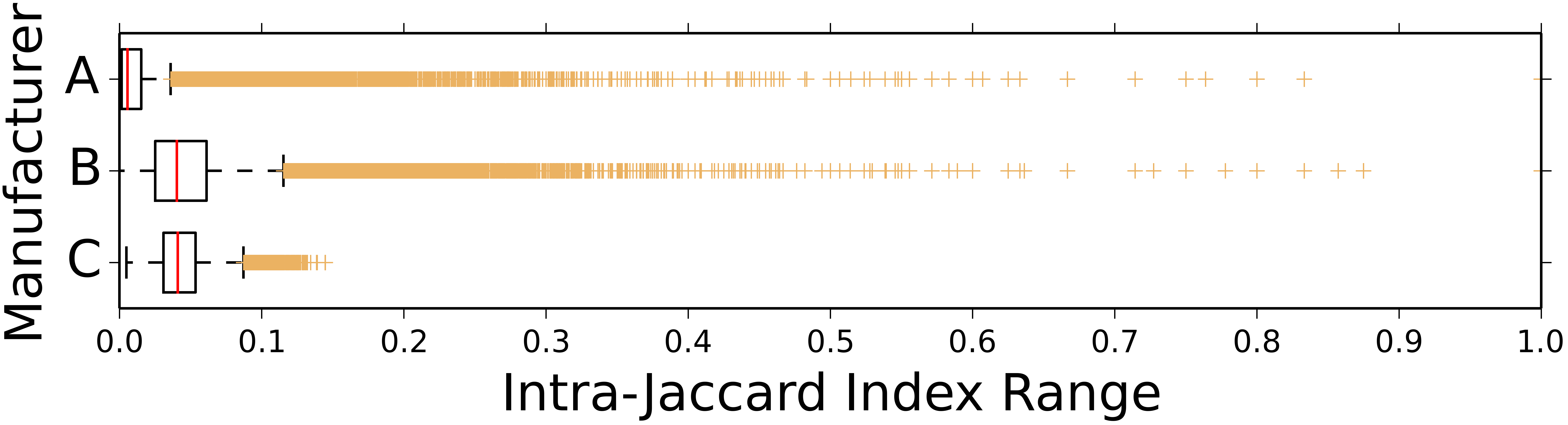} 
    \caption{Distribution of the Intra-Jaccard index range values calculated between many PUF responses that a PUF memory segment generates over a 30-day period.\vspace{5pt}}
    \label{fig:key_idea_time} 
\end{figure}

Figure~\ref{fig:key_idea_time} shows the distribution of \emph{Intra-Jaccard
index ranges} across our memory segments with box-and-whisker
plots\footnote{The box is bounded by the first quartile (i.e., the median of
the first half of the ordered set of Intra-Jaccard index ranges) and third
quartile (i.e., the median of the second half of the ordered set of
Intra-Jaccard index ranges). The median is marked by a red line within the
bounding box. The \emph {inter-quartile range} (IQR) is defined as the
difference between the third and first quartiles. The whiskers are drawn out to
extend an additional $1.5 \times IQR$ above the third quartile and $1.5 \times
IQR$ below the first quartile. Outliers are shown as orange crosses indicating
data points outside of the range of whiskers.} for each of the three
manufacturers. We observe that the Intra-Jaccard index ranges are quite low,
i.e., less than 0.1 on average for all manufacturers. Thus, we conclude that
the vast majority of memory segments across all manufacturers exhibit very high
repeatability over long periods of time.

In order to show that every chip has a significant proportion of memory
segments that exhibit high reliability over time, we analyze per-chip
Intra-Jaccard index range properties.
Table~\ref{Tab:mem_segs_long_term_per_manufacturer} shows the
\jkfive{\emph{median [minimum, maximum]} of the fraction} of memory segments
per chip that are observed to have Intra-Jaccard index ranges below 0.1 and
0.2. Over 90\% of all segments \emph{in each chip} are suitable for PUF
evaluation for Intra-Jaccard index ranges below 0.1, and over 97\% for
Intra-Jaccard index ranges below 0.2.  This means that each chip has a
significant number of memory segments that are viable for DRAM latency PUF
evaluation.  Furthermore, the distributions are very narrow, which indicates
that different chips show similar behavior. We conclude that every chip has a
significant number of PUF memory segments that exhibit high repeatability
across time. We show in Section~\ref{subsec:char_and_enroll} how we can use a
simple characterization step to identify these viable memory segments quickly
and reliably.

\begin{table}[h!]
\footnotesize
\begin{center}
\begin{tabular}{ c|c|c| } 
\cline{2-3} 
& \multicolumn{2}{c|}{\%Memory Segments per Chip} \\ 
\cline{2-3}
 & Intra-Jaccard index range <0.1 & Intra-Jaccard index range <0.2 \\
\cline{2-3}
A & 100.00 [99.08, 100.00] & 100.00 [100.00, 100.00] \\ 
B & 90.39 [82.13, 99.96] & 96.34 [95.37, 100.00] \\
C & 95.74 [89.20, 100.00] & 96.65 [95.48, 100.00] \\

\cline{2-3}
\end{tabular} 
\caption{Percentage of PUF memory segments per chip with Intra-Jaccard index \jkfive{ranges <0.1 or 0.2 over a 30-day period. Median [minimum, maximum] values are shown.}} 
\label{Tab:mem_segs_long_term_per_manufacturer} 
\end{center} 
\end{table} 
\vspace{-8pt}

\begin{figure}[h]
    \centering
    \includegraphics[width=0.7\linewidth]{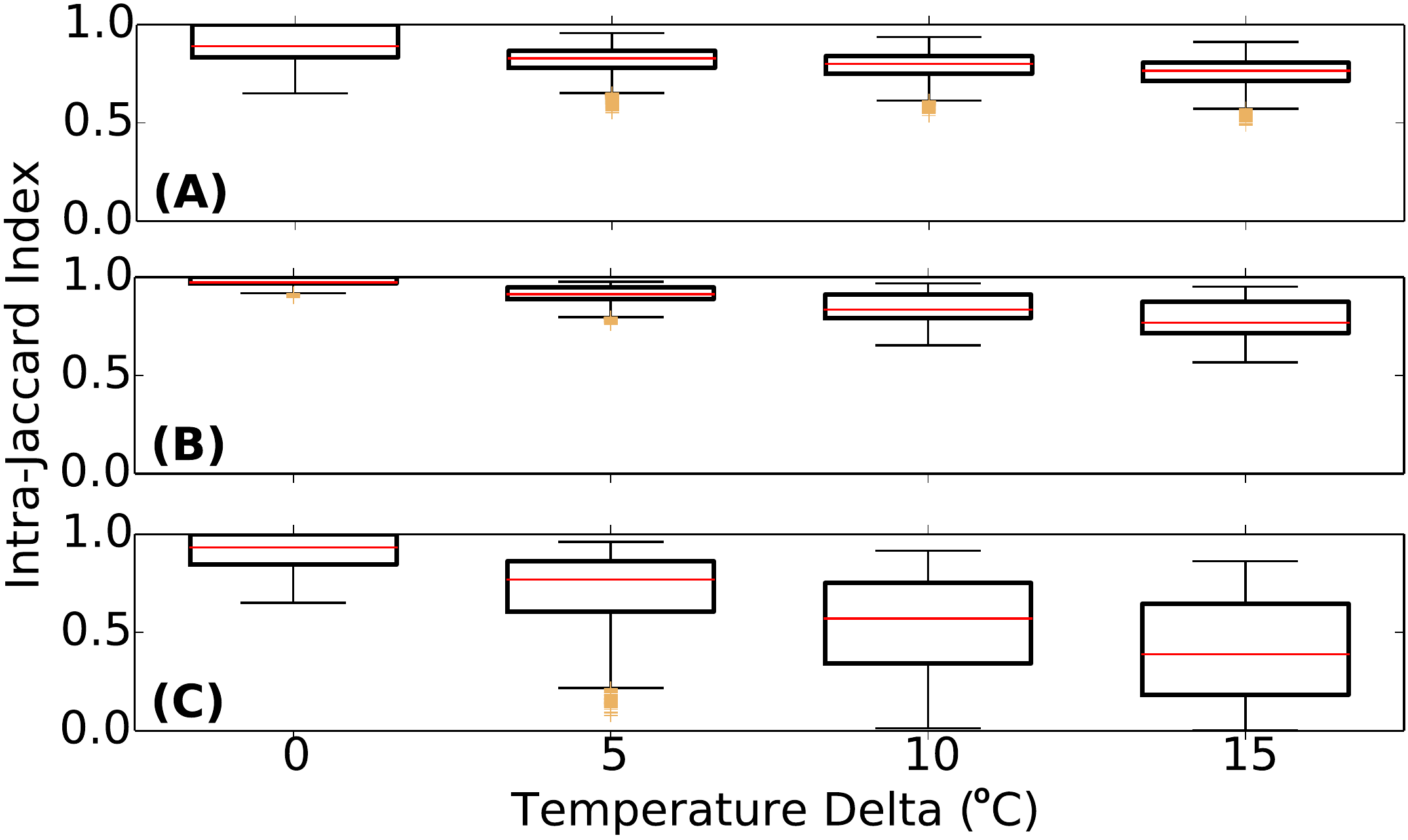} 
    \caption{\jksix{DRAM latency PUF repeatability vs. temperature.}} 
    \label{fig:key_idea_repeatability_temperature} 
\end{figure}

\textbf{Temperature Effects.} To demonstrate how changes in
temperature affect PUF evaluation, we evaluate \jkfive{the DRAM latency} PUF 10 times for each
of the memory segments in Table~\ref{Tab:mem_segs_long_term} at each
5$^{\circ}$C increment throughout our testable temperature range
(55$^{\circ}$C-70$^{\circ}$C).
Figure~\ref{fig:key_idea_repeatability_temperature} shows the distributions of
Intra-Jaccard indices calculated between every possible pair of PUF responses
generated by the \emph{same} memory segment. The deltas between the operating
temperatures at the time of PUF evaluation are denoted in the x-axis
(\emph{temperature delta}). Since we test at four evenly-spaced temperatures,
we have four distinct temperature deltas. The y-axis marks the Jaccard indices
calculated between the PUF responses. The distribution of Intra-Jaccard
indices found for a given temperature delta is shown using a box-and-whisker
plot. 

Figure~\ref{fig:key_idea_repeatability_temperature} subdivides the
distributions for each of the three manufacturers as indicated by A, B, and C.
\jkfive{Two observations are in order. 1)} Across all three manufacturers, the
distribution of Intra-Jaccard indices strictly shifts towards zero as the
temperature delta increases. 2) The Intra-Jaccard distribution of PUF
responses from chips of manufacturer C are the most sensitive to changes in
temperature as reflected in the large distribution shift in
Figure~\ref{fig:key_idea_repeatability_temperature}(C). \jkfive{Both
observations show} that evaluating a PUF at a temperature different from the
temperature during enrollment affects the quality of the PUF response and
reduces repeatability.  However, 1) for small temperature deltas \jkfive{(e.g.,
5$^{\circ}$)}, PUF repeatability is not significantly affected, and 2) we
discuss in Section~\ref{subsec:char_and_enroll} how we can ameliorate this
effect during device enrollment.


\subsection{Runtime-Accessible PUF Metrics Evaluation}
\label{section:runtime_metrics_evaluation}

Throughout the remainder of this section, we show 1) how the DRAM latency PUF
satisfies the characteristics of a \emph{runtime-accessible} PUF (i.e., low
latency and low system interference) discussed in
Section~\ref{subsection:ideal_runtime_puf}, and 2) that the DRAM latency PUF
significantly outperforms the DRAM retention PUF \jkfive{in terms of both
evaluation time and system interference.} 

\subsubsection{Low Latency}\hspace{2pt}
\label{subsubsec:low_latency} 
The DRAM latency PUF consists of two key phases: 1) inducing latency
failures, and 2) filtering the PUF segment, which improves PUF repeatability
(to be discussed in Section~\ref{subsection:failure_repeatability}).
During Phase~1, we induce latency failures multiple times (i.e., for
multiple \emph{iterations}) over the PUF memory segment and count the
failures in a separate buffer for additional bookkeeping (we discuss this in
further detail in Section~\ref{section:generating_reliable_lpuf}). The
execution time of this phase depends directly on three factors:
\begin{enumerate}
    \itemsep0em

    \item The value of the $t_{RCD}$ timing parameter. \jkfour{A smaller $t_{RCD}$
    value causes each read to have a shorter latency.} 

    \item The size of the PUF memory segment. A larger memory segment requires
    more DRAM read requests per iteration. In our devices, we observe
    that latency failures are induced at a granularity of 32 bytes with each
    read request, so we can find the total number of required DRAM reads by 
    dividing the size of the memory segment by 32 bytes.

    \item The number of iterations used to induce latency failures. \jksix{More
    iterations lead to a longer evaluation time. } 
\end{enumerate} 
Increasing any one of these factors independently of the others directly
results in an increase in PUF evaluation time. We experimentally find that
\jkfive{a single} low-$t_{RCD}$ access to DRAM, along with its associated
bookkeeping and \jkfour{memory barrier}, takes 3.4$\mu$s. Because the value of
$t_{RCD}$ is on the scale of tens of nanoseconds~\cite{2014lpddr4}, changing its
value negligibly affects the time for each low-$t_{RCD}$ access. \jkfour{Thus,
we use a constant 3.4$\mu$s for each read regardless of the $t_{RCD}$ value to
find a} good estimate of the PUF evaluation time in
Equation~\ref{eq:eval_time}.  We experimentally show that Phase~2 has
negligible runtime ($<0.1\%$ \jkfive{of total DRAM latency PUF evaluation
time}) compared with Phase~1, so we omit Phase~2 in our PUF evaluation time
estimation.\jksix{We express PUF evaluation time estimation as:} 
\vspace{-5pt}
\begin{equation} \label{eq:eval_time} 
T_{PUF\_eval} = (N_{iters}) \times [(size_{mem\_seg}) / (32~\mathrm{bytes})] \times 3.4\mu s
\vspace{-5pt}
\end{equation} 
where $N_{iters}$ is the number of times we induce latency failures on each 32
byte block of the memory segment, and $size_{mem\_seg}$ is the size
of the memory segment used to evaluate the PUF. For our final chosen
configuration (discussed in detail in Section~\ref{impl_challenges}), we use
the parameters $size_{mem\_seg}=8$KiB
(Section~\ref{subsection:among_memseg_variation}), $t_{RCD}=9.8$ns
(Section~\ref{subsec:changing_timing}), and $N_{iters}=100$
(Section~\ref{subsection:failure_repeatability}). Using
Equation~\ref{eq:eval_time}, we expect this configuration to result in an
evaluation time of approximately 87ms.

In order to \jkfour{experimentally verify Equation~\ref{eq:eval_time}, we measure the
evaluation time of the DRAM latency PUF for 10000 evaluations across chips from} 
all three manufacturers at 55$^{\circ}$C. We find that evaluation times are
normally distributed per-manufacturer according to
$\mathcal{N}_A(\mu=89.1\mathrm{ms},~\sigma=0.0132\mathrm{ms})$,
$\mathcal{N}_B(\mu=88.2\mathrm{ms},~\sigma=0.0135\mathrm{ms})$, and
$\mathcal{N}_C(\mu=87.2\mathrm{ms},~\sigma=0.0102\mathrm{ms})$. These
distribution parameters show that evaluation times have very similar means and
are extremely tightly distributed (i.e., $<0.0002$ relative standard
deviation). This is expected \jkfive{because}, for any particular
configuration, DRAM latency PUF evaluation essentially \jkfive{requires} a
\emph{constant} number of DRAM accesses. Therefore, any variation in PUF
evaluation time comes from variations in code execution (e.g., multitasking,
interrupts, \jkfive{DRAM refresh}, etc.) rather than any characteristics of the
PUF itself. In order to compare these runtime distributions with the result of
Equation~\ref{eq:eval_time}, we take take the mean of the mixture distribution
of the three per-manufacturer distributions (i.e.,
$\mathcal{N}_{ABC}(\mu=88.2\mathrm{ms},~\sigma=0.716\mathrm{ms})$) and find
that the 87ms estimate from Equation~\ref{eq:eval_time} results in only 1.4\%
error.

Figure~\ref{fig:raw_runtime} provides a comparison of DRAM latency PUF
evaluation time with retention PUF evaluation time across our testable
temperature range (i.e., 55$^{\circ}$C-70$^{\circ}$C). We find that the DRAM
latency PUF significantly outperforms the DRAM retention PUF for an
equivalent DRAM capacity overhead of 64KiB (i.e., 8KiB latency PUF memory
segment + 56KiB counter buffer), providing an average (minimum, maximum)
speedup of 152x (109x, 181x) at 70$^{\circ}$C and 1426x (868x, 1783x) at
55$^{\circ}$C. By increasing the memory segment size from 64KiB to
64MiB, we can evaluate a DRAM retention PUF in 1.05s (1.01s, 1.09s) at
70$^{\circ}$C (Section~\ref{subsection:retention_time_optimization}). However,
the DRAM latency PUF still outperforms this configuration \emph{without} an
increase in DRAM capacity overhead \jkfour{(i.e., still with an 8KiB memory
segment),} providing a speedup of 12.1x (11.6x, 12.5x).

Similarly to prior work on DRAM latency reduction~\cite{chang2016understanding,
lee2015adaptive}, we experimentally find that inducing latency failures is
minimally affected by changes in temperature. Importantly, since our method of
inducing latency failures does \emph{not} change with
temperature~(Section~\ref{section:generating_reliable_lpuf}), DRAM latency PUF
evaluation time remains reliably short across \emph{all} operating
temperatures. We conclude that the DRAM latency PUF 1) \jkfour{can be evaluated
at speeds that are orders of magnitude faster than the DRAM retention PUF, and
2) overcomes the temperature dependence of the DRAM retention PUF} and
maintains a low evaluation time across all temperatures.





\subsubsection{Low System Interference} \hspace{2pt}
\label{subsubsec:system_interference} 
The DRAM latency PUF exhibits two major sources of system
interference: 1) requiring exclusive DRAM \jkfour{rank/bank} access throughout
PUF evaluation, and 2) using a region in a separate DRAM rank to count 
latency failures (Section~\ref{section:generating_reliable_lpuf}).

\jkfour{First, because DRAM timing parameters can only be manipulated for the
coarse granularity of a DRAM rank, any other access to the same rank containing
the PUF memory segment must be blocked during PUF evaluation. Such blocking
prevents other accesses from obeying the same reduced timing parameters and
corrupting the data. For this reason, DRAM latency PUF evaluation requires
exclusive access to a full DRAM rank for the entire duration of PUF evaluation.
Fortunately, the DRAM latency PUF's quick evaluation time (i.e., 88.2ms on
average) guarantees that the DRAM rank will be unavailable only for a short
period of time. This is \jkfive{in stark} contrast with the DRAM retention PUF,
which 1) blocks rank/bank access for much longer periods of time (e.g., on the
order of \emph{minutes} \jkfive{or seconds}), and 2) requires the memory
controller to issue a large number of refresh operations to rows in the
rank/bank outside of the PUF memory segment \jkfive{for the same period of
time}~\cite{xiong2016run}.} 

Second, the DRAM latency PUF algorithm (described in detail in
Section~\ref{section:generating_reliable_lpuf}) requires a \jkfour{small
\emph{counter buffer} (e.g., a 56KiB buffer for an 8KiB PUF memory segment)
which stores counters for each bit of the PUF memory segment. This comes at the
cost of both DRAM capacity overhead and additional memory traffic penalty.}
However, given that the DRAM capacity overhead is small (e.g., <0.003\% for a
2GB DRAM using an 8KiB memory segment) and the additional bandwidth consumed is
extremely low (e.g., on the order of 100MB/s using an 8KiB memory segment) in
the context of total DRAM bandwidth \jkfour{(e.g., 8GB/s),} we conclude that
the additional system interference induced by the counter buffer is
insignificant.  In practice, we expect system caches to (fully) hold
the counter buffer, further reducing the required DRAM bandwidth.

\section{Design Considerations} 
\label{impl_challenges} 

As we experimentally showed in Section~\ref{section:key_idea}, utilizing
DRAM latency failures is a viable method for evaluating runtime-accessible PUFs in
commodity, unmodified DRAM chips. However, due to variation across DRAM cells
and chips, there are \jkfive{various important} design considerations that must
be made in the implementation of the DRAM latency PUF. In this section, we
discuss these considerations for implementing the DRAM latency PUF. 

\subsection{Repeatability of Cell Latency Failures} 
\label{subsection:failure_repeatability} 

Due to many underlying factors (e.g., process variation, temperature), each
DRAM cell fails with a different probability when read with a timing parameter
reduced beyond the manufacturer specification~\cite{lee-sigmetrics2017,
lee2015adaptive, chang2016understanding, lee2016reducing, chang2017thesis}.  We
define a \emph{latency-weak cell} as a cell that has a significant probability
of failure when read with a reduced timing parameter. Our DRAM latency PUFs are
comprised of the locations of latency-weak cells because such cells can be
repeatably found. In order to repeatably find the set of latency-weak cells, we
employ \emph{many} iterations (e.g., on the order of 100) of inducing latency
failures at the PUF memory segment. This improves the chances of a PUF
evaluation to find a significant proportion of the latency-weak cells.  Because
we assume that any given cell has a static probability \mpseven{($p$)} to fail
when accessed with reduced latency, we can model the number of \mpseven{times}
that a cell must be accessed before observing a latency failure as a
\emph{geometric random variable} \mpseven{with a success probability of $p$.
The geometric distribution with parameter $p$ has a mean value of
$\frac{1}{p}$. By sampling cells over $x$ iterations during DRAM latency PUF
evaluation, we expect to find all cells that fail with a probability greater
than or equal to $\frac{1}{x}$}.  There is a chance that a cell with a failure
probability below the threshold fails during an instance of the PUF evaluation,
reducing the similarity of the PUF responses across evaluations and thus the
repeatability of the PUF. To mitigate this issue, we apply a \emph{filter} (see
Section~\ref{section:generating_reliable_lpuf}) that removes cells that we
observe to fail in only a small proportion of the $x$ iterations. We
empirically find that removing cells that fail in less than 10\% of the
iterations results in the \jkfive{highest Intra-Jaccard indices across PUF
responses.} 

In order to determine how many iterations to induce latency failures for during
latency PUF evaluation, we generate PUF responses across our devices using a
varying number of iterations between 1 and 1024. For each set of PUF responses
generated with a given number of iterations, we calculate the box-and-whisker
plots for both Inter- and Intra-Jaccard distributions (not shown). We find that
for PUF responses from chips across all manufacturers, the Inter-Jaccard
\emph{and} Intra-Jaccard distributions have strictly the same or increasing
medians, \jkfive{first and third quartiles}, and whiskers, for an
increasing number of iterations. 

Higher Intra-Jaccard index distribution values represent a more repeatable PUF
since the distribution directly reflects the similarities of PUF responses from
the same memory segment.  We find that the Intra-Jaccard index distribution's
median and bottom whiskers increase by 0.0025 and 0.0054, respectively, for
every doubling of the number of iterations. On the other hand, higher
Inter-Jaccard index distribution values represent higher similarity across
distinct memory segments. Such higher values would limit the PUF's ability to
identify many unique devices.  We find that the Inter-Jaccard index
distribution's median and top whiskers increase by 0.0012 and 0.0011,
respectively, for every doubling of the number of iterations. \jkfive{Based on
our experimental analyses of these tradeoffs,} we choose to induce latency
failures for 100 iterations during each DRAM latency PUF evaluation.  We next
discuss in detail \jkfive{our algorithm for} evaluating DRAM latency PUFs with
high repeatability. 

\subsection{DRAM Latency PUF Evaluation Algorithm}
\label{section:generating_reliable_lpuf}

We provide an implementable algorithm for evaluating a repeatable DRAM latency
PUF at a given memory segment.  While we focus on evaluating DRAM latency PUFs
with $t_{RCD}$-induced failures, Algorithm~\ref{alg:lat_puf} works with any other
timing parameter capable of inducing failures. We first initialize the
PUF memory segment indicated by \jkfive{$Segment[seg\_id]$} by setting every
bit \jksix{in the memory segment} to ``1'' (line~2). We then attempt to find
the reliable set of failures as fast as possible in the memory segment
(lines~4-11). Because DRAM RD commands require a $t_{RCD}$ delay only after the
activation of a previously closed DRAM row, $t_{RCD}$ failures can only be
observed when issuing a read request to a \emph{closed} DRAM row. The key idea
is to iterate over each row sequentially such that each read request goes to a
different row (i.e., perform column order accesses through the memory segment
of interest) as shown in lines 7-9.  Before inducing failures across the PUF
memory segment, we must first obtain exclusive access to the rank containing
the PUF memory segment (line~4)\jkfive{, due to the rank-level granularity of
changing DRAM timing parameters (Section~\ref{section:out_of_spec}). We} then
reduce the value of $t_{RCD}$ for the entire rank containing the PUF memory
segment (line~5).  During the iterations of inducing $t{RCD}$ failures (lines
6-11), we issue a memory barrier (line~10) after each read. This ensures that
1) \emph{only one} memory instruction is in flight at a given time and, thus,
improves repeatability by simplifying the logic required by the memory
controller when issuing memory accesses, and 2) read requests do \emph{not} get
reordered by the memory controller to exploit row buffer
locality~\cite{rixner2000memory, rixner2004memory, mutlu2007stall,
kim2010atlas, mutlu2008parallelism, subramanian2016bliss, usui2016dash, kim2010thread}.
Instead, each access activates a new row, while obeying the $t_{RCD}$ timing
parameter.  We find that the instruction order indicated by lines 6-11 is the
fastest method for finding a reliable set of latency failures in a memory
segment. For every read, the $t_{RCD}$ failure locations are determined and
their failures are counted in a separate rank for bookkeeping (line~11). After
all iterations of inducing $t_{RCD}$ failures, we must reset the $t_{RCD}$
value to the default (line~12), filter the PUF segment (line~13; see
\emph{Filtering Mechanism}), release exclusive access to the rank containing
the PUF memory segment (line~14), and finally return the PUF response, i.e.,
the resulting \emph{error pattern} from the PUF evaluation at the PUF memory
segment (line~15).

\begin{algorithm}\footnotesize
    \SetAlgoNlRelativeSize{0.7} 
    \SetAlgoNoLine
    \DontPrintSemicolon
    \SetAlCapHSkip{0pt}
	\caption{Evaluate DRAM latency PUF} 
    \label{alg:lat_puf} 


    \textbf{evaluate\_DRAM\_latency\_PUF($seg\_id$):} \par
	~~~~write known data \jksix{(all 1's)} to Segment[$seg\_id$] \par
    ~~~~$rank\_id$ $\gets$ DRAM rank containing $seg\_id$ \par	
	~~~~obtain exclusive access to Rank[$rank\_id$] \par
	~~~~set low $t_{RCD}$ for Rank[$rank\_id$] \par
	~~~~\textbf{for} $i~=~1~to~num\_iterations:$\par
    ~~~~~~~~\textbf{for all} $col$ in Segment[$seg\_id$] \par
    ~~~~~~~~~~~~\textbf{for all} $row$ in Segment[$seg\_id$]: \textcolor{gray}{~~~~~~~~~~~~~\hspace{1pt}// column-order reads} \par 
    ~~~~~~~~~~~~~~~~$read()$ \textcolor{gray}{~~~~~~~~~~~~~~~~~~~~~~~~~~~~~~~~~~~~~~~// induce read failures} \par
	~~~~~~~~~~~~~~~~$memory\_barrier()$ \textcolor{gray}{~~~~~~~~~~~~~~~~~~~~~~\hspace{1pt}// one access at a time}\par
    ~~~~~~~~~~~~~~~~$count\_failures()$ \textcolor{gray}{~~~~~~~~~~~~~~~~~~~~~~~~~// record in another rank} \par 
	~~~~set default $t_{RCD}$ for Rank[$rank\_id$] \par
    ~~~~filter the PUF memory segment \textcolor{gray}{~~~~~~~~~~~~~~~~// See \emph{Filtering Mechanism}} \par 
	~~~~release exclusive access to Rank[$rank\_id$] \par
	~~~~\textbf{return} error pattern at Segment[$seg\_id$]
\end{algorithm} 

\textbf{Filtering Mechanism.} In order to improve the repeatability of
the DRAM latency PUF, we employ a \emph{filtering mechanism} which removes the
cells with low failure probability from the PUF response (as shown on
line~13 in Algorithm~\ref{alg:lat_puf}). The key idea is to count, for each bit
location in the PUF memory segment, the number of iterations in which the
location fails and then use that count to determine whether the bit location
should be \emph{set} (``1'') or \emph{cleared} (``0'') in the final PUF
response.  Every bit in the DRAM PUF memory segment has a corresponding counter
that we store in the \emph{counter buffer}\jkfive{, a data structure we
allocate in a DRAM rank separate from the one containing the PUF memory
segment.} This is to ensure that read/write requests to the counter buffer follow
manufacturer-specified timing parameters and do not induce latency failures. 

After each reduced-latency read request in the PUF memory segment, we find all
bit locations in the read data that resulted in a latency failure, and
increment their corresponding counters in the counter buffer. After all
iterations of inducing latency failures are completed, we compare every counter
of each bit location in the PUF memory segment against a threshold. If a
counter holds a value greater than the threshold (i.e., the counter's
corresponding bit location failed more than $n$ times, where $n$ is the
threshold), we set the corresponding bit location. Otherwise, we clear it.

\textbf{Memory Footprint.} Equation~\ref{eq:mem_usage} provides the memory
footprint required by PUF evaluation: 
\vspace{-5pt}
\begin{equation}
\label{eq:mem_usage} 
mem_{total}~=~(size_{mem\_seg})~+~(size_{counter\_buffer}) 
\vspace{-5pt}
\end{equation} 
where $size_{mem\_seg}$ is the size of the PUF memory segment and
$size_{counter\_buffer}$ is the size of the counter buffer. The size of the counter buffer
can be calculated using Equation~\ref{eq:counter_buffer_size}: 
\vspace{-5pt}
\begin{equation} 
\label{eq:counter_buffer_size} 
size_{counter\_buffer}~=~(size_{mem\_seg})\times\lceil\log_{2} N_{iters}\rceil
\vspace{-5pt}
\end{equation} 
where $size_{mem\_seg}$ is the size of the PUF memory segment and $N_{iters}$
is the number of iterations that we want to induce latency failures for. Since
we require one counter per bit in the memory segment, we must multiply this
quantity by the size of each counter. Since the counter must be able to store
up to the value of $N_{iters}$ (e.g., in the case of a cell that fails every
iteration), each counter must be $\lceil\log_{2} N_{iters}\rceil$ bits wide.
For a memory segment size of 8KiB, we find that \jkfour{the DRAM latency PUF's
total} memory footprint is 64KiB. From this, we conclude that DRAM latency PUFs
have insignificant DRAM capacity overhead.

\subsection{Variation Among PUF Memory Segments}
\label{subsection:among_memseg_variation} 

We observe a variation in latency failure rates across different memory
segments, which make some DRAM memory segments more desirable to evaluate DRAM
latency PUFs with than others. Because we want to find 512 bits that fail per
PUF memory segment (Section~\ref{subsection:rpufgt}), we consider only those
memory segments that have at least 512 failing bits as \emph{good} memory
segments. In order to determine the best size of the memory segment to evaluate
the DRAM latency PUF on, we study the effect of varying memory segment size on
1) DRAM capacity overhead, 2) PUF evaluation time, and 3) fraction of good
memory segments per device.  As the memory segment size increases, both the
DRAM capacity overhead and the PUF evaluation time increase linearly.  The
number of possible PUF memory segments for a DRAM device with a DRAM latency
PUF is obtained by counting the number of contiguous PUF memory segments across
all of DRAM (i.e., dividing the DRAM size by the PUF memory segment size).
Thus, larger PUF memory segments result in fewer possible PUF memory segments
for a DRAM device. From an experimental analysis of the associated tradeoffs of
varying the PUF memory segment size (not shown), we choose a PUF memory
segment size of 8KiB.\footnote{We will provide details in a technical
report/extended version for all other results that we cannot provide detail
for in the submission.}

In Table~\ref{Tab:varying_reliability}, we represent the distribution of the
percentage of good memory segments per chip with a \emph{median [minimum,
maximum]} across each of the three manufacturers. The left column shows the
number of chips tested, the right column shows the representation of the
distribution, and the rows indicate the different manufacturers of the chips.
We see that an overwhelming majority of memory segments from manufacturers A
and B are good for PUF evaluation. Memory segments from chips of manufacturer C
were observed to exhibit less latency failures, but across each of our chips we
could find at least 19.4\% of the memory segments to be good for PUF
evaluation.  Of the total number of PUF memory segments tested (shown in
Table~\ref{Tab:mem_segs_long_term}), we experimentally find that 100\%,
64.06\%, and 19.37\% of memory segments are \emph{good} (i.e., contain enough
failures to be considered for PUF evaluation) in the worst-case chips from
manufacturers A, B, and C. We conclude that there are plenty of PUF memory
segments that are good enough for DRAM latency PUF evaluation.


\begin{table}[h!]
\footnotesize
\begin{center}
\begin{tabular}{ c|c|c| } 
\cline{2-3}
 & \#Chips & Good Memory Segments per Chip (\%) \\ 
\cline{2-3}
A & 19 & 100.00 [100.00, 100.00] \\ 
B & 12 & 100.00 [64.06, 100.00] \\ 
C & 14 & 30.86 [19.37, 95.31] \\ 
\cline{2-3}
\end{tabular} 
\caption{Percentage of \emph{good} memory segments per chip across manufacturers. Median [min, max] values are shown.}
\label{Tab:varying_reliability} 
\end{center} 
\end{table} 
\vspace{-8pt}

\subsection{Support for Changing Timing Parameters}
\label{subsec:changing_timing}

In order to induce latency failures, the manufacturer-specified DRAM
timing parameters must be changed. Some existing
processors~\cite{lee2015adaptive, AMD_opteron, bkdg_amd2013} enable software to
directly manipulate DRAM timing parameters. These processors can trivially
implement and evaluate a DRAM latency PUF with \emph{minimal} changes to the
software \jkfive{and no changes to hardware}. However, for other processors
that cannot directly manipulate DRAM timing parameters, we would need to simply
enable software to programmatically modify memory controller registers which
indicate the DRAM timing parameters that a memory access must observe.

We find that we can reliably induce latency failures when we reduce the
value of $t_{RCD}$ from a default value of 18ns to between 6ns and 13ns. Given
this wide range of failure-inducing $t_{RCD}$ values, most memory controllers
should be able to issue read requests with a $t_{RCD}$ value within this
range.

\subsection{Device Enrollment}
\label{subsec:char_and_enroll}

Device enrollment is a one-time process consisting of evaluating all possible
PUFs from across the entire challenge-response space and securely storing the
evaluated PUFs in a trusted database such that they can be later queried for
authentication~\cite{katzenbeisserpufs,sutar2016d,xiong2016run}. Since the
goal of PUF authentication is to ensure that a challenge-response is difficult
to replicate without access to the original device, enrollment must be done
securely so that the full set of all possible challenge-response pairs is known
only to the trusted database and can be created \emph{only} by the device owner.

Similar to prior works' approach to DRAM
PUFs~\cite{xiong2016run,tehranipoor2015dram,sutar2016d}, we assume that a
trusted third party (e.g., the DRAM manufacturer) performs device enrollment prior
to making the system available to the end consumer. This ensures that the
complete set of all possible challenge responses are known only by the trusted
third party. After the device is in the field, even the trusted third party
\emph{cannot} regenerate the enrollment data. Thus, allowing the trusted third
party to both characterize and enroll the responses makes it extremely
difficult for a malicious attacker to obtain the full set of possible response
pairs without first compromising the trusted third party.

Because DRAM latency PUF responses vary depending on the temperature of DRAM
during evaluation time (see Section~\ref{subsubsec:repeatability}), we must
enroll multiple golden keys at varying temperature intervals. This enables a
PUF response to match at least one golden key during authentication regardless
of the temperature during evaluation time. We find that some chips generate PUF
responses with less variation across a range of temperatures than other chips.
Chips with less variation can enroll golden keys for temperatures at larger
intervals than chips with more variation.

\subsection{In-DRAM Error Correcting Codes}
\label{subsec:ind_ecc}

Some new DRAM chips utilize in-DRAM \emph{error-correcting codes} (ECC), which
perform single-bit error correction per \emph{word} (i.e., typically 64 data
bits) invisibly to the system~\cite{nair2016xed, kang2014co, oh20153}, to overcome the
reliability challenges of DRAM technology scaling~\cite{mutlu2013memory,
mutlu2017rowhammer, kim2014flipping, raidr, mutlu2014research,
meza2015revisiting, schroeder2009dram}. When such chips are used for DRAM
latency PUFs,  ECC words with only one error appear to be error-free to the
memory controller, leaving fewer total errors available for PUF response.
\jksix{We note that} error correction \emph{deterministically} transforms DRAM
error patterns.  Thus, a DRAM PUF (on a chip with in-DRAM ECC) that repeatably
induces the same error pattern prior to ECC correction, would repeatably result
in a different but \emph{consistent error pattern} after ECC correction.

In order to support PUF evaluation in a system using DRAM chips with in-DRAM
ECC, we would need to evaluate PUFs with \jkfive{a higher} \emph{raw bit error
rate} (i.e., the error rate before ECC is performed) relative to non-ECC DRAMs.
A higher raw bit error rate would produce enough observable failures (after ECC
is performed) for a PUF. The DRAM latency PUF can achieve this higher
raw bit error rate by simply reducing the latency parameter value further. Such
reduction would \jksix{also ideally reduce PUF evaluation time and system
interference. Therefore, we expect the DRAM latency PUF to be evaluated even
faster in chips with built-in ECC.}\footnote{\jksix{In contrast,} DRAM
retention PUF \emph{must} be evaluated with a longer refresh interval to
increase the raw bit error rate \jksix{in a DRAM chip with in-DRAM ECC}.  This
leads to a significantly longer DRAM retention PUF evaluation time when in-DRAM
ECC is used.} As stronger ECC mechanisms are used, i.e., ECC can correct DRAM
words containing more than 1 error, we expect even lower evaluation latencies
and lower system interference with the DRAM latency PUF. 

\subsection{Effect of High-Temperature} 
\label{subsec:high_temp_runtime}

Our evaluation of the DRAM latency and retention PUFs is limited to the
70$^{\circ}$C maximum DRAM temperature. We clearly show in
Section~\ref{subsection:rpufgt} that DRAM latency PUFs are much faster than
DRAM retention PUFs at 70$^{\circ}$C and lower (Figure~\ref{fig:raw_runtime}).
However, at higher temperatures (e.g., $>85^\circ$C), DRAM retention PUFs could
become faster than DRAM latency PUFs.\footnote{Note that the DDR protocol
specifies that every cell must be refreshed at least every 32ms for LPDDR3/4 or
64ms for DDR3/4 below $85^\circ$C and at even higher rates at higher
temperatures.}

If a DRAM retention PUF is faster than the DRAM latency PUF at a very
high temperature, it is easy to envision a mechanism that dynamically switches
between DRAM latency PUFs and DRAM retention PUFs based on the device operating
temperature at the time of evaluation. This mechanism could exploit the
strengths of each type of DRAM PUF in order to allow the fastest possible PUF
evaluation time. By exploiting in-DRAM temperature sensors that already exist
in modern DRAM chips~\cite{2014lpddr4, lpddr3, lee2015adaptive, ddr4}, this
mechanism could potentially be implemented with no additional hardware overhead
beyond what is already required for DRAM retention PUFs and DRAM latency PUFs
individually.

Such a mechanism would require challenge-response pairs from both the DRAM
retention PUF and DRAM latency PUF to be enrolled. Furthermore, since retention
failure rates vary significantly across different
chips~\cite{liu2013experimental, khan2014efficacy, patel2017reaper}, each chip
will have a different temperature at which the mechanism switches between the
two DRAM PUFs. This could considerably impact enrollment time and complicate
the device authentication process. Ultimately, it is up to the system architect
to decide whether such a mechanism is worth the evaluation runtime benefits
\jksix{at very high temperatures} (which is likely to be on the order of tens
of milliseconds). We leave a full exploration of this hybrid DRAM
latency-retention PUF mechanism to future work.

\section{Related Work} 
\label{related}

To our knowledge, this is the first work to: 1) introduce the idea of
violating DRAM read latency parameters to create a fast, runtime-accessible
DRAM PUF without modifying commodity DRAM devices, 2) introduce an
effective DRAM PUF that is runtime-accessible at all operating temperatures,
3) demonstrate a wide variety of tradeoffs in DRAM PUFs, based on extensive new
experimental data from 223 state-of-the-art LPDDR4 DRAM chips, 4) demonstrate
the prohibitively slow evaluation times of the DRAM retention PUF, the
previously fastest DRAM PUF suitable for commodity devices. 

In this section, we discuss prior works that propose \jkfive{DRAM} PUFs and
PUFs based on other substrates. The proliferation of recent works on DRAM PUFs
reflects the growing importance of \jkfour{DRAM PUFs given DRAM's} near
ubiquity in modern systems and large address space. 

\noindent
\textbf{DRAM Retention PUFs.} We have already described the basics of DRAM
retention PUFs in Section~\ref{ret_puf_analysis} \jkfour{and
extensively evaluated them in Section~\ref{ret_puf_analysis}}. We briefly
explain the differences \jkfour{between} prior proposals. Keller et
al.~\cite{keller2014dynamic} is the first to propose using DRAM retention
failures as unique identifiers, shortly followed by Xiong et
al.~\cite{xiong2016run} and D-PUF~\cite{sutar2016d}, both of which enable
runtime-accessible DRAM retention PUFs. Other works propose further
optimizations for improving the quality of DRAM retention
PUFs~\cite{rahmati2015probable, tang2017dram, tehranipoor2017investigation}.
As our experimental evaluations across 223 LPDDR4 DRAM chips show, DRAM
retention PUFs take very long to evaluate at \jkfour{common-case operating}
temperatures; \jksix{they are orders of magnitude slower than our proposal (see
Section~\ref{ret_puf_analysis}).} \jky{A work~\cite{muelich2019channel}
published after the DRAM Latency PUF, \jkx{demonstrates} highly reliable debiasing
techniques for PUFs generated with retention failures.}

\noindent
\textbf{Other DRAM PUFs.} Hashemian et al.~\cite{hashemian2015robust} propose
adding a delay generator to the DRAM write-circuitry to induce failures.
However, this requires additional hardware and cannot be applied to existing
DRAM designs. \jky{Tehranipoor et al.~\cite{tehranipoor2015dram,
karimian2019generate}} suggest using DRAM start-up values for PUFs, but this
precludes runtime evaluation by requiring a DRAM power cycle for every
authentication. In a work published after this work, Talukder et
al.,~\cite{talukder2019prelatpuf} propose evaluating a PUF by inducing
precharge latency failures and show that they can provide PUFs orders of
magnitude faster, since precharge latency failures occur at larger
granularities than activation latency failures.

\noindent
\textbf{PUFs Based on Other Substrates.} Many PUFs have been proposed for
various other substrates, including other memory technologies and customized
hardware designs. We categorize them into delay-based and memory-based PUFs. 

\textbf{Delay-based PUFs} include \jkfour{1)} arbiter
PUFs~\cite{lee2004technique, lim2005extracting, majzoobi2008testing,
ruhrmair2009foundations, ozturk2008physical, ozturk2008towards,
hammouri2008unclonable, ye2015opuf}, which rely on process variation to extract
the unique behavior of two identical competing circuit paths, \jkfour{2)} ring
oscillator PUFs~\cite{gassend2003physical, gassend2002silicon, suh2007physical,
liu2017acro}, which rely on frequencies of oscillating signals from chained
inverters, and \jkfour{3)} Current Mirror Array (CMA)
PUFs~\cite{wang2017current}, which rely on the manufacturing process variation
in a customized circuit typically used for machine learning tasks.  These works
rely on customized hardware \emph{not} present in commodity systems.
FPGA-based PUFs~\jky{\cite{kumar2008butterfly, guajardo2007physical,
majzoobi2010fpga, maiti2011impact, guajardo2007fpga, gu2015ultra,
tian2020fingerprinting, ge2019fpga}} overcome the need for hardware changes.
However, they are not as common as DRAM in computer systems. 

\textbf{Memory-based PUFs} include SRAM PUFs, which rely on SRAM start up
values~\cite{guajardo2007fpga, holcomb2007initial, holcomb2009power,
cortez2013adapting, bhargava2012reliability, zheng2013resp, xiao2014bit,
sutar2017memory} and voltage reduction induced
failures~\cite{bacha2015authenticache}; butterfly PUFs, which mimic the behavior
of SRAM cells with cross-coupled data latches~\cite{kumar2008butterfly}; latch
PUFs, which cross-couple two NOR-gates~\cite{su20071}; flip-flop PUFs, which
exploit the power up behavior of regular flip-flops~\cite{maes2008intrinsic,
van2010hardware}; and PUFs for emerging memory
technologies~\jky{\cite{iyengar2014dwm, koeberl2013memristor, rose2013foundations,
vatajelu2015stt, che2014non, chen2015exploiting, liu2015experimental,
das2015mram, zhang2015highly, zhang2015optimizating, vatajelu2016stt,
beckmann2017performance, pang2017optimization, kamal2019mixed}}. These prior
works either require additional customized hardware or usage of SRAM, which has
a small address space compared to DRAM, and thus cannot accommodate a large
number of challenge-response pairs.

\section{\jky{Limitations}} 

\jkz{Although the DRAM Latency PUF can be used immediately in certain systems,
the DRAM Latency PUF has a few limitations that prevent its immediate
deployment in all real systems at its full potential. }

\jkz{First, the DRAM Latency PUF has a linear challenge-response space which
results in its categorization as a weak PUF. While weak PUFs have many use
cases in the field, a strong PUF (i.e., a PUF with an exponential
challenge-response space) is more practical and provides added value to a
system. We believe that studying the interactions between many DRAM timing
parameters may help to identify effects of parameters that can be combined to
build a strong PUF with an exponential challenge-response space. }

\jkz{Second, the DRAM Latency PUF requires a flexible memory controller that
can interleave DRAM accesses with varying timing parameters in order to
minimize the overhead that evaluating this PUF would cause on a standard system
that likely has higher overhead in changing timing parameters.} 

\jkz{Third, as the DRAM bus is considered an insecure communication channel,
the current interface for evaluating the PUF over the DRAM bus is prone to
interference and eavesdropping. As the DRAM Latency PUF can be evaluated purely
on the DRAM chip itself, we believe that developing simple \jky{logic nearby
memory to facilitate the evaluation of the DRAM latency PUF on DRAM} without
relying on an insecure channel may help to further improve the \jky{security}
of the PUF.}

\jkz{Fourth, our aging studies are limited to a \jky{30-day} period. While we
demonstrate that PUF responses do not change significantly over 30 days,
long-term aging may have stronger effects. If \jky{long-term} effects change
PUF responses such that they are incomparable to the golden PUF responses,
occasional re-profiling may be required to reliably employ the DRAM Latency
PUF. }
 
\section{Summary} 

We introduce and analyze the DRAM latency PUF, a new DRAM PUF suitable for
runtime authentication. The DRAM latency PUF intentionally violates
\jkfour{manufacturer-specified} DRAM timing parameters in order to provide
\jkfour{many highly repeatable, unique, and unclonable} PUF responses with low
latency.  Through experimental evaluation using 223 state-of-the-art LPDDR4
DRAM devices, we show that the DRAM latency PUF reliably generates PUF
responses at runtime-accessible speeds (i.e., 88.2ms on average) at all
operating temperatures.  We show that the DRAM latency PUF achieves an average
speedup of 152x/1426x at 70$^{\circ}$C/55$^{\circ}$C when compared with a DRAM
retention PUF of \jkfour{the same} DRAM capacity overhead, and it achieves even
greater speedups at lower temperatures.  We conclude that the DRAM latency PUF
enables a fast and effective substrate for runtime device authentication across
all operating temperatures, and we hope that the advent of runtime-accessible
PUFs like the DRAM latency PUF \jkfour{and the detailed experimental
characterization data we provide on modern DRAM devices} will enable security
architects to develop even more secure systems for future devices.

\chapter{D-RaNGe: Using Commodity DRAM Devices to Generate True Random Numbers with Low Latency and High Throughput} 
\label{ch5-drange} 

We propose a new DRAM-based true random number generator (TRNG) that
leverages DRAM cells as an entropy source. \hhtwo{The key idea is to}
intentionally \hhtwo{violate} the \hhtwo{DRAM} access timing parameters and use
the resulting errors as the source of randomness. Our technique specifically
decreases the DRAM row activation latency (timing parameter $t_{RCD}$) below
manufacturer-recommended specifications, to induce read errors, or activation
failures, that exhibit true random behavior. We then aggregate the resulting
data from multiple cells to obtain a TRNG capable of providing a
\jkfour{high throughput} of random numbers at low latency.

To demonstrate that our TRNG design is viable using commodity DRAM chips, we
rigorously characterize the behavior of activation failures in 282
state-of-the-art LPDDR4 \mpt{devices from three major DRAM manufacturers. We
verify our observations using four additional DDR3 DRAM devices from the same
manufacturers}. Our results show that many cells in each device produce random data
that remains robust over both time and temperature variation. We use our
observations to develop \mechanism, a methodology for extracting true random
numbers from commodity DRAM devices with high throughput and low latency by
deliberately violating \hhtwo{the} read access timing parameters. We evaluate the quality
of our TRNG using the commonly-used NIST statistical test suite for randomness
and find that \mechanism: 1) successfully passes each test, and 2) generates
true random numbers with over two orders of magnitude higher throughput
than the previous highest-throughput DRAM-based TRNG. 

\section{True Random Number Generators (TRNGs)} 
\label{subsec:trng}

A \emph{true random number generator (TRNG)} requires physical processes (e.g.,
radioactive decay, thermal noise, Poisson noise) to construct a bitstream of
random data. Unlike pseudo-random number generators, the random numbers
generated by a TRNG do \emph{not} depend on the \hh{previously-generated} numbers
and \emph{only} depend on the random noise obtained from physical processes.
TRNGs are usually validated using statistical tests such as
NIST~\cite{rukhin2001statistical} or DIEHARD~\cite{marsaglia2008marsaglia}. A
TRNG typically consists of 1) an \emph{entropy source}, 2) a \emph{randomness
extraction technique}, and sometimes 3) a \emph{post-processor}, which improves
the randomness of the extracted data often at the expense of throughput. These
three components are typically used to reliably generate true random
numbers\hh{~\cite{stipvcevic2014true, sunar2007provably}}.

\textbf{Entropy Source.} The entropy source is a critical component of a random
number generator, as its amount of entropy affects the unpredictability and the
throughput of the generated random data. Various physical phenomena can be
used as entropy sources. In the domain of electrical circuits, thermal and
Poisson noise, jitter, and circuit metastability have been proposed as processes
that have high entropy~\cite{wang2012flash, ray2018true, holcomb2007initial,
holcomb2009power, van2012efficient, mathew20122, brederlow2006low,
tokunaga2008true, bhargava2015robust}. To ensure robustness, the entropy source
should not be visible or modifiable by an adversary.  Failing to satisfy that
requirement would result in generating predictable data, and thus put the
system into a state susceptible to security attacks.

\textbf{Randomness Extraction Technique.} The randomness extraction
technique harvests random data from an entropy source. A good randomness
extraction technique should have two key properties. First, it should have
high throughput, i.e., extract as much as randomness possible in a short amount
of time~\cite{kocc2009cryptographic, stipvcevic2014true}\hh{, especially
important for applications that require high-throughput random number
generation (e.g., security applications~\cite{gutterman2006analysis, von2007dual, kim2017nano,
drutarovsky2007robust, kwok2006fpga, cherkaoui2013very, zhang2017high,
quintessence2015white, bagini1999design, rock2005pseudorandom, ma2016quantum,
stipvcevic2014true, barangi2016straintronics, tao2017tvl, botha2005gammaray,
mathew20122, yang201416}, scientific simulation~\cite{ma2016quantum,
botha2005gammaray})}. Second, it should not disturb the physical
process~\cite{kocc2009cryptographic, stipvcevic2014true}. Affecting the entropy
source during the randomness extraction process would make the harvested data
predictable, lowering the reliability of the TRNG.

\textbf{Post-processing.} Harvesting randomness from a physical phenomenon
\emph{may} produce bits that are biased or
correlated~\cite{kocc2009cryptographic, rahman2014ti}. In such a case, a
post-processing step, which is also known as \emph{de-biasing}, is applied to
eliminate the bias and correlation. The post-processing step also provides
protection against environmental changes and adversary
tampering~\cite{kocc2009cryptographic, rahman2014ti, stipvcevic2014true}.
\hhtwo{Well-known} post-processing techniques are the von Neumann
corrector~\cite{jun1999intel} and cryptographic hash functions such as
SHA-1~\cite{eastlake2001us} or MD5~\cite{rivest1992md5}. These post-processing
steps work well, but generally result in decreased throughput (e.g., up to
80\%~\cite{kwok2011comparison}).

\section{Motivation and Goal} 
\label{sec:motivation}

True random numbers sampled from physical phenomena have a number of real-world
applications from system security~\hh{\cite{bagini1999design,
rock2005pseudorandom, stipvcevic2014true}} to recreational
entertainment~\hh{\cite{stipvcevic2014true}}. As user data
privacy becomes a \emph{highly-sought} commodity in Internet-of-Things (IoT)
and mobile devices, enabling primitives that provide security on such systems
becomes \jk{critically important}~\cite{lin2017survey, ponemon2017sec, zhang2017high}.
Cryptography is one typical method for securing systems against various attacks
by encrypting the system's data with keys generated with true random values.
Many cryptographic algorithms require random values to generate keys in many
standard protocols (e.g., TLS/SSL/RSA/VPN keys) to either 1) encrypt network
packets, file systems, and data, 2) select internet protocol sequence numbers
(TCP), or 3) generate data padding values~\cite{gutterman2006analysis,
von2007dual, kim2017nano, drutarovsky2007robust, kwok2006fpga,
cherkaoui2013very, zhang2017high, quintessence2015white}. TRNGs are also
commonly used in authentication protocols and in countermeasures against
hardware attacks~\cite{cherkaoui2013very}, in which psuedo-random number
generators (PRNGs) \hh{are shown to be insecure}~\cite{von2007dual,
cherkaoui2013very}. To keep up with the \emph{ever-increasing} rate of
secure data creation, especially with the growing number of commodity
data-harvesting devices (e.g., IoT and mobile devices), the ability to generate
true random numbers with \emph{high throughput and low latency} becomes ever
more relevant to maintain user data privacy. In addition,
\emph{high-throughput} TRNGs are \hh{already} \emph{essential} components of
various important applications such as scientific
simulation~\hh{\cite{ma2016quantum, botha2005gammaray}}, industrial testing,
statistical sampling, randomized algorithms, and recreational
entertainment~\cite{bagini1999design, rock2005pseudorandom, ma2016quantum,
stipvcevic2014true, barangi2016straintronics, tao2017tvl, botha2005gammaray,
zhang2017high, mathew20122, yang201416}. 

A \emph{\hh{widely-available,} high-throughput, low-latency} TRNG will enable
all previously mentioned applications that rely on TRNGs, including improved
\jktwo{security and privacy} in most systems that are known to be vulnerable to
attacks~\cite{lin2017survey, ponemon2017sec, zhang2017high}, as well as enable
research that we may not anticipate at the moment. One such direction is using
a one-time pad \hhthree{(i.e., a private key used to encode and decode only a single message)} with quantum key
distribution, which requires at least \emph{4Gb/s} \hh{of true random number
generation throughput}~\cite{wang2016theory, clarke2011robust, lu2015fpga}.
Many \emph{high-throughput} TRNGs have been recently
proposed~\cite{kwok2006fpga, tsoi2007high, zhang201568, cherkaoui2013very,
barangi2016straintronics, zhang2017high, ning2015design, quintessence2015white,
wang2016theory, mathew20122, yang201416, bae20173, fischer2004high,
gyorfi2009high}, and the availability of these high-throughput TRNGs can enable
\hh{a wide range of new applications with improved security} \jktwo{and privacy}.

DRAM offers a promising substrate for developing an effective and
widely-available TRNG due to the prevalence of DRAM throughout all modern
computing systems ranging from microcontrollers to supercomputers. A
high-throughput DRAM-based TRNG would help enable widespread adoption of
applications that are today limited to only select architectures equipped with
dedicated high-performance TRNG engines. Examples of such applications include
high-performance scientific simulations and cryptographic applications for
securing devices and communication protocols, both of which would run
\hhthree{much} more
\jk{efficiently on \hhthree{mobile devices, embedded devices,} or microcontrollers with the availability
of higher-throughput TRNGs \hhthree{in} the system.} 



In terms of the CPU architecture itself, a high-throughput DRAM-based TRNG
\jkfour{could} help the memory controller to improve scheduling
decisions~\cite{usui2016dash, mutlu2008parallelism, ausavarungnirun2012staged,
subramanian2016bliss, subramanian2014blacklisting, kim2010thread,
mutlu2007stall, subramanian2013mise, subramanian2015application} and
\jkfour{enable the implementation} a truly-randomized version of
PARA~\cite{kim2014flipping} (i.e., a protection mechanism against the RowHammer
\jktwo{vulnerability~\cite{kim2014flipping, mutlu2017rowhammer}}).
\jktwo{Furthermore, a DRAM-based TRNG would likely have additional hardware and
software applications as system designs become more capable and increasingly
security-critical.} 

In addition to traditional computing paradigms, DRAM-based TRNGs can
benefit processing-in-memory (PIM) architectures~\cite{ghose2018enabling, mutlu2019processing,
seshadri2017simple}, which co-locate logic \hhthree{within or near} memory \jktwo{to overcome the
large bandwidth and energy bottleneck caused by} the memory bus and leverage
the \emph{significant} data parallelism available within the DRAM chip itself.
Many prior works provide primitives for PIM or exploit PIM-enabled systems for
workload acceleration~\cite{ahn2015scalable, ahn2015pim, lee2015simultaneous,
seshadri2015fast, seshadri2013rowclone, seshadri2015gather, seshadri2017ambit,
liu2017concurrent, seshadri2017simple, pattnaik2016scheduling,
babarinsa2015jafar, farmahini2015nda, gao2015practical, gao2016hrl,
hassan2015near, hsieh2016transparent, morad2015gp, sura2015data, zhang2014top,
hsieh2016accelerating, boroumand2017lazypim, boroumand2019conda, chang2016low,
kim2018grim, ghose2018enabling, mutlu2019processing, boroumand2018google,
seshadri2016simple}. A low-latency, high-throughput DRAM-based TRNG can enable
PIM applications to source random values \emph{directly within the memory
itself}, thereby enhancing the overall potential, \jktwo{security, and
privacy,} of PIM-enabled architectures. For example, \hhf{in applications that
require true random numbers, a DRAM-based TRNG can enable large contiguous code
segments to execute in memory, which would reduce communication with the CPU,
and thus improve system efficiency. A DRAM-based TRNG can also enable security
tasks to run completely in memory.  This would remove the dependence of
PIM-based security tasks on an I/O channel and would increase \jkfive{overall}
system security.}

We posit, based on analysis done in prior works~\cite{kocc2009cryptographic,
jun1999intel, schindler2002evaluation}, that an \emph{effective} TRNG must
satisfy \emph{six} key properties: \jktwo{it} must 1) have \hh{low} implementation
cost, 2) be fully non-deterministic such that it is impossible to predict the
next output given complete information about how the mechanism operates, 3)
provide a continuous stream of true random numbers with high throughput, 4)
provide true random numbers with low latency, 5) exhibit low system
interference, i.e., not significantly \hh{slow down} concurrently-running
applications, and 6) generate random values with low energy overhead.

To this end, our \textbf{goal} in this work, is to provide a
\hh{widely-available} TRNG for DRAM
devices that satisfies all six key properties of an effective TRNG. 


\section{Testing Environment}
\label{dlrng:sec:methodology} 

In order to test our hypothesis that DRAM cells are an effective source of
entropy when accessed with reduced DRAM timing parameters, we developed an
infrastructure to characterize modern LPDDR4 DRAM chips.
We also use an infrastructure for DDR3 DRAM chips,
SoftMC~\cite{hassan2017softmc, softmc-safarigithub}, to demonstrate
empirically that our proposal is applicable beyond the LPDDR4
technology. Both testing environments give us precise control over DRAM
commands and DRAM timing parameters as verified with a logic analyzer probing
the command bus.

We perform all tests, unless otherwise specified, using a total of 282 2y-nm
LPDDR4 DRAM chips from three major manufacturers in a thermally-controlled
chamber held at 45$^{\circ}$C. For consistency across results, we
precisely stabilize the ambient temperature using heaters and fans
controlled via a microcontroller-based proportional-integral-derivative (PID)
loop to within an accuracy of 0.25$^{\circ}$C and a reliable range of
40$^{\circ}$C to 55$^{\circ}$C.  We maintain DRAM temperature at 15$^{\circ}$C
above ambient temperature using a separate local heating source. We use
temperature sensors to smooth out temperature variations caused by self-induced
heating.

We also use a separate infrastructure, based on open-source
SoftMC~\cite{hassan2017softmc, softmc-safarigithub}, to validate our mechanism
on 4 DDR3 DRAM chips from a single manufacturer. SoftMC enables precise control
over timing parameters, and we house the DRAM chips inside \jktwo{another}
temperature chamber to maintain a stable ambient testing temperature (with the
same temperature range as the temperature chamber \jktwo{used for the LPDDR4
devices}).  

To explore the various effects of temperature, short-term aging, and
circuit-level interference (in Section~\ref{sec:dlrng_characterization}) on
activation failures, we reduce the $t_{RCD}$ parameter from the default
$18ns$ to $10ns$ for all experiments, unless otherwise stated.
Algorithm~\ref{dlrng:alg:testing_lat} explains the general testing methodology we
use to induce activation failures. First, we write a data pattern to the
    \begin{algorithm}[tbh]\footnotesize
        \SetAlgoNlRelativeSize{0.7}
        \SetAlgoNoLine
        \DontPrintSemicolon
        \SetAlCapHSkip{0pt}
        \caption{DRAM Activation Failure Testing}
        \label{dlrng:alg:testing_lat} 

        \textbf{DRAM\_ACT\_failure\_testing($data\_pattern$, $DRAM\_region$):} \par 
        ~~~~write $data\_pattern$ (e.g., solid 1s) into all cells in $DRAM\_region$ \par 
        ~~~~set low $t_{RCD}$ for ranks containing $DRAM\_region$ \par 
        ~~~~\textbf{foreach} $col$ in $DRAM\_region$: \par 
        ~~~~~~~~\textbf{foreach} $row$ in $DRAM\_region$: \par 
        ~~~~~~~~~~~~$activate(row)$ \textcolor{gray}{~~~~// fully refresh cells } \par 
        ~~~~~~~~~~~~$precharge(row)$ \textcolor{gray}{~// ensure next access activates the row} \par
        ~~~~~~~~~~~~$activate(row)$ \par 
        ~~~~~~~~~~~~$read(col)$ \textcolor{gray}{~~~~~~~~~~~// induce activation failure on col} \par
        ~~~~~~~~~~~~$precharge(row)$ \par
        ~~~~~~~~~~~~record activation failures to storage \par 
        ~~~~set default $t_{RCD}$ for DRAM ranks containing $DRAM\_region$ \par 
    \end{algorithm}
region of DRAM under test (Line~2). Next, we reduce the $t_{RCD}$
parameter to begin inducing activation failures (Line~3). We then access
the DRAM region in column order (Lines~4-5) in order to ensure that each DRAM
access is to a closed DRAM row and thus requires an activation. This enables
each access to induce activation failures in DRAM. Prior to each
reduced-latency read, we first refresh the target row such that each cell has
the same amount of charge each time it is accessed with a reduced-latency read.
We effectively refresh a row by issuing an activate (Line~6) followed by a
precharge (Line~7) to that row. We then induce the activation failures by
issuing consecutive activate (Line~8), read (Line~9), and precharge (Line~10)
commands. Afterwards, we record any activation failures that we observe
(Line~11).  We find that this methodology enables us to quickly induce
activation failures across \emph{all} of DRAM, and \jktwo{minimizes} testing
time. 

\section{Activation Failure Characterization} 
\label{sec:dlrng_characterization} 

To demonstrate the viability of using DRAM cells as an entropy source for
random data, we explore and characterize DRAM failures when employing a
reduced DRAM activation latency ($t_{RCD}$) across 282 LPDDR4 DRAM chips. We
also compare our findings against \mpx{those of} prior works that study an
older generation of DDR3 DRAM chips~\cite{chang2016understanding,
lee-sigmetrics2017, lee2015adaptive, kim2018solar} to \jk{cross-}validate our
infrastructure. \jk{To understand the effects of changing environmental
conditions on a DRAM cell that is used as a source of entropy, we rigorously
characterize DRAM cell behavior as we vary four environmental conditions.
First, we study the effects of DRAM array design-induced variation (i.e., the
spatial distribution of activation failures in DRAM). Second, we study data
pattern dependence (DPD) effects on DRAM cells. Third, we study the effects of
temperature variation on DRAM cells. Fourth, we study a DRAM cell's activation
failure probability over time.} We present several key observations that
support the viability of a mechanism that generates random numbers by
accessing DRAM cells with a reduced $t_{RCD}$.  In
Section~\ref{section:DLRNG_mechanism}, we discuss a mechanism to effectively
\jk{sample DRAM cells to extract true random numbers while minimizing the effects
of environmental condition variation (presented in this section) on the DRAM
cells.} 

\subsection{Spatial Distribution of Activation Failures}
\label{subsec:spatial_char} 

\begin{figure}[h] 
    \centering \includegraphics[width=0.6\linewidth]{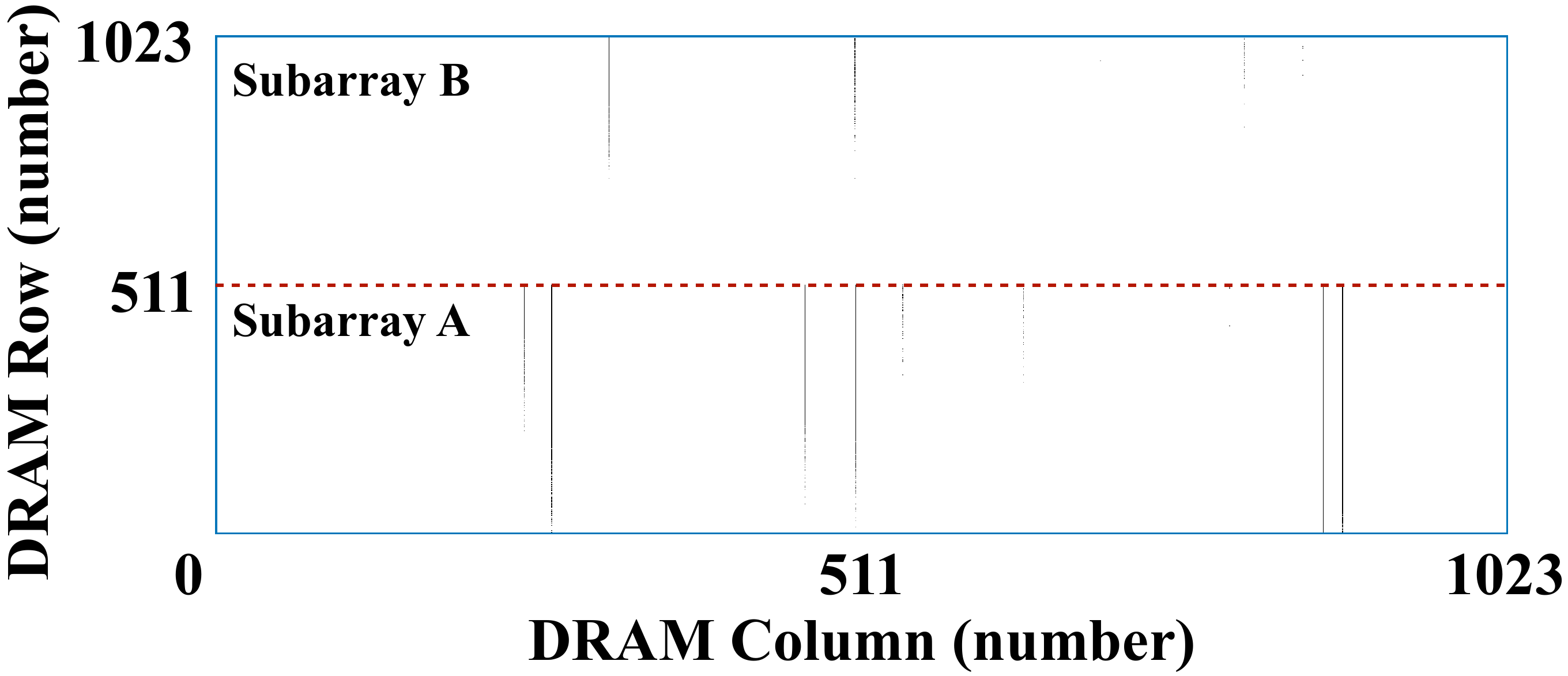} 
    \caption{Activation failure bitmap in $1024\times1024$ cell array.}
    \label{fig:spatial_failures} 
\end{figure}

To study \mpx{which} regions of DRAM \mpx{are better suited to} generating
random data, we first visually inspect the spatial distributions of activation
failures \mpx{both} across DRAM chips and within each chip \mpx{individually}. 
Figure~\ref{fig:spatial_failures} plots the spatial distribution of activation
failures in a \emph{representative} $1024\times1024$ array of DRAM cells
\mpx{taken} from a single DRAM chip\mpx{. Every} observed activation
failure is marked in black. We make two observations. First, we observe that
each contiguous region of 512 DRAM rows\footnote{\jk{We note that subarrays
have either 512 or 1024 (not shown) rows depending on the manufacturer of the
DRAM device.}} \mpx{consists} of \mpx{repeating} rows with the same set (or
subset) of \jk{column bits} that are prone to activation failures. As shown in
the figure, rows 0 to 511 have the same 8 (or a subset of the 8) \jk{column
bits} failing in the row, and rows 512 to 1023 have the same 4 (or a subset of
the 4) \jk{column bits} failing in the row. We hypothesize that these
contiguous regions reveal the DRAM subarray architecture \mpx{as a result of
variation across} the local sense amplifiers in the subarray. \jkthree{We
indicate the two subarrays in Figure~\ref{fig:spatial_failures} as Subarray A
and Subarray B.} A ``weaker'' local sense amplifier results in cells that share
\mpx{its} respective \emph{local bitline} in the subarray \mpx{having} an
increased probability of failure. For this reason, we observe \mpx{that}
activation failures \mpx{are} localized to a few columns within a DRAM subarray
as shown in Figure~\ref{fig:spatial_failures}. Second, we observe that within a
subarray, the activation failure probability \mpx{increases} across rows (i.e.,
activation failures are \emph{more} likely to occur in \jk{higher-numbered}
rows in the subarray and \mpx{are} \emph{less} likely in \jk{lower-numbered}
rows in the subarray). This can be seen \jkthree{from} the fact that more cells
fail in \jk{higher-numbered} rows in the subarray (i.e., there are more black
marks higher in each subarray). We hypothesize that the \jkthree{failure
probability of a cell attached to a local bitline} correlates \mpx{with} the
distance between the row and the local sense amplifiers, and further rows have
less time to amplify their data due to the signal propagation delay in a
bitline. These observations are similar to \mpx{those made in} prior
studies~\cite{lee-sigmetrics2017, chang2016understanding, lee2015adaptive,
kim2018solar} on DDR3 devices.


We next study the granularity at which \mpx{we can induce} activation failures
when accessing a row. We observe (not shown) that activation failures
\mpx{occur} \emph{only} within the \mpx{first} cache line that is \mpx{accessed}
 immediately following an activation. No subsequent access to an
\jk{already} \emph{open} row results in activation failures. This is because
cells within the \emph{same} row have a longer time to restore their cell
charge (Figure~\ref{fig:dram_op}) when they are accessed after the row
\mpx{has already been opened}.  We draw two key conclusions: 1) the region
\emph{and} bitline of DRAM being accessed affect the number of observable
activation failures, and 2) different \mpx{DRAM} subarrays \emph{and}
different local bitlines exhibit varying levels of entropy.

\subsection{Data Pattern Dependence} 
\label{subsec:dpd} 

To understand the data pattern dependence of activation failures and DRAM cell
entropy, we study \mpy{how effectively we can discover failures using
different data patterns across multiple rounds of testing.} Our \emph{goal} in
this \mpy{experiment} is to determine \mpy{which} data pattern results in the
highest entropy such that we can generate random values with high throughput.
Similar to prior works~\cite{patel2017reaper, liu2013experimental} that
extensively describe the data patterns, we analyze a total of 40 unique data
patterns: solid 1s, checkered, row stripe, column stripe, 16 walking 1s,
\emph{and} the inverses of all 20 aforementioned data patterns.

\begin{figure}[h] 
    \centering \includegraphics[width=0.7\linewidth]{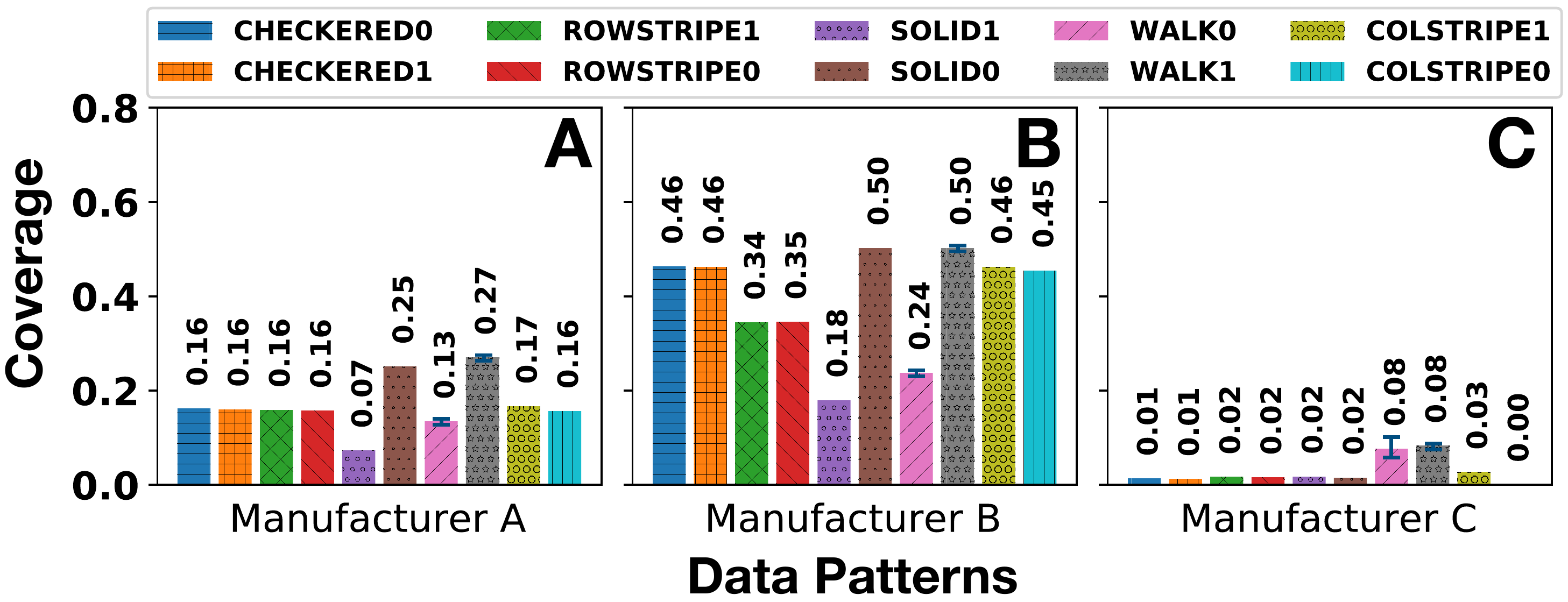} 
    \caption{Data pattern dependence of DRAM cells prone to activation failure over 100 iterations} 
    \label{fig:dpd_coverage}
\end{figure}

Figure~\ref{fig:dpd_coverage} plots the \mpy{ratio of activation failures
discovered by a particular data pattern after 100 iterations of
Algorithm~\ref{dlrng:alg:testing_lat} relative to the \emph{total} number of
failures discovered by \emph{all} patterns for a representative chip from each
manufacturer}. \mpy{We call this metric \emph{coverage} because it
indicates the effectiveness of a single data pattern to identify all possible
DRAM cells that are prone to activation failure. We show results for each
pattern individually except for the WALK1 and WALK0 patterns, for which we
show the mean (bar) and minimum/maximum (error bars) coverage across all 16
iterations of each walking pattern.}

\mpy{We make three key observations from this experiment.} \mpy{First}, we
find that testing with different data patterns identifies different subsets of
the total set of possible activation failures.  This indicates that \jk{1)
different data patterns cause different DRAM cells to fail and 2) specific
data patterns induce more activation failures than others. Thus,} certain data
patterns may extract more entropy from a DRAM cell array than other data
patterns. \mpy{Second}, we find that, of \emph{all} 40 tested data patterns,
each of the 16 \emph{walking 1s}, for a given device, provides a \jk{similarly
high} coverage, regardless of the manufacturer. This high coverage is
similarly provided by only one other data pattern per manufacturer: solid 0s
for manufacturers A and B, and walking 0s for manufacturer C. \mpy{Third, if
we repeat this experiment (i.e., Figure~\ref{fig:dpd_coverage}) while varying
the number of iterations of Algorithm~\ref{dlrng:alg:testing_lat}, the \emph{total
failure count} across all data patterns \emph{increases} as we increase the
number of iterations of Algorithm~\ref{dlrng:alg:testing_lat}.} This indicates that
\mpy{not all DRAM cells fail deterministically when accessed with a reduced
$t_{RCD}$, providing a potential source of entropy for random number
generation.}

We next analyze \mpy{each cell's probability of failing when accessed with a
reduced $t_{RCD}$ (i.e., its  \emph{activation failure probability}) to
determine which data pattern most effectively identifies cells that provide
high entropy}. We note that DRAM cells with an activation failure probability
$F_{prob}$ of 50\% provide high entropy when accessed many times. With the
same data used to produce Figure~\ref{fig:dpd_coverage}, we study the
different data patterns with regard to the number of cells they cause to fail
50\% of the time. Interestingly, we find that the data pattern that
\jkfour{induces} the most failures overall does not necessarily find the most
number of cells that fail 50\% of the time. In fact, when searching for cells
with an $F_{prob}$ between 40\% and 60\%, \jk{we observe that} the data
\mpy{patterns that find} the \jk{highest number of cells} \mpy{are} solid 0s,
checkered 0s, and solid 0s for manufacturers A, B, and C, respectively.  We
conclude that: 1) due to manufacturing and design variation across DRAM
devices from different manufacturers, different data patterns \mpy{result in}
different failure probabilities in our DRAM devices, and 2) to provide high
entropy \mpy{when} accessing DRAM cells with a reduced $t_{RCD}$, we should
use the respective data pattern that finds the most number of cells with an
$F_{prob}$ of 50\% for DRAM devices from a given manufacturer.

Unless otherwise stated, in the rest of this chapter, we use \mpy{the} solid 0s,
checkered 0s, and solid 0s data patterns for manufacturers A, B, and C,
respectively, to analyze $F_{prob}$ at the granularity of a single cell
\mpy{and} to study the effects of temperature and time on our sources of
entropy.

\subsection{Temperature Effects} 
\label{subsec:temp_effects}

\begin{figure}[h] 
    \centering \includegraphics[width=0.7\linewidth]{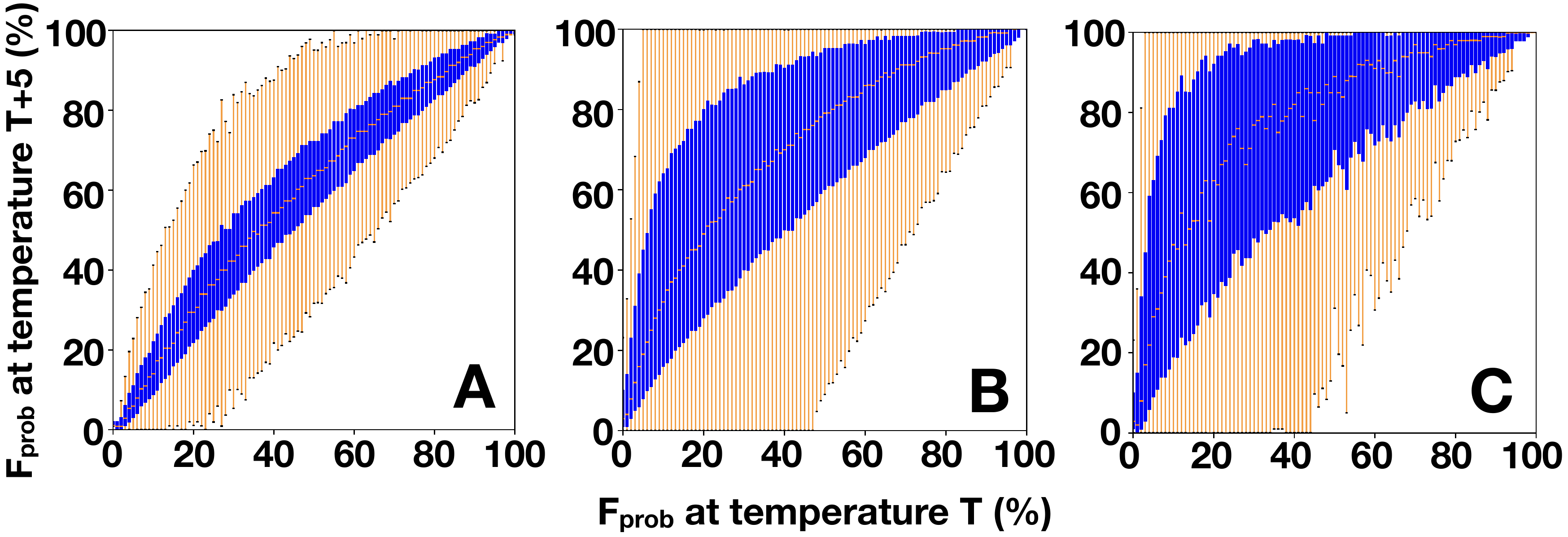} 
    \caption{Effect of temperature variation on failure \hhf{probability}} 
    \label{fig:temperature_effects} 
\end{figure}

In this section, we study whether temperature \mpy{fluctuations affect} a DRAM
cell's activation failure probability and thus the entropy that can be
extracted from the DRAM cell. To analyze \mpy{temperature effects}, we record
the $F_{prob}$ of cells \mpy{throughout} our DRAM devices \mpy{across 100
iterations of Algorithm~\ref{dlrng:alg:testing_lat} at 5$^{\circ}$C increments}
between 55$^{\circ}$C and 70$^{\circ}$). Figure~\ref{fig:temperature_effects}
aggregates \mpy{results} across 30 DRAM modules from each DRAM manufacturer.
\jk{Each point in the figure represents how the $F_{prob}$ of a DRAM cell
changes as the temperature changes (i.e., ${\Delta}F_{prob}$). The x-axis
shows the $F_{prob}$ of a single cell at temperature $T$ (i.e., the baseline
temperature), and the y-axis shows the $F_{prob}$ of the same cell at
temperature $T+5$ (i.e., 5$^{\circ}$C above the baseline temperature).}
Because we test each cell at each temperature across 100 iterations, the
granularity \mpy{of} $F_{prob}$ on both the x- and y-axes is 1\%. For a given
$F_{prob}$ at temperature $T$ \jk{(x\% on the x-axis)}, we aggregate
\emph{all} respective $F_{prob}$ points at temperature $T+5$ \jk{(y\% on the
y-axis)} with box-and-whiskers plots\footnote{A box-and-whiskers plot
emphasizes the important metrics of a dataset's distribution. The box is
lower-bounded by the first quartile (i.e., the median of the first half of the
ordered set of data points) and upper-bounded by the third quartile (i.e., the
median of the second half of the ordered set of data points). The median falls
within the box. The \emph{inter-quartile range} (IQR) is the distance between
the first and third quartiles (i.e., box size).  Whiskers extend an additional
$1.5 \times IQR$ on either sides of the box.  We indicate outliers, or data
points outside of the range of the whiskers, with pluses.} to show how the
given $F_{prob}$ is affected by the increased DRAM temperature.  The
\emph{box} is drawn in blue and contains the \emph{median} drawn in red.  The
\emph{whiskers} are drawn in gray, and the \emph{outliers} are indicated with
orange pluses. 

We observe that $F_{prob}$ at temperature $T+5$ tends to be higher than
$F_{prob}$ at temperature $T$\jkfour{,} \mpy{as shown by} the blue region of the figure
(i.e., the boxes of the box-and-whiskers plots) \mpy{lying} above the $x~=~y$
line. However, \mpy{fewer} than 25\% of all data points fall below the $x~=~y$
line, indicating that a portion of cells have a lower $F_{prob}$ as
temperature is increased.

We observe that DRAM devices from different manufacturers are affected by
temperature differently. DRAM cells of manufacturer A have the \emph{least}
variation of ${\Delta}F_{prob}$ when temperature is increased since the boxes
of the box-and-whiskers plots are strongly correlated with the $x~=~y$ line.
\jk{How a DRAM cell's activation failure probability changes in DRAM devices
from \emph{other} manufacturers is unfortunately \emph{less} predictable
under} temperature change (i.e., a DRAM cell from manufacturers B or C has
higher variation in $F_{prob}$ change), but \mpy{the data still shows a strong
positive correlation between temperature and $F_{prob}$}.  We conclude that
temperature affects \jk{cell failure probability ($F_{prob}$)} to different
degrees depending on the manufacturer of the DRAM device, \mpy{but increasing
temperature generally increases the activation failure probability.}

\subsection{Entropy Variation over Time}
\label{subsec:short_term_variation}

To determine whether the failure probability of a DRAM cell \jk{changes over
time}, we complete 250 \emph{rounds} of recording the activation failure
probability of DRAM cells over the span of 15 days.  Each round consists of
accessing every cell in DRAM 100 times with a reduced $t_{RCD}$ value and
recording the failure probability for each individual cell (out of 100
iterations). We \mpy{find} that a DRAM cell's activation failure probability
does \emph{not} change significantly over time. This means that, \mpy{once we
identify} a DRAM cell that exhibits high entropy, we can rely on the cell to
maintain its high entropy over time.  We hypothesize that this is because 
\jkfour{a DRAM cell fails with high entropy} when process manufacturing
variation in peripheral and DRAM cell circuit elements combine such that, when
we read the cell using a reduced $t_{RCD}$ value, we induce a metastable state
resulting from the cell voltage falling between the reliable sensing margins
(i.e., falling close to $\frac{V_{dd}}{2}$)~\cite{chang2016understanding}.
Since manufacturing variation is fully determined at manufacturing time, a
\jkfour{DRAM cell's activation failure probability} is stable over time given
the same experimental conditions. In Section~\ref{subsec:cell_selection}, we
discuss our \mpy{methodology} for selecting DRAM cells for extracting stable
entropy, such that we can preemptively avoid longer-term aging effects that we
do not study in this chapter.

\delete{\subsection{Effects of Reduced $t_{RCD}$ on DRAM Writes}} 
\delete{We rely on inducing activation failures on DRAM cells having the same data
pattern each time. This means that between every reduced $t_{RCD}$ access,
we must restore, i.e., rewrite, the data pattern in the recently accessed DRAM
word.  Therefore, we would like to study the effects that a reduced $t_{RCD}$
has on DRAM write accesses to see whether we need to reset $t_{RCD}$ to its
default value before restoring the data pattern in the DRAM cells. To test
whether DRAM writes are affected by a reduced $t_{RCD}$ value, we run two
experiments with our set of DRAM devices. First, we sweep the value of
$t_{RCD}$ between $10ns$ and $18ns$ and write a known data pattern across DRAM. We
then subsequently check the DRAM array for unexpected data and we observe no
failures in this range of $t_{RCD}$. } 
\delete{In our second experiment, we would like to verify that a shortened write
command does not result in side-effects such as lower voltage levels in the
written DRAM cells. Thus, for a range of $t_{RCD}$ values, we run the following
test.  For a $t_{RCD}$ value, we write a known data pattern into the DRAM
array, and disable DRAM refresh, so the charge in every cell will drain over
time. We then compare the failure rates across the DRAM array for a set
interval with refresh disabled. We note highly similar failure rates regardless
of the value of $t_{RCD}$ and conclude that writing to DRAM with a
significantly reduced $t_{RCD}$ has no impact on the quality of the writes.
This enables DRAM to service application writes between reduced $t_{RCD}$
accesses for true random number generation to minimize system interference.} 


\section{\mechanism: A DRAM-based TRNG} 
\label{section:DLRNG_mechanism} 

\jkfour{Based on} our rigorous analysis of \mpy{DRAM} activation failures
(presented in Section~\ref{sec:dlrng_characterization}), we propose \mechanism,
a flexible \mpy{mechanism} that provides high-throughput DRAM-based true random
number \mpy{generation} (TRNG) by sourcing entropy from a subset of DRAM cells
\jk{and is built fully within the memory controller.} \mechanism~is \mpy{based
on the \textbf{key observation}} that DRAM cells fail \mpy{probabilistically}
when accessed with reduced DRAM timing parameters\jk{, and this probabilistic
failure mechanism can be used as a source of true random numbers}.  While there
are many other timing parameters that we could \mpy{reduce} to induce failures
in DRAM~\cite{chang2016understanding, lee-sigmetrics2017, lee2015adaptive,
kim2018solar, lee2016reducing, chang2017thesis}, we focus specifically on
reducing $t_{RCD}$ below manufacturer-recommended values to study the resulting
activation failures.\footnote{We believe that reducing other timing parameters
could be used to generate true random values, but we leave their exploration to
future work.}

Activation failures occur as a result of reading the value from a DRAM cell
\emph{too soon} after sense amplification. This results in reading the value
at the sense amplifiers before the bitline voltage is amplified to an
\jk{I/O-readable} voltage level. The probability of reading incorrect data
from the DRAM cell therefore depends largely \jk{on the bitline's voltage} at
the time of reading the sense amplifiers. Because there is significant process
variation across the DRAM cells and I/O
circuitry~\cite{chang2016understanding, lee-sigmetrics2017, lee2015adaptive,
kim2018solar}, we observe a wide \mpy{variety} of failure probabilities for
different DRAM cells (as discussed in
Section~\ref{sec:dlrng_characterization}) for a given $t_{RCD}$ value\mpy{,
ranging from 0\% probability to 100\% probability}.

\textbf{We discover that in each DRAM chip, a subset of cells fail at
\mpy{$\jkfour{\sim}$}50\% probability, and a subset of these cells fail
randomly with high entropy (shown in Section~\ref{subsec:diff_chips}).} In this
section, we first discuss our method of identifying such cells, \jk{which} we
refer to as \emph{RNG cells} (in Section~\ref{subsec:cell_selection}). Second,
we describe the mechanism with which D-RaNGe \emph{samples} RNG cells to
extract random data (Section~\ref{subsec:sampling_rng_cells}). Finally, we
discuss a potential design for integrating D-RaNGe \jk{in} a full system
(Section~\ref{subsec:full_system_integration}).

\subsection{RNG Cell Identification} 
\label{subsec:cell_selection} 

Prior to generating random data, we must first identify cells that are capable
of producing truly random output (i.e., RNG cells).  Our process of
identifying RNG cells involves reading every cell in the DRAM array 1000 times
with a \emph{reduced} $t_{RCD}$ and approximating each cell's Shannon
entropy~\cite{shannon1948mathematical} by counting the occurrences of 3-bit
symbols across its 1000-bit stream. We identify cells that \mpy{generate an
approximately equal} number of every \mpy{possible} 3-bit symbol ($\pm10\%$ of
the number of expected symbols) as RNG cells. 

We find that \mpy{RNG cells provide unbiased output, meaning that a}
post-processing step (described in Section~\ref{subsec:trng}) is \emph{not}
necessary \mpy{to provide sufficiently high entropy for random number
generation}. We also find that RNG cells \emph{maintain high entropy across
system reboots}. In order to account for our observation that entropy from an
RNG cell changes depending on the DRAM temperature
\jk{(Section~\ref{subsec:temp_effects})}, we identify reliable RNG cells at
each temperature and store their locations in the memory controller. Depending
on the DRAM temperature \jk{at the time an application requests random
values,} D-RaNGe samples the appropriate RNG cells. To ensure that DRAM aging
does not negatively impact the reliability of RNG cells, we require
re-identifying the set of RNG cells at regular intervals. From our observation
that entropy does not change significantly over a tested 15 day period of
sampling RNG cells
\jk{(Section~\ref{subsec:short_term_variation})}, we expect the interval of
re-identifying RNG cells to be at least 15 days long.  Our RNG cell
identification process is effective at identifying cells that are reliable
entropy sources for random number generation, and we quantify their randomness
using the NIST test suite for randomness~\cite{rukhin2001statistical} in
Section~\ref{subsec:nist}. 

\subsection{Sampling RNG Cells for Random Data} 
\label{subsec:sampling_rng_cells} 

Given the availability of these RNG cells, we use our observations in
Section~\ref{sec:dlrng_characterization} to design a high-throughput TRNG that
quickly and repeatedly samples RNG cells with reduced DRAM timing parameters.
Algorithm~\ref{alg:lat_rng} demonstrates the key components of \mechanism~that
enable us to generate random numbers with high throughput. 
    \begin{algorithm}[tbh]\footnotesize
        \SetAlgoNlRelativeSize{0.7}
        \SetAlgoNoLine
        \DontPrintSemicolon
        \SetAlCapHSkip{0pt}
        \caption{\mechanism:~A DRAM-based TRNG} 
        \label{alg:lat_rng}

        \textbf{\mechanism($num\_bits$):} \textcolor{gray}{~~~// \emph{num\_bits}: number of random bits requested} \par 
        ~~~~$DP$: a known data pattern that results in high entropy \par 
        ~~~~select 2 DRAM words with RNG cells in distinct rows in each bank \par
        ~~~~write $DP$ to chosen DRAM words and their neighboring cells \par
        ~~~~get exclusive access to rows of chosen DRAM words and nearby cells \par 
        ~~~~set low $t_{RCD}$ for DRAM ranks containing chosen DRAM words \par 
        ~~~~\textbf{for} each bank: \par 
        ~~~~~~~~read data in $DW_1$ \textcolor{gray}{~~// induce activation failure} \par 
        ~~~~~~~~write the read value of $DW_1$'s RNG cells to $bitstream$ \par
        ~~~~~~~~write original data value back into $DW_1$ \par 
        ~~~~~~~~memory barrier \textcolor{gray}{~~~~// ensure completion of write to $DW_1$} \par 
        ~~~~~~~~read data in $DW_2$ \textcolor{gray}{~// induce activation failure} \par 
        ~~~~~~~~write the read value of $DW_2$'s RNG cells to $bitstream$ \par 
        ~~~~~~~~write original data value back into $DW_2$ \par 
        ~~~~~~~~memory barrier \textcolor{gray}{~~~// ensure completion of write to $DW_2$} \par 
        ~~~~~~~~\textbf{if} $bitstream_{size}$$~\geq~num\_bits$: \par 
        ~~~~~~~~~~~~break \par 
        ~~~~set default $t_{RCD}$ for DRAM ranks of the chosen DRAM words \par 
        ~~~~release exclusive access to rows of chosen words and nearby cells \par 
    \end{algorithm}
\mechanism~ takes in $num\_bits$ as an argument, which is defined as the
number of random bits desired (Line~1). \mechanism~then prepares to generate
random numbers in Lines~2-6 by first selecting DRAM words (i.e., the
granularity at which \jk{a DRAM module} is accessed) containing known RNG cells
for generating random data (Line~3).  To maximize \jkfour{the} throughput of
random number generation, \mechanism~\mpy{chooses} DRAM words with the highest
density of RNG cells in each bank (to exploit DRAM parallelism). Since each
DRAM access can induce activation failures \emph{only} in the accessed DRAM
word, the density of RNG cells per DRAM word determines the number of random
bits D-RaNGe can generate per access.    
For each available DRAM bank, \mechanism~selects two DRAM words (in
distinct DRAM rows) containing RNG cells. The purpose of selecting two DRAM
words in \emph{different} rows is to \emph{repeatedly} cause \emph{bank conflicts}, or
\jk{issue} requests to \emph{closed} DRAM rows so that every read request will
\emph{immediately} follow an activation. This is done by alternating accesses
to the chosen DRAM words in different DRAM rows. After selecting DRAM words
for generating random values, \mechanism~ writes a known data pattern that
results in high entropy to each chosen DRAM word and its neighboring cells
(Line~4) and gains exclusive access to rows containing the two chosen DRAM
words as well as their neighboring cells (Line~5).\footnote{Ensuring
exclusive access to DRAM rows can be done by remapping rows to 1) redundant
DRAM rows or 2) buffers in the memory controller so that these rows are hidden
from the \jkfour{system software} and only accessible by the memory controller for
generating random numbers. }  This ensures that the data pattern surrounding
the RNG cell and the original value of the RNG cell stay constant prior to each
access such that the failure probability of each RNG cell remains reliable (as
observed to be necessary in Section~\ref{subsec:dpd}).  To begin generating
random data (i.e., sampling RNG cells), \mechanism~reduces the value of
$t_{RCD}$ (Line~6).  From every available bank \jk{(Line~7)},
\mechanism~generates random values in parallel (Lines~8-15).  Lines 8 and 12
indicate the commands to alternate accesses to two DRAM words in distinct rows
of a bank to both 1) induce activation failures and 2) precharge the
\hhtwo{recently-accessed} row.  After inducing activation failures in a DRAM
word, \mechanism~extracts the value of the RNG cells within the DRAM word
(Lines~9 and 13) to use as random data and restores the DRAM word to its
original data value (Lines~10 and 14) to maintain the original data pattern.
Line~15 ensures that writing the original data value is complete before
attempting to \jk{sample} the DRAM words again. Lines~16 and 17 simply end the
loop if enough random bits of data have been harvested.  Line~18 sets the
$t_{RCD}$ timing parameter back to its default value, so other applications can
access DRAM without corrupting data. Line~19 releases exclusive access to the
rows containing the chosen DRAM words and their neighboring rows. 

We find that this methodology maximizes the opportunity for activation failures
in DRAM, thereby maximizing the rate of generating random data from RNG
cells.

\subsection{Full System Integration} 
\label{subsec:full_system_integration} 

In this work, we focus on developing a flexible substrate for sampling RNG
cells fully from within the memory controller. D-RaNGe generates random
numbers using a simple firmware routine running entirely within the memory
controller.  The firmware executes the sampling algorithm
(Algorithm~\ref{alg:lat_rng}) whenever \mpx{an application requests random
samples and there is available DRAM} bandwidth (i.e., DRAM is not servicing other
\jk{requests} or maintenance commands). In order to minimize latency between
requests for samples and their \mpx{corresponding} responses, a small queue of
already-harvested random data may be maintained in the memory controller for
use by the system. Overall performance overhead can be minimized by tuning
\mpx{both 1) the queue size and 2) how the memory
controller prioritizes requests for random numbers relative to normal memory
requests}.

In order to integrate \mechanism~with the rest of the system, the system
designer needs to decide how to best expose an interface by which an
application can \jk{leverage \mechanism~to generate true random numbers} on
their system. There are many ways to achieve this, including, but not limited
to:
\begin{itemize}
\item \mpx{Providing a simple \texttt{REQUEST} and \texttt{RECEIVE} interface
for applications to request and receive the random numbers using memory-mapped
configuration status registers (CSRs)~\cite{wolrich2004mapping} or other existing I/O
datapaths (e.g., x86 \texttt{IN} and \texttt{OUT} opcodes, Local Advanced
Programmable Interrupt Controller (LAPIC configuration \cite{intel32intel}).}



\item \mpx{Adding a new ISA instruction (e.g., Intel
\texttt{RDRAND}~\cite{hamburg2012analysis}) that retrieves random numbers from
the memory controller and stores them into processor registers.}
\end{itemize} 
The operating system may then expose \mpx{one or more of these interfaces to
user applications through standard kernel-user interfaces (e.g., system calls,
file I/O, operating system APIs). The system designer has complete freedom to
choose between these (and other) mechanisms that expose an interface for user
applications to interact with D-RaNGe. We expect that the best option will be
\hhtwo{system specific}, depending both on the desired D-RaNGe use cases and the ease
with which the design can be implemented.}



\section{\mechanism~Evaluation} 
\label{dlrng:sec:evaluation} 

We evaluate \mpy{three key aspects of} \mechanism. \mpy{First, we} show that
the random data \jkthree{obtained from RNG cells identified by D-RaNGe}
\mpy{passes} all of the tests in the NIST test suite for randomness
(Section~\ref{subsec:nist}). Second, we analyze the
\jkthree{existence} of RNG cells across 59 LPDDR4 and 4 DDR3 DRAM chips (due to long
testing time) randomly sampled from the overall population of DRAM chips across
all three major DRAM manufacturers (Section~\ref{subsec:diff_chips}). Third, we
evaluate \mechanism~in terms of the six key properties of an ideal TRNG as
explained in Section~\ref{sec:motivation}
(Section~\ref{subsec:trng_key_char_eval}). 


\subsection{NIST Tests} 
\label{subsec:nist} 


First, we identify RNG cells using our RNG cell \jkthree{identification}
process (Section~\ref{subsec:cell_selection}). Second, we sample \jkfour{\emph{each}}
identified RNG \hhf{cell} \jkthree{one million times} to generate large amounts of
random data (\mpy{i.e.,} 1~Mb \emph{bitstreams}). Third, we evaluate the
entropy of the bitstreams from the identified RNG cells with the NIST test
suite for randomness~\cite{rukhin2001statistical}. Table~\ref{Tab:NIST} shows
the average results of 236 1~Mb bitstreams\footnote{We test data obtained from
4 RNG cells from each of 59 DRAM chips, to maintain a reasonable NIST testing
time and \jkfour{thus} show that RNG cells across all \jkfour{tested} DRAM
chips reliably generate random values.} across the 15 tests of the full NIST
\begin{table}[h!]
\footnotesize
\begin{center}
\begin{tabular}{ |c||c|c|c }
\cline{1-3}
\textbf{NIST Test Name} & \textbf{P-value} & \textbf{Status} \\
\cline{1-3}
\hhline{|=|=|=|}
monobit & 									0.675  & PASS \\ 
frequency\_within\_block & 					0.096  & PASS \\ 
runs & 										0.501  & PASS \\ 
longest\_run\_ones\_in\_a\_block & 			0.256  & PASS \\ 
binary\_matrix\_rank & 						0.914  & PASS \\ 
dft & 										0.424  & PASS \\ 
non\_overlapping\_template\_matching &	 	>0.999 & PASS \\ 
overlapping\_template\_matching & 			0.624  & PASS \\ 
maurers\_universal & 						0.999  & PASS \\ 
linear\_complexity & 						0.663  & PASS \\ 
serial & 									0.405  & PASS \\ 
approximate\_entropy & 						0.735  & PASS \\ 
cumulative\_sums & 							0.588  & PASS \\ 
random\_excursion & 						0.200  & PASS \\ 
random\_excursion\_variant & 				0.066  & PASS \\ 
\cline{1-3}
\end{tabular}
\caption{\hhf{\mechanism\ results} with NIST randomness test suite. \vspace{-10pt}}
\label{Tab:NIST} 
\end{center}
\end{table}
test suite for randomness. P-values are calculated for each test,\footnote{A
p-value close to 1 indicates that we must accept the null hypothesis, while a
p-value close to 0 and below a small threshold, e.g., $\alpha=0.0001$
(\jkthree{recommended by the NIST Statistical Test Suite
documentation~\cite{rukhin2001statistical}}), indicates that we must reject the
null hypothesis.} where the null hypothesis for each test is that a perfect
random number generator would \emph{not} have produced random data with
\emph{better} characteristics for the given test than the tested
sequence~\cite{marton2015interpretation}. Since the resulting P-values for each
test in the suite are greater than our chosen level of significance,
$\alpha=0.0001$, we accept our null hypothesis for each test. We note that all
236 bitstreams pass all 15 tests with similar P-values.  Given our
$\alpha=0.0001$, our proportion of passing sequences (1.0) falls within the
range of acceptable proportions of sequences that pass each test
(\jkthree{[0.998,1]} calculated by \mpy{the} NIST statistical test suite using
$(1-\alpha)\pm3\sqrt{\frac{\alpha(1-\alpha)}{k}}$, where $k$ is the number of
tested sequences). This \emph{strongly} indicates that \mechanism~can generate
unpredictable, \mpy{truly} random values. Using the proportion of 1s and 0s
generated from each RNG cell, we calculate \mpy{Shannon}
entropy~\cite{shannon1948mathematical} and find the \emph{minimum} entropy
across all RNG cells to be 0.9507.

\subsection{RNG Cell Distribution} 
\label{subsec:diff_chips} 

\begin{figure}[h] 
    \centering \includegraphics[width=0.7\linewidth]{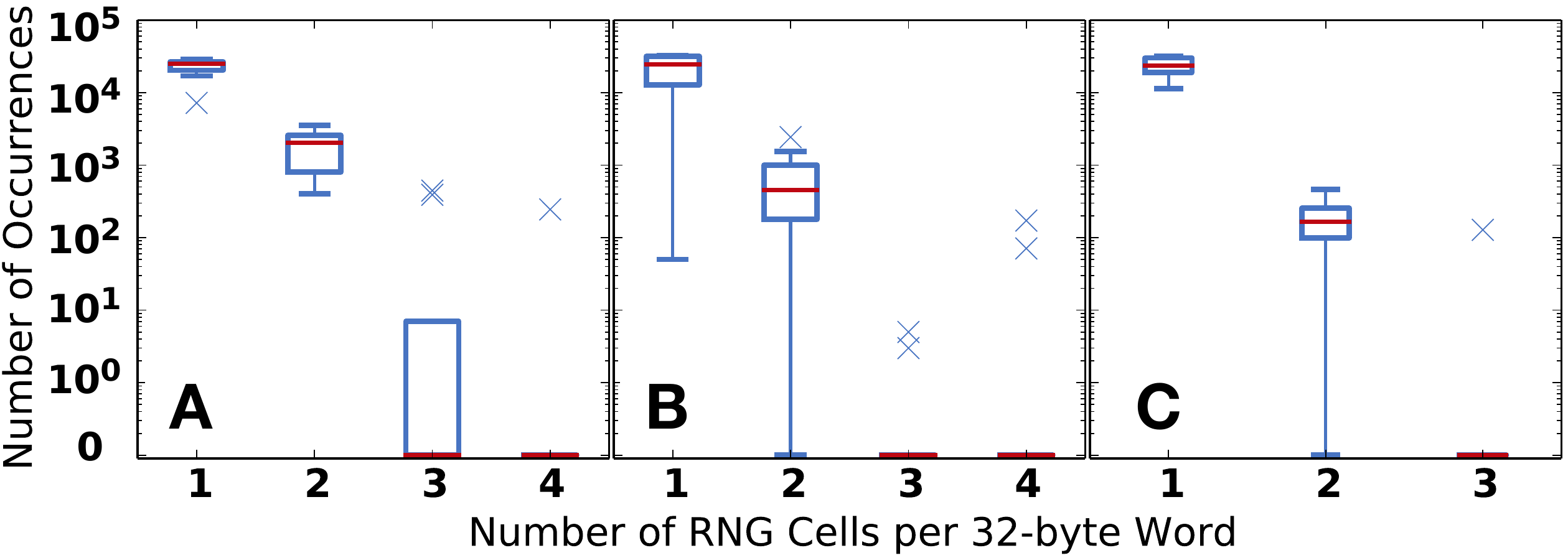} 
    \caption{Density of RNG cells in DRAM words per bank.} 
    \label{fig:bits_per_word} 
\end{figure}

The throughput at which \mechanism~generates random numbers is a function of
the 1) density of RNG cells per DRAM word and 2) bandwidth \jkthree{with which we
can access} DRAM words when using our methodology for inducing activation
failures. Since each DRAM access can induce activation failures \emph{only} in
the accessed DRAM word, the density of RNG cells per DRAM word indicates the
number of random bits D-RaNGe can sample per access.  We first study the
density of RNG cells per word across DRAM \jkthree{chips}.
Figure~\ref{fig:bits_per_word} plots the distribution of the number of words
containing x RNG cells (\jkthree{indicated by the value on the} x-axis) per
\emph{bank} across 472 banks from 59 DRAM devices from \jkthree{all manufacturers}.
The distribution is presented as a box-and-whiskers plot where the y-axis
\jkthree{has} a logarithmic scale with a 0 point. The three plots respectively
show the distributions for DRAM devices from the three manufacturers
(indicated at the bottom left corner of each plot). 

We make three key observations. First, RNG cells are \emph{widely available in
every bank} across many chips. This means that we can use the available DRAM
access parallelism that multiple banks offer and sample RNG cells from each
DRAM bank in parallel to improve random number generation throughput.  Second,
\emph{every} bank that we \mpy{analyze} has \emph{multiple DRAM words}
containing at least one RNG cell. The DRAM bank with the smallest
\mpy{occurrence} of RNG cells has 100 DRAM words containing only 1 RNG cell
(manufacturer B). Discounting this point, the distribution of the number of
DRAM words containing only 1 RNG cell is tight with a high number of RNG cells
(e.g., tens of thousands) in each bank, regardless of the manufacturer. Given
our random sample of DRAM chips, we expect that the existence of RNG cells in
DRAM banks will hold true for all DRAM chips. Third, we observe that a single
DRAM word can contain as many as 4 RNG cells. Because the throughput of
accesses to DRAM is fixed, the number of RNG cells in the accessed words
essentially acts as a multiplier for the throughput of random numbers
generated (e.g., accessing DRAM words containing 4 RNG cells results in
\emph{4x} the throughput of random numbers compared to accessing DRAM words
containing 1 RNG cell).

\subsection{TRNG Key Characteristics Evaluation} 
\label{subsec:trng_key_char_eval} 

We now evaluate \mechanism~in terms of the six key properties of an effective
TRNG as explained in Section~\ref{sec:motivation}. 

\textbf{Low Implementation Cost.} To induce activation failures, we must be
able to reduce the DRAM timing parameters below manufacturer-specified values.
Because memory controllers issue memory accesses according to the timing
parameters specified in a set of internal registers, D-RaNGe requires
\jkfour{simple} software support to be able to programmatically modify the
memory controller's registers.  Fortunately, there exist some
processors~\cite{lee2015adaptive, opteron_amd, bkdg_amd2013, ram_overclock}
that \emph{already} enable software to directly change memory controller
register values, i.e., the DRAM timing parameters. These processors can easily
generate random numbers with \mechanism. 

All other processors that do \emph{not} currently support direct changes to
memory controller registers require \emph{minimal} software changes to expose
an interface for changing the memory controller
registers~\cite{ARM2016memrefman, samsung2014refman, hassan2017softmc,
softmc-safarigithub}. To \jkthree{enable} a more efficient implementation, the
memory controller could be programmed such that it issues DRAM accesses with
distinct timing parameters on a per-access granularity to reduce the overhead
in 1) changing the DRAM timing parameters and 2) allow concurrent DRAM
accesses by other applications. In the rare case where these registers are
unmodifiable by even the hardware, the hardware changes necessary to enable
register modification \mpy{are} minimal and \mpy{are} simple to
implement~\cite{lee2015adaptive, hassan2017softmc, softmc-safarigithub}.

We experimentally find that we can induce activation failures \mpy{with}
$t_{RCD}$ between $6ns$ and $13ns$ (\mpy{reduced} from the default of $18ns$). Given
this wide range of failure-inducing $t_{RCD}$ values, most memory controllers
should be able to adjust their timing parameter registers to a value within
this range.

\textbf{Fully Non-deterministic.} As we have shown in
Section~\ref{subsec:nist}, the bitstreams extracted from the
\jkthree{D-RaNGe-identified} RNG cells pass \emph{all} 15 NIST tests\jkthree{. We have}
full reason to believe that we are inducing a metastable state of the sense
amplifiers (as hypothesized by~\cite{chang2016understanding}) such that we are
effectively sampling random physical phenomena to extract unpredictable random
values.  

\begin{figure}[h]
\centering \includegraphics[width=0.7\linewidth]{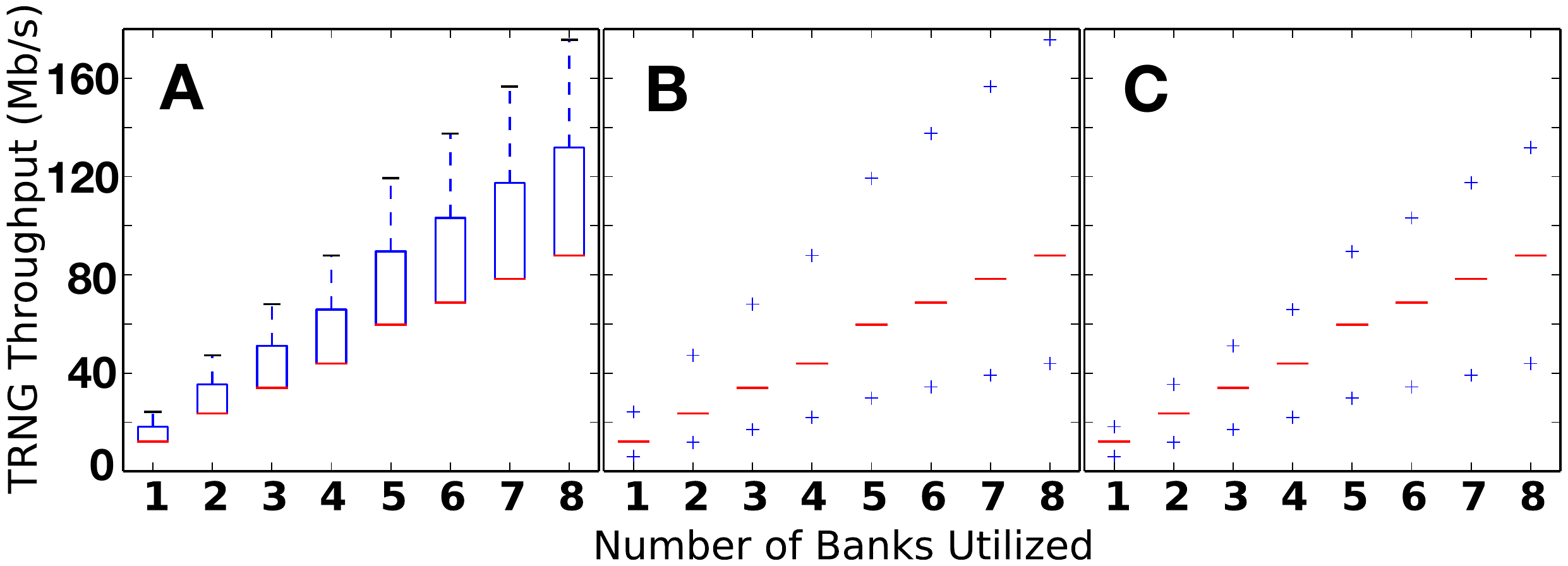} 
\caption{Distribution of TRNG throughput across chips.} 
\label{fig:trng_throughput}
\end{figure}

\textbf{High Throughput of Random Data.} Due to the various use cases of random
number generation discussed in Section~\ref{sec:motivation}, different
applications have different throughput requirements for random number
generation, and applications may tolerate a reduction in performance so
that \mechanism~can quickly generate true random numbers. Fortunately,
\mechanism~provides flexibility to tradeoff between the \textit{system
interference} it causes, i.e., the slowdown experienced by concurrently running
applications, and the random number generation throughput it provides. To
demonstrate this flexibility, Figure~\ref{fig:trng_throughput} plots the TRNG
throughput of \mechanism~when using varying numbers of banks (\emph{x} banks on
the x-axis) across the three DRAM manufacturers (indicated at the top left
corner of each plot). For each number of banks used, we plot the distribution
of TRNG throughput that we observe \emph{real} DRAM devices to provide\jkthree{. The
available density of RNG cells in a DRAM device (provided in
Figure~\ref{fig:bits_per_word}) dictates the TRNG throughput that the DRAM
device can provide. We plot each distribution} as
a box-and-whiskers plot. For each number of banks used, we select x
banks with the greatest sum of RNG cells across each banks' two DRAM words with
the highest density of RNG cells (that are \emph{not} in the same DRAM row).
We select two DRAM words per bank because we must alternate accesses between
two DRAM rows (as shown in \jkfour{Lines 8 and 12} of Algorithm~\ref{alg:lat_rng}).
The sum of the RNG cells available across the two selected DRAM words for each
bank is considered each bank's \emph{TRNG data rate}, and we use this value to
obtain \mechanism's throughput. We use Ramulator~\cite{ramulatorgithub,
kim2016ramulator} to obtain the rate at which we can execute the core loop of
Algorithm~\ref{alg:lat_rng} with varying numbers of banks. We obtain the random
number generation throughput for x banks with the following equation:   \vspace{-4pt}
\begin{equation} \label{eq:x_bank_throughput}
TRNG\_Throughput_{x\_Banks} = \sum_{n=1}^{x} \frac{TRNG\_data\_rate_{Bank\_n}}{Alg2\_Runtime_{x\_banks}} 
\end{equation} 
where $TRNG\_data\_rate_{Bank\_n}$ is the TRNG data rate for the selected bank,
and $Alg2\_Runtime_{x\_banks}$ is the runtime of the core loop of
Algorithm~\ref{alg:lat_rng} when using $x$ Banks. We note that because we
observe small variation in the density of RNG cells per word (between 0 and 4),
we see that \jkthree{TRNG throughput across different chips is} generally very similar.
For this reason, we see that the box and whiskers are condensed into a single
point for distributions of manufacturers B and C.  We find that when
\emph{fully} using \emph{all} 8 banks in a single DRAM channel, every
device can produce \emph{at least} 40 Mb/s of random data regardless of
manufacturer.  The highest throughput we observe from devices of manufacturers
A/B/C respectively are \changes{179.4/134.5/179.4 Mb/s}. On average, across all
manufacturers, we find that \mechanism~can provide a throughput of
\changes{108.9 Mb/s.} 

We draw two key conclusions. First, due to the \emph{parallelism} of multiple
banks, the throughput of random number generation increases linearly as we
use more banks. Second, there is variation of TRNG throughput across
different DRAM devices, but the medians across manufacturers are very 
similar. 

We note that any throughput sample point on this figure can be multiplied by
the number of available channels in a memory hierarchy for a better TRNG
throughput estimate for a system with multiple DRAM channels.  For an example 
memory hierarchy comprised of 4 DRAM channels, \mechanism~results in a maximum
(average) throughput of 717.4 Mb/s (435.7 Mb/s). 

\textbf{Low Latency.} \jkfour{Since D-RaNGe's sampling mechanism consists of a
single DRAM access, the latency of generating random values is directly related
to the DRAM access latency. Using the timing parameters specified in the JEDEC
LPDDR4 specification~\cite{2014lpddr4}, we calculate D-RaNGe's latency to
generate a 64-bit random value. To \hhf{calculate the} \emph{maximum latency} for D-RaNGe,
we assume that 1) each DRAM access provides only 1 bit of random data (i.e.,
each DRAM word contains \emph{only} 1 RNG cell) and 2) we can use only a single
bank within a single channel to generate random data. We find that D-RaNGe can
generate 64 bits of random data with a \emph{maximum} latency of $960 ns$. If
D-RaNGe takes full advantage of DRAM's \hhf{channel- and bank-level}
parallelism in a system with 4 DRAM channels \jkfive{and 8 banks per channel,} D-RaNGe
can generate 64 bits of random data by issuing 16 DRAM accesses per channel in
parallel. This results in a latency of $220 ns$.  To \hhf{calculate the}
\emph{empirical minimum latency} for D-RaNGe, we fully parallelize D-RaNGe
\hhf{across banks in all 4 channels} while \hhf{also} assuming that each DRAM
access provides 4 bits of random data, since we find a maximum density of 4 RNG
cells per DRAM word in the LPDDR4 DRAM devices that we characterize
(Figure~\ref{fig:bits_per_word}).  We find the empirical minimum latency to be
\emph{only} $100 ns$ in our tested devices.}


\textbf{Low System Interference.} The flexibility of using a different number
of banks across the available channels in a system's memory hierarchy allows
\mechanism~to \jkthree{cause} varying levels of system interference at the
expense of TRNG throughput. \jkfour{This enables application developers to
generate random values with D-RaNGe at varying tradeoff points} depending on
the running applications' \jkfour{memory access} requirements. We analyze
\mechanism's system interference with respect to DRAM storage overhead and DRAM
latency. 

In terms of storage overhead, \mechanism~simply requires exclusive access
rights to \jkfour{six DRAM rows} per bank, consisting of the two rows
containing the RNG cells and each row's two \jkfour{physically-adjacent} DRAM
rows containing the chosen data pattern.\footnote{As in prior
work~\cite{kim2014flipping, bains2015method, mutlu2017rowhammer}, we argue that
manufacturers can disclose \jkfour{which rows are physically adjacent to each
other.}} This results in an insignificant 0.018\% DRAM storage overhead cost. 

\jkthree{To evaluate} \mechanism's effect on \mpy{the} DRAM \jkthree{access
latency of regular memory requests}, we present one implementation of
\mechanism. For a single DRAM channel, which is the granularity at which DRAM
timing parameters are applied, \mechanism~can alternate between using a
reduced $t_{RCD}$ and the default $t_{RCD}$. When using a reduced $t_{RCD}$,
\mechanism~\mpy{generates} random numbers across every bank in the channel. On
the other hand, when using the default $t_{RCD}$, memory requests from running
applications \mpy{are} serviced to ensure application progress.  The length of
these time intervals (with default/reduced $t_{RCD}$) can both be adjusted
according to the applications' random number \jkthree{generation}
requirements.  Overall,
\mechanism~\jkthree{provides} significant flexibility in \jkthree{trading off} its system
overhead \jkthree{with its TRNG} throughput. \jkthree{However,} it is
up to the system designer to use and exploit the flexibility for their
requirements.  To show the potential throughput of \mechanism~without impacting
\jkthree{concurrently-running} applications, we run simulations with
\jkfour{the SPEC CPU2006}~\cite{spec2006} workloads, and calculate the idle DRAM bandwidth
available that we can use to issue D-RaNGe commands. We find \mpy{that,}
across all workloads, we can obtain an average (maximum, minimum) random-value
throughput of 83.1 (98.3, 49.1) Mb/s \mpy{with \emph{no} significant impact on
overall system performance}. 


\textbf{Low Energy Consumption.} To evaluate the energy consumption of D-RaNGe,
we use DRAMPower~\cite{drampowergithub} to analyze the output traces of
Ramulator~\cite{ramulatorgithub, kim2016ramulator} when DRAM is (1)
generating random numbers (Algorithm~\ref{alg:lat_rng}), and (2) idling and not
servicing memory requests. We subtract quantity (2) from (1) to
obtain the estimated energy consumption of D-RaNGe. We then divide the value by
the total number of random bits found during execution and find that, on
average, D-RaNGe finds random bits at the cost of 4.4 nJ/bit.


\section{Comparison with Prior DRAM TRNGs} 
\label{comparison}

To our knowledge, D-RaNGe is the highest-throughput TRNG \hh{\emph{for
commodity DRAM devices}} that works by exploiting activation failures as a
sampling mechanism for observing entropy in DRAM cells. There are a number of
proposals to construct TRNGs using commodity DRAM devices, which \jkthree{we
summarize} in Table~\ref{tab:prior_dram_works} based on their entropy sources.
In this section, we compare each of these works with \mechanism. \jkthree{We}
show how \mechanism\ fulfills the six key properties of an ideal TRNG
(Section~\ref{sec:motivation}) better than any prior DRAM-based TRNG proposal.
We group our comparisons by the entropy source of each prior DRAM-based TRNG
proposal.

\begin{table*}[h!]
\begin{adjustwidth}{-1.2cm}{}
\tiny 
\begin{center}
\begin{tabular}{ |c||c|c|c|c|c|c|c| }
\hline 
      \textbf{Proposal}
	& \textbf{Year}
	& \begin{tabular}{@{}c@{}}\textbf{Entropy} \\ \textbf{Source}\end{tabular} 
	& \begin{tabular}{@{}c@{}}\textbf{True} \\ \textbf{Random}\end{tabular} 
	& \begin{tabular}{@{}c@{}}\textbf{Streaming} \\ \textbf{Capable}\end{tabular} 
	& \begin{tabular}{@{}c@{}}\textbf{\hhf{64-bit TRNG}} \\ \textbf{Latency}\end{tabular} 
	& \begin{tabular}{@{}c@{}}\textbf{Energy} \\ \textbf{Consumption}\end{tabular} 
	& \begin{tabular}{@{}c@{}}\textbf{Peak} \\ \textbf{Throughput}\end{tabular} \\ 
\hline \hline
Pyo+~\cite{pyo2009dram}
	& 2009
	& Command Schedule
	& \xmark  
	& \cmark  
    & $18{\mu}s$ 
    & N/A 
	& $3.40 Mb/s$ \\
\hline
Keller+~\cite{keller2014dynamic}          
	& 2014
	& Data Retention
	& \cmark  
	& \cmark  
    & $40 s$
    & $6.8mJ/bit$
	& $0.05 Mb/s$  \\
\hline
Tehranipoor+~\cite{tehranipoor2016robust} 
	& 2016
	& Startup Values
	& \cmark  
	& \xmark  
    & $>60ns$ \jkfour{(optimistic)} 
    & $>245.9 pJ/bit$ \jkfour{(optimistic)} 
	& N/A  \\
\hline
Sutar+~\cite{sutar2018d}                  
	& 2018
	& Data Retention
	& \cmark  
	& \cmark  
    & $40 s$ 
    & $6.8 mJ/bit$
	& $0.05 Mb/s$ \\
\hline
\textbf{\mechanism} 
	& 2018
	& Activation Failures
	& \cmark  
	& \cmark  
    & $100 ns < x < 960 ns$ 
    & $4.4 nJ/bit$
	& $717.4 Mb/s$ \\
\hline
\end{tabular}
\caption{Comparison to previous DRAM-based TRNG proposals. \vspace{-10pt}} 
\label{tab:prior_dram_works}
\end{center}
 \end{adjustwidth}
\end{table*}

\subsection{DRAM Command Scheduling} 

Prior work~\cite{pyo2009dram} proposes using non-determinism in DRAM command
scheduling for true random number generation. In particular, since pending
access commands contend with regular refresh operations, the latency of a DRAM
access is hard to predict and is useful for random number generation. 

Unfortunately, this method fails to satisfy two important properties of an
ideal TRNG. First, it harvests random numbers from the instruction and DRAM
command scheduling decisions made by the processor and memory controller, which
does \emph{not} constitute a fully non-deterministic entropy source. Since the
quality of the harvested random numbers depends directly on the quality of the
processor and memory controller implementations, the entropy source is visible
to and potentially modifiable by an adversary \hh{(e.g., by simultaneously
running a memory-intensive workload on another processor
core~\cite{moscibroda2007memory})}. Therefore, this method does not meet our
design goals as it does not securely generate random numbers. 

Second, although this technique has a higher throughput than those based on
DRAM data retention (Table~\ref{tab:prior_dram_works}), \mechanism\ still
outperforms this method in terms of throughput by 211x (maximum) and 128x
(average) because a single byte of random data requires a \emph{significant}
amount of time to generate. Even if we scale the throughput results provided
by~\cite{pyo2009dram} to a modern day system (e.g., $5 GHz$ processor, 4 DRAM
channels\footnote{The authors do not provide their DRAM configuration, so we
optimistically assume that they evaluate their proposal using one DRAM channel.
We also assume that by utilizing 4 DRAM channels, the authors can harvest four
times the entropy, which gives the benefit of the doubt
to~\cite{pyo2009dram}.}), the theoretical maximum throughput of \jkthree{Pyo et
al.'s} approach\footnote{We base our estimations on \cite{pyo2009dram}'s claim
that they can harvest one byte of random data every 45000 cycles. However,
using these numbers along with the authors' stated processor configuration
(i.e., $2.8 GHz$) leads to a discrepancy between our calculated maximum
throughput ($\approx 0.5 Mb/s$) and that reported in~\cite{pyo2009dram}
($\approx 5 Mb/s$). We believe our estimation methodology and calculations are
sound. In our work, we compare \mechanism's peak throughput against that of
\cite{pyo2009dram} using a more modern system configuration (i.e., $5 GHz$
processor, 4 DRAM channels) than used in the original work, which gives the
benefit of the doubt to~\cite{pyo2009dram}.} is \emph{only} $3.40 Mb/s$ as
compared with the maximum (average) throughput of $717.4 Mb/s$ ($435.7 Mb/s$)
for \mechanism. To calculate the latency of generating random values, we assume
the same system configuration with \hh{\cite{pyo2009dram}'s} claimed number of
cycles \hh{45000} to generate random bits. To provide 64 bits of random data,
\cite{pyo2009dram} takes 18$\mu$s, which is significantly higher than D-RaNGe's
\jkthree{minimum/maximum latency of $100ns/960ns$}. Energy
consumption for ~\cite{pyo2009dram} depends heavily on the entire system that
it is running on, so we do not compare against this \hh{metric}.


\subsection{DRAM Data Retention} 

Prior works~\cite{keller2014dynamic, sutar2018d} propose using DRAM data
retention failures to generate random numbers. Unfortunately, this approach is
\emph{inherently too slow} for high-throughput operation due to the long wait
times required to induce \jkthree{data retention failures in DRAM}. While the
failure rate can be increased by increasing the operating temperature, a wait
time on the order of seconds is required to induce \jkthree{enough}
failures~\cite{liu2013experimental, khan2014efficacy, qureshi2015avatar,
patel2017reaper, kim2018dram} to achieve \jkfour{high-throughput} random number
generation, which is orders of \jkfour{magnitude} slower than D-RaNGe. 

Sutar et al.~\cite{sutar2018d} report that they are able to generate {256-bit}
random numbers using a hashing algorithm (e.g., {SHA-256}) on a $4~MiB$ DRAM
block that contains \mpo{data retention errors resulting from having disabled
DRAM refresh} for 40~seconds. \mpf{Optimistically assuming a large DRAM
capacity of $32~GiB$ and ignoring the time required to read out and hash the
erroneous data, a waiting time of 40~seconds to induce data retention errors
allows for an estimated maximum random number throughput of $0.05~Mb/s$. This
throughput is already far \hh{smaller} than \mechanism's measured maximum
throughput of \changes{$717.4 Mb/s$}, and it would decrease linearly with
DRAM capacity.  Even if we were able to induce a large number of
data retention errors by waiting only 1 second, the \hh{maximum} random number
generation throughput would be $2~Mb/s$, i.e., orders of \jkfour{magnitude} 
\hh{smaller} than \jkthree{that of} \mechanism.}

Because \cite{sutar2018d} requires a wait time of 40~seconds before
producing any random values, its latency for random number generation is
extremely high \jkfour{(40s)}. \hh{In contrast,} D-RaNGe can produce random values
very quickly since it generates random values \jkthree{potentially with} each DRAM access
(10s of nanoseconds). D-RaNGe therefore has a latency many orders of \jkfour{magnitude} 
lower than \hh{Sutar et al.'s mechanism~\cite{sutar2018d}.} 

We estimate the energy consumption of retention-time \jkthree{based TRNG} mechanisms
with Ramulator\hh{~\cite{kim2016ramulator, ramulatorgithub}} and
DRAMPower\hh{~\cite{drampowergithub, chandrasekar2011improved}}. We model first
\mpy{writing data to} a 4MiB DRAM \jkfour{region} (to constrain the energy consumption
estimate to the region of interest), waiting for 40 seconds, and then reading
from that region. We then divide the energy consumption of these operations by
the number of bits found (256 bits). \hh{We find that the energy consumption
is} around $6.8 mJ$ per bit\hh{, which} is orders of \jkfour{magnitude} more costly
than \jkthree{that of} \mechanism, which provides random \jkthree{numbers} at
$4.4 nJ$ per bit.

\subsection{DRAM Startup Values} 

Prior work~\cite{tehranipoor2016robust, eckert2017drng} proposes using DRAM
startup values as random numbers. Unfortunately, this method is unsuitable for
continuous high-throughput operation since it requires a DRAM power cycle in
order to obtain random data. We are unable to accurately model the latency of
this mechanism since it relies on the startup time of DRAM (i.e., bus frequency
\hh{calibration}, temperature \hh{calibration}, timing register
initialization~\cite{ddr4operationhynix}). This \jkthree{heavily depends} on
the implementation of the system and \jkfour{the} DRAM device in use. Ignoring
these components, \hh{we estimate the throughput of generating random numbers
using startup values by taking into account only the latency of a single} DRAM
read (\emph{after} all initialization is complete), which \jkthree{is $60ns$}.
We model energy consumption ignoring the initialization phase as well, by
modeling the energy to read a MiB of DRAM and \hh{dividing} that quantity by
\cite{tehranipoor2016robust}'s claimed number of random bits found in that
region (420Kbit). \hh{Based on this calculation, we estimate energy consumption
as} $245.9pJ$ per bit. While \hh{the energy consumption of
\cite{tehranipoor2016robust}} is smaller than the energy cost of \mechanism, we
note that \hh{our energy estimation for \cite{tehranipoor2016robust}} does
\emph{not} account for the energy consumption required for initializing DRAM to
be able to read out the random values.  Additionally,
~\cite{tehranipoor2016robust} requires a full system reboot which is often
impractical for applications and for effectively providing a \emph{steady
stream of random values}. \cite{eckert2017drng} suffers from the same issues
since it uses the same mechanism as \cite{tehranipoor2016robust} to
generate random numbers and is strictly worse since \cite{eckert2017drng}
results in $31.8x$ less entropy. 

\subsection{Combining DRAM-based TRNGs} 

We note that \mechanism's method for sampling random values from DRAM is
entirely distinct from prior DRAM-based TRNGs that we have discussed in this
section. This makes it possible to combine \mechanism~with prior work to
produce random values at an even higher throughput.

\section{Other Related Works} 
\label{dlrng:related}


In this work, we focus on the design of a DRAM-based hardware mechanism to
implement a TRNG, which makes the focus of our work orthogonal to those that
design PRNGs. In contrast to prior DRAM-based TRNGs discussed in
Section~\ref{comparison}, we propose using \emph{activation failures} as an
entropy source. Prior works characterize activation failures in order to
exploit the resulting error patterns for overall DRAM latency
reduction~\hh{\cite{chang2016understanding, kim2018solar, lee-sigmetrics2017,
lee2015adaptive}} and to implement physical unclonable functions
(PUFs)~\cite{kim2018dram}.  However, none of these works measure the randomness
inherent in activation failures or propose using them to generate random
numbers.  

\hh{Many} TRNG designs have been proposed that exploit sources of entropy
\mpf{that are \emph{not}} based on DRAM. \mpf{Unfortunately, these proposals
either 1) require custom hardware modifications that preclude their application
to commodity devices, or 2) do not sustain continuous (i.e., constant-rate)
high-throughput operation}. We briefly discuss different entropy sources with
examples. 

\textbf{Flash Memory Read Noise.} Prior proposals use random telegraph noise in
flash memory devices as an entropy source (up to 1 Mbit/s)~\cite{wang2012flash,
ray2018true}. Unfortunately, flash memory is orders of magnitude slower than
DRAM, making flash unsuitable for high-throughput \jkthree{and low-latency}
operation. \jky{A more recent work~\cite{chakraborty2020true} demonstrates that
random numbers can be extracted from the variability of write and erase latency
in flash memory devices.} \jkx{However this technique suffers from low
throughput (i.e., up to 0.25 Kb/s) as it depends on long flash memory
latencies.} 

\textbf{SRAM-based Designs.} SRAM-based TRNG designs exploit
randomness in startup values \cite{ 
	  holcomb2007initial, holcomb2009power, 
	van2012efficient}. 
Unfortunately, these proposals are unsuitable for continuous, high-throughput
operation since they require a power cycle.

\textbf{GPU- and FPGA-Based Designs.}  Several works harvest random numbers
from GPU-based (up to 447.83 Mbit/s)~\cite{chan2011true, tzeng2008parallel,
teh2015gpus} and FPGA-based (up to 12.5 Mbit/s)~\jky{\cite{majzoobi2011fpga,
wieczorek2014fpga, chu1999design, hata2012fpga, fischer2002true}} entropy
sources. \hh{These} \mpf{proposals do not require modifications to commodity
GPUs or FPGAs\hh{.  Yet,} GPUs and FPGAs are not as prevalent as DRAM in
commodity devices today.}

\textbf{Custom Hardware.} Various works propose TRNGs based in part or fully on
non-determinism \jkthree{provided by custom hardware designs} (\jkfour{with
TRNG throughput} up to 2.4 Gbit/s)~\cite{ 
	amaki2015oscillator, 
	yang2016all, 
	bucci2003high, 
	bhargava2015robust, 
	petrie2000noise, 
	mathew20122, brederlow2006low, tokunaga2008true, 
	kinniment2002design, holleman20083, holcomb2009power, 
	pareschi2006fast, 
	stefanov2000optical}.  
Unfortunately, \jkthree{the need for custom hardware limits the widespread use of such
proposals in} commodity hardware devices (today).
\delete{\textbf{System-Level RNGs.} Operating systems often provide interfaces
for harvesting entropy from devices running on a system.  Unfortunately, the
entropy source of a system-level RNG depends on 1) the particular devices
attached to the system (e.g., I/O peripherals, disk drives) and 2) the quality
of the drivers that directly harvest entropy from these devices. This means
that a system-level RNG is not guaranteed to be harvesting random numbers from
a fully non-deterministic entropy source; therefore, system-level RNGs do not
meet our design goals.}

\section{\jky{Limitations}} 

\jkz{While D-RaNGe cam be immediately deployed in certain available systems
today, D-RaNGe has limitations in implementation that limits its full
potential even in these systems.} 

\jkz{First, D-RaNGe requires a flexible memory controller \jky{that has} the
ability to issue DRAM commands with varying timing parameters on the fly.
Without such a memory controller, D-RaNGe is limited to existing systems with
tunable timing parameters and may have high overhead in switching latencies for
D-RaNGe accesses and regular memory accesses. We believe that a flexible memory
controller has many use cases for \jky{the system, including Solar-DRAM, the
DRAM Latency PUF,} and D-RaNGe, and we expect future work to develop such a
memory controller.} 

\jkz{Second, D-RaNGe requires an intelligent memory access scheduler that can
interleave D-RaNGe accesses with regular accesses. The scheduler \jky{may} have
to predict and account for many parameters including 1) future memory idle
time, 2) priority levels for D-RaNGe accesses and concurrently running
applications, 3) anticipated random number \jky{throughput and latency}
requirements, and 4) amount of random numbers saved in the buffer. Developing
such a memory controller would enable systems to satisfy random number
requirements of running applications with minimal interference from D-RaNGe.}

\section{Summary} 
\label{sec:conclusion} 

We propose \mechanism, a mechanism for extracting true random numbers with high
throughput from unmodified commodity DRAM devices on any system that allows
manipulation of DRAM timing parameters in the memory controller.  \mechanism\
harvests fully non-deterministic random numbers from DRAM row activation
failures, which are \jk{bit errors} induced by intentionally \jk{accessing DRAM
with lower latency than required for correct row activation.} Our TRNG is based
on two key observations: 1) activation failures can be induced quickly and 2)
repeatedly accessing certain DRAM cells with reduced activation latency results
in reading true random data. We validate the quality of our TRNG with the
commonly-used NIST statistical test suite for randomness. Our evaluations show
that \mechanism\ significantly outperforms the previous highest-throughput
DRAM-based TRNG by up to 211x \jkthree{(128x on average)}. We conclude that
DRAM row activation failures can be effectively exploited to efficiently
generate true random numbers with high throughput on a wide range of devices
that use commodity DRAM chips.

\chapter{Revisiting RowHammer: An Experimental Analysis of Modern Devices and Mitigation Techniques} 
\label{ch6-rh} 

RowHammer is a circuit-level DRAM vulnerability, first rigorously analyzed
and introduced in 2014, where repeatedly accessing data in a DRAM row can cause
bit flips in nearby rows. The RowHammer vulnerability has since garnered
significant interest in both computer architecture and computer security
research communities because it stems from physical circuit-level interference
effects that worsen with continued DRAM density scaling. As
DRAM manufacturers primarily depend on density scaling to increase DRAM
capacity, future DRAM chips will likely be more vulnerable to RowHammer than
those of the past. Many RowHammer mitigation mechanisms have been proposed by
both industry and academia, but it is unclear whether these mechanisms will
remain viable solutions for future devices, as their overheads increase
with DRAM's vulnerability to RowHammer. 

In order to shed more light on how RowHammer affects modern and future devices
at the circuit-level, we first present an experimental characterization of
RowHammer on 1580 DRAM chips (408$\times$ DDR3, 652$\times$ DDR4, and
520$\times$ LPDDR4) from 300 DRAM modules (60$\times$ DDR3, 110$\times$
DDR4, and 130$\times$ LPDDR4) with RowHammer protection mechanisms disabled,
spanning multiple different technology nodes from across each of the three
major DRAM manufacturers.  Our studies definitively show that newer DRAM chips
are more vulnerable to RowHammer: as device feature size reduces, the
number of activations needed to induce a RowHammer bit flip also reduces, to as
few as $9.6k$ (4.8k to two rows each) in the most vulnerable chip we tested.  

We evaluate five state-of-the-art RowHammer mitigation mechanisms using
cycle-accurate simulation in the context of real data taken from our chips to
study how the mitigation mechanisms scale with chip vulnerability. We find
that existing mechanisms either are not scalable or suffer from prohibitively
large performance overheads in projected future devices
given our observed trends of RowHammer vulnerability. Thus, it is critical to
research more effective solutions to RowHammer. 

\section{RowHammer: DRAM Disturbance Errors} 

Modern DRAM devices suffer from \emph{disturbance errors} that
occur when a high rate of accesses to a single DRAM row unintentionally
flip the values of cells in nearby rows. This phenomenon is known as
\emph{RowHammer}~\cite{kim2014flipping}. It inherently stems from
electromagnetic interference between nearby cells. RowHammer is exacerbated by
reduction in process technology node size because adjacent DRAM cells
become both smaller and closer to each other.  Therefore, as DRAM manufacturers
continue to increase DRAM storage density, a chip's vulnerability to RowHammer
bit flips increases~\cite{kim2014flipping, mutlu2017rowhammer,
mutlu2019rowhammer}.

RowHammer exposes a system-level security vulnerability that has been studied
by many prior works both from the attack and defense perspectives. Prior
works demonstrate that RowHammer can be used to mount system-level attacks for
privilege escalation (e.g.,~\cite{cojocar2019exploiting,
gruss2018another, gruss2016rowhammer, lipp2018nethammer, qiao2016new,
razavi2016flip, seaborn2015exploiting, tatar2018defeating, van2016drammer,
xiao2016one, frigo2020trrespass, ji2019pinpoint}),
leaking confidential data (e.g.,~\cite{kwong2020rambleed}), and
denial of service (e.g.,~\cite{gruss2018another, lipp2018nethammer}). These
works effectively demonstrate that a system must provide protection against
RowHammer to ensure
robust (i.e., reliable and secure) execution.

Prior works propose defenses against RowHammer attacks both at the
hardware (e.g.,~\cite{kim2014flipping, lee2019twice, ryu2017overcoming,
son2017making, ghasempour2015armor, you2019mrloc, seyedzadeh2018cbt,
kang2020cat, kim2014architectural, gomez2016dram, bains2015rowref, bains14d, bains14c, greenfield14b, bains2016row, bains2015row, hassan2019crow, fisch2017dram}) and
software (e.g.,~\cite{rh-apple, aweke2016anvil, brasser2016can,
kim2014flipping, konoth2018zebram, li2019detecting, van2018guardion, rh-lenovo,
rh-hp, irazoqui2016mascat, bu2018srasa, rh-cisco, wu2019protecting,
bock2019rip, kim2019effective, wang2019reinforce, chakraborty2019deep,
wang2019detect}) levels. DRAM manufacturers themselves employ in-DRAM RowHammer
prevention mechanisms such as \emph{Target Row Refresh (TRR)}~\cite{ddr4},
which internally performs proprietary operations to reduce the vulnerability of
a DRAM chip against potential RowHammer attacks, although these solutions have
been recently shown to be vulnerable~\cite{frigo2020trrespass}. Memory
controller and system manufacturers have also included defenses such as
increasing the refresh rate~\cite{rh-apple, aweke2016anvil, rh-lenovo} and
Hardware RHP~\cite{intel2017cannon, omron2019ny, tq2020tq,
versalogic2019blackbird}.  For a detailed survey of the RowHammer problem, its
underlying causes, characteristics, exploits building on it, and mitigation
techniques, we refer the reader to~\cite{mutlu2019rowhammer}.

\section{Motivation and Goal} 
\label{rh:sec:motivation}



Despite the considerable research effort expended towards understanding and
mitigating RowHammer, scientific literature still lacks rigorous
experimental data on how the RowHammer vulnerability is changing with the
advancement of DRAM designs and process technologies. In general, important
practical concerns are difficult to address with existing data in
literature. For example: 
\begin{itemize}
\item How vulnerable to RowHammer are future DRAM chips expected to be at the circuit level?
\item How well would RowHammer mitigation mechanisms prevent or mitigate RowHammer in future devices?
\item What types of RowHammer solutions would cope best with increased circuit-level vulnerability due to continued technology node scaling?
\end{itemize} 
While existing experimental characterization
studies~\cite{kim2014flipping, park2016statistical, park2016experiments} take
important steps towards building an overall understanding of the RowHammer
vulnerability, they are too scarce and collectively do not provide a holistic
view of RowHammer evolution into the modern day. To help overcome this lack of
understanding, we need a unifying study of the RowHammer vulnerability of a
broad range of DRAM chips spanning the time since the original RowHammer
paper was published in 2014~\cite{kim2014flipping}.


To this end, \textbf{our goal} in this chapter is to evaluate and understand
how the RowHammer vulnerability of real DRAM chips at the circuit level
changes across different chip types, manufacturers, and process technology
node generations. Doing so enables us to predict how the RowHammer
vulnerability in DRAM chips will scale as the industry continues to increase
storage density and reduce technology node size for future chip designs. To
achieve this goal, we perform a rigorous experimental characterization study of
DRAM chips from three different DRAM types (i.e., DDR3, DDR4, and LPDDR4),
three major DRAM manufacturers, and at least two different process technology
nodes from each DRAM type. We show how different chips from different DRAM
types and technology nodes (abbreviated as ``type-node'' configurations)
have varying levels of vulnerability to RowHammer. We compare the chips'
vulnerabilities against each other and
project how they will likely scale when reducing the technology node size even
further (Section~\ref{rh:sec:characterization}). Finally, we study how effective
existing RowHammer mitigation mechanisms will be, based on our observed
and projected experimental data on the RowHammer vulnerability
(Section~\ref{sec:implications}).

\section{Experimental Methodology}
\label{sec:methodology}

We describe our methodology for characterizing DRAM chips for RowHammer. 

\subsection{Testing Infrastructure}
\label{subsec:methodology:infrastructure}

In order to characterize the effects of RowHammer across a broad range of
modern DRAM chips, we experimentally study DDR3, DDR4, and LPDDR4 DRAM chips
across a wide range of testing conditions. To achieve this, we use two
different testing infrastructures: (1) the SoftMC
framework~\cite{softmc-safarigithub, hassan2017softmc} capable of testing DDR3
and DDR4 DRAM modules in a temperature-controlled chamber and (2) an in-house
temperature-controlled testing chamber capable of testing LPDDR4 DRAM chips.

\begin{figure}[h] \centering
    \includegraphics[width=0.7\linewidth]{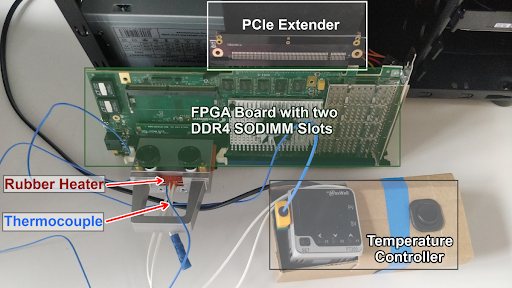}
    \caption{Our SoftMC infrastructure~\cite{softmc-safarigithub, hassan2017softmc} for testing DDR4 DRAM chips.}
    \label{fig:softmc_ddr4}
\end{figure}

\textbf{SoftMC.} Figure~\ref{fig:softmc_ddr4} shows our SoftMC setup for
testing DDR4 chips. In this setup, we use an FPGA board with a Xilinx Virtex
UltraScale 95 FPGA~\cite{xilinx_virtex_ultrascale}, two DDR4 SODIMM slots, and
a PCIe interface. To open up space around the DDR4 chips for temperature
control, we use a vertical DDR4 SODIMM riser board to plug a DDR4 module into the
FPGA board.  We heat the DDR4 chips to a target temperature using silicone
rubber heaters pressed to both sides of the DDR4 module. We control the temperature
using a thermocouple, which we place between the rubber heaters and the DDR4
chips, and a temperature controller. To enable fast data transfer between the
FPGA and a host machine, we connect the FPGA to the host machine using PCIe via a 30~cm
PCIe extender. We use the host machine to program the SoftMC hardware and
collect the test results. Our SoftMC setup for testing DDR3 chips is similar
but uses a Xilinx ML605 FPGA board~\cite{xilinx_ml605}. Both infrastructures
provide fine-grained control over the types and timings of DRAM commands sent to
the chips under test and provide precise temperature control at typical
operating conditions. 

\textbf{LPDDR4 Infrastructure.} Our LPDDR4 DRAM testing infrastructure uses
industry-developed in-house testing hardware for package-on-package LPDDR4
chips. The LPDDR4 testing infrastructure is further equipped with cooling
and heating capabilities that also provide us with precise temperature
control at typical operating conditions.  

\subsection{Characterized DRAM Chips}
\label{subsec:methodology:devices}

Table~\ref{tab:devices} summarizes the DRAM chips that we test using both
infrastructures. We have chips from all of the three major DRAM manufacturers
spanning DDR3, DDR4, and two known technology nodes of LPDDR4. We refer to the
DRAM type (e.g., LPDDR4) and technology node of a DRAM chip as a \emph{DRAM
type-node configuration} (e.g., LPDDR4-1x). For DRAM chips whose technology node we do
not exactly know, we identify their node as \emph{old} or \emph{new}.
\begin{table}[h!]
\footnotesize 
\begin{center}
\caption{Summary of DRAM chips tested.} 
\begin{tabular}{ lrrrr }
\toprule
\multicolumn{1}{c}{\textbf{DRAM}} & \multicolumn{4}{c}{\textbf{Number of Chips (Modules) Tested}} \\
\multicolumn{1}{c}{\textbf{type-node}} & \textbf{\textit{Mfr. A}} & \textbf{\textit{Mfr. B}} & \textbf{\textit{Mfr. C}} & \textbf{\textit{Total}} \\
\toprule
DDR3-old & 56 (10) & 88 (11) & 28 (7) & \textbf{172 (28)} \\
DDR3-new & 80 (10) & 52 (9) & 104 (13) & \textbf{236 (32)} \\
\Xhline{0.5\arrayrulewidth}
DDR4-old & 112 (16) & 24 (3) & 128 (18) & \textbf{264 (37)} \\
DDR4-new & 264 (43) & 16 (2) & 108 (28) & \textbf{388 (73)} \\
\Xhline{0.5\arrayrulewidth}
LPDDR4-1x & 12 (3) & 180 (45) & N/A & \textbf{192 (48)} \\
LPDDR4-1y & 184 (46) & N/A & 144 (36) & \textbf{328 (82)} \\ 
\toprule
\end{tabular}
\label{tab:devices}
\vspace{-3mm} 
\end{center}
\end{table}


\textbf{DDR3 and DDR4.} Among our tested DDR3 modules, we identify two distinct
batches of chips based on their manufacturing date, datasheet publication date,
purchase date, and RowHammer characteristics. We categorize
DDR3 devices with a manufacturing date earlier than 2014 as DDR3-old chips, and
devices with a manufacturing date including and after 2014 as DDR3-new chips.
Using the same set of properties, we identify two distinct batches of devices
among the DDR4 devices. We categorize DDR4 devices with a manufacturing date
before 2018 or a datasheet publication date of 2015 as DDR4-old chips and
devices with a manufacturing date including and after 2018 or a datasheet
publication date of 2016 or 2017 as DDR4-new chips. Based on our observations
on RowHammer characteristics from these chips, we expect that DDR3-old/DDR4-old
chips are manufactured at an older date with an older process technology
compared to DDR3-new/DDR4-new chips, respectively. This enables us to directly
study the effects of shrinking process technology node sizes in DDR3 and DDR4
DRAM chips. 

\textbf{LPDDR4.} For our LPDDR4 chips, we have two known distinct
generations manufactured with different technology node sizes, 1x-nm and 1y-nm,
where 1y-nm is smaller than 1x-nm.  Unfortunately, we are missing data from
some generations of DRAM from specific manufacturers (i.e., LPDDR4-1x from
manufacturer C and LPDDR4-1y from manufacturer B) since we did not have access
to chips of these manufacturer-technology node combinations due to
confidentiality issues. Note that while we know the external technology node
values for the chips we characterize (e.g., 1x-nm, 1y-nm), these values are
\emph{not} standardized across different DRAM manufacturers and the actual
values are confidential.  This means that a 1x chip from one manufacturer is
not necessarily manufactured with the same process technology node as a 1x chip
from another manufacturer. However, since we do know relative process node
sizes of chips from the \emph{same} manufacturer, we can directly observe how
technology node size affects RowHammer on LPDDR4 DRAM chips. 

\subsection{Effectively Characterizing RowHammer}
\label{subsec:methodology:characterizing_rowhammer}

In order to characterize RowHammer effects on our DRAM chips at the
circuit-level, we want to test our chips at the worst-case RowHammer
conditions. We identify two conditions that our tests must satisfy to
effectively characterize RowHammer at the circuit level: our testing
routines must both: 1) run without interference (e.g., without DRAM refresh or
RowHammer mitigation mechanisms) and 2) systematically test each DRAM row's
vulnerability to RowHammer by issuing the \emph{worst-case sequence of DRAM
accesses} for that particular row. 

\noindent\textbf{Disabling Sources of Interference.} To directly observe
RowHammer effects at the circuit level, we want to minimize the external
factors that may limit 1) the effectiveness of our tests or 2) our ability
to effectively characterize/observe circuit-level effects of RowHammer on our
DRAM chips.  First, we want to ensure that we have control over how our
RowHammer tests behave without disturbing the desired access pattern in any way. Therefore, during
the core loop of each RowHammer test (i.e., when activations are issued at a
high rate to induce RowHammer bit flips), we disable all DRAM
self-regulation events such as refresh and calibration, using control registers
in the memory controller. This guarantees consistent testing without
confounding factors due to intermittent events (e.g., to avoid the
possibility that a victim row is refreshed during a RowHammer test routine such
that we observe fewer RowHammer bit flips).  Second, we want to directly
observe the circuit-level bit flips such that we can make conclusions about
DRAM's vulnerability to RowHammer at the circuit technology level rather than
the system level. To this end, to the best of our knowledge, we disable all
DRAM-level (e.g., TRR~\cite{2014lpddr4, ddr4, frigo2020trrespass}) and
system-level RowHammer mitigation mechanisms (e.g.,
pTRR~\cite{aichinger2015ddr}) along with all forms of rank-level
error-correction codes (ECC), which could obscure RowHammer bit flips.
Unfortunately, all of our LPDDR4-1x and LPDDR4-1y chips use on-die
ECC~\cite{micron2017whitepaper, kwak2017a, kang2014co, patel2019understanding,
kwon2017an} (i.e., an error correcting mechanism that corrects single-bit
failures entirely within the DRAM chip~\cite{patel2019understanding}), which we
cannot disable. Third, we ensure that the core loop of our RowHammer test
runs for less than 32 ms (i.e., the lowest refresh interval specified by
manufacturers to prevent DRAM data retention failures across our tested
chips~\cite{patel2017reaper, liu2013experimental, khan2014efficacy, jedec2012,
ddr4, 2014lpddr4}) so that we do not conflate retention failures with RowHammer
bit flips. 

\noindent\textbf{Worst-case RowHammer Access Sequence.} We leverage
\emph{three} key observations from prior work~\cite{kim2014flipping,
aweke2016anvil, gruss2018another, xiao2016one, cojocar2020we} in order to craft
a worst-case RowHammer test pattern. First, a repeatedly accessed row (i.e.,
\emph{aggressor row}) has the greatest impact on its immediate
physically-adjacent rows (i.e., repeatedly accessing physical row $N$ will
cause the highest number of RowHammer bit flips in physical rows $N+1$ and
$N-1$). Second, a \emph{double-sided hammer} targeting physical victim row $N$
(i.e., repeatedly accessing physical rows $N-1$ and $N+1$) causes the
\emph{highest} number of RowHammer bit flips in row $N$ compared to any other
access pattern. Third, increasing the rate of DRAM activations (i.e., issuing
the same number of activations within shorter time periods) results in an
increasing number of RowHammer bit flips. This rate of activations is limited
by the DRAM timing parameter $t_{RC}$ (i.e., the time between two successive
activations) which depends on the DRAM clock frequency and the DRAM type: DDR3
(52.5ns)~\cite{jedec2012}, DDR4 (50ns)~\cite{ddr4}, LPDDR4
(60ns)~\cite{2014lpddr4}. Using these observations, we test each row's
worst-case vulnerability to RowHammer by repeatedly accessing the two directly
physically-adjacent rows as fast as possible.

To enable the quick identification of physical rows $N-1$ and $N+1$ for a
given row $N$, we reverse-engineer the \emph{undocumented} and
\emph{confidential} logical-to-physical DRAM-internal row address
remapping. To do this, we exploit RowHammer's key observation that
repeatedly accessing an arbitrary row causes the two directly
physically-adjacent rows to contain the \emph{highest} number of RowHammer bit
flips~\cite{kim2014flipping}. By repeating this analysis across rows throughout
the DRAM chip, we can deduce the address mappings for each type of chip that we
test. We can then use this mapping information to quickly test RowHammer
effects at worst-case conditions. We note that for our LPDDR4-1x chips from
Manufacturer B, when we repeatedly access a single row within two consecutive
rows such that the first row is an even row (e.g., rows 2 and 3) in the logical
row address space as seen by the memory controller, we observe 1) no RowHammer
bit flips in either of the two consecutive rows and 2) a near equivalent number
of RowHammer bit flips in each of the four immediately adjacent rows: the two
previous consecutive rows (e.g., rows 0 and 1) and the two subsequent
consecutive rows (e.g., rows 4 and 5).  This indicates a row address remapping
that is internal to the DRAM chip such that every pair of consecutive rows
share the same internal wordline.  To account for this DRAM-internal row
address remapping, we test each row $N$ in LPDDR4-1x chips from manufacturer B
by repeatedly accessing physical rows $N-2$ and $N+2$.

\noindent\textbf{Additional Testing Parameters.} To investigate RowHammer
characteristics, we explore two testing parameters at a stable ambient
temperature of $50^{\circ}C$:  
\begin{enumerate}
    \item \textbf{Hammer count ($HC$).} We test the effects of changing the number of times we access (i.e., activate) a victim row's physically-adjacent rows (i.e., aggressor rows). We count each pair of activations to the two neighboring rows as one \emph{hammer} (e.g., one activation each to rows $N-1$ and $N+1$ counts as one hammer). We sweep the hammer count from 2k to 150k (i.e., 4k to 300k activations) across our chips so that the hammer test runs for less than 32ms. 
    \item \textbf{Data pattern ($DP$).} We test several commonly-used DRAM data patterns where every byte is written with the same data: Solid0 (SO0: 0x00), Solid1 (SO1: 0xFF), Colstripe0 (CO0: 0x55), Colstripe1 (CO1: 0xAA)~\cite{liu2013experimental, patel2017reaper, khan2014efficacy}. In addition, we test data patterns where each byte in every other row, including the row being hammered, is written with the same data, Checkered0 (CH0: 0x55) or Rowstripe0 (RS0: 0x00), and all other rows are written with the inverse data, Checkered1 (CH1: 0xAA) or Rowstripe1 (RS1: 0xFF), respectively.
\end{enumerate} 

\noindent\textbf{RowHammer Testing Routine.}
Algorithm~\ref{alg:rowhammer_char} presents the general testing methodology we
use to characterize RowHammer on DRAM chips. For different data patterns
($DP$) (line~2) and hammer counts ($HC$) 
(line~8), the test individually targets each row in DRAM (line~4) as
a victim row (line~5).  For each victim row, we identify the two
physically-adjacent rows ($aggressor\_row1$ and $aggressor\_row2$) as aggressor
rows (lines~6 and 7).  Before beginning the core loop of our RowHammer test
(Lines 11-13), two things happen: 1) the memory controller disables DRAM
refresh (line~9) to ensure no interruptions in the core loop of our test due to
refresh operations, and 2) we refresh the victim row (line~10) so that we
begin inducing RowHammer bit flips on a fully-charged row, which ensures that
bit flips we observe are not due to retention time violations. The core loop of our
RowHammer test (Lines 11-13) induces RowHammer bit flips in the victim row
    \begin{algorithm}[tbh]\footnotesize
        \setstretch{0.6} 
        \SetAlgoNlRelativeSize{0.7}
        \SetAlgoNoLine
        \DontPrintSemicolon
        \SetAlCapHSkip{0pt}
        \caption{DRAM RowHammer Characterization}
        \label{alg:rowhammer_char}
        \textbf{DRAM\_RowHammer\_Characterization():} \par
		\quad\quad \textbf{foreach} $DP$ in [Data Patterns]: \par
        \quad\quad\quad\quad write $DP$ into all cells in $DRAM$ \par
		\quad\quad\quad\quad\textbf{foreach} $row$ in $DRAM$: \par
		\quad\quad\quad\quad\quad\quad set $victim\_row$ to $row$ \par
		\quad\quad\quad\quad\quad\quad set $aggressor\_row1$ to $victim\_row - 1$ \par
		\quad\quad\quad\quad\quad\quad set $aggressor\_row2$ to $victim\_row + 1$ \par
		\quad\quad\quad\quad\quad\quad \textbf{foreach} $HC$ in [$HC$ sweep]: \par
		\quad\quad\quad\quad\quad\quad\quad\quad Disable DRAM refresh \par
		\quad\quad\quad\quad\quad\quad\quad\quad Refresh $victim\_row$ \par
		\quad\quad\quad\quad\quad\quad\quad\quad \textbf{for} $n = 1 \rightarrow HC$: \textcolor{gray}{// core test loop} \par
		\quad\quad\quad\quad\quad\quad\quad\quad\quad\quad activate $aggressor\_row1$ \par
		\quad\quad\quad\quad\quad\quad\quad\quad\quad\quad activate $aggressor\_row2$ \par
		\quad\quad\quad\quad\quad\quad\quad\quad Enable DRAM refresh \par
        \quad\quad\quad\quad\quad\quad\quad\quad Record RowHammer bit flips to storage \par
		\quad\quad\quad\quad\quad\quad\quad\quad Restore bit flips to original values \par 
    \end{algorithm}
by first activating $aggressor\_row1$ then $aggressor\_row2$, $HC$ times. After
the core loop of our RowHammer test, we re-enable DRAM refresh (line~14) to
prevent retention failures and record the observed bit flips to secondary
storage (line~15) for analysis (presented in
Section~\ref{rh:sec:characterization}). Finally, we prepare to test the next $HC$
value in the sweep by restoring the observed bit flips to their original values
(Line 16) depending on the data pattern ($DP$) being tested. 

\textbf{Fairly Comparing Data Across Infrastructures.} Our
carefully-crafted RowHammer test routine allows us to compare our test results
between the two different testing infrastructures.
This is because, as we described earlier, we 1) reverse engineer the row address mappings of each DRAM
configuration such that we effectively test double-sided RowHammer on every
single row, 2) issue activations as fast as possible for each chip, such that
the activation rates are similar across infrastructures, and 3) disable all
sources of interference in our RowHammer tests. 

 
\section{RowHammer Characterization}
\label{rh:sec:characterization}

\newcounter{obscount}

In this section, we present our comprehensive characterization of RowHammer on
the 1580 DRAM chips we test.\footnote{We list our full set of chips in Appendix A of our
extended technical report~\cite{kim2020revisitingrh}.}

\subsection{RowHammer Vulnerability}
\label{subsec:vulnerability}

We first examine which of the chips that we test are susceptible to RowHammer.
Across all of our chips, we sweep the hammer count ($HC$) between 2K and
150K (i.e., 4k and 300k activates for our double-sided RowHammer test) and
observe whether we can induce any RowHammer bit flips at all in each chip.  We
find that we can induce RowHammer bit flips in all chips except many
DDR3 chips. Table~\ref{tab:ddr3_failing_devices} shows the fraction of DDR3
chips in which we \emph{can} induce RowHammer bit flips (i.e., \emph{RowHammerable}
chips). 

\begin{table}[h!]
\footnotesize
\begin{center}
\caption{Fraction of DDR3 DRAM chips vulnerable to RowHammer when $\bm{HC}\, \mathbf{<150k}$.} 
\begin{tabular}{ lrrr}
\toprule
\multicolumn{1}{c}{\textbf{DRAM}} & \multicolumn{3}{c}{\textbf{RowHammerable chips}} \\
\multicolumn{1}{c}{\textbf{type-node}} & \textbf{\textit{Mfr. A}} & \textbf{\textit{Mfr. B}} & \textbf{\textit{Mfr. C}} \\
\toprule
DDR3-old & 24/88 & 0/88 & 0/28 \\
DDR3-new & 8/72 & 44/52 & 96/104 \\
\toprule
\end{tabular}
\label{tab:ddr3_failing_devices}
\vspace{-3mm} 
\end{center}
\end{table}

\stepcounter{obscount} \textbf{Observation \arabic{obscount}.} \emph{Newer
DRAM chips appear to be more vulnerable to RowHammer based on the increasing
fraction of RowHammerable chips from DDR3-old to DDR3-new DRAM chips of
manufacturers B and C.} 

We find that the fraction of manufacturer A's chips that are RowHammerable
decreases from DDR3-old to DDR3-new chips, but we also note that the number of
RowHammer bit flips that we observe across each of manufacturer A's chips is
very low ($<20$ on average across RowHammerable chips) compared to the number
of bit flips found in manufacturer B and C's DDR3-new chips ($87k$ on average
across RowHammerable chips) when $HC=150K$. Since DDR3-old chips of all
manufacturers and DDR3-new chips of manufacturer A have very few to no bit
flips, we refrain from analyzing and presenting their characteristics in many
plots in Section~\ref{rh:sec:characterization}. 

%
%


\subsection{Data Pattern Dependence}
\label{rh:subsec:dpd}

To study data pattern effects on observable RowHammer bit flips, we
test our chips using Algorithm~\ref{alg:rowhammer_char} with $hammer\_count \
(HC)=150k$ at $50^{\circ}C$, sweeping the 1) $victim\_row$ and 2)
$data\_pattern$ (as described in
Section~\ref{subsec:methodology:characterizing_rowhammer}).\footnote{Note
that for a given data pattern ($DP$), the same data is always written to
$victim\_row$. For example, when testing Rowstripe0, every byte in
$victim\_row$ is always written with 0x00 and every byte in the two
physically-adjacent rows are written with 0xFF.} 

\begin{figure}[h] \centering
    \includegraphics[width=0.7\linewidth]{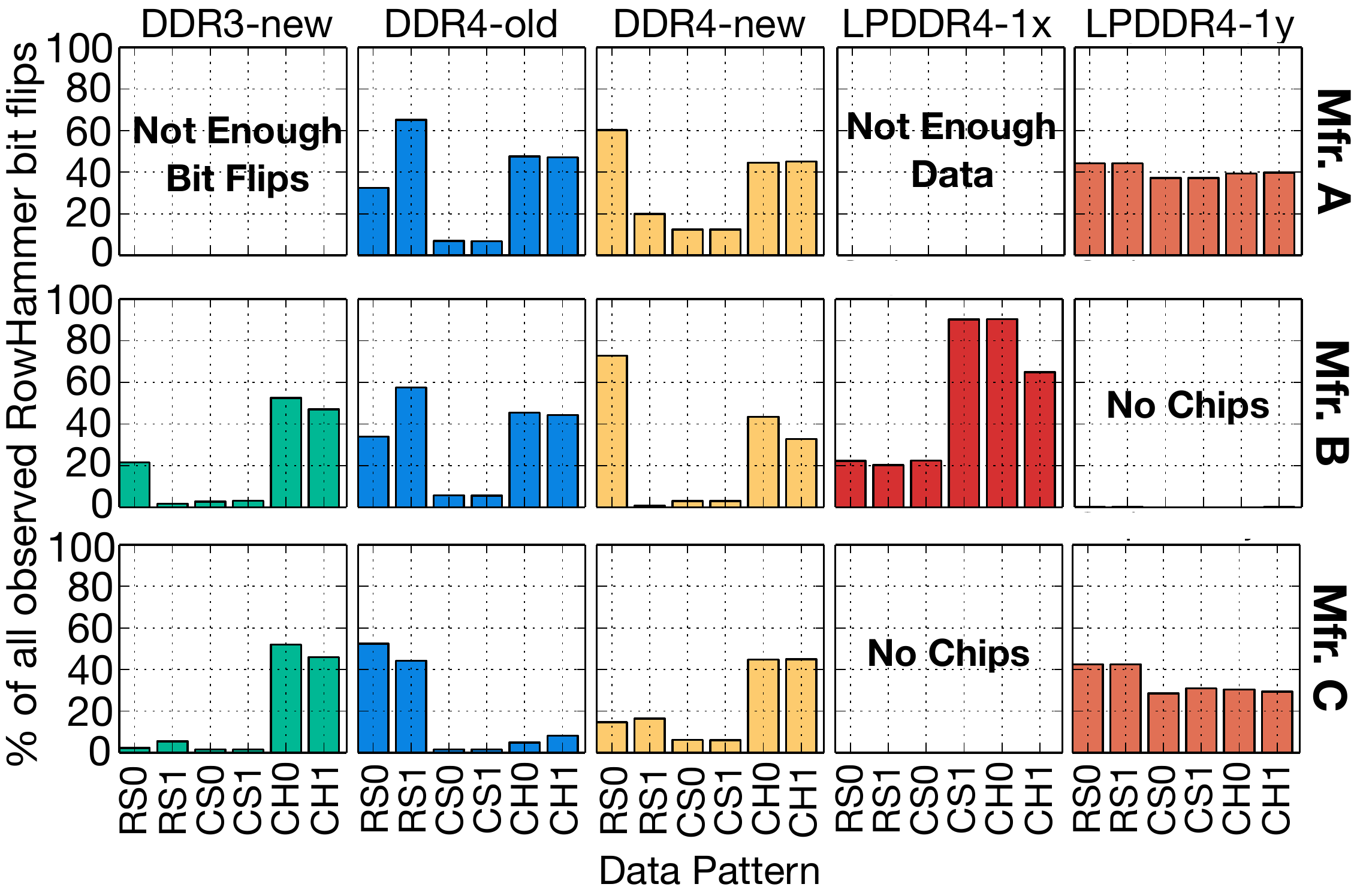}
	\caption{RowHammer bit flip coverage of different data patterns (described in Section~\ref{subsec:methodology:characterizing_rowhammer}) for a single representative DRAM chip of each type-node configuration.} 
    \label{fig:char:dpd_distribution}
\end{figure}

We first examine the set of all RowHammer bit flips that we observe when
testing with different data patterns for a given $HC$.
For each data pattern, we run our RowHammer test routine ten times. We then
aggregate all unique RowHammer bit flips per data pattern.
We combine all unique RowHammer bit flips found by all data patterns and
iterations into a full set of observable bit flips. Using the combined data,
we calculate the fraction of the full set of observable bit flips that each
data pattern identifies (i.e., the data pattern's \emph{coverage}).
Figure~\ref{fig:char:dpd_distribution} plots the coverage (y-axis) per individual data pattern (shared x-axis) for a
single representative DRAM chip from each DRAM type-node
configuration that we test. Each row of subplots shows the coverages for chips of the same
manufacturer (indicated on the right y-axis), and the columns show the
coverages for chips of the same DRAM type-node configuration (e.g., DDR3-new).

\stepcounter{obscount} \textbf{Observation \arabic{obscount}.}
\emph{Testing with different data patterns is essential for
comprehensively identifying RowHammer bit flips because no individual data
pattern achieves full coverage alone.} 

\stepcounter{obscount} \textbf{Observation \arabic{obscount}.}
\emph{The worst-case data pattern (shown in
Table~\ref{tab:dpd}) is consistent across chips of the same manufacturer and
DRAM type-node configuration.}\footnote{We do not consider the true/anti cell
pattern of a chip~\cite{liu2013experimental, kim2014flipping,
frigo2020trrespass} and agnostically program the data pattern accordingly into
the DRAM array. More RowHammer bit flips can be induced by considering the
true/anti-cell pattern of each chip and devising corresponding data patterns to
exploit this knowledge~\cite{frigo2020trrespass}.} 

\begin{table}[h]
\footnotesize
\begin{center}
\caption{Worst-case data pattern for each DRAM type-node configuration at $\mathbf{50^{\circ}C}$ split into different manufacturers.}
\begin{tabular}{ lrrr}
\toprule
\multicolumn{1}{c}{\textbf{DRAM}} & \multicolumn{3}{c}{\textbf{Worst Case Data Pattern at $\mathbf{50^{\circ}C}$}} \\
\multicolumn{1}{c}{\textbf{type-node}} & \textbf{\textit{Mfr. A}} & \textbf{\textit{Mfr. B}} & \textbf{\textit{Mfr. C}} \\
\toprule
DDR3-new & N/A & Checkered0 & Checkered0 \\
\Xhline{0.2\arrayrulewidth}
DDR4-old & RowStripe1 & RowStripe1 & RowStripe0 \\
DDR4-new & RowStripe0 & RowStripe0 & Checkered1 \\
\Xhline{0.2\arrayrulewidth}
\ap{LPDDR4-1x} & Checkered1 & Checkered0 & N/A \\
\ap{LPDDR4-1y} & RowStripe1 & N/A & RowStripe1 \\
\toprule
\end{tabular}
\label{tab:dpd}
\vspace{-3mm} 
\end{center}
\end{table}

We believe that different data patterns induce the most RowHammer bit flips in
different chips because DRAM manufacturers apply a variety of proprietary
techniques for DRAM cell layouts to maximize the cell density for different
DRAM type-node configurations. For the remainder of this
chapter, we characterize each chip using \emph{only} its worst-case data
pattern.\footnote{We use the worst-case data pattern to 1) minimize the
extensive testing time, 2) induce many RowHammer bit flips, and 3) experiment
at worst-case conditions. A diligent attacker would also try to find the
worst-case data pattern to maximize the probability of a successful
RowHammer attack.}

\subsection{Hammer Count ($HC$) Effects} 
\begin{figure}[htbp] \centering
    \includegraphics[width=0.95\linewidth]{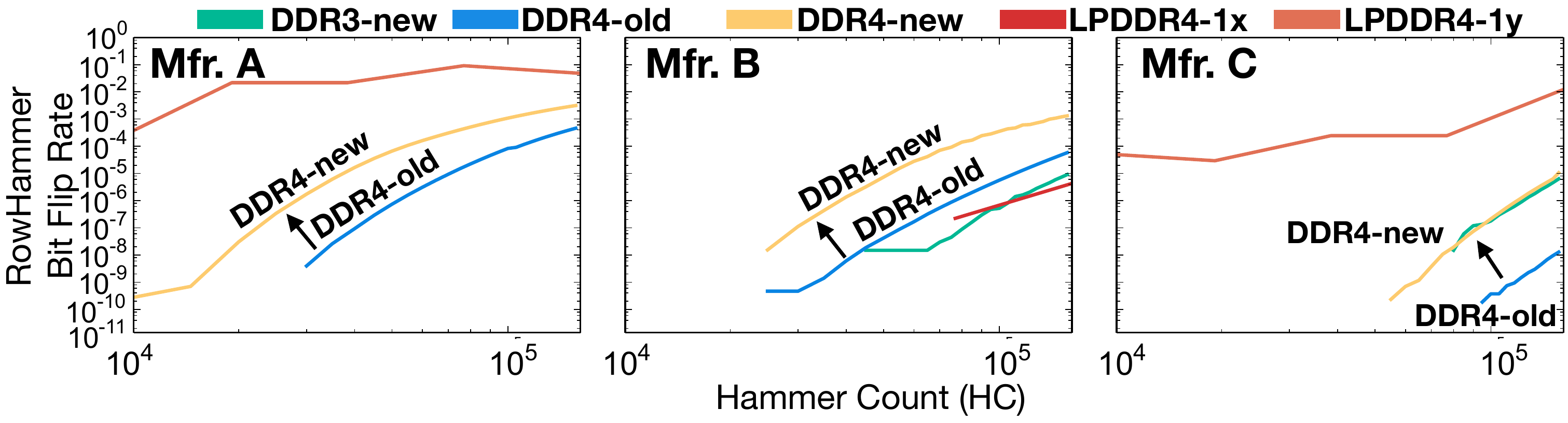}
    \caption{Hammer count ($HC$) vs. RowHammer bit flip rate across DRAM type-node configurations.} 
    \label{fig:char:refresh_sweep}
\end{figure}

We next study the effects of increasing the hammer count on the number of
observed RowHammer bit flips across our chips.
Figure~\ref{fig:char:refresh_sweep} plots the effects of increasing the number
of hammers on the RowHammer bit flip rate\footnote{We define the RowHammer bit
flip rate as the number of observed RowHammer bit flips to the total number of
bits in the tested DRAM rows.} for our tested DRAM chips of various DRAM
type-node configurations across the three major DRAM manufacturers. For all
chips, we hammer each row, sweeping $HC$ between 10,000 and 150,000.
For each $HC$ value, we plot the average rate of observed
RowHammer bit flips across all chips of a DRAM type-node configuration.

\stepcounter{obscount} \textbf{Observation \arabic{obscount}.} \emph{The log of
the number of RowHammer bit flips has a linear relationship with the log of
$HC$.}\footnote{Our observation is consistent with prior
work~\cite{park2016statistical}.}

We observe this relationship between $HC$ and RowHammer bit flip rate because
more accesses to a single row results in more cell-to-cell interference,
and therefore more charge is lost in victim cells of nearby rows. 

We examine the effects of DRAM technology node on the RowHammer bit flip rate
in Figure~\ref{fig:char:refresh_sweep}.  We observe that the bit flip rate
curve shifts \emph{upward} and \emph{leftward} when going from DDR4-old to DDR4-new chips,
indicating respectively, 1) a higher rate of bit flips for the same $HC$ value
and 2) occurrence of bit flips at lower $HC$ values, as technology node size
reduces from DDR4-old to DDR4-new.  

\stepcounter{obscount} \textbf{Observation \arabic{obscount}.} \emph{Newer
DDR4 DRAM technology nodes show a clear trend of increasing RowHammer bit flip
rates: the \emph{same} $HC$ value causes an increased average RowHammer bit
flip rate from DDR4-old to DDR4-new DRAM chips of all DRAM manufacturers.}

We believe that due to increased density of DRAM chips from older to newer
technology node generations, cell-to-cell interference increases and
results in DRAM chips that are more vulnerable to RowHammer bit flips.


\subsection{RowHammer Spatial Effects}
\label{subsec:spatial_effects} 

\begin{figure}[H] \centering
    \includegraphics[width=0.91\linewidth]{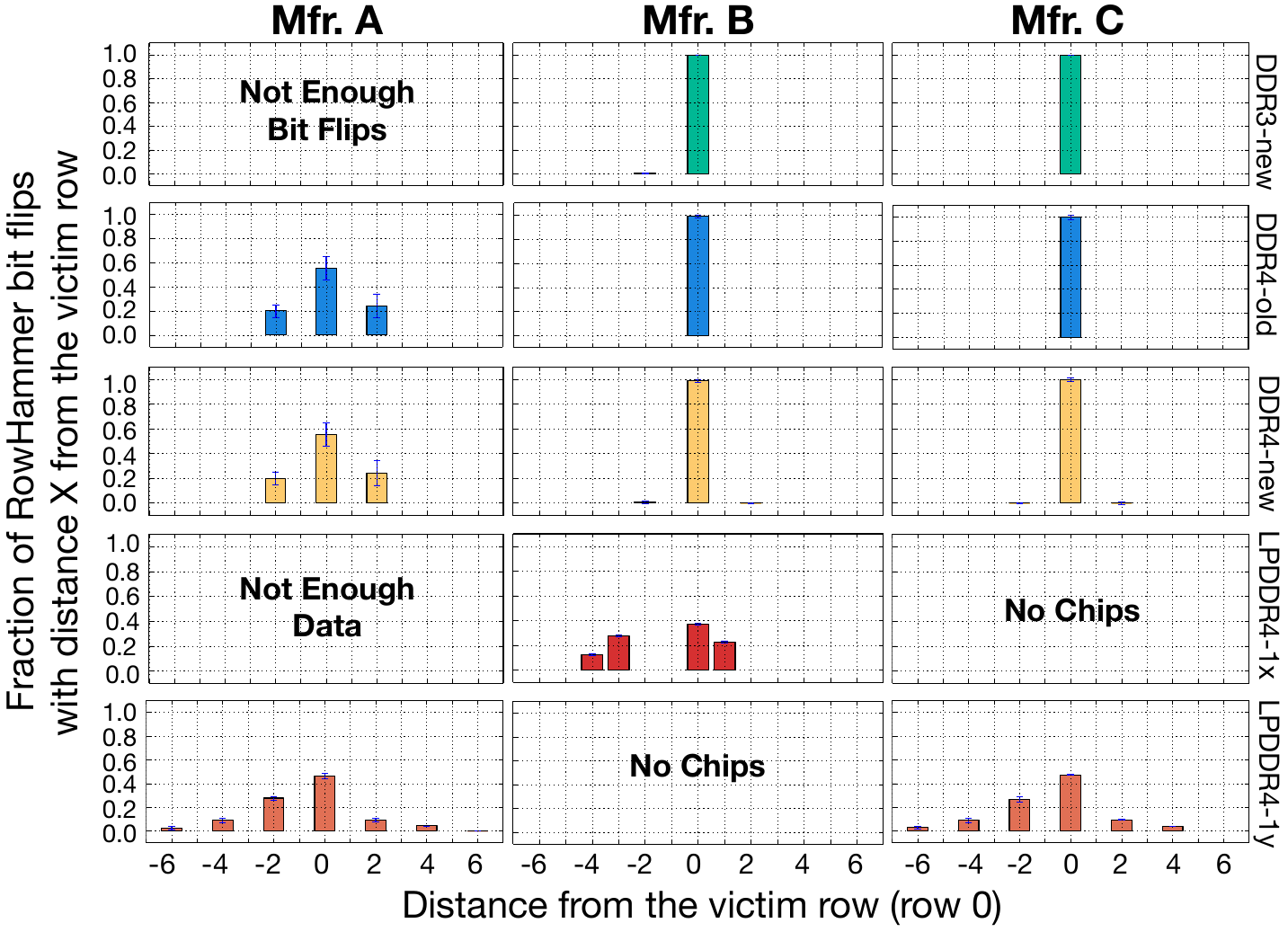}
    \caption{Distribution of RowHammer bit flips across row offsets from the victim row.} 
    \label{fig:char:distances}
\end{figure}

We next experimentally study the spatial distribution of RowHammer bit flips
across our tested chips. In order to normalize the RowHammer effects that we
observe across our tested chips, we first take each DRAM chip and use a hammer
count specific to that chip to result in a RowHammer bit flip rate of
$10^{-6}$.\footnote{We choose a RowHammer bit flip rate of $10^{-6}$ since we
are able to observe this bit flip rate in most chips that we characterize with
$HC<150k$.} For each chip, we analyze the spatial distribution of bit flips throughout the chip.
Figure~\ref{fig:char:distances} plots the fraction of RowHammer bit flips
that occur in a given row offset from the $victim\_row$ out of all observed
RowHammer bit flips. Each column of subplots shows the distributions for chips
of different manufacturers and each row of subplots shows the distribution for
a different DRAM type-node configuration. The error bars show the
standard deviation of the distribution across our tested chips. Note that the 
repeatedly-accessed rows (i.e., \emph{aggressor rows}) are
at $x=1$ and $x=-1$ for all plots except in LPDDR4-1x chips from
manufacturer B, where they are at $x=-2$ and $x=2$ (due to the internal address
remapping that occurs in these chips as we describe in
Section~\ref{subsec:methodology:characterizing_rowhammer}).  Because an access
to a row essentially refreshes the data in the row, repeatedly accessing
aggressor rows during the core loop of the RowHammer test prevents any bit
flips from happening in the aggressor rows. Therefore, there are no RowHammer
bit flips in the aggressor rows across each DRAM chip in our plots (i.e., $y=0$
for $x=[-2,-1,2,3]$ for LPDDR4-1x chips from manufacturer B and for $x=1$ and
$x=-1$ for all other chips). 

We make three observations from Figure~\ref{fig:char:distances}. First, we
observe a general trend across DRAM type-node configurations of a given DRAM
manufacturer where newer DRAM technology nodes have an increasing number of
rows that are susceptible to RowHammer bit flips that are \emph{farther} from
the victim row. For example, in LPDDR4-1y chips, we observe RowHammer bit flips
in as far as 6 rows from the victim row (i.e., $x=-6$), whereas in DDR3 and
DDR4 chips, RowHammer bit flips only occur in as far as 2 rows from the victim
row (i.e., $x=-2$). We believe that this effect could be due to 1) an
increase in DRAM cell density, which leads to cell-to-cell interference
extending farther than a single row, with RowHammer bit flips occurring in rows
increasingly farther away from the aggressor rows (e.g., 5 rows away) for
higher-density chips, and 2) more shared structures internal to
the DRAM chip, which causes farther (and multiple) rows to be affected by
circuit-level interference. 

\stepcounter{obscount} \textbf{Observation \arabic{obscount}.} \emph{For a
given DRAM manufacturer, chips of newer DRAM technology nodes can exhibit
RowHammer bit flips 1) in more rows and 2) farther away from the victim row.} 

Second, we observe that rows containing RowHammer bit flips that are
farther from the victim row have fewer RowHammer bit flips than rows closer to
the victim row. Non-victim rows adjacent to the aggressor rows ($x=2$ and
$x=-2$) contain RowHammer bit flips, and these bit flips demonstrate the
effectiveness of a single-sided RowHammer attack as only one of their adjacent
rows are repeatedly accessed. As discussed earlier
(Section~\ref{subsec:methodology:characterizing_rowhammer}), the single-sided
RowHammer attack is not as effective as the double-sided RowHammer attack, and
therefore we find fewer bit flips in these rows. In rows farther 
away from the victim row, we attribute the diminishing number of RowHammer bit
flips to the diminishing effects of cell-to-cell interference with distance. 

\stepcounter{obscount} \textbf{Observation \arabic{obscount}.} \emph{The number
of RowHammer bit flips that occur in a given row decreases as the distance from
the victim row increases.} 

Third, we observe that only even-numbered offsets from the victim row
contain RowHammer bit flips in all chips except LPDDR4-1x chips from
Manufacturer B.  However, the rows containing RowHammer bit flips in
Manufacturer B's LPDDR4-1x chips would be even-numbered offsets if we translate
all rows to physical rows based on our observation in
Section~\ref{subsec:methodology:characterizing_rowhammer} (i.e., divide each
row number by 2 and round down).  While we are uncertain why we observe
RowHammer bit flips only in physical even-numbered offsets from the victim row,
we believe that it may be due to the internal circuitry layout of DRAM rows.  

\begin{figure}[H] \centering
    \includegraphics[width=0.8\linewidth]{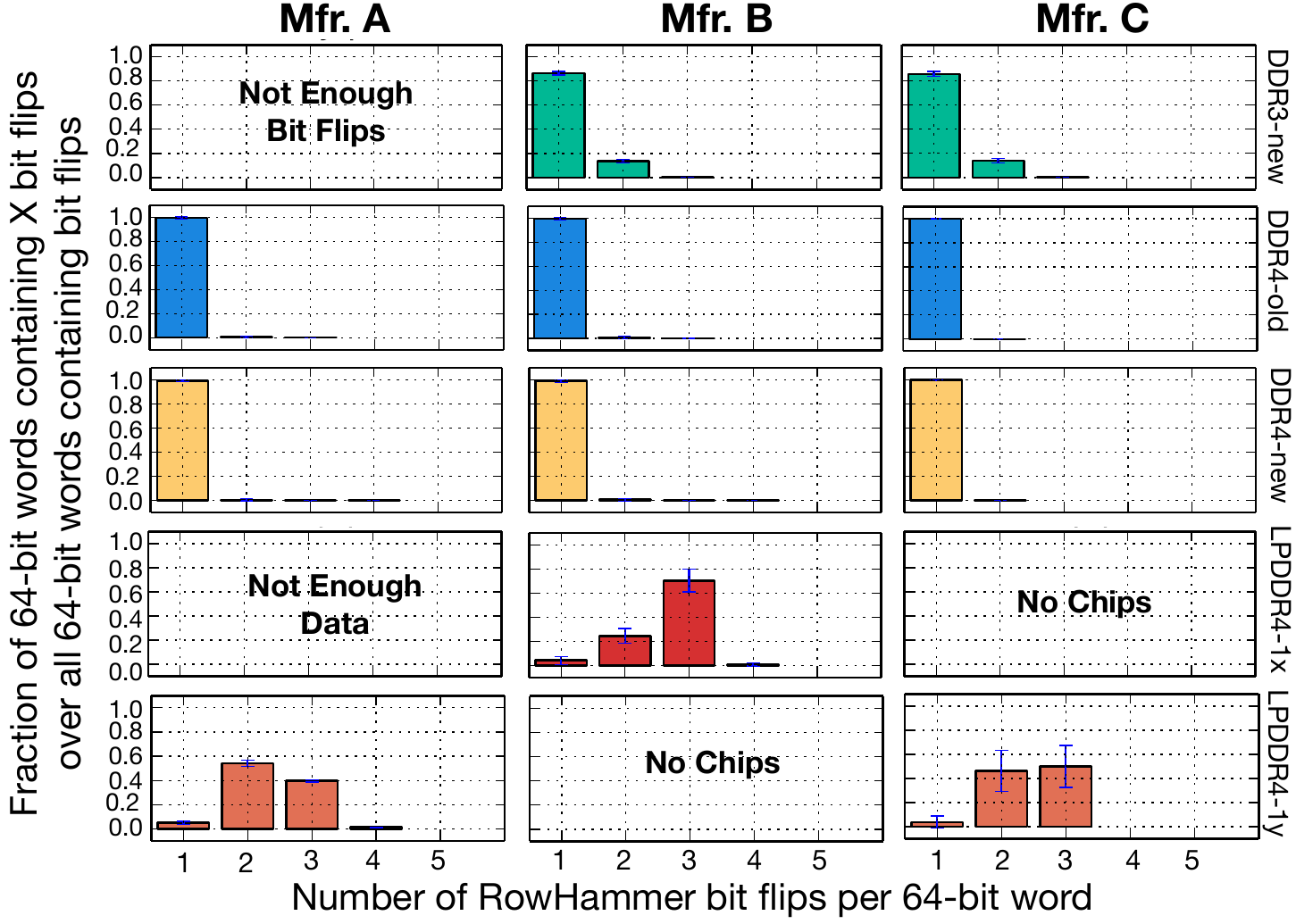}
    \caption{Distribution of the number of RowHammer bit flips per 64-bit word for each DRAM type-node configuration.} 
    \label{fig:char:density}
\end{figure}

We next study the spatial distribution of RowHammer-vulnerable DRAM cells in a
DRAM array using the same set of RowHammer bit flips.
Figure~\ref{fig:char:density} shows the distribution of 64-bit words containing
x RowHammer bit flips across our tested DRAM chips. We find the proportion of
64-bit words containing x RowHammer bit flips out of all 64-bit words in each
chip containing any RowHammer bit flip and plot the distribution as a
bar chart with error bars for each x value.

\stepcounter{obscount} \textbf{Observation \arabic{obscount}.} \emph{At a
RowHammer bit flip rate of $10^{-6}$, a single 64-bit value can contain up to
four RowHammer bit flips.} 

Because ECC~\cite{kang2014co, micron2017whitepaper, oh20153,
patel2019understanding} is typically implemented for DRAM at a 64-bit
granularity (e.g., a single-error correcting code would only protect a 64-bit
word if it contains at most one error), observation \arabic{obscount}
indicates that even at a relatively low bit flip rate of $10^{-6}$, a DRAM
chip can only be protected from RowHammer bit flips with a strong ECC code
(e.g., 4-bit error correcting code), which has high hardware overhead. 

\stepcounter{obscount} \textbf{Observation \arabic{obscount}.} \emph{The
distribution of RowHammer bit flip density per word changes significantly in
LPDDR4 chips compared to other DRAM types.} 

We find DDR3 and DDR4 chips across all manufacturers to exhibit an
exponential decay curve for increasing RowHammer bit flip densities with most
words containing only one RowHammer bit flip. However, LPDDR4 chips across
all manufacturers exhibit a much smaller fraction of words containing a
single RowHammer bit flip and significantly larger fractions of words
containing two and three RowHammer bit flips compared to DDR3 and DDR4 chips.
We believe this change in the bit flip density distribution is due to the
on-die ECC that manufacturers have included in LPDDR4 chips~\cite{kang2014co,
micron2017whitepaper, oh20153, patel2019understanding}, which is a 128-bit
single-error correcting code that corrects and hides \emph{most} single-bit
failures within a 128-bit ECC word using redundant bits (i.e.,
\emph{parity-check bits}) that are hidden from the system.  

With the failure rates at which we test, many ECC words contain
several bit flips. This exceeds the ECC's correction strength and causes the
ECC logic to behave in an undefined way. The ECC logic may 1) correct one of
the bit flips, 2) do nothing, or 3) introduce an \emph{additional} bit flip by
corrupting an error-free data bit~\cite{son2015cidra, patel2019understanding}.
On-die ECC makes single-bit errors rare because 1) any true single-bit error
is immediately corrected and 2) a multi-bit error can \emph{only} be reduced
to a single-bit error when there are no more than two bit flips within the
data bits \emph{and} the ECC logic's undefined action happens to change the
bit flip count to exactly one. In contrast, there are many more scenarios that
yield two or three bit-flips within the data bits, and a detailed experimental
analysis of how on-die ECC affects DRAM failure rates in LPDDR4 DRAM chips can
be found in~\cite{patel2019understanding}.



\subsection{First RowHammer Bit Flips} 

\begin{figure}[h] \centering
    \includegraphics[width=0.8\linewidth]{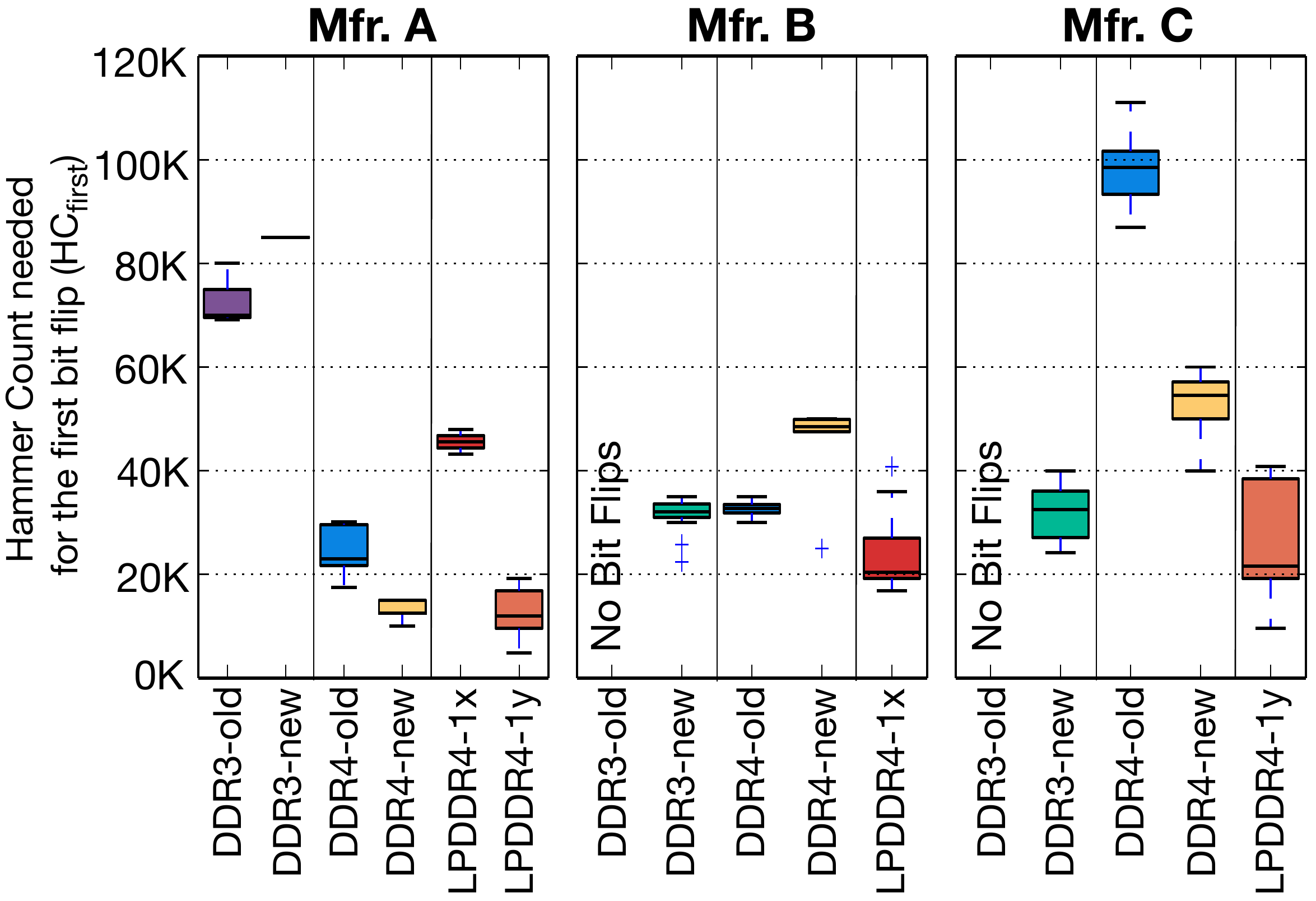}
    \caption{Number of hammers required to cause the first RowHammer bit flip ($HC_{first}$) per chip across DRAM type-node configurations.}
    \label{fig:char:first_fail}
\end{figure}

We next study the vulnerability of each chip to RowHammer. One critical
component of vulnerability to the double-sided RowHammer
attack~\cite{kim2014flipping} is identifying the weakest cell, i.e., the DRAM cell
that fails with the fewest number of accesses to physically-adjacent rows.
In order to perform this study, we sweep $HC$ at a fine granularity and record 
the $HC$ that results in the first RowHammer bit flip in the chip
({\hcfirst}).  Figure~\ref{fig:char:first_fail} plots the distribution of
{\hcfirst} across all tested chips as box-and-whisker plots.\footnote{A
box-and-whiskers plot emphasizes the important metrics of a dataset's
distribution. The box is lower-bounded by the first quartile (i.e., the median
of the first half of the ordered set of data points) and upper-bounded by the
third quartile (i.e., the median of the second half of the ordered set of data
points). The median falls within the box. The \emph{inter-quartile range} (IQR)
is the distance between the first and third quartiles (i.e., box size).
Whiskers extend an additional $1.5 \times IQR$ on either sides of the box.  We
indicate outliers, or data points outside of the range of the whiskers, with
pluses.} The subplots contain the distributions of each tested DRAM type-node 
configuration for the different DRAM manufacturers. The x-axis organizes the
distributions by DRAM type-node configuration in order of age (older on the left to younger
on the right).  We further subdivide the subplots for chips of the same DRAM
type (e.g., DDR3, DDR4, LPDDR4) with vertical lines. Chips of the same DRAM
type are colored with the same color for easier visual comparison across DRAM
manufacturers.

\stepcounter{obscount} \textbf{Observation \arabic{obscount}.} \emph{Newer
chips from a given DRAM manufacturer appear to be more vulnerable to RowHammer
bit flips. This is demonstrated by the clear reduction in {\hcfirst} 
values from old to new DRAM generations (e.g., LPDDR4-1x to LPDDR4-1y in
manufacturer A, or DDR4-old to DDR4-new in manufacturers A and C).} 

We believe this observation is due to DRAM technology process scaling
wherein both 1) DRAM cell capacitance reduces and 2) DRAM cell density
increases as technology node size reduces. Both factors together lead to more
interference between cells and likely faster charge leakage from the DRAM
cell's smaller capacitors, leading to a higher vulnerability to RowHammer. We find
two exceptions to this trend (i.e., a general increase in {\hcfirst} from
DDR3-old to DDR3-new chips of manufacturer A and from DDR4-old to DDR4-new
chips of manufacturer B), but we believe these potential
anomalies may be due to our inability to identify explicit manufacturing
dates and correctly categorize these particular chips. 

\stepcounter{obscount} \textbf{Observation \arabic{obscount}.} \emph{In
LPDDR4-1y chips from manufacturer A, there are chips whose weakest cells
fail after only 4800 hammers.} 

This observation has serious implications for the future as DRAM
technology node sizes will continue to reduce and {\hcfirst} will only get
smaller. We discuss these implications further in
Section~\ref{sec:implications}.  Table~\ref{tab:device_Nth} shows the
lowest observed {\hcfirst} value for any chip within a DRAM type-node
configuration (i.e., the minimum values of each distribution in
Figure~\ref{fig:char:first_fail}).

\begin{table}[h]
\footnotesize
\begin{center}
\renewcommand{\arraystretch}{0.9}
\renewcommand{\aboverulesep}{0ex}
\renewcommand{\belowrulesep}{0ex}
\caption{Lowest $\bm{HC_{first}}$ values ($\mathbf{\times1000}$) across all chips of each DRAM type-node configuration.} 
\begin{tabular}{l R{1.45cm} R{1.45cm} R{1.2cm}}
\toprule
\multicolumn{1}{c}{\textbf{DRAM}} & \multicolumn{3}{c}{\textbf{$\bm{HC_{first}}$ (Hammers until first bit flip) $\mathbf{\times1000}$}} \\
\multicolumn{1}{c}{\textbf{type-node}} & \quad \textbf{\textit{Mfr. A}} & \quad \textbf{\textit{Mfr. B}} & \quad \textbf{\textit{Mfr. C}} \\
\toprule
DDR3-old & 69.2 & 157 & 155 \\ 
DDR3-new & 85 & 22.4 & 24 \\
\Xhline{0.2\arrayrulewidth}
DDR4-old & 17.5 & 30 & 87  \\
DDR4-new & 10 & 25 & 40 \\
\Xhline{0.2\arrayrulewidth}
\ap{LPDDR4-1x} & 43.2 & 16.8 & N/A \\
\ap{LPDDR4-1y} & 4.8 & N/A & 9.6 \\
\toprule
\end{tabular}
\label{tab:device_Nth}
\vspace{-3mm} 
\end{center}
\end{table}
\textbf{Effects of ECC.} The use of error correcting codes (ECC) to
improve the reliability of a DRAM chip is common practice, with most
system-level~\cite{balasubramonian2019innovations, cojocar2019exploiting,
gong2018duo, kim2015bamboo} or on-die~\cite{micron2017whitepaper, kwak2017a,
kang2014co, patel2019understanding, kwon2017an} ECC mechanisms providing single
error correction capabilities at the granularity of 64- or 128-bit words.
We examine 64-bit ECCs since, for the same correction capability (e.g.,
single-error correcting), they are stronger than 128-bit ECCs. In order
to determine the efficacy with which ECC can mitigate RowHammer effects on real
DRAM chips, we carefully study three metrics across each of our chips: 1)
the lowest $HC$ required to cause the first RowHammer bit flip (i.e.,
{\hcfirst}) for a given chip (shown in Figure~\ref{fig:char:first_fail}), 2)
the lowest $HC$ required to cause at least two RowHammer bit flips (i.e.,
{\hcsecond}) within any 64-bit word, and 3) the lowest $HC$ required to cause
at least three RowHammer bit flips (i.e., {\hcthird}) within any 64-bit word.
These quantities tell us, for ECCs of varying strengths (e.g.,
single-error correction code, double-error correction code), at which
$HC$ values the ECC can 1) mitigate RowHammer bit flips and 2) no
longer reliably mitigate RowHammer bit flips for that particular chip.

\begin{figure}[h] \centering
    \includegraphics[width=0.99\linewidth]{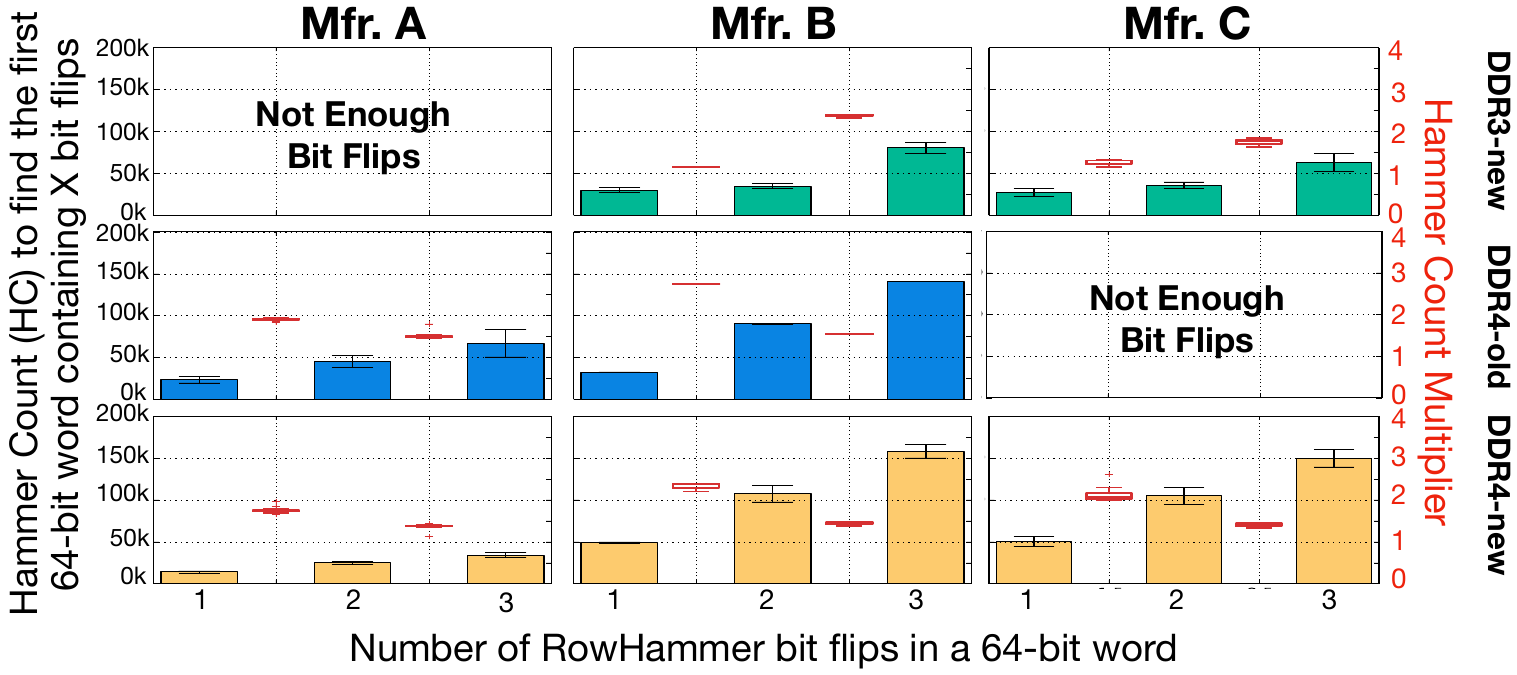}
    \caption{Hammer Count (left y-axis) required to find the first 64-bit word containing one, two, and three RowHammer bit flips. Hammer Count Multiplier (right y-axis) quantifies the $HC$ difference between every two points on the x-axis (as a multiplication factor of the left point to the right point).}
    \label{fig:ECC_efficacy} 
\end{figure}

Figure~\ref{fig:ECC_efficacy} plots as a bar graph the $HC$ (left y-axis)
required to find the first 64-bit word containing one, two, and three RowHammer
bit flips (x-axis) across each DRAM type-node configuration.  The error bars
represent the standard deviation of $HC$ values across all chips tested. On the
same figure, we also plot with red boxplots, the increase in $HC$ (right
y-axis) between the $HC$s required to find the first 64-bit word containing one
and two RowHammer bit flips, and two and three RowHammer bit flips. These
multipliers indicate how {\hcfirst} would change in a chip if the chip uses
single-error correcting ECC or moves from a single-error correcting to a
double-error correcting ECC.  Note that we 1) leave two plots (i.e., Mfr.  A
DDR3-new and Mfr. C DDR4-old) empty since we are unable to induce enough
RowHammer bit flips to find 64-bit words containing more than one bit flip in
the chips and 2) do not include data from our LPDDR4 chips because they already
include on-die ECC~\cite{micron2017whitepaper, kwak2017a, kang2014co,
patel2019understanding, kwon2017an}, which obfuscates errors potentially
exposed to any other ECC mechanisms~\cite{patel2019understanding}.  

\stepcounter{obscount} \textbf{Observation \arabic{obscount}.} \emph{A
single-error correcting code can significantly improve {\hcfirst} by up to
$2.78\times$ in DDR4-old and DDR4-new DRAM chips, and $1.65\times$ in
DDR3-new DRAM chips.} 


\stepcounter{obscount} \textbf{Observation \arabic{obscount}.}
\emph{Moving from a double-error correcting code to a triple-error
correcting code has diminishing returns in DDR4-old and DDR4-new DRAM
chips (as indicated by the reduction in the $HC$ multiplier) compared to when
moving from a single-error correcting code to a double-error
correcting code. However, using a triple-error correcting code in
DDR3-new DRAM chips continues to further improve the {\hcfirst} and thus
reduce the DRAM chips' vulnerability to RowHammer.}

\subsection{Single-Cell RowHammer Bit Flip Probability} 

We examine how the failure probability of a single RowHammer bit flip changes as $HC$
increases.  We sweep $HC$ between 25k to 150k with a step size of 5k and hammer
each DRAM row over 20 iterations. For each $HC$ value, we identify each
cell's bit flip probability (i.e., the number of times we observe a RowHammer
bit flip in that cell out of all 20 iterations). We then observe how each
cell's bit flip probability changes as $HC$ increases.  We expect that by
exacerbating the RowHammer conditions (e.g., increasing the hammer count),
the exacerbated circuit-level interference effects should result in an
increasing RowHammer bit flip probability for each individual cell.  Out of the
full set of bits that we observe \emph{any} RowHammer bit flips in,
Table~\ref{tab:sensitivity} lists the percentage of cells that have a strictly
monotonically increasing bit flip probability as we increase $HC$.

\begin{table}[h]
\footnotesize
\begin{center}
\caption{Percentage of cells with monotonically increasing RowHammer bit flip probabilities as $\bm{HC}$ increases.}
\begin{tabular}{ lR{2cm}R{2cm}R{2cm}}
\toprule
\multicolumn{1}{c}{\multirow{3}{*}{\begin{tabular}{c}\textbf{DRAM} \\ \textbf{type-node}\end{tabular}}} & \multicolumn{3}{c}{\textbf{Cells with monotonically increasing}} \\
 & \multicolumn{3}{c}{\textbf{ RowHammer bit flip probabilities (\%)}} \\
 & \textbf{\textit{Mfr. A}} & \textbf{\textit{Mfr. B}} & \textbf{\textit{Mfr. C}} \\
\toprule
DDR3-new & $97.6\pm0.2$ & 100 & 100 \\
\Xhline{0.2\arrayrulewidth}
DDR4-old & $98.4\pm0.1$ & 100 & 100 \\
DDR4-new & $99.6\pm0.1$ & 100 & 100 \\
\Xhline{0.2\arrayrulewidth}
LPDDR4-1x & $50.3\pm1.2$ & $52.4\pm1.4$ & N/A \\
LPDDR4-1y & $47.0\pm0.8$ & N/A & $54.3\pm5.7$ \\
\toprule
\end{tabular}
\label{tab:sensitivity}
\vspace{-3mm} 
\end{center}
\end{table}

\stepcounter{obscount} \textbf{Observation \arabic{obscount}.} \emph{For DDR3
and DDR4 chips, an overwhelming majority (i.e., more than 97\%) of the cells
tested have monotonically increasing RowHammer bit flip probabilities for
DDR3 and DDR4 chips.} 

This observation indicates that exacerbating the RowHammer
conditions by increasing $HC$ increases the probability that a DRAM cell 
experiences a RowHammer bit flip. However, we find that the proportion of cells
with monotonically increasing RowHammer bit flip probabilities as $HC$
increases is around only 50\% in the LPDDR4 chips that we test.
We believe that this decrease is due to the addition of on-die ECC in LPDDR4
chips, which can obscure the probability of observing a RowHammer bit flip from
the system's perspective in two ways. First, a RowHammer bit flip at bit X can
no longer be observable from the system's perspective if another RowHammer bit
flip at bit Y occurs within the same ECC word as a result of increasing $HC$,
and the error correction logic corrects the RowHammer bit flip at bit X.
Second, the system may temporarily observe a bit flip at bit X at
a specific $HC$ if the set of real RowHammer bit flips within an ECC word
results in a miscorrection at bit X. Since this bit flip is a result of the ECC
logic misbehaving rather than circuit-level interference, we do not observe the
expected trends for these transient miscorrected bits.

%



\section{Implications for Future Systems} 
\label{sec:implications}

Our characterization results have major implications for continued DRAM
technology scaling since DRAM's increased vulnerability to RowHammer means that
systems employing future DRAM devices will likely need to handle
significantly elevated failure rates. While prior works propose a wide variety
of RowHammer failure mitigation techniques (described in
Sections~\ref{subsec:implications:existing_mechanisms} and \ref{sec:related}),
these mechanisms will need to manage increasing failure rates going forward and
will likely suffer from high overhead (as we show in
Section~\ref{subsec:implications:evaluation}).

While DRAM and system designers currently implement several RowHammer
mitigation mechanisms (e.g., \emph{pseudo Target Row Refresh}
(pTRR)~\cite{kaczmarski2014thoughts}, \emph{Target Row Refresh}
(TRR)~\cite{lin2017handling})\footnote{Frigo et al.~\cite{frigo2020trrespass} 
recently demonstrated that these mechanisms do \emph{not} prevent \emph{all}
RowHammer bit flips from being exposed to the system, and an attacker can still
take over a system even with these mechanisms in place.}, the designers make a
number of unknown implementation choices in these RowHammer mitigation
mechanisms that are not discussed in public documentation. Therefore,
we cannot fairly evaluate how their performance overheads scale as DRAM
chips become more vulnerable to RowHammer. Instead, we evaluate five
state-of-the-art academic proposals for RowHammer mitigation
mechanisms~\cite{kim2014flipping, lee2019twice, son2017making, you2019mrloc} as
well as an ideal refresh-based mitigation mechanism.  

We evaluate each RowHammer mitigation mechanism in terms of two major
challenges that they will face going forward as they will need to support DRAM
chips more vulnerable to RowHammer: design scalability and system performance
overhead. We first qualitatively explain and discuss the five state-of-the-art
mitigation mechanisms and how they can potentially scale to support DRAM
chips that are more vulnerable to RowHammer. We then quantitatively evaluate
their performance overheads in simulation as {\hcfirst} decreases. In
order to show the opportunity for reducing performance overhead in RowHammer
mitigation, we also implement and study an \emph{ideal refresh-based mechanism}
that prevents RowHammer by refreshing a DRAM row \emph{only} immediately before
it is about to experience a bit flip. 

\subsection{RowHammer Mitigation Mechanisms}
\label{subsec:implications:existing_mechanisms}

There is a large body of work (e.g., \cite{aweke2016anvil, van2018guardion,
konoth2018zebram, brasser2016can, wu2019protecting, bock2019rip,
kim2019effective, wang2019reinforce, chakraborty2019deep, li2019detecting,
ghasempour2015armor, gruss2018another, irazoqui2016mascat}) that proposes
software-based RowHammer mitigation mechanisms. Unfortunately, many of these
works have critical weaknesses (e.g., inability to track all DRAM activations)
that make them vulnerable to carefully-crafted RowHammer attacks, as
demonstrated in some followup works (e.g., \cite{gruss2018another}).
Therefore, we focus on evaluating six mechanisms (i.e., five state-of-the-art
hardware proposals and one \emph{ideal} refresh-based mitigation
mechanism), which address a strong threat model that assumes an attacker
can cause row activations with precise memory location and timing information. We briefly
explain each mitigation mechanism and how its design scales for DRAM chips with
increased vulnerability to RowHammer (i.e., lower {\hcfirst} values). 

\textbf{Increased Refresh Rate~\cite{kim2014flipping}.} The original
RowHammer study~\cite{kim2014flipping} describes increasing the overall
DRAM refresh rate such that it is impossible to issue enough activations
within one refresh window (i.e., the time between two consecutive refresh
commands to a single DRAM row) to any single DRAM row to induce a RowHammer bit
flip. The study notes that this is an undesirable mitigation mechanism due to
its associated performance and energy overheads. In order to reliably mitigate
RowHammer bit flips with this mechanism, we scale the refresh rate such that
the refresh window (i.e., $t_{REFW}$; the time interval between consecutive
refresh commands to a single row) equals the number of hammers until the first
RowHammer bit flip (i.e., {\hcfirst}) multiplied by the activation latency
$t_{RC}$. Due to the large number of rows that must be refreshed within a
refresh window, this mechanism inherently does not scale to {\hcfirst} values
below 32k.

\textbf{PARA~\cite{kim2014flipping}.} Every time a row is opened and closed,
PARA (Probabilistic Adjacent Row Activation) refreshes one or more of the row's
adjacent rows with a low probability $p$.  Due to PARA's simple approach, it is
possible to easily tune $p$ when PARA must protect a DRAM chip with a lower
{\hcfirst} value. In our evaluation of PARA, we scale $p$ for different values
of {\hcfirst} such that the bit error rate (BER) does not exceed 1e-15 per hour
of continuous hammering.\footnote{We adopt this BER from typical consumer
memory reliability targets~\cite{micheloni2015apparatus, jedec2003failure,
patel2017reaper, cai2012error, luo2018heatwatch, luo2018improving, cai2017flashtbd,
luo2016enabling, cai2017errors}.}

\textbf{ProHIT~\cite{son2017making}.} ProHIT maintains a history of DRAM
activations in a set of tables to identify any row that may be activated
{\hcfirst} times. ProHIT manages the tables probabilistically to
minimize the overhead of tracking frequently-activated DRAM rows.
ProHIT~\cite{son2017making} uses a pair of tables labeled "Hot" and "Cold" to
track the victim rows. When a row is activated, ProHIT checks whether each
adjacent row is already in either of the tables. If a \emph{row} is not in
either table, it is inserted into the cold table with a probability $p_i$. If
the table is full, the least recently inserted entry in the cold table is then
evicted with a probability $(1-p_{e}) + p_{e}/(\#cold\_entries)$ and the other
entries are evicted with a probability $p_{e}/(\#cold\_entries)$. If the row
already exists in the cold table, the row is promoted to the highest-priority
entry in the hot table with a probability $(1 - p_t) + p_t/(\#hot\_entries)$
and to other entries with a probability $p_t/(\#hot\_entries)$. If the row
already exists in the hot table, the entry is upgraded to a higher priority
position. During each refresh command, ProHIT simultaneously refreshes the row
at the top entry of the hot table, since this row has likely experienced the
most number of activations, and then removes the entry from the table. 

For ProHIT~\cite{son2017making} to effectively mitigate RowHammer with
decreasing {\hcfirst} values, the size of the tables and the probabilities for
managing the tables (e.g., $p_i$, $p_e$, $p_t$) must be adjusted. Even though
Son et al. show a low-cost mitigation mechanism for a specific {\hcfirst} value
(i.e., 2000), they do \emph{not} provide models for appropriately setting these
values for arbitrary {\hcfirst} values and how to do so is not intuitive.
Therefore, we evaluate ProHIT only when {\hcfirst}$\,= 2000$. 

\textbf{MRLoc}~\cite{you2019mrloc}. MRLoc refreshes a victim row using a
probability that is dynamically adjusted based on each row's access history.
This way, according to memory access locality, the rows that have been recorded
as a victim more recently have a higher chance of being refreshed. MRLoc uses a
queue to store victim row addresses on each activation. Depending on the time
between two insertions of a given victim row into the queue, MRLoc adjusts the
probability with which it issues a refresh to the victim row that is present in
the queue.

MRLoc's parameters (the queue size and the parameters used to calculate the
probability of refresh) are tuned for {\hcfirst}$\,= 2000$. You et
al.~\cite{you2019mrloc} choose the values for these parameters
empirically, and there is no concrete discussion on how to adjust these
parameters as {\hcfirst} changes.  Therefore we evaluate MRLoc for only
{\hcfirst}$\,= 2000$. 

As such, even though we quantitatively evaluate both
ProHIT~\cite{son2017making} and MRLoc~\cite{you2019mrloc} for completeness and
they may seem to have good overhead results at one data point, we are unable to
demonstrate how their overheads scale as DRAM chips become more vulnerable to
RowHammer. 

\textbf{TWiCe}~\cite{lee2019twice}. TWiCe tracks the number of times a
victim row's aggressor rows are activated using a table of counters and
refreshes a victim row when its count is above a threshold such that RowHammer
bit flips cannot occur. TWiCe uses two counters per entry: 1) a lifetime
counter, which tracks the length of time the entry has been in the table, and
2) an \emph{activation counter}, which tracks the number of times an aggressor
row is activated. The key idea is that TWiCe can use these two counters to
determine the rate at which a row is being hammered and can quickly prune
entries that have a low rate of being hammered. TWiCe also minimizes its table
size based on the observation that the number of rows that can be
activated enough times to induce RowHammer failures within a refresh window is
bound by the DRAM chip's vulnerability to RowHammer. 

When a row is activated, TWiCe checks whether its adjacent rows are already in
the table. If so, the activation count for each row is incremented. Otherwise,
new entries are allocated in the table for each row. Whenever a row's
activation count surpasses a threshold $t_{RH}$ defined as {\hcfirst}$/4$,
TWiCe refreshes the row. TWiCe also defines a pruning stage that 1) increments
each lifetime counter, 2) checks each row's hammer rate based on both counters,
and 3) prunes entries that have a lifetime hammer rate lower than a
\emph{pruning threshold}, which is defined as $t_{RH}$ divided by the
number of refresh operations per refresh window (i.e.,
$t_{RH}/(t_{REFW}/t_{REFI})$). TWiCe performs pruning operations during refresh
commands so that the latency of a pruning operation is hidden behind the DRAM
refresh commands.

If $t_{RH}$ is lower than the number of refresh intervals in a refresh window
(i.e., $8192$), a couple of complications arise in the design. TWiCe either 1) cannot
prune its table, resulting in a very large table size since every row that is
accessed at least once will remain in the table until the end of the refresh
window or 2) requires floating point operations in order to calculate
thresholds for pruning, which would significantly increase the latency of the
pruning stage.  Either way, the pruning stage latency would increase
significantly since a larger table also requires more time to check each entry,
and the latency may no longer be hidden by the refresh command. 

As a consequence, TWiCe does \emph{not} support $t_{RH}$ values lower than the
number of refresh intervals in a refresh window ($\sim8k$ in several DRAM
standards, e.g., DDR3, DDR4, LPDDR4). This means that in its current form, we
\emph{cannot} fairly evaluate TWiCe for {\hcfirst} values below $32k$, as
$t_{RH} =\,${\hcfirst}$/4$. However, we do evaluate an ideal version of
TWiCe (i.e., \emph{TWiCe-ideal}) for {\hcfirst} values below $32k$ assuming
that TWiCe-ideal solves \emph{both} issues of the large table size and the
high-latency pruning stage at lower {\hcfirst} values. 

\textbf{Ideal Refresh-based Mitigation Mechanism.} We implement an ideal
refresh-based mitigation mechanism that tracks all activations to every row in
DRAM and issues a refresh command to a row only right before it can potentially 
experience a RowHammer bit flip (i.e., when a physically-adjacent row has
been activated {\hcfirst} times). 

\subsection{Evaluation of Viable Mitigation Mechanisms}
\label{subsec:implications:evaluation}

We first describe our methodology for evaluating the five state-of-the-art
RowHammer mitigation mechanisms (i.e., increased refresh
rate~\cite{kim2014flipping}, PARA~\cite{kim2014flipping},
ProHIT~\cite{son2017making}, MRLoc~\cite{you2019mrloc},
TWiCe~\cite{lee2019twice}) and the ideal refresh-based mitigation mechanism.

\subsubsection{Evaluation Methodology}~
\label{subsec:implications:methodology}
We use Ramulator~\cite{ramulatorgithub, kim2016ramulator}, a cycle-accurate
DRAM simulator with a simple core model and a system configuration as listed in
Table~\ref{tab:ramulator_config}, to implement and evaluate the RowHammer
mitigation mechanisms. To demonstrate how the performance overhead of each
mechanism would scale to future devices, we implement, to the best of our
ability, parameterizable methods for scaling the mitigation mechanisms to DRAM
chips with varying degrees of vulnerability to RowHammer (as described in
Section~\ref{subsec:implications:existing_mechanisms}). 

\begin{table}[ht]
	\setlength\tabcolsep{1.5pt} 
    \footnotesize 
    \centering
    \caption{System configuration for simulations.} 
    \begin{tabular}{m{3cm}m{6cm}}
    \toprule
    \textbf{Parameter}     & \textbf{Configuration}  \\\toprule 
    Processor          & 4GHz, 8-core, 4-wide issue, 128-entry instr. window \\\hline
    Last-level Cache   & 64-Byte cache line, 8-way set-associative, 16MB \\\hline
    Memory Controller  & 64 read/write request queue, FR-FCFS~\cite{rixner2000memory, zuravleff1997controller} \\\hline
	Main Memory        & DDR4, 1-channel, 1-rank, 4-bank groups, 4-banks per bank group, 16k rows per bank\\ \toprule
    \end{tabular}
    \label{tab:ramulator_config}
	\vspace{-3mm} 
\end{table}

\textbf{Workloads.} We evaluate 48 8-core workload mixes drawn randomly from
the full SPEC CPU2006 benchmark suite~\cite{spec2006} to demonstrate the
effects of the RowHammer mitigation mechanisms on systems during typical use
(and \emph{not} when a RowHammer attack is being mounted).  The set of
workloads exhibit a wide range of memory intensities. The workloads' MPKI
values (i.e., last-level cache misses per kilo-instruction) range from 10 to
740. This wide range enables us to study the effects of RowHammer mitigation
on workloads with widely varying degrees of memory intensity. We note that there could
be other workloads with which mitigation mechanisms exhibit higher performance
overheads, but we did not try to maximize the overhead experienced by workloads
by biasing the workload construction in any way. 
We simulate each workload until each core executes at least 200 million
instructions. For all configurations, we initially warm up the caches by
fast-forwarding 100 million instructions.

\textbf{Metrics.} Because state-of-the-art RowHammer mitigation
mechanisms rely on additional DRAM refresh operations to prevent RowHammer, we
use two different metrics to evaluate their impact on system performance.
First, we measure \emph{DRAM bandwidth overhead}, which quantifies the
fraction of the total system DRAM bandwidth consumption coming from the
RowHammer mitigation mechanism. Second, we measure overall workload performance
using the \emph{weighted speedup} metric~\cite{eyerman2008system,
snavely2000symbiotic}, which effectively measures job throughput for multi-core
workloads~\cite{eyerman2008system}. We normalize the weighted speedup to
its baseline value, which we denote as 100\%, and find that when using
RowHammer mitigation mechanisms, most values appear below the baseline.
Therefore, for clarity, we refer to normalized weighted speedup as
\emph{normalized system performance} in our evaluations.



\subsubsection{Evaluation of Mitigation Mechanisms}
\begin{figure}\centering
    \includegraphics[width=0.95\linewidth]{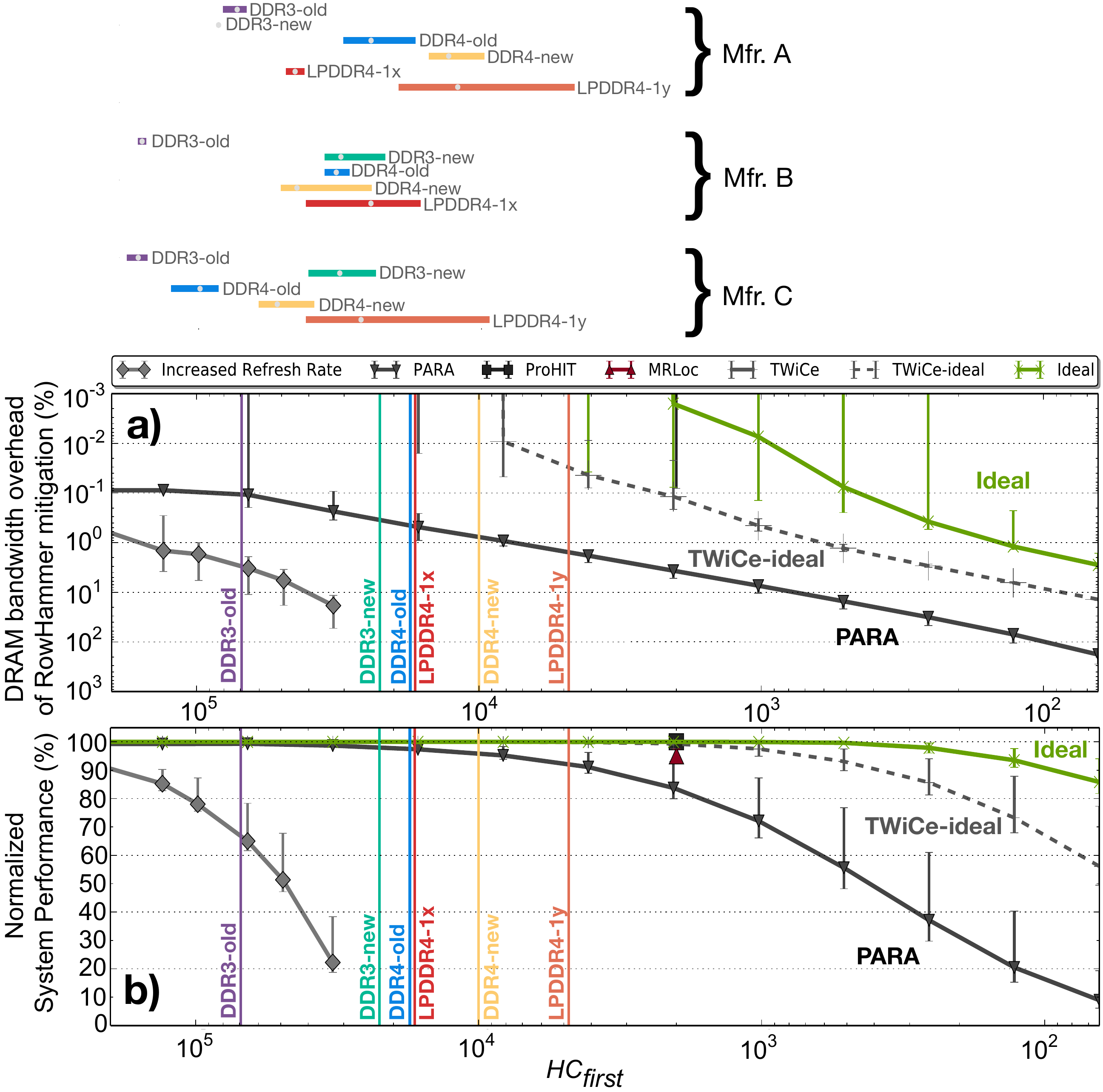} 
    \caption{Effect of RowHammer mitigation mechanisms on a) DRAM bandwidth overhead (note the inverted log-scale y-axis) and b) system performance, as DRAM chips become more vulnerable to RowHammer (from left to right).} 
    \label{fig:rh_mitigation_overhead_perf}
	\vspace{-2mm} 
\end{figure}

~
Figure~\ref{fig:rh_mitigation_overhead_perf} shows the results of our
evaluation of the RowHammer mitigation mechanisms (as described in
Section~\ref{subsec:implications:existing_mechanisms}) for chips of varying
degrees of RowHammer vulnerability (i.e., $200k \geq\,${\hcfirst}$\,\geq 64$)
for our two metrics: 1) DRAM bandwidth overhead in
Figure~\ref{fig:rh_mitigation_overhead_perf}a and 2) normalized system
performance in Figure~\ref{fig:rh_mitigation_overhead_perf}b. Each data
point shows the average value across 48 workloads with minimum and maximum
values drawn as error bars. 

For each DRAM type-node configuration that we characterize, we plot the minimum
{\hcfirst} value found across chips within the configuration (from
Table~\ref{tab:device_Nth}) as a vertical line to show how each RowHammer
mitigation mechanism would impact the overall system when using a DRAM chip of
a particular configuration. Above the figures (sharing the x-axis with
Figure~\ref{fig:rh_mitigation_overhead_perf}), we draw horizontal lines
representing the ranges of {\hcfirst} values that we observe for every tested
DRAM chip per DRAM type-node configuration across manufacturers. We color the
ranges according to DRAM type-node configuration colors in the figure, and
indicate the average value with a gray point.  Note that these lines directly
correspond to the box-and-whisker plot ranges in
Figure~\ref{fig:char:first_fail}. 

We make \emph{five} key observations from this figure. First, DRAM 
bandwidth overhead is highly correlated with normalized system performance,
as DRAM bandwidth consumption is the main source of system interference caused by 
RowHammer mitigation mechanisms. We note that several points (i.e., 
ProHIT, MRLoc, and TWiCe and Ideal evaluated at higher {\hcfirst} values) are not visible
in Figure~\ref{fig:rh_mitigation_overhead_perf}a since we are plotting an
inverted log graph and these points are very close to zero.  Second, in the
latest DRAM chips (i.e., the LPDDR4-1y chips), only PARA, ProHIT, and MRLoc 
are viable options for mitigating RowHammer bit flips with reasonable average
normalized system performance: 92\%, 100\%, and 100\%, respectively. 
Increased Refresh Rate and TWiCe do not scale to such degrees of RowHammer vulnerability
(i.e., {\hcfirst}$\,= 4.8k$), as discussed in
Section~\ref{subsec:implications:existing_mechanisms}.  Third, only PARA's
design scales to low {\hcfirst} values that we may see in future DRAM chips,
but has very low average normalized system performance (e.g., 72\% when
{\hcfirst}$\,=1024$; 47\% when {\hcfirst}$\,=256$; 20\% when {\hcfirst}$\,=128$).
While TWiCe-ideal has higher normalized system performance over PARA (e.g.,
98\% when {\hcfirst}$\,=1024$; 86\% when {\hcfirst}$\,=256$; 73\% when
{\hcfirst}$\,=128$), there are significant practical limitations in enabling
TWiCe-ideal for such low {\hcfirst} values (discussed in
Section~\ref{subsec:implications:existing_mechanisms}). Fourth, 
ProHIT and MRLoc both exhibit high normalized system performance at their single data
point (i.e., 95\% and 100\%, respectively when {\hcfirst}$\,= 2000$), but
these works do not provide models for scaling their mechanisms to lower
{\hcfirst} values and how to do so is not intuitive (as described in
Section~\ref{subsec:implications:existing_mechanisms}). Fifth, the ideal
refresh-based mitigation mechanism is \emph{significantly} and increasingly
better than any existing mechanism as {\hcfirst} reduces below $1024$.
This indicates that there is still significant opportunity for developing a
refresh-based RowHammer mitigation mechanism with low performance overhead that
scales to low {\hcfirst} values.  However, the ideal mechanism affects system
performance at very low {\hcfirst} values (e.g., 99.96\% when
{\hcfirst}$\,=1024$; 97.91\% when {\hcfirst}$\,=256$; 93.53\% when
{\hcfirst}$\,=128$), indicating the potential need for a better approach to
solving RowHammer in future ultra-dense DRAM chips. 

We conclude that while existing mitigation mechanisms may exhibit
reasonably small performance overheads for mitigating RowHammer bit flips in
modern DRAM chips, their overheads do \emph{not} scale well in future DRAM
chips that will likely exhibit higher vulnerability to RowHammer. Thus, we need
new mechanisms and approaches to RowHammer mitigation that will scale to DRAM
chips that are highly vulnerable to RowHammer bit flips.





\subsection{RowHammer Mitigation Going Forward}
\label{sec:profiling}


DRAM manufacturers continue to adopt smaller technology nodes to improve DRAM
storage density and are forecasted to reach 1z and 1a technology nodes within
the next couple of years~\cite{DRAMtechroadmap}. Unfortunately, our findings
show that future DRAM chips will likely be increasingly vulnerable to
RowHammer. This means that, to maintain market competitiveness without
suffering factory yield loss, manufacturers will need to develop
effective RowHammer mitigations for coping with increasingly
vulnerable DRAM chips.

\subsubsection{Future Directions in RowHammer Mitigation} 
\label{subsec:implications:mitigation_solutions} 

~RowHammer mitigation mechanisms have been proposed across the computing
stack ranging from circuit-level mechanisms built into the DRAM chip itself to
system-level mechanisms that are agnostic to the particular DRAM chip that the
system uses.  Of these solutions, our evaluations in
Section~\ref{subsec:implications:existing_mechanisms} show that, while the
ideal refresh-based RowHammer mitigation mechanism, which inserts the
minimum possible number of additional refreshes to prevent RowHammer bit flips,
scales reasonably well to very low {\hcfirst} values (e.g., only 6\%
performance loss when {\hcfirst} is 128), existing RowHammer mitigation
mechanisms either \emph{cannot} scale or cause severe system performance
penalties when they scale. 






To develop a scalable and low-overhead mechanism that can prevent
RowHammer bit flips in DRAM chips with a high degree of RowHammer vulnerability
(i.e., with a low {\hcfirst} value), we believe it is essential to explore
all possible avenues for RowHammer mitigation. Going forward, we identify two
promising research directions that can potentially lead to new
RowHammer solutions that can reach or exceed the scalability of the
ideal refresh-based mitigation mechanism: (1) DRAM-system cooperation and (2)
profile-guided mechanisms. The remainder of this section briefly discusses
our vision for each of these directions.

\noindent
\textbf{DRAM-System Cooperation.} Considering either DRAM-based or
system-level mechanisms alone ignores the potential benefits of addressing the 
RowHammer vulnerability from both perspectives together. While the root causes
of RowHammer bit flips lie within DRAM, their negative effects are observed at the
system-level.  Prior work~\cite{mutlu2014research, mutlu2013memory} stresses
the importance of tackling these challenges at all levels of the stack, and we
believe that a holistic solution can achieve a high degree of protection at
relatively low cost compared to solutions contained within either domain alone.

\noindent
\textbf{Profile-Guided Mechanisms.} The ability to accurately profile for
RowHammer-susceptible DRAM cells or memory regions can provide a powerful
substrate for building targeted RowHammer solutions that efficiently
mitigate RowHammer bit flips at low cost. Knowing (or effectively predicting)
the locations of bit flips before they occur in practice could lead to a large
reduction in RowHammer mitigation overhead, providing new information that no
known RowHammer mitigation mechanism exploits today. For example, within the
scope of known RowHammer mitigation solutions, increasing the refresh rate can
be made far cheaper by only increasing the refresh rate for known-vulnerable
DRAM rows. Similarly, ECC or DRAM access counters can be used only for
known-vulnerable cells, and even a software-based mechanism can be adapted
to target only known-vulnerable rows (e.g., by disabling them or remapping
them to reliable memory).

Unfortunately, there exists no such effective RowHammer error profiling
methodology today. Our characterization in this work essentially follows the
na{\"i}ve approach of individually testing each row by attempting to induce
the worst-case testing conditions (e.g., $HC$, data pattern, ambient
temperature etc.). However, this approach is extremely time consuming due to
having to test each row individually (potentially multiple times with various 
testing conditions). Even for a relatively small DRAM module of 8GB with 8KB
rows, hammering each row only once for only one refresh window of 64ms requires
over 17 hours of continuous testing, which means that the na{\"i}ve approach to
profiling is infeasible for a general mechanism that may be used in a
production environment or for online operation. We believe that developing a
fast and effective RowHammer profiling mechanism is a key research challenge,
and we hope that future work will use the observations made in this study
and other RowHammer characterization studies to find a solution.

\section{Related Work}
\label{sec:related}


Although many works propose RowHammer attacks and mitigation mechanisms, only
three works~\cite{kim2014flipping, park2016statistical, park2016experiments}
provide detailed failure-characterization studies that examine how RowHammer
failures manifest in real DRAM chips. However, none of these studies
show how the number of activations to induce RowHammer bit flips is
changing across modern DRAM types and generations, and the original RowHammer
study~\cite{kim2014flipping} is already six years old and limited to DDR3
DRAM chips only. This section highlights the most closely related prior works
that study the RowHammer vulnerability of older generation chips or examine
other aspects of RowHammer.

\textbf{Real Chip Studies.} Three key studies (i.e., the pioneering
RowHammer study~\cite{kim2014flipping} and two subsequent
studies~\cite{park2016statistical, park2016experiments}) perform extensive
experimental RowHammer failure characterization using older DDR3 devices.
However, these studies are restricted to only DDR3 devices and do not provide
a scaling study of hammer counts across DRAM types and generations. In
contrast, our work provides the first rigorous experimental study showing how
RowHammer characteristics scale across different DRAM generations and how DRAM
chips designed with newer technology nodes are increasingly vulnerable to
RowHammer. Our work complements and furthers the analyses provided in prior
studies.


\textbf{Simulation Studies.} Yang et al.~\cite{yang2019trap} use device-level
simulations to explore the root cause of the RowHammer vulnerability. While
their analysis identifies a likely explanation for the failure mechanism
responsible for RowHammer, they do not present experimental data taken from
real devices to support their conclusions.






\textbf{RowHammer Mitigation Mechanisms.} Many prior works~\cite{you2019mrloc,
son2017making, aweke2016anvil, konoth2018zebram, van2018guardion,
brasser2016can, kim2014flipping, kim2014architectural, irazoqui2016mascat,
gomez2016dram, lee2019twice, bu2018srasa, bains2015rowref, bains14d, bains14c,
greenfield14b, bains2016row, bains2015row, rh-apple, rh-hp, rh-lenovo,
rh-cisco, hassan2019crow, wu2019protecting, bock2019rip, kim2019effective,
wang2019reinforce, fisch2017dram, chakraborty2019deep, li2019detecting,
wang2019detect} propose RowHammer mitigation techniques. Additionally, several
patents for RowHammer prevention mechanisms have been filed~\cite{bains2015row,
bains14d, bains14c, bains2016row, bains2015rowref, greenfield14a}. However,
these works do not analyze how their solutions will scale to future DRAM
generations and do not provide detailed failure characterization data from
modern DRAM devices. Similar and other related works on RowHammer can be found
in a recent retrospective~\cite{mutlu2019rowhammer}.

 
\section{\jky{Limitations}} 

\jkz{Due to limited resources and available testing time, we were unable to
study several aspects of RowHammer that could help to further understand it as
a phenomenon. }

\jkz{First, while we demonstrate a scaling study on RowHammer as DRAM process
technology node size scales, we are unsure of the exact process technology node
sizes for several chip generations. Therefore, we could only provide a relative
study on how the vulnerability to RowHammer in DRAM chips across technology
node generations changes. Given the exact sizes of the technology nodes, we
could provide an analytical model such that we could predict the progression of
RowHammer vulnerability in future chips. } 

\jkz{Second, we did not account for chip aging in our RowHammer study. We
believe that understanding how aging affects the RowHammer vulnerability in
chips is critical to understand whether a system will become more vulnerable
over time as its DRAM chip ages in use. Furthermore, if aging can affect the
RowHammer vulnerability of a chip to the point that mitigation mechanisms can
no longer effectively mitigate RowHammer bit flips, it is important to
understand when and how often a DRAM chip should be swapped out to maintain a
\jky{RowHammer-free} system. }

\section{Summary} 

We provide the first rigorous experimental RowHammer failure characterization
study that demonstrates how the RowHammer vulnerability of modern DDR3, DDR4,
and LPDDR4 DRAM chips scales across DRAM generations and technology nodes.
Using experimental data from 1580 real DRAM chips produced by the three
major DRAM manufacturers, we show that modern DRAM chips that use smaller
process technology node sizes are significantly more vulnerable to RowHammer
than older chips. Using simulation, we show that existing RowHammer mitigation 
mechanisms 1) suffer from prohibitively large performance overheads at
projected future hammer counts and 2) are still far from an \emph{ideal}
selective-refresh-based RowHammer mitigation mechanism. Based on our study, we
motivate the need for a scalable and low-overhead solution to RowHammer and
provide two promising research directions to this end. We hope that the results
of our study will inspire and aid future work to develop efficient solutions
for the RowHammer bit flip rates we are likely to see in DRAM chips in the near
future.

\chapter{Putting It All Together} 

\jkz{Each of these four works individually provide mechanisms that improve
system performance, security or reliability. However, these works are all
orthogonal in that a subset or all of them can be implemented simultaneously on
the same system to provide each individual benefit with little interference.
Furthermore, since each mechanism does not require any changes to DRAM, the
implementation should be \jky{constrained to the memory controller}. There are
many ways to potentially combine each individual mechanism, but we provide a
simple high-level example.} 

\section{Implementing \jky{All Proposed Techniques} on the Same System} 

\jkz{Solar-DRAM, The DRAM Latency PUF, and D-RaNGe all rely on latency failure
profiles for their specific cases on a given DRAM chip. Once the profiles are
generated via characterization, the storage overhead of the profiles is simply
the combined size of each profile. Depending on which mechanism is issuing a
DRAM access, the access latency is set according to the respective profile.} 

\jkz{As Solar-DRAM improves system performance with its faster DRAM accesses,
the latency of regular DRAM accesses should be set according to the Solar-DRAM
profile. In the relatively infrequent event that the \jky{system} requests a
PUF response or a true random value, firmware for The DRAM Latency PUF or
D-RaNGe mechanism will override regular DRAM accesses and issue DRAM accesses
with latencies adjusted according to their respective profiles.} 

\jkz{While Solar-DRAM purely provides performance benefits, evaluating PUFs
(via DRAM Latency PUF) and generating true random values (via D-RaNGe) come at
the cost of additional DRAM accesses interleaved within regular DRAM accesses.
These additional DRAM accesses likely will increase the system performance
overhead depending on the implementations (as discussed in
Sections~\ref{subsec:trng_key_char_eval}
and~\ref{section:runtime_metrics_evaluation}).  Fortunately, D-RaNGE and The
DRAM Latency PUF both enable flexible implementations and can minimize their
overheads depending on the user needs and the importance of \jky{and the need
for} timely PUF evaluations and true random values.} 

\jkz{While utilizing these mechanisms in conjunction may have compound overheads
in terms of both performance and storage, we believe that enabling these
mechanisms on a system simultaneously will provide system performance,
security, and reliability benefits that outweigh the combined overheads of the
mechanisms.} 

\section{\jky{Cost-Benefit} Analysis}  

\jkz{While each individual mechanism provides benefits to the system, whether
performance, security, or reliability, each mechanism has associated costs in
deployment. These costs come in the form of 1) performance or energy
\jky{overheads}, 2) profiling time prior to deployment \jky{(or during online
operation)}, or 3) minor system changes. These costs must be considered when
deploying each mechanism on various systems according to the manufacturer and
user constraints (e.g., fiscal, physical, time, \jky{and other resource}
limitations), and requirements (e.g., performance, reliability, security
service level agreements). However, due to the flexibility and orthogonality of
these mechanisms, costs can be reduced by implementing any subset of these
works on a system according to the constraints and requirements. }

\jkz{Since our works mainly focus on demonstrating the \jky{new} fundamental
ideas rather than present an optimal implementation for each of these works, we
cannot \jky{provide a quantitative cost-benefit} analysis of each of our
mechanisms.  However, given the benefits that each mechanism provides with the
relatively low overheads in the \jky{relatively non-optimized} solutions
offered in the works, we do foresee future work in supporting
\jky{near-optimal} deployment strategies and implementations for each mechanism
that would significantly reduce the associated costs and improve the \jky{ease
of} for implementation in real systems.  }

\chapter{Conclusions and Future Directions} 

\section{Conclusions} 

In this dissertation, we present a number of novel observations on DRAM, via
characterization studies of real DRAM chips, that we exploit to develop
mechanisms that improve system performance and enhance system security and
reliability.

First, we \jky{introduce} 1) a rigorous characterization of activation failures
across 282 \emph{real state-of-the-art LPDDR4} DRAM \jks{modules}, 2)
Solar-DRAM, whose key idea is to exploit our observations and issue DRAM
accesses with variable latency depending on the target DRAM location's
propensity to fail with reduced access latency, and 3) an evaluation of
Solar-DRAM and its three individual components, with comparisons to the
state-of-the-art~\cite{chang2016understanding}.  We find that Solar-DRAM
provides significant performance improvement over the state-of-the-art DRAM
latency reduction mechanism across a wide variety of workloads,
\jkf{\emph{without} requiring any changes to DRAM chips or software.}

Second, we propose \mechanism, a mechanism for extracting true random numbers with high
throughput from unmodified commodity DRAM devices on any system that allows
manipulation of DRAM timing parameters in the memory controller.  \mechanism\
harvests fully non-deterministic random numbers from DRAM row activation
failures, which are \jk{bit errors} induced by intentionally \jk{accessing DRAM
with lower latency than required for correct row activation.} Our TRNG is based
on two key observations: 1) activation failures can be induced quickly and 2)
repeatedly accessing certain DRAM cells with reduced activation latency results
in reading true random data. We validate the quality of our TRNG with the
commonly-used NIST statistical test suite for randomness. Our evaluations show
that \mechanism\ significantly outperforms the previous highest-throughput
DRAM-based TRNG by up to 211x \jkthree{(128x on average)}. We conclude that
DRAM row activation failures can be effectively exploited to improve the
security of a wide range of systems that use commodity DRAM chips via a
high-throughput true random number generator, which can enable a number of
security applications such as cryptography. 

Third, we introduce and analyze the DRAM latency PUF, a new DRAM PUF suitable for
runtime authentication. The DRAM latency PUF intentionally violates
\jkfour{manufacturer-specified} DRAM timing parameters in order to provide
\jkfour{many highly repeatable, unique, and unclonable} PUF responses with low
latency.  Through experimental evaluation using 223 state-of-the-art LPDDR4
DRAM devices, we show that the DRAM latency PUF reliably generates PUF
responses at runtime-accessible speeds (i.e., 88.2ms on average) at all
operating temperatures.  We show that the DRAM latency PUF achieves an average
speedup of 152x/1426x at 70$^{\circ}$C/55$^{\circ}$C when compared with a DRAM
retention PUF of \jkfour{the same} DRAM capacity overhead, and it achieves even
greater speedups at lower temperatures.  We conclude that the DRAM latency PUF
enables a fast and effective substrate for runtime device authentication across
all operating temperatures, and we hope that the advent of runtime-accessible
PUFs like the DRAM latency PUF \jkfour{and the detailed experimental
characterization data we provide on modern DRAM devices} will enable security
architects to develop even more secure systems for future devices.

Finally, we provide the first rigorous experimental RowHammer failure
characterization study that demonstrates how the RowHammer vulnerability of
modern DDR3, DDR4, and LPDDR4 DRAM chips scales across DRAM generations and
technology nodes.  Using experimental data from 1580 real DRAM chips produced
by the three major DRAM manufacturers, we show that modern DRAM chips that use
smaller process technology node sizes are significantly more vulnerable to
RowHammer than older chips. Using simulation, we show that existing RowHammer
mitigation mechanisms 1) suffer from prohibitively large performance overheads
at projected future hammer counts and 2) are still far from an \emph{ideal}
selective-refresh-based RowHammer mitigation mechanism. Based on our study, we
motivate the need for a scalable and low-overhead solution to RowHammer and
provide two promising research directions to this end. We hope that the results
of our study will inspire and aid future work to develop efficient solutions
for the RowHammer bit flip rates we are likely to see in DRAM chips in the near
future.

\section{Future Research Directions}

Our four preliminary works, Solar-DRAM, the DRAM Latency PUF, D-RaNGe, and
Revisiting RowHammer demonstrate that novel observations via DRAM
characterization can be exploited to improve latency, security, and reliability
aspects of DRAM by developing  mechanisms that exploit the DRAM characteristics
when they are accessed with reduced DRAM timing parameters. To explore other
methods for improving these aspects of DRAM, our proposal for future work is to
1) characterize various additional timing parameters and circuit-level aspects,
and 2) propose mechanisms based on our prior observations. 

\subsection{Reducing DRAM Latency by Exploiting Different Timing Parameters} 

In Chapter~\ref{ch3-solar}, we have shown that we can reduce the DRAM timing
parameter, tRCD, to substantially improve DRAM access latency. However, we
believe that there are a number of other DRAM timing parameters (e.g., tRP,
tWR, tRTP) that could also have substantial impact on overall system
performance when reduced below manufacturer-specified values. 

The key challenge in demonstrating a viable mechanism for reliably reducing
DRAM timing parameters lies in rigorously characterizing DRAM to demonstrate
exploitable trends for efficiently reducing DRAM timing parameters. While
Chapter~\ref{ch3-solar} discusses exploitable spatial distributions for
reducing tRCD (i.e., failures constrained to local bitlines), we expect that
each timing parameter, when reduced, will result in various exploitable
spatial distributions. Using this knowledge gained from characterization, we
can then safely and reliably reduce DRAM timing parameters for system
performance improvement depending on the region of DRAM being accessed. To
further exploit findings from this type of characterization, we believe the
next step would be to determine the behavior of interactions when reducing
multiple DRAM timing parameters simultaneously, such that we can further reduce
DRAM access latencies reliably. 

We identify two further challenges that are necessary to overcome for enabling
this direction.  First, it is critical to determine an \emph{efficient} method
for profiling DRAM chips (e.g., determining the set of reliable timing
parameters).  Due to the large search space in parameters (e.g., temperature,
data pattern, timing parameter, timing parameter value), a comprehensive
characterization for every chip is far too expensive. We believe that searching
for correlations between failures of cells with different parameters, could
help to reduce the characterization phase in such a mechanism. For example, if
we find that tRCD failures are correlated with tWR failures, then we can simply
characterize one of the parameters and set the other value accordingly. Second,
it is essential to develop a memory controller and scheme that determines when
an access can be issued with reduced timing parameters and dynamically adjusts
timing parameters to minimize DRAM access latency while maintaining data
correctness during operation.

\subsection{Improving Security Primitives for DRAM Chips} 

In Chapters~\ref{ch4-dlpuf} and \ref{ch5-drange}, we introduced two security
primitives, a Physical Unclonable Function (PUF) and a True Random Number
Generator (TRNG), for commodity DRAM chips. We believe that there is
significant room to improve the security and reliability guarantees of each of
our proposed mechanisms: The DRAM Latency PUF and D-RaNGe. 

First, as discussed in Chapter~\ref{ch4-dlpuf}, a PUF with a larger challenge
response space can uniquely identify a larger set of devices and improve the
authentication process to minimize the probability for misidentification. In
the ideal case, the challenge response space would grow exponentially with its
input parameters.  A PUF with an exponential challenge response space is
classified as a strong PUF which has substantially more use cases compared to a
weak PUF. By rigorous characterization of varying DRAM timing parameters, we
can determine whether it is feasible to increase the challenge response space
simply by using DRAM timing parameters as an additional dimension to the input
challenges.  Ideally, reducing multiple DRAM timing parameters would result in
uncorrelated failure locations such that the input space would grow
exponentially with the number of additional DRAM timing parameters used. We
believe that following an experimental methodology similar to the one described
in Chapter~\ref{ch4-dlpuf} can demonstrate whether this direction is viable. 

Second, we believe it is important to demonstrate practical end-to-end
implementations of both The DRAM Latency PUF and D-RaNGe for adoption in real
devices. We identify various challenges for the DRAM Latency PUF and D-RaNGe.
For a practical implementation of The DRAM Latency PUF, it is critical to
investigate 1) methods for supporting dynamically changing temperatures at
runtime such that a PUF response's variations due to ambient temperature do not
prohibit correct authentication, and 2) a DRAM memory controller that can
schedule PUF evaluation accesses with custom DRAM timing parameters and
minimize its overhead on running workloads. For a practical implementation of
the random number generator, D-RaNGe, we identify three directions that are
critical to investigate. First, it is critical to demonstrate a process for
quickly identifying RNG cells that work under varying conditions (e.g.,
temperature). Second, it is critical to investigate a smarter DRAM memory
controller that can intelligently schedule D-RaNGe accesses with custom DRAM
timing parameters for minimizing the overhead that would come with directly
implementing D-RaNGe on any existing memory controller today. Third, it is
important to further increase the throughput and decrease the latency. We
believe that by combining the varying existing DRAM-based TRNGs (discussed in
Chapter~\ref{related}), we can invoke the different TRNGs (or a combination of
them) depending on the running workloads to maximize the utility of idle DRAM
time.

\subsection{RowHammer Mitigation Going Forward} 

DRAM manufacturers continue to adopt smaller technology nodes to improve DRAM
storage density and are forecasted to reach 1z and 1a technology nodes within
the next couple of years~\cite{DRAMtechroadmap}. Unfortunately, our findings
show that future DRAM chips will likely be increasingly vulnerable to
RowHammer. This means that, to maintain market competitiveness without
suffering factory yield loss, manufacturers will need to develop
effective RowHammer mitigation solutions for coping with increasingly
vulnerable DRAM chips.

RowHammer-mitigation solutions range from circuit-level mechanisms built into
the DRAM chip itself to system-level mechanisms that are agnostic to the
particular DRAM chip that the system uses. Of these solutions, our evaluations
in Section~\ref{subsec:implications:existing_mechanisms} show that\jk{, while
the ideal refresh-based RowHammer-mitigation mechanism that inserts the minimum
possible number of additional refreshes to prevent RowHammer bit flips scales
well to relatively low $HC_{first}$ values (e.g., only 6\% performance
degradation when $HC_{first}$ is 128), currently existing refresh-based
mechanisms fall short of the ideal mechanism's benefits (e.g., at least 28\%
performance degradation when $HC_{first}$ is 128).}

To achieve the scalability of the ideal refresh-based mechanism
while maintaining an efficient implementation, we believe it is essential to
explore all possible avenues for RowHammer mitigation. Going forward, we
identify two promising research directions that can potentially lead to
new RowHammer-mitigation solutions that meet or exceed the scalability of
the ideal refresh-based mitigation mechanism: (1) DRAM-system cooperation and
(2) a profile-guided mechanism. The remainder of this section briefly discusses
our vision for each of these directions.

\noindent
\textbf{DRAM-System Cooperation.} Considering either DRAM-based or
system-level mechanisms alone ignores the potential benefits of addressing the 
RowHammer vulnerability from both perspectives together. While the root causes
of RowHammer bit flips lie within DRAM, their negative effects are observed at the
system-level.  Prior work~\cite{mutlu2014research} stresses the importance of
tackling these challenges at all levels of the stack, and we believe that a
holistic solution can achieve a high degree of protection at relatively low
cost compared to solutions contained within either domain alone.

\noindent
\textbf{Profile-Guided Mechanisms\jky{.}} The ability to accurately profile for
RowHammer-susceptible DRAM cells or memory regions can provide a powerful
substrate for building targeted RowHammer-mitigation solutions that
efficiently mitigate RowHammer bit flips at low cost. Knowing (or effectively predicting)
the locations of bit flips before they occur in practice could lead to a large
reduction in RowHammer-mitigation overhead, providing new information that no
known RowHammer-mitigation mechanism exploits today. For example, within the
scope of known RowHammer-mitigation solutions, increasing the refresh rate can
be made far cheaper by only increasing the refresh rate for known vulnerable
DRAM rows. Similarly, ECC and DRAM access counters can be used only for cells
that are known to be vulnerable, and even software-based mechanism can be
adapted to target only known vulnerable rows (e.g., disabling them, remapping
them to reliable memory).

Unfortunately, there exists no such effective RowHammer error profiling
methodology today. Our characterization in this work essentially follows the
na{\"i}ve approach of individually testing each row by attempting to induce
the worst-case testing conditions (e.g., $HC$, data pattern, ambient
temperature etc.). However, this approach is extremely time consuming due to
having to test each row individually (potentially multiple times with various 
testing conditions). Even for a relatively small DRAM module of 8GB with 8KB
rows, hammering each row only once for only one refresh window of 64ms requires
over 17 hours of continuous testing, which means that the na{\"i}ve approach to
profiling is infeasible for a general mechanism that may be used in a
production environment or for online operation. We believe that
developing a fast and effective RowHammer profiling mechanism is a key
research challenge, and we hope that future work will use the observations made
in this and other RowHammer characterization studies to find a solution.

\section{Final Concluding Remarks} 

In this dissertation, we demonstrated that by rigorously understanding and
exploiting DRAM device characteristics, we can significantly improve system
performance and enhance system security and reliability. We have presented four
characterization-based works \jkz{that show by understanding per-chip error
characteristics using a profiling mechanism, we can develop mechanisms that
exploit the chip-dependent error profiles to improve system performance or
enhance system security and reliability}: 1) Solar-DRAM, which exploits our
experimental characterization on latency variation within a chip to reduce
latency \jkz{with a profile that identifies regions that can be reliably accessed
with lower latencies,} 2) The DRAM Latency PUF, which exploits our observation
from characterization that latency failures are unique to a DRAM chip due to
process manufacturing variation and \jkz{a profile of these error
characteristics can be used generate reliable and unique identifiers}, 3)
D-RaNGe, which demonstrates via characterization \jkz{that a profile of error
characteristics can identify specific cells in DRAM that can be repeatedly
accessed with reduced latency to result in random failures} and can be used as
an efficient true random number generator, a feature often used in security
applications, and 4) Revisiting RowHammer, which demonstrates via
characterization that the DRAM-based RowHammer vulnerability is getting worse
as technology node size scales and existing RowHammer mitigation mechanisms
either do not scale or have prohibitively high overheads to mitigate RowHammer
bit flips in future DRAM chips. We conclude and hope that the proposed
characterization-based studies of DRAM chips and novel observations will pave
the way for new research that can develop new mechanisms to improve system
performance, energy efficiency, system security, or reliability of future
memory systems.

\section*{Other Works of the Author}

Throughout the course of my Ph.D. study, I have worked on several different
topics with many fellow graduate students from Carnegie Mellon University, ETH
Zurich, and other institutions. In this chapter, I would like to acknowledge
these works. 

I have worked on a number of other projects on DRAM. In collaboration with
Minesh Patel, we have developed REAPER~\cite{patel2017reaper}, a profiling
mechanism for retention failures to maintain DRAM reliability while reducing
the refresh overhead. We also characterized and studied DRAM chips with on-die
ECC~\cite{patel2019understanding}, demonstrating how on-die ECC affects the
DRAM's properties, and we develop Error-correction Inference (EIN), a
statistical inference methodology that infers pre-correction error rates of
DRAM with on-die ECC. \jky{In a follow-up work called
BEER~\cite{patel2020beer}, \jkx{we propose} a new methodology for
determining the full DRAM on-die ECC function (i.e., its parity-check matrix)
without hardware support, hardware intrusion, or access to error syndromes and
parity information.} In collaboration with Hasan Hassan, we developed
CROW~\cite{hassan2019crow}, a flexible substrate that we use to lower DRAM
activation latency to frequently accessed rows and reduce the overhead of DRAM
refresh operations. In collaboration with Yaohua Wang, we developed
CAL~\cite{wang2018reducing}, a mechanism that predicts the next access time of
a given row, and only partially restores charge to a row that will be accessed
soon, and \jky{a substrate which can perform cache-block level data relocation
within a DRAM bank at a \jkx{distance-independent} latency~\cite{wang2020reducing}.} I
have also contributed to a retrospective survey~\cite{mutlu2019rowhammer} on
the DRAM-based security vulnerability, RowHammer, which was first rigorously
analyzed in a paper~\cite{kim2014flipping} that I contributed to before my
Ph.D. study. In collaboration with Lucian Cojocar, Stefan Saroiu, and Alec
Wolman at Microsoft Research and others, we developed a methodology for testing
the RowHammer vulnerability in server nodes and identify an instruction
sequence that results in the highest rate of DRAM activations (i.e., hammers).
I have also contributed to Ambit~\cite{seshadri2017ambit}, an accelerator for
bulk bitwise operations in memory and a positioning paper describing important
domains of work toward the practical construction and widespread adoption of
Processing-in-Memory (PIM) architectures. I have also contributed to a
positioning paper~\cite{ghose2019processing}, which describes the challenges that
remain for the widespread adoption of PIM.

I have also contributed to FLIN~\cite{tavakkol2018flin}, a lightweight I/O
request scheduling mechanism that provides fairness among requests from
different applications, SysScale~\cite{hajyahya2020sysscale}, a multi-domain
power management technique that improves the energy efficiency of mobile SoCs,
\jky{and FlexWatts~\cite{hajyahya2020power}, a hybrid adaptive \jkx{Power
Delivery Network (PDN)} for modern high-end client processors whose goal is to
maintain high energy-efficiency across the processor’s wide spectrum of power
\jkx{consumption} and workloads}. 

Another topic that I have developed an interest and worked on was
bioinformatics. I authored GRIM-Filter~\cite{kim2018grim}, a fast seed location
filtering algorithm for DNA read mapping and AirLift~\cite{kim2019airlift}, a
methodology for quickly and comprehensively mapping a set of reads from one
reference to another reference. I collaborated with Damla Senol on a survey of
Nanopore sequencing technologies~\cite{senol2019nanopore} and
\jky{GenASM~\cite{senol2020genasm}, \jkx{a flexible} approximate string
matching acceleration framework.} In addition, I also worked with Can Firtina
on Apollo~\cite{firtina2020apollo}, a sequencing-technology-independent
assembly polishing algorithm, and with Hongyi Xin on LEAP~\cite{xin2017leap},
an algorithm for sequence alignment. 

\begin{singlespace}
\small
\bibliography{refs}
\bibliographystyle{plain}
\end{singlespace}

\includegraphics[scale=0.7]{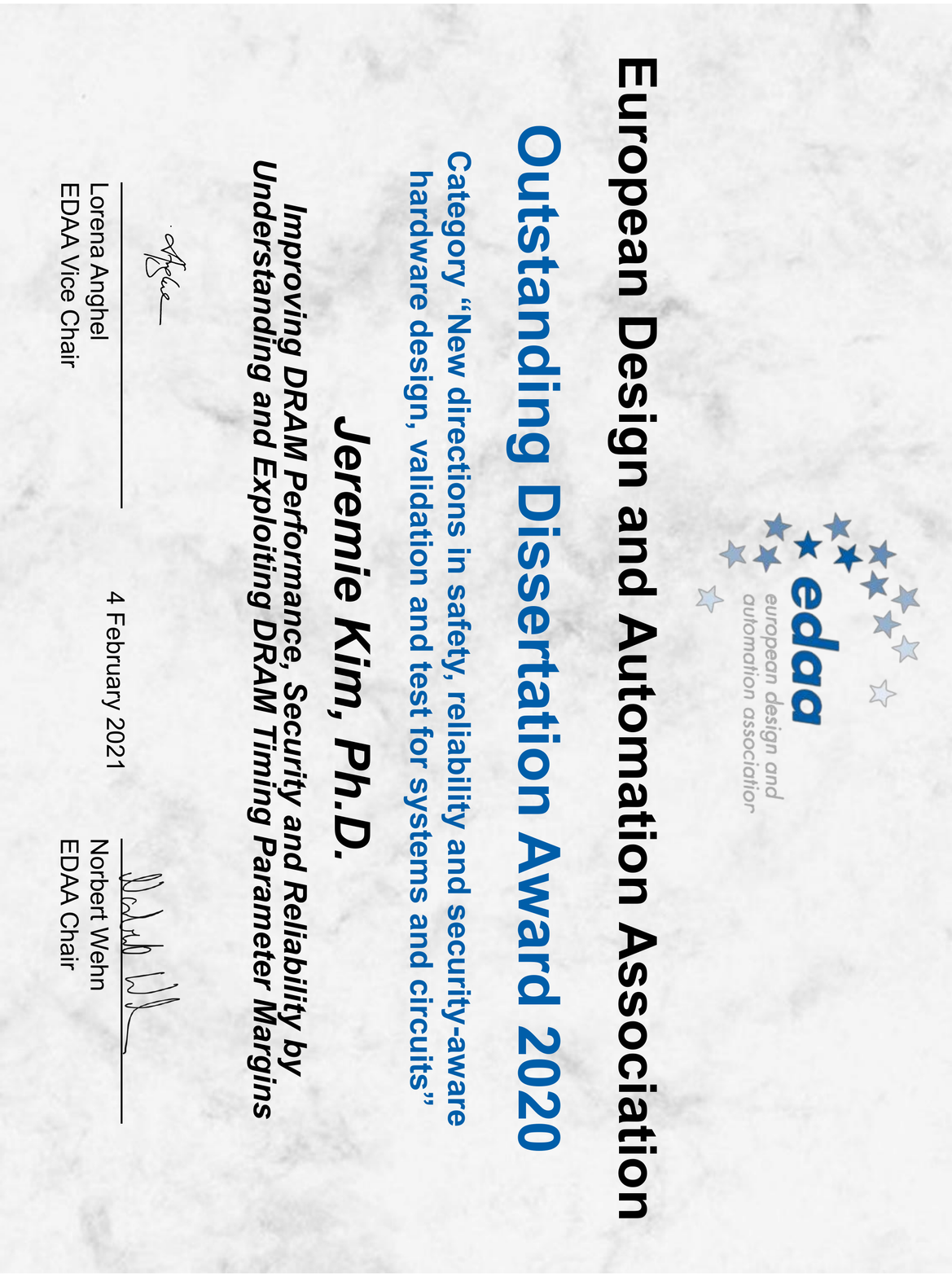}

\end{document}